# Characteristics of planetary candidates observed by *Kepler*, II: Analysis of the first four months of data

William J. Borucki[0,1], David G. Koch[1], Gibor Basri[2], Natalie Batalha[3], Timothy M. Brown[5], Stephen T. Bryson[1], Douglas Caldwell[6], Jørgen Christensen-Dalsgaard[7], William D. Cochran[8], Edna DeVore[6], Edward W. Dunham[9], Thomas N. Gautier III[11], John C. Geary[10], Ronald Gilliland[12], Alan Gould[13], Steve B. Howell[14], Jon M. Jenkins[6], David W. Latham[10], Jack J. Lissauer[1], Geoffrey W. Marcy[2], Jason Rowe[1], Dimitar Sasselov[10], Alan Boss[4], David Charbonneau[10], David Ciardi[22], Laurance Doyle[6], Andrea K. Dupree[10], Eric B. Ford[16], Jonathan Fortney[17], Matthew J. Holman[10], Sara Seager[18], Jason H. Steffen[19], Jill Tarter[6], William F. Welsh[20], Christopher Allen[21], Lars A. Buchhave[10], Jessie L. Christiansen[6], Bruce D. Clarke[6], Santanu Das[23], Jean-Michel Désert[10], Michael Endl[8], Daniel Fabrycky[17], Francois Fressin[10], Michael Haas[1], Elliott Horch[24], Andrew Howard[2], Howard Isaacson[2], Hans Kjeldsen[7], Jeffery Kolodziejczak[25], Craig Kulesa[15], Jie Li[6], Philip W. Lucas[28], Pavel Machalek[6], Donald McCarthy[15], Phillip MacQueen[8], Søren Meibom[10,], Thibaut Miquel[27]Andrej Prsa[26], Samuel N. Quinn[10],Elisa V. Quintana[6], Darin Ragozzine[10], William Sherry[14], Avi Shporer[5], Peter Tenenbaum[6], Guillermo Torres[10], Joseph D. Twicken[6], Jeffrey Van Cleve[6], and Lucianne Walkowicz[2]

[1]NASA Ames Research Center, Moffett Field, CA 94035, USA
[2]University of California, Berkeley, CA, 94720, USA
[3]San Jose State University, San Jose, CA, 95192, USA
[4]Carnegie Institution of Washington, Washington, DC 20015 USA
[5]Las Cumbres Observatory Global Telescope, Goleta, CA 93117, USA
[6]SETI Institute, Mountain View, CA, 94043, USA
[7]Aarhus University, Aarhus, Denmark
[8]McDonald Observatory, University of Texas at Austin, Austin, TX, 78712, USA
[9]Lowell Observatory, Flagstaff, AZ, 86001, USA
[10]Harvard-Smithsonian Center for Astrophysics, Cambridge, MA, 02138, USA
[11]Jet Propulsion Laboratory, Calif. Institute of Technology, Pasadena, CA, 91109, USA
[12]Space Telescope Science Institute, Baltimore, MD, 21218, USA
[13] Lawrence Hall of Science, Berkeley, CA 94720, USA
[14]NOAO, Tucson, AZ 85719 USA
[15]University of Arizona, Steward Observatory, Tucson, AZ 85721, USA
[16]Univ. of Florida, Gainesville, FL, 32611 USA
[17]Univ. of Calif., Santa Cruz, CA 95064 USA
[18]MIT, Cambridge, MA 02139 USA
[19]Fermilab, Batavia, IL 60510 USA
[20]San Diego State Univ., San Diego, CA 92182 USA
[21]Orbital Sciences Corp., Mountain View, CA 94043 USA
[22]Exoplanet Science Institute/Caltech, Pasadena, CA 91125 USA
[23]University Affiliated Research Center, University of California, Santa Cruz, CA 95064 USA
[24]Southern Connecticut State University, New Haven, CT 06515 USA
[25]MSFC, Huntsville, AL 35805 USA
[26]Villanova University, Villanova, PA 19085 USA
[27]CNES, Toulouse, France
[28]Centre for Astrophysics Research, Science & Technology Research Institute, University of Hertfordshire, Hatfield, UK

[0]Correspondence should be addressed to: William Borucki, William.J.Borucki@nasa.gov

**Abstract.** On 1 February 2011 the *Kepler* Mission released data for 156,453 stars observed from the beginning of the science observations on 2 May through 16 September 2009. There are 1235 planetary candidates with transit like signatures detected in this period. These are associated with 997 host stars. Distributions of the characteristics of the planetary candidates are separated into five class-sizes; 68 candidates of approximately Earth-size ($R_p < 1.25$ $R_\oplus$), 288 super-Earth size (1.25 $R_\oplus < R_p <$ 2 $R_\oplus$), 662 Neptune-size (2 $R_\oplus$, $< R_p <$ 6 $R_\oplus$), 165 Jupiter-size (6 $R_\oplus < R_p <$ 15 $R_\oplus$), and 19 up to twice the size of Jupiter (15 $R_\oplus < R_p <$ 22 $R_\oplus$). In the temperature range appropriate for the habitable zone, 54 candidates are found with sizes ranging from Earth-size to larger than that of Jupiter. Six are less than twice the size of the Earth. Over 74% of the planetary candidates are smaller than Neptune. The observed number versus size distribution of planetary candidates increases to a peak at two to three times Earth-size and then declines inversely proportional to area of the candidate. Our current best estimates of the intrinsic frequencies of planetary candidates, after correcting for geometric and sensitivity biases, are 5.4% for Earth-size candidates, 6.8% for super-Earth size candidates, 19.3% for Neptune-size candidates, 2.4% for Jupiter-size candidates, and 0.15% for very-large candidates; a total of 0.341 candidates per star. Multi-candidate, transiting systems are frequent; 17% of the host stars have multi-candidate systems, and 33.9% of all the candidates are part of multi-candidate systems.

## 1. Introduction

*Kepler* is a Discovery-class mission designed to determine the frequency of Earth-size planets in and near the habitable zone (HZ) of solar-type stars. Details of the *Kepler* Mission and instrument can be found in Koch *et al*. (2010a), Jenkins *et al*. (2010c), and Caldwell *et al*. (2010). All data through 16 September 2009 are now available through the Multi-Mission Archive (MAST[1]) at the Space Telescope Science Institute for analysis by the community.

Based on the first 43 days of data, five exoplanets with sizes between 0.37 and 1.6 Jupiter radii and orbital periods from 3.2 to 4.9 days were recognized and then confirmed by radial velocity observations during the 2009 observing season (Borucki *et al*. 2010, Koch *et al*. 2010b, Dunham *et al*. 2010, Jenkins *et al*. 2010a, and Latham *et al*. 2010). Ten more planets orbiting a total of 3 stars have subsequently been announced (Holman et al., 2010, Torres et al. 2011, Batalha et al. 2011, Lissauer et al. 2011a).

Because of great improvements to the data-processing pipeline, many more candidates are much more visible than in the data used for the papers published in early 2010. When *Kepler's* first major exoplanet data release occurred on 15 June 2010, 706 targets stars had candidate exoplanets (Borucki et al. 2011). In this data release we identify 997 stars with a total of 1235 planetary candidates that show transit-like signatures in the first 132 days of data. A list of false positive events found in the released data is also included in Table 4 with a brief note explaining the reason for classification as a false positive. All false positives are also archived at the MAST. A total of 1202 planetary candidates are discussed herein.

[1] http://archive.stsci.edu/*Kepler*/data_search/search.php

The algorithm that searches for patterns of planetary transits also finds stars with multiple planet candidates. A separate paper presents an analysis of five of these candidates (Steffen *et al*. 2010). Data and search techniques capable of finding planetary transits are also very sensitive to eclipsing binary (EB) stars, and indeed the number of EBs discovered with *Kepler* exceeds the number of planetary candidates. With more study, some of the current planetary candidates might also be shown to be EBs and some planetary candidates or planets might be discovered orbiting some of the EBs. Prsa *et al*. (2011) present a list of EBs with their basic system parameters that have been detected in these early data.

## 2. Description of the Data

Data for all stars are recorded at a cadence of one per 29.4244 minutes (hereafter, long cadence, or LC). Data for a subset of up to 512 stars are also recorded at a cadence of one per 58.85 seconds (hereafter, short cadence or SC), sufficient to conduct asteroseismic observations needed for measurements of the stars' sizes, masses, and ages. The results presented here are based only on LC data. For a full discussion of the LC data and their reduction, see Jenkins *et al*. (2010b, 2010c). See Gilliland *et al*. (2010) for a discussion of the SC data.

The results discussed in this paper are based on three data segments; the first segment (labeled Q0) started JD 2544953.53 and ended on 2454963.25 and was taken during commissioning operations; the second data segment (labeled Q1) taken at the beginning of science operations that started on JD 2454964.50 and finished on JD 2454997.99 and a third segment (labeled Q2) starting on JD 2455002.51 and finishing on JD 2455091.48. The durations of the segments are; 9.7, 33.5, and 89.0 days, respectively. The observations span a total period of 137.95 days including the gaps. A total of 156,097 LC targets in Q1 and 166247 LC and 1492 SC targets were observed in Q2. The stars observed in Q2 were mainly a superset of those observed in Q1. These data have been processed with Science Operations Center (SOC) pipeline version 6.2 and archived at the MAST. Originally, the bulk of these data were scheduled for release on 15 June 2011, but the exoplanet targets are being released early, so 165470 LC and 1478 SC targets will be publically available to the public on 1 February 2011. The remaining few targets have a proprietary user other than the *Kepler* science team (e.g., guest observers). Data for these targets will become public by 15 June 2011. The current release date and the proprietary owner for each target are posted at MAST as soon as the data enter the archive, which occurs about four months after data acquisition for the quarter in question is complete.

The results reported here are for the LC observations of 153,196 stars observed during Q2. Other stars were giants or super-giants, did not have valid parameter values, or were in some way inappropriate to the discussion of the exoplanet search. The enlarged set of stars observed in Q2 included most of the stars observed in Q1and additional stars due to the more efficient use of the available pixels. The selected stars are primarily main sequence dwarfs chosen from the *Kepler* Input Catalog[27] (KIC). Targets were chosen to maximize the number that were both bright and small enough to show detectable transit signals for small planets in and near the habitable zone (HZ) (Gould *et al.* 2003, Batalha *et al*. 2010a). Most stars were in the *Kepler* magnitude range 9 < Kp < 16. The *Kepler* passband covers both the V and R photometric passbands (Figure 1 in Koch et al. 2010a). See the discussion in Batalha et al. (2010b).

[27] http://archive.stsci.edu/*Kepler*/*Kepler*_fov/search.php

*2.1 Noise Sources in the Data*

The *Kepler* photometric data contain a wide variety of both random and systematic noise sources. These sources and others are discussed in Jenkins *et al*. (2010b) and Caldwell *et al*. (2010). Work is underway to improve the mitigation and flagging of the affected data. Stellar variability over the periods similar to transit durations is also a major source of noise.

Because of the complexity of the various small effects that are important to the quality of the *Kepler* data, prospective users of *Kepler* data are strongly urged to study the data release notes (available at the MAST) for the data sets they intend to use. Note that the *Kepler* data analysis pipeline was designed to perform differential photometry to detect planetary transits, so other uses of the data products require caution.

*2.2 Distinguishing Planetary Candidates from False Positive Events*

The search for planets starts with a search of the time series of each star for a pattern that exceeds a detection threshold commensurate with a non-random event. Observed patterns of transits consistent with those from a planet transiting its host star are labeled "planetary candidates." (In a few cases, a single drop in brightness that had a high SNR and was of the form of a transit was sufficient to identify a planetary candidate.) Those that were at one time considered to be planetary candidates but subsequently failed some consistency test are labeled "false positives". After passing all consistency tests described below, and only after a review of all the evidence by the entire *Kepler* Science Team, does the candidate become a *confirmed* or *validated* exoplanet. Steps such as high-precision radial velocity (RV) measurements (Borucki *et al*. 2010, Koch *et al*. 2010b, Dunham *et al*. 2010, Jenkins *et al*. 2010a, and Latham *et al*. 2010), or transit timing variations (Holman et al 2010, Lissauer et al. 2011a) are used when practical. When such methods cannot be used to *confirm* an exoplanet, an extensive analysis of spacecraft and ground-based data may allow *validation* of an exoplanet by showing that the planetary interpretation is at least 100 times as probable as a false positive (Torres et al. 2011, Lissauer et al. 2011a). This paper does not attempt to promote the candidates discussed herein to validated or confirmed exoplanets, but rather documents the full set of current candidates and the many levels of steps toward eventual validation, or in some cases, rejection as a planet that have been taken.

There are two general causes of false positive events in the *Kepler* data that must be evaluated and excluded before a candidate planet can be considered a valid discovery: 1) statistical fluctuations or systematic variations in the time series, and 2) astrophysical phenomena that produce similar signals. A sufficiently high detection threshold (i.e., 7.1 $\sigma$) was chosen such that the totality of data from Q0 thru Q5 (end date JD 2455371.170) provides an expectation of fewer than one false positive event due to statistical fluctuations over the ensemble of all stars for entire mission duration. Similarly, systematic variations in the data have been interpreted in a conservative manner and should result in false positives only rarely. However, astrophysical phenomena that produce transit-like signals are common.

*2.2.1 Search for False Positives in the Output of the Data Pipeline*

The Transiting Planet Search (TPS) pipeline searches through each systematic error-corrected flux time series for periodic sequences of negative pulses corresponding to transit signatures. The approach is a wavelet-based, adaptive matched filter that characterizes the power spectral density (PSD) of the background process yielding the observed light curve and uses this time-variable PSD estimate to realize a pre-whitening filter and whiten the light curve (Jenkins 2002, Jenkins et al. 2010c,d). TPS then convolves a transit waveform, whitened by the same pre-whitening filter as the data, with the whitened data to obtain a time series of single event statistics. These represent the likelihood that a transit of that duration is present at each time step. The single event statistics are combined into multiple event statistics by folding them at trial orbital periods ranging from 0.5 days to as long as one quarter (~93 days) of a spacecraft year. Every quarter year, the spacecraft must be rotated 90 degrees to keep the solar panels pointed at the Sun. This rotation put the images of the stars on a different set of detectors and resets the photometric values. Automated identification of candidates with periods longer than one quarter will be done by the pipeline in the coming months, but is currently done by *ad hoc* methods. The ad hoc methods produced many of the Kepler-Object-of-Interests (KOI) with numbers larger than 1000, but might cause a bias against candidates with periods longer than one quarter. For a more comprehensive discussion of the data analysis, see Wu et al (2010) and Batalha et al (2010b).

After automatic identification with TPS or *ad hoc* detection of longer period candidates, the light curves of potential planet candidates were modeled and examined by eye to determine the gross viability of the candidate. If the potential candidate was not an obvious variable star or eclipsing binary showing significant ellipsoidal variation the candidate was elevated to Kepler Object of Interest (KOI) status, given a KOI number (see section 3.1) and was subjected to tests described in the next paragraphs. After passing these tests, the KOI is forwarded to the Follow-up Observation Program (FOP) for various types of observations and additional analysis. See the discussion in Gautier et al. (2010) and Bryson et al. (2011).

Using these estimates and information about the star from the KIC, tests are performed to search for a difference in even- and odd-numbered event depths. If a significant difference exists, this suggests that a comparable-brightness EB has been found for which the true period is twice the period initially determined due to the presence of primary and secondary eclipses. Similarly, a search is conducted for evidence of a secondary eclipse or a possible planetary occultation roughly halfway between the potential transits. If a secondary eclipse is seen, then this could indicate that the system is an EB with the period assumed. However, the possibility of a self-luminous planet (as with HAT-P-7; Borucki *et al.* 2009) must be considered before dismissing a candidate as a false positive.

Many false positives due to background eclipsing binaries (BGEBs) are not detected by the pipeline techniques described above, for example if their secondary transit signals are so weak that they are lost in the noise. The term "eclipsing binaries", as distinct from BGEBs, are gravitationally-bound, multi-star targets and are usually detected by the secondary eclipse or RV observations. To detect BGEBs, a very sensitive validation technique is used on all candidates to determine the relative position of the image centroid during and outside of the transit epoch. The shift in the centroid position of the target star measured in and out of the transits must be consistent with that predicted from the fluxes and locations of the target and nearby stars. (See Bryson et al. 2011.) In particular, a post-processing examination uses an average difference image formed by subtracting the pixels during transit from the pixels out of transit. A pixel response function fit to this difference image provides a direct sub-pixel measurement of the transit source location on the sky (Torres et al. 2011). When the measured position of the transit source does not coincide with the target star the most common cause will be a BGEB false positive, although for strongly blended targets in the direct image further analysis is necessary to support this rejection. This analysis of centroid motion is capable of identifying BGEBs as close as about 1 arcsecond to the target star in favorable circumstances, even with *Kepler*'s 4-arcsecond pixel scale.

Centroid analysis is conducted for each candidate that is unsaturated in the *Kepler* observations and follow-up observations by AO and speckle imaging of the area near the target star are carried out for many candidates. adaptive optics (AO) observations in the infrared were conducted at the 5-m at Palomar Observatory and the 6.5 m at the MMT with ARIES; speckle observations were obtained at the WIYN 3.5m telescope. However, the area behind and immediately surrounding the star, can conceal a BGEB that could imitate a candidate signature. The area that could conceal an EB varies with brightness of the target star because of photon noise limitations to AO and speckle searches, but is of order 1 square arc sec. Model estimates of the *a priori* probability that an EB is present in the magnitude range that could mimic the transit signal range from $10^{-6}$ to $10^{-4}$. Thus the estimated number of target star locations that might have an EB too close to the star to be detected by AO or speckle imaging is 0.1 to 10 based on observations of 150,000 stars.

A much more comprehensive and intensive analysis has been done for the candidates listed here than was done for the data released in June 2010 (Borucki et al. 2011). Consequently the fraction of the candidates that are false positives in the active candidate list should be substantially smaller than the earlier estimate.

### *2.2.2 Estimate of false positive rate*

While many of the candidates have been vetted through the steps described above, the process of determining the residual false positive fraction for *Kepler* candidates at various stages in the validation process has not proceeded far enough to make good quantitative statements about the expected true planet fraction, or reliability, of the released list. However, we can make rough estimates of the quality of the vetting that the KOIs have had. Several groups of KOIs in Table 2 are distinguished by the FOP ranking flag. These groups have had different levels of scrutiny for false positives and will therefore have different expectations for reliability.

KOIs with ranking of 1 are validated and published planets with expected reliability above 98%. We are reluctant to state a higher reliability since unforeseen issues have led to retractions of apparently well-established planets in other planet detection programs.

KOIs with rankings of 2 and 3 have been subject to thorough analysis of their light curves to look for signs of eclipsing binary origin, analysis of centroid motion to detect BGEBs confused with their target stars, and varying degrees of spectroscopic and imaging follow-up observation from ground and space based observatories. These analyses and follow-up observations are generally sufficient to eliminate many stellar mass objects at or near the location of the target star as the source of the transit signal. A ranking of 2 means that none of the results argued against the planet interpretation. A ranking of 3 means that some of the results were suspicious enough to warrant caution but did not unambiguously rule out the planet interpretation. The criteria are subjective and are not meant to be quantitative. The main sources of unreliability, false positives among the rank 2 and 3 KOIs are likely to be from BGEBs with angular separation from the target star too small to be detected by our centroid motion analysis, grazing eclipses in binary systems, and eclipsing stars in hierarchical multiple systems where transits by stellar companions and giant planets dilute the light of other system components. Note that spectroscopy, even at low signal-to-noise such as the reconnaissance spectra we are pursuing, easily rules out grazing eclipsing binaries, as they would show RV variations of tens of km/s. However, those KOIs in Table 2 without a flag=1 in the FOP column did not have such spectroscopy, leaving open the possibility of such grazing eclipsing binaries.

For bright unsaturated stars with $Kp \leq 11.5$ and transit depths strong enough to provide overall detection significances of $20\sigma$ and more, the minimum angular separation for the current centroid motion analysis is about 1 arcsec. This limit becomes significantly larger for fainter stars and/or low-amplitude transit signals associated with smaller planets. For these signal levels, the transit significance of $\sim 10\sigma$ supports a centroid motion analysis constraint on the inner detection limit of about 3 arcsec. These minimum detection angles of 1 to 3 arcsecs are quoted as $3\sigma$ angles beyond which high confidence of discriminating against BGEBs exists. High resolution imaging provided additional reduction of the effective the minimum detection angle for about 100 of the rank 2

KOIs. We expect 10% of the BGEBs to remain in the rank 2 list. KOIs were given a rank of 3 when the centroid motion analysis or follow-up spectroscopy was ambiguous so that the KOI could not be definitely declared a false positive. We estimate that as many as 30% false positives could remain among the rank 3 KOIs.

About 12% of star systems in the solar neighborhood are found to be triple, or of higher multiplicity, hierarchical systems (Raghavan et al. 2010), so a similar fraction is expected to appear in the *Kepler* target list. Only a small percentage of the hierarchical systems will produce eclipses that are seen by *Kepler* and many of these signals can be identified as binary star eclipses by examination of their light curves. From the rare occurrence rate of EBs and the also rare occurrence rate of triple star systems, the fraction of KOIs that are triple-star systems with an EB is expected to be less than 5%.

A potentially more frequent type of misidentification in a hierarchical system is a planet transiting in a binary system. If the double nature of the star system is not identified, dilution of the planetary transit by the second star will result in miscalculation of the planet size. Raghavan et al. (2010) give the binary star system fraction as 34%, but little is yet known about the frequency of planets in binary systems and, again, only a small fraction of planets in binary systems will transit because the orbital planes of the planets are expected to be coplanar with the orbital plane of the stars. Adopting Raghavan et al.'s occurrence rate of binary stars, and assuming that the typical number of planets per star system doesn't depend on the multiplicity of the system, we expect that up to 34% of the KOIs represent planets of larger radius than indicated in Table 2. The distribution of the amounts of dilution cannot be easily determined as it depends on two effects, namely the distribution of the ratio of star brightnesses and the distribution of planet sizes that transit one (or the other) of the two stars in the binary system. Estimating these planet-transit effects in binary systems requires knowledge of the systematic dependence of planet size on orbital distance, a chicken-and-egg problem that we cannot easily resolve at present. For binaries in which the transiting planet orbits the primary star, the dilution will be less than 50% flux. But for binaries in which the transiting object (planet or star) orbits the fainter secondary star, the transiting object's radius can be arbitrarily larger than that stated in Table 2.

Considering all sources of remaining false positives we expect the list of rank 2 KOIs to be >80% reliable and the rank 3 list to be >60% reliable. A careful assessment of false positive scenarios, especially background and gravitationally bound eclipsing binaries and planets, suggests that 90% to 95% of the Kepler planet candidates are indeed true planets (Morton & Johnson 2011). This agrees with our best estimates.

Rank 4 KOIs have had scant examination of their light curves and no follow-up observation and were therefore subject only to centroid motion analysis. We expect the reliability of rank 4 KOIs to be similar to that of rank 3.

### *2.2.3 Development of a model to estimate the probability of an EB near the position of a candidate.*

Low-mass planets, especially those in long-period orbits within the habitable zone, have low amplitude RV signal levels that are often too small to be confirmed by current Doppler observation capabilities. Consequently, validation must be accomplished by the series of steps outlined above. An estimate is also made of the probability that an EB is present that is too near the target star to detect by AO, speckle imaging, or centroid motion. The area number density (number per solid angle) of EBs is calculated based on the assumption that the number of EBs to the number of background stars is constant near the position of each target star. Because the area number density varies rapidly with Galactic latitude and because the *Kepler* field-of-view (FOV) covers over 10° of latitude, predictions of the EB density also vary greatly over the FOV. Consequently, a model was constructed to estimate the probability per square arcsec that an EB is present in the magnitude range that would provide a signal with an amplitude similar to that of the candidate and at the position of each target star. The model is based on the fraction of stars observed by *Kepler* to be binary (Prsa et al. 2011), and it uses the number and magnitude distributions of stars from the Besancon model after correction from the V band to the *Kepler* passband. The value of the probability that there is a BGEB at the location of the target star is listed in Table 2 for each candidate.

## 3. Results

The characteristics of the host stars and the candidates are summarized in Tables 1 and 2, respectively. A total of 1235 KOIs were found in the Q0 through Q2 data. Table 3 provides short notes on many of these KOIs. Table 4 lists the 511 candidates considered to be false positives; comments are included. The false positives have been removed from the list of candidates in Table 2 and are not used in the distributions discussed here. The 15 candidates with a diameter over twice that of Jupiter, and thus larger than late M dwarf stars, were also removed from discussion. This leaves a total of 1235 -18 single-transit candidates -15 candidates greater than twice the size of Jupiter = 1202 candidates for consideration in this discussion.

To provide the most accurate predictions for future observations, the values for the epoch and orbital period given in Table 2 are derived from all data currently available to the Kepler team; i.e., data obtained through Q5 (from JD 2455276.481 through JD 2455371.170) were used. For some candidates, reconnaissance spectra were taken with moderate exposures to look for double- and single-lined binaries. They are most useful in finding outliers for the stellar temperatures and log*g* listed in the KIC. Adaptive optics and speckle observations were taken to check for the presence of faint nearby stars that could be BGEBs or that could dilute the signal level. Flags also indicate the particularly interesting candidates for which radial velocity (RV) measurements of extremely high precision (~ 2 m/s) or high precision (~ 10 m/s) observations were obtained. The last column of Table 2 indicates whether a note is available about that candidate in Table 3. For consistency, all values of the stellar parameters are derived from the KIC.

### *3.1 Naming Convention*

To avoid confusion in naming the target stars, host stars, planetary candidates, and confirmed/validated planets, the following naming convention has been used. Kepler stars are referred to as KIC NNNNNNN (with a space between the "KIC" and the number), where the integer refers to the ID in the *Kepler* Input Catalog archived at MAST. Confirmed planets are named Kepler followed by a hyphen, a number for the planetary system, and a letter designating the first, second, etc. confirmed planet as "b", "c", etc., for example Kepler-4b. Candidates are labeled *Kepler* Object of Interest ("KOI") followed by a decimal number. The two digits beyond the decimal provide identification of the candidates when more than one is found for a given star, e.g., KOI NNN.01, KOI NNN.02, KOI NNN.03, etc. For example KOI 377.03, the third transit candidate identified around star KOI 377, became Kepler-9d after validation as a planet (Torres et al. 2011). KOI numbers are always cross-referenced to a KIC ID. For a multi-candidate system these digits beyond the decimal indicate the order in which the candidates were identified by the analysis pipelines and are not necessarily in order of orbital period. It should be noted that the KOI list is not contiguous and not all integers have an associated KOI.

*3.2 Statistical Properties of Planet Candidates*

We conducted a statistical analysis of the 1202 candidates to investigate the general trends and initial indications of the characteristics of the planetary candidates. The list of candidates was augmented with known planets in the field of view. In particular, TrES-2, HAT-P7b, HAT-P11b, (Kepler-1b, -2b, -3b, respectively), Kepler-4b-8b (Borucki *et al*. 2010, Koch *et al*. 2010b, Dunham *et al*. 2010, Latham *et al*. 2010, and Jenkins *et al*. 2010). Kepler-9bcd (Holman et al. 2010, Torres et al. 2011), Kepler-10b (Batalha et al. 2011), and Kepler-11b-g (Lissauer et al. 2011a) were included. However one candidate identified by a guest observer (KOI 824.01) is included in the list of candidates but is not used in the graphs and statistics because it wasn't in the range of parameters chosen for the search. As noted above, not all candidates appearing in Table 2 were used in the statistical analysis or in the graphical associations shown in the figures: specifically, candidates greater than twice the size of Jupiter, those that showed only one transit in the Q0/Q2 data but no others in the succeeding observations, and those orbiting stars larger than 10 solar radii or with temperatures in excess of 9500 K were excluded. Comparisons are limited to orbital periods of ≤ 138 days. The figures are indicative of the properties and associations of candidates with various parameters, but are not meant to be definitive.

The readers are cautioned that the sample is affected by many poorly quantified biases. Obviously some of the released candidates could be false positives, but other characteristics such as stellar radius, magnitude, noise spectrum, and analysis protocols can all play significant roles in the statistical results. Nevertheless, the large number of candidates provides interesting, albeit tentative, associations with stellar properties. No correction is made to the frequency plots due to the linearly decreasing probability of a second transit occurring during the Q0 through Q2 period. This correction is not needed because data for following quarters were used to calculate the epochs and periods for all candidates that showed at least one transit in the Q0 through Q2 period and at least one in the subsequent observations.. In the figures below, the distributions of various parameters are plotted and compared with values in the literature and those selected from the Extrasolar Planets Encyclopedia[2] (EPE; values as of 7 December 2010). We consulted the literature to identify those planets discovered by the RV method and excluded those discovered by the transit method. This step avoids biasing the RV-discovered planets with the short-period planets that are often found by the transit method.

The results discussed here are primarily based on the observations of stars with Kp < 16, with effective temperature below 9500 K, and with size less than ten times the solar radius. The latter condition is imposed because the photometric precision is insufficient to find Jupiter-size and smaller planets orbiting stars with 100 times the area of the Sun. Stellar parameters are based on KIC data. The function of the KIC was to provide a target sample with a high fraction of dwarf stars that are suitable for transit work, and to provide a first estimate of stellar parameters that is intended to be refined spectroscopically for KOI targets at a later time. Although post-identification reconnaissance spectroscopic observations have been made for more than half of the stars with candidates, it is important to recognize that some of the characteristics listed for the stars are still uncertain, especially surface gravity (i.e., log *g*) and metallicity ([M/H]). The errors in the stellar diameters can reach 25%, with proportional changes to the estimated diameter of the candidates.

In Figure 1, the stellar distributions of magnitude and effective temperature are given for reference. In later figures, the association of the candidates with these properties is examined.

---

[2] Extrasolar Planet Encyclopedia; http://exoplanet.eu/

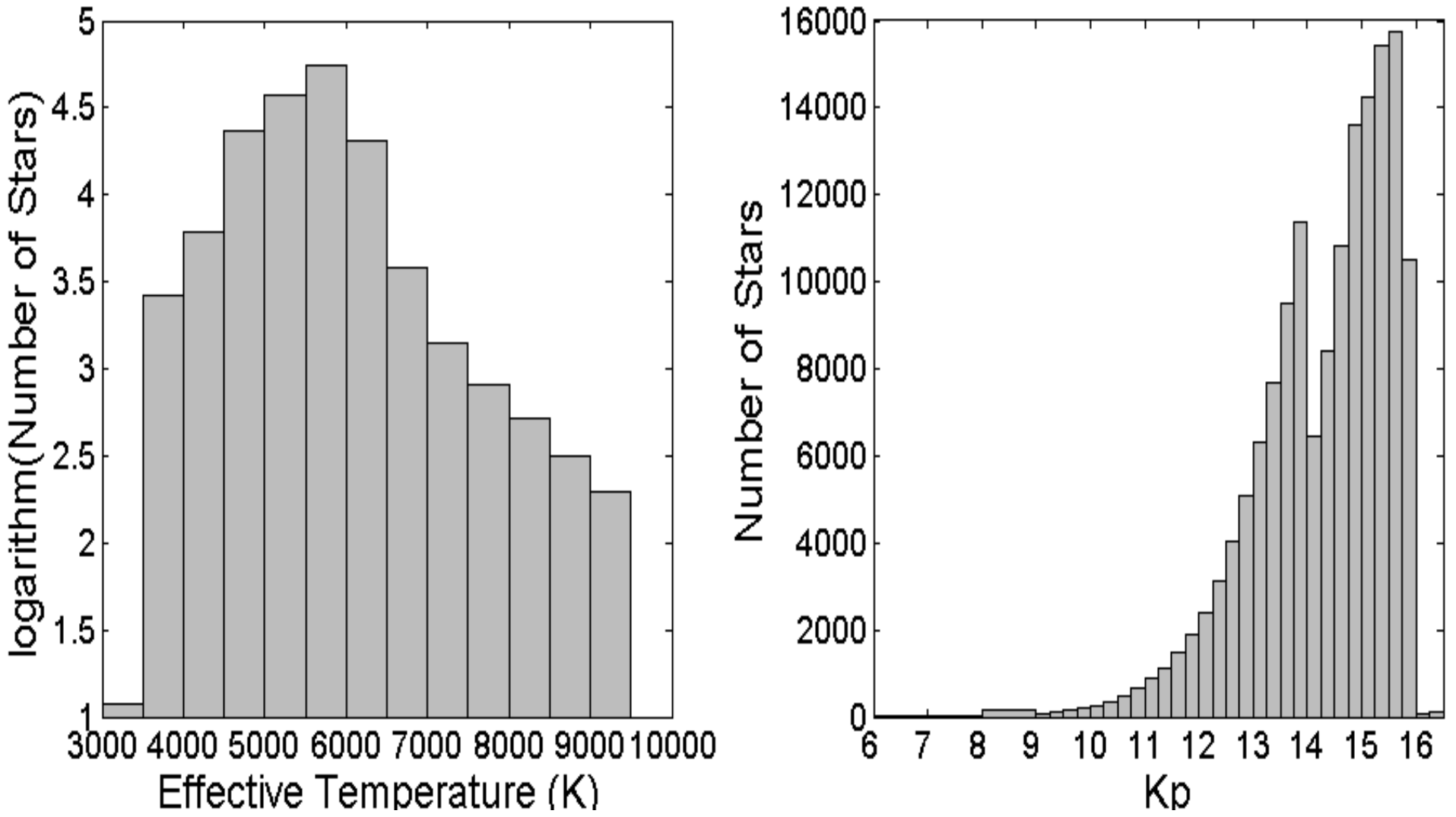

**Figure 1.** Distributions of effective temperature and magnitude for the stars observed during Q2 and considered in this study. Bin size for left panel is 500 K. The bin size for right hand panel is one magnitude from 6 to 9 and 0.25 mag from 9 to 16.5.

It is clear from the left panel in Figure 1 that most of the stars monitored by *Kepler* have temperatures between 4000 and 6500 K; they are mostly late F, G and K spectral types. Because of their faintness, only 2510 stars cooler than 4000 K (i.e., dwarf stars of spectral type M) were monitored. Although cooler stars are more abundant, hotter stars are the most frequently seen for a magnitude-limited survey of dwarfs.

The selection of target stars was purposefully skewed to enhance the detectability of Earth-size planets by choosing those stars with an effective temperature and magnitude that maximized the transit signal-to-noise ratio (SNR) (Batalha et al. 2010b). The step decrease seen in the right hand panel of Figure 1 at Kepler magnitude (Kp) equal 14.0 and the turnover near Kp = 15.5, seen in the right hand panel of figure 1, are due to the selection of only those stars in the FOV that are bright enough and small enough to show terrestrial-size planets. After all available bright dwarf stars were chosen for the target list, many target slots remained, but only stars fainter than Kp=14 were available (Batalha *et al*. 2010b). From the fainter stars the smallest stars are given preference. At the lower left of the right hand chart, the bin size has been increased to show the small number of candidates brighter than Kp = 9. *In the following figures, the bias introduced by the selection of stellar size- and magnitude distributions must always be considered.*

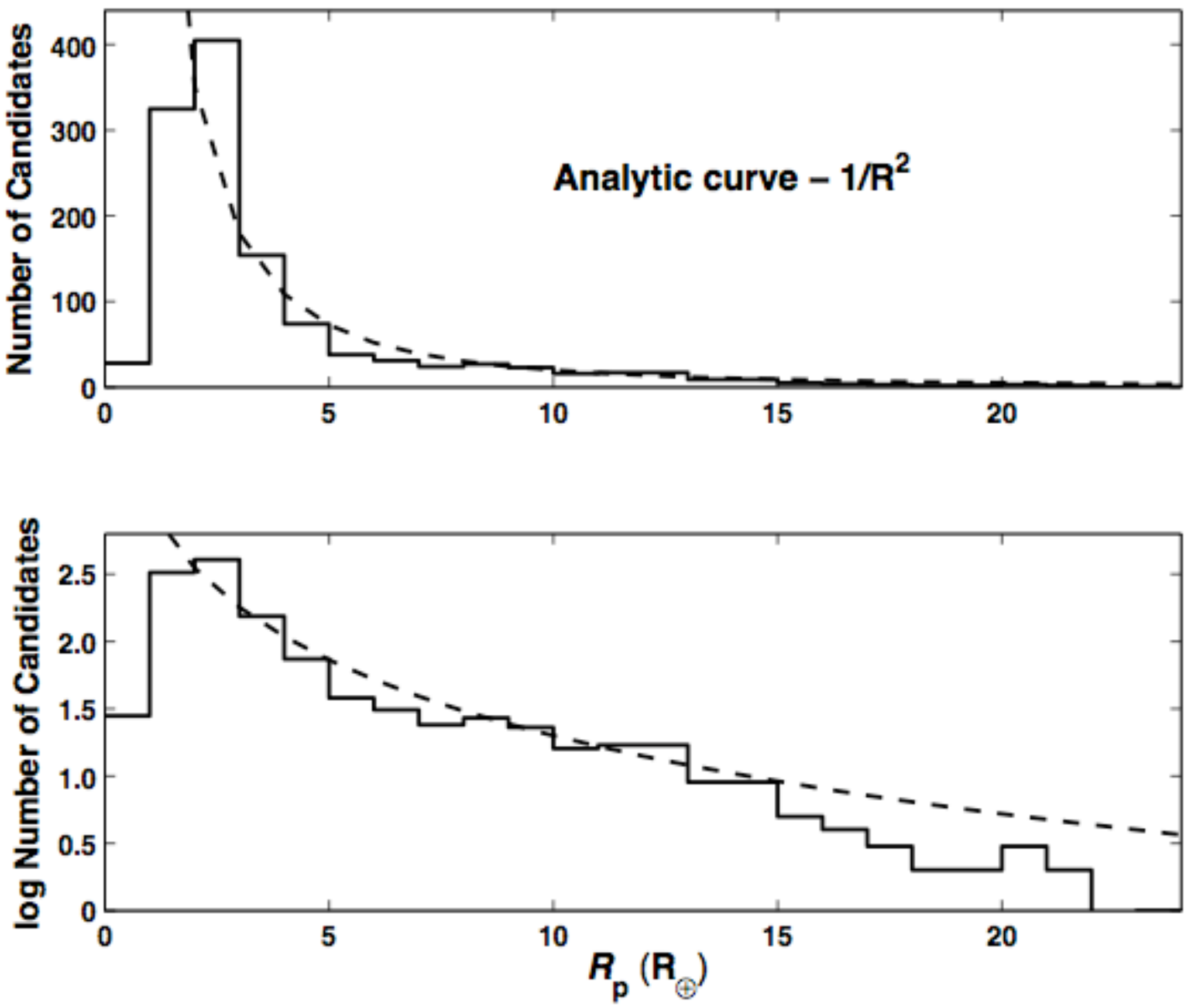

**Figure 2.** Size distribution of the number of *Kepler* candidates vs. planet radius ($R_p$) (upper panel). The logarithm of the number of candidates is presented in the lower panel to better show the tail of the distribution. Bin sizes in both panels are 1 $R_\oplus$.

As noted in Borucki et al. (2011), the results shown in Figure 2 imply that small candidate planets are much more common than large candidate planets. Of the 1202 candidates considered for the analysis, 74% are smaller than Neptune ($R_p$= 3.8 $R_\oplus$). Table 6 shows the observed distribution and the definition of sizes used throughout the paper for these 1202 candidates.

The dashed curve in both panels of Figure 2 represents a 1/($R_p^{-2}$ dependence of the number of candidates on candidate radius; i.e., dN/dr scales as $R_p^{-3}$ for 2$R_\oplus$ <Rp <15 $R_\oplus$. The data shown here are restricted to orbital periods ≤ 138 days. Because it is much easier to detect larger candidates than smaller ones, this result implies that the frequency of candidates decreases with the area of the candidate, assuming that the false positive rate, completeness, and other biases are independent of candidate size for candidates larger than 2 Earth radii. However, the current survey is not complete, especially for the fainter stars, smallest candidates, and long orbital periods, and further observations could influence the distribution.

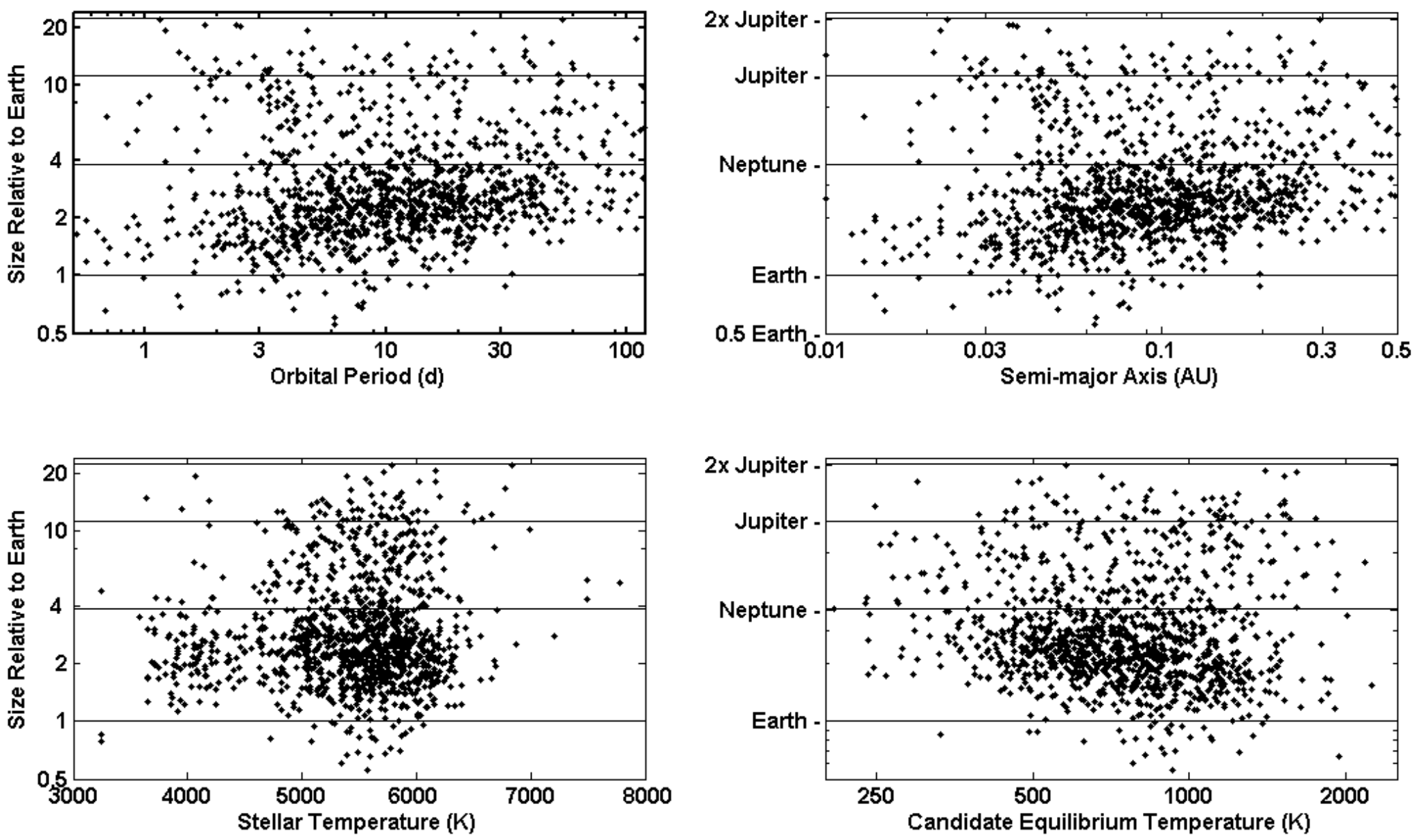

**Figure 3.** Candidate size versus orbital period, semi-major axis, stellar temperature, and candidate equilibrium temperature[3]. Uncertainties in candidate size are mostly due to the uncertainty in stellar sizes, i.e., approximately 25%. Horizontal lines mark ratios of candidate sizes for Earth-size, Neptune-size , and Jupiter-size relative to Earth-size.

Figure 3 presents scatter plots showing the observed relative size of individual candidates versus orbital period, semi-major axis, stellar temperature, and candidate temperature. The values on the abcissa are limited to show only the most populous range. Outliers can be found in Table 2. The upper left panel shows a concentration (in log-log space) of candidates for orbital periods between 3 and 30 days and sizes between 1 and 4 $R_\oplus$. The upper right panel shows a similar concentration. Both of them show a nearly empty area to the lower right that likely represents the lack of small candidates caused by the lower detectability of small candidates in long period orbits.

All panels in Figure 3 show a scarcity of candidates with radius $R_p$ smaller than 1 $R_\oplus$. The paucity of small candidates at even the shortest orbital periods could be due to incompleteness for the smaller signals, coupled with analysis of only a portion of the eventually expected *Kepler* data, and higher than expected noise levels. These effects could mask a real dependence of number on size. The modestly higher noise levels than those anticipated are thought to follow primarily from an underestimate of intrinsic stellar noise and are the topic of an on-going study.

---

[3] Teq was derived by assuming an even distribution of heat from the day to night side of the planet(e.g., a planet with an atmosphere or a planet with rotation period shorter than the orbital period) and the planet and star actas blackbodies in equilibrium; \begin{equation}\label{eq:teq} T_{eq}=T_* (R_*/2a)^{1/2} [f(1-A_B)]^{1/4}, \end{equation} where $T_*$ and $R_*$ are the effective temperature and radius of the host star, the planet at distance $a$ with a Bond albedo of $A_B$ and $f$ is a proxy for atmospheric thermal circulation. The Bond albedo, $A_B$, is the fraction of total power incident on a body scattered back into space which we assume to be 30\% and $f=1$ indicates full thermal circulation.

Figure 4 expands that portion of the lower right panel to emphasize those candidates with estimated radiative equilibrium temperatures in the range of liquid water at a pressure of 1 bar.

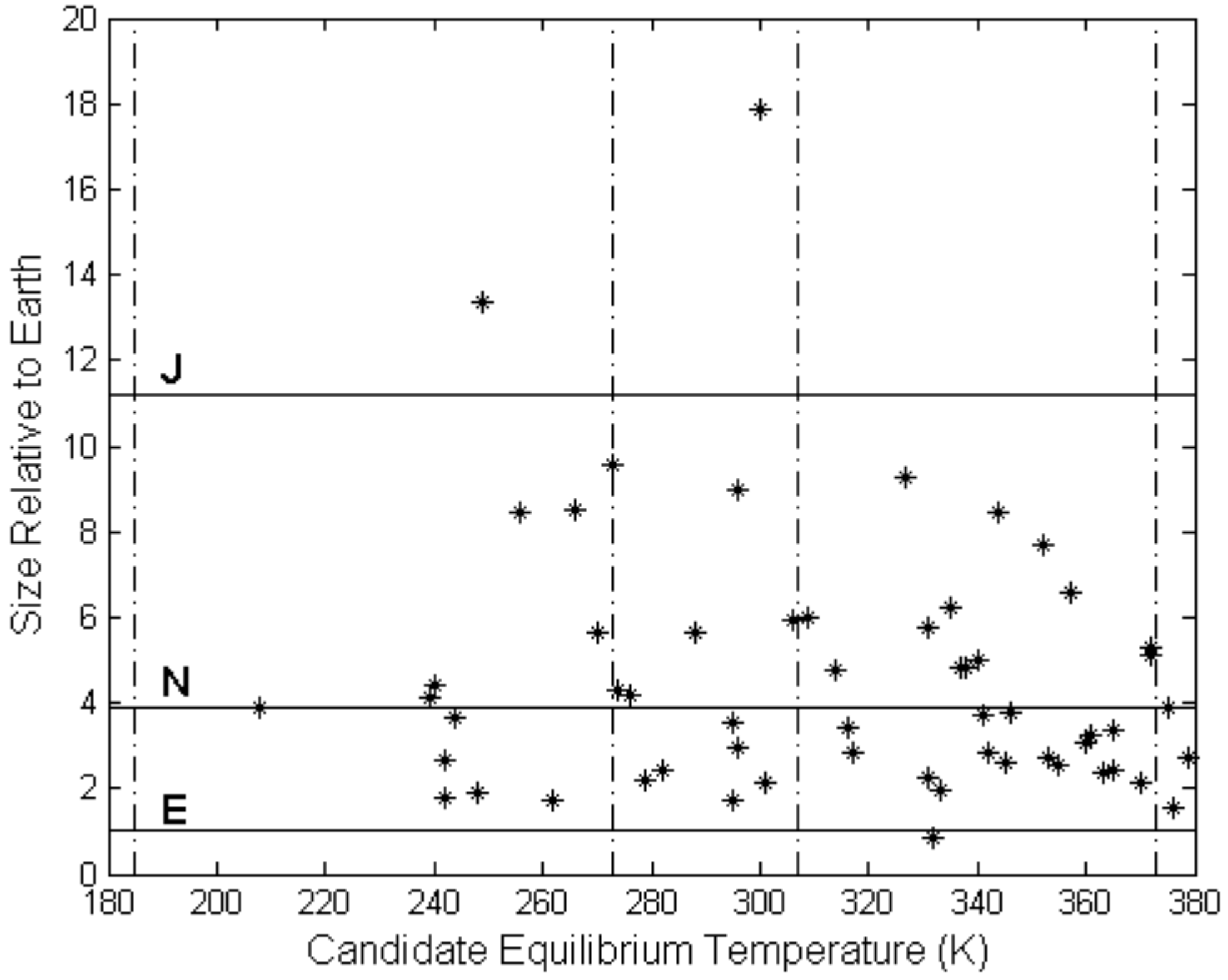

**Figure 4.** Candidate sizes and estimated radiative equilibrium temperatures ($T_{eq}$) centered on the habitable zone temperature range. The dotted lines bracket the range of temperatures allowing water to exist as a liquid at one atmosphere of pressure. Uncertainties are discussed in the text.

The habitable zone (HZ) is often defined to be that region around a star where a rocky planet with an Earth-like atmosphere could have a surface temperature between the freezing point and boiling point of water, or analogously the region receiving roughly the same insolation as the Earth from the Sun (Kasting et al.1993,  Rampino and Caldeira 1994, Heath et al. 1999, Joshi 2003, Tarter et al. 2007). The surface temperature range for habitable zones is likely to include radiative equilibrium temperatures well below 273 K because of warming by any atmosphere that might be present. For example, the greenhouse effect raises the Earth's surface temperature by 33 K and that of Venus by approximately 500 K. Further, the spectral characteristics of the stellar flux vary strongly with $T_{eff}$ and affect both the atmospheric composition and the chemistry of photosynthesis (Heath et al. 1999, Segura et al. 2005). Consequently, Figure 4 shows temperatures well below the freezing point of water. The vertical lines at 183 and 307 K delineate the radiative temperature range for which the surface temperature of a rocky planet with an atmosphere similar to that of the Earth is expected to be within the freezing and boiling point of water (Jim Kasting, private communication, 2/28/2011).

The calculated equilibrium temperatures shown in Figure 4 are for grey-body spheres without atmospheres. The calculations assume a Bond albedo of 0.3, emissivity of 0.9, and a uniform surface temperature. The uncertainty in the computed equilibrium temperatures is approximately 22% (see Appendix) because of uncertainties in the stellar size, mass, and temperature as well as the planetary albedo. For planets with an atmosphere, the surface temperature would be higher than the radiative equilibrium temperature.

Within this temperature range, there are 54 candidates are present with sizes ranging from Earth-size to larger than that of Jupiter. Table 5 lists the candidates in the HZ. The detection of Earth-size candidates depends on the signal level, which in turn depends on the size of the candidate relative to the size of the star, the number of transits observed, and the combined noise of the star and the instrument. It is important to recognize that the size of the star is generally not well characterized until spectroscopic studies and analysis are completed. In particular, some of the cooler stars could be nearly double the size shown in Table 1 and that some of the candidates could prove to be false positives.
As can be seen in Table 5, there are two candidates with $R_p$ < 1.5 $R_\oplus$ (KOI 314.02 and KOI 326.01) present in the list. The uncertainty in the sizes of these candidates is approximately 25% to 35% due to the uncertainty in size of stars and of the transit depth.

The predicted semi-amplitudes of the RV signals for small candidates such as KOI 314.02 and 326.01 are 1.2 m/s and 0.5 m/s, respectively. These RV amplitudes follow from assuming a circular orbit and a density of 5.5 g/cm$^3$ for both candidates. RV semi-amplitudes of 1.0 m/s are at the very limit of what might currently be possible to detect with the largest telescopes and best spectrometers. In principle, RV amplitudes under 1 m/s could be detected, but there are many impediments to achieving such precision including the surface velocity fields (turbulence) and spots on the rotating surface. In addition, stars with one transiting planet may well harbor multiple additional planets that do not transit, causing additional RV variations. Moreover, these two stars have V-band magnitudes of 14, making it very difficult to acquire sufficient photons in a high resolution spectrum to achieve the required Doppler precision. Of course, for all of these small planets RV measurements can place firm upper limits to their masses and densities.

**Table 6.** Number of Candidates versus Size.

| Candidate Label | Candidate Size ($R_\oplus$) | Number of Candidates plus known planets |
|---|---|---|
| Earth-size | $R_p \leq 1.25$ | 68 |
| super-Earth-size | $1.25 < R_p \leq 2.0$ | 288 |
| Neptune-size | $2.0 < R_p \leq 6.0$ | 662 |
| Jupiter-size | $6.0 < R_p \leq 15$ | 165 |
| very-Large-size | $15.0 < R_p \leq 22.4$ | 19 |
| Not considered | $R_p > 22.4$ | 15 |

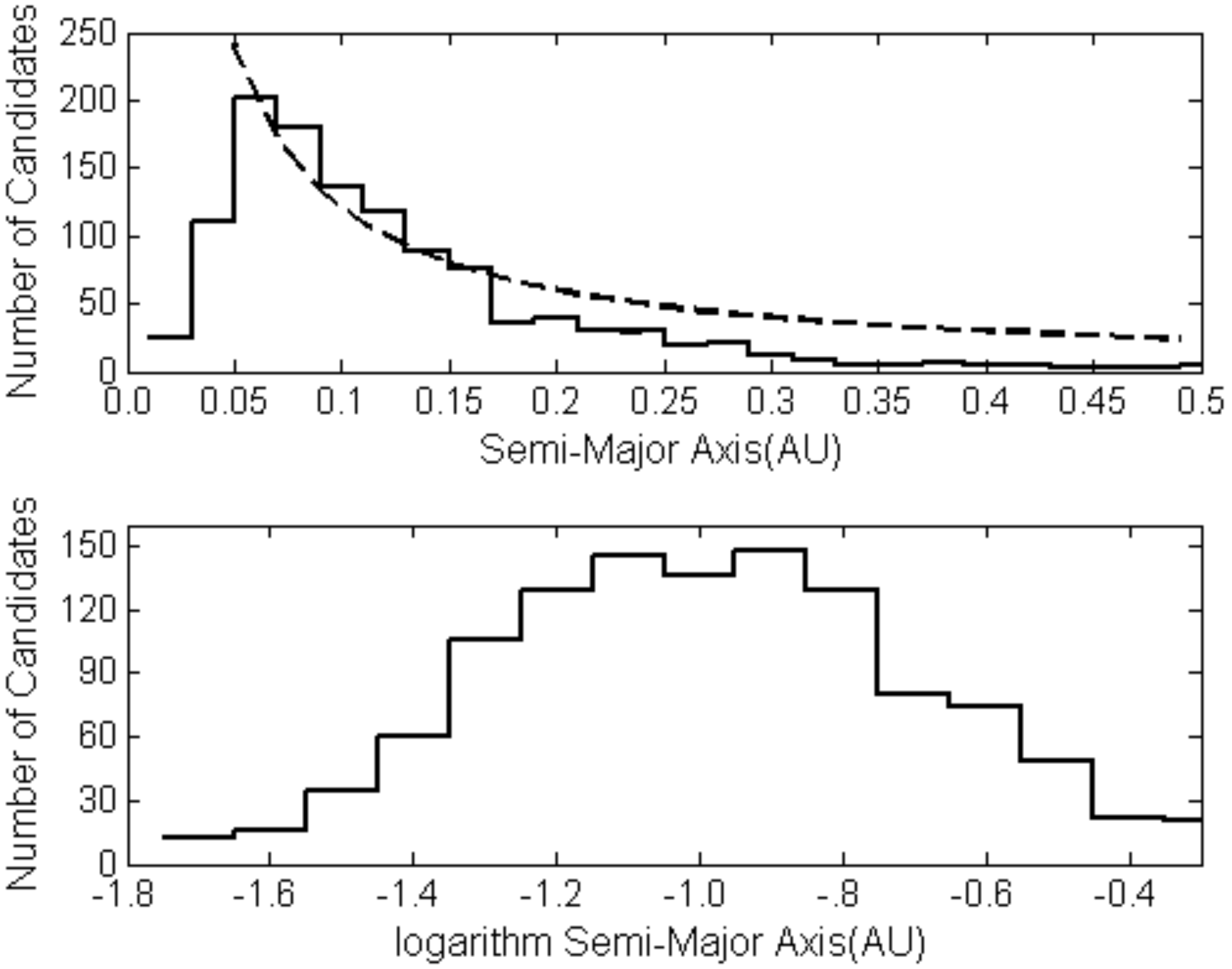

**Figure 5.** Upper panel: Historgrams of the observed number of candidates vs. linear intervals in the semi-major axis. The dashed line shows the relative effect of geometricical probability of alignment Lower panel: The number of candidates vs. logarithmic intervals of the semi-major axis. Bin size is 0.02 AU in the upper panel and 0.1 in the lower panel.

In Figure 5, the dependence of the number of candidates on the semi-major axis is examined. For $a$ less than 0.04 AU, it is evident that the distribution is severely truncated. As is evident in Figure 5, this feature is present in each of the candidate size groups. In the upper panel of Figure 5, an analytic curve shows the expected reduction in the number in each interval due to the decreasing geometrical probability that orbits are aligned with the line-of-sight. It has been scaled over the range of semi-major axis from 0.04 to 0.5 AU, corresponding to orbital periods from 3 days to 138 days for a solar-mass star. The fit is fair-to-poor implying that the intrinsic distribution is not constant with semi-major axis after a correction only for the alignment probability.

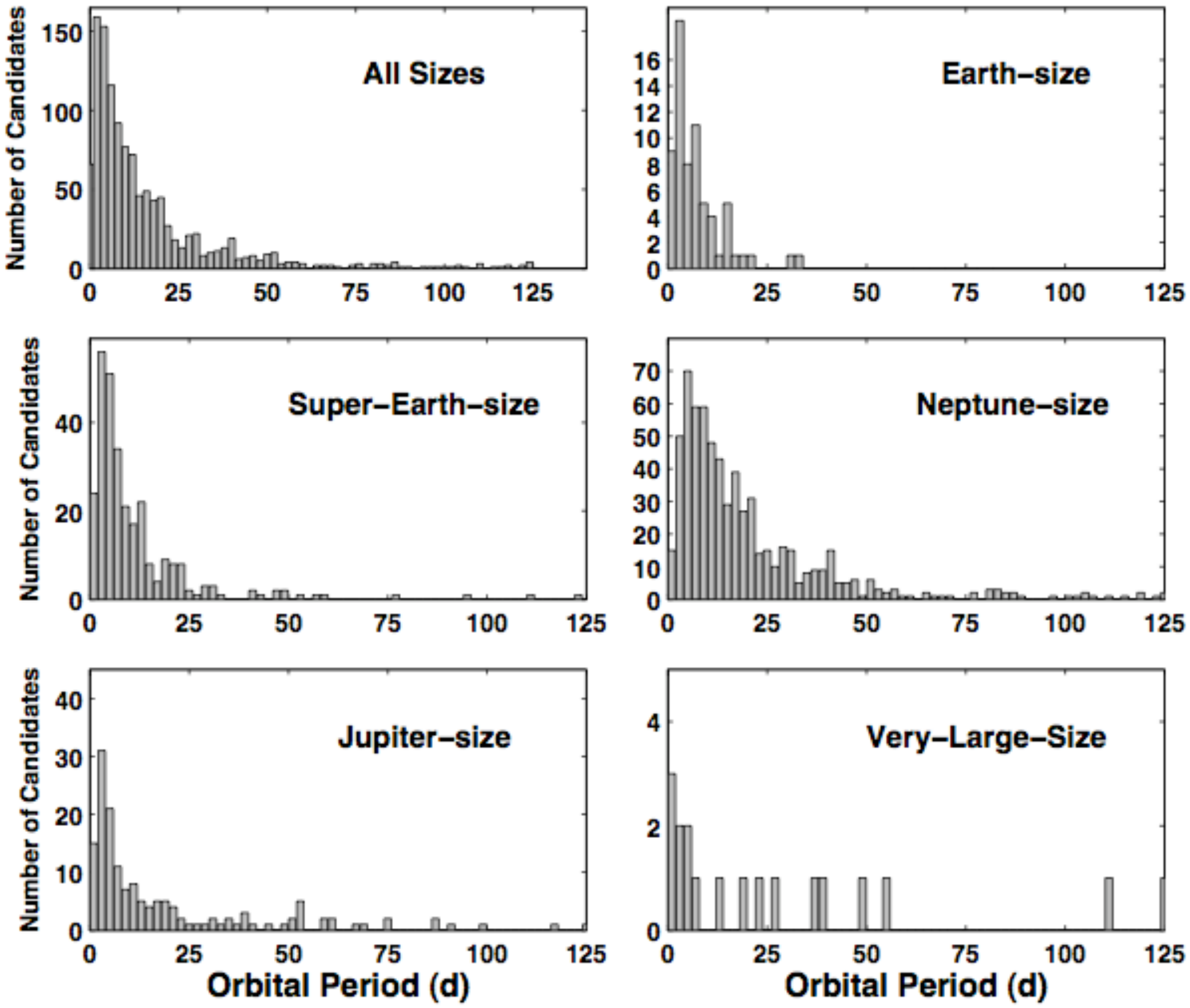

**Figure 6.** Number of candidates vs orbital period for several choices of candidate size. Bin size is 2 days. Refer to Table 6 for the definition of each size category.

The panels in Figure 6 show that the period distribution of Neptune-size candidates has a less steep slope compared to Jupiter-size candidates in the period range from one week to one month. Because of the large numbers in both samples and the ease of detecting such large candidates, the difference in the dependence of number on semi-major axis is likely to be real. All show maxima in the number of candidates for orbital periods between 2 to 5 days for all sizes and a narrow dip at periods shorter than two days. (The small number of very-Large candidates might be the reason for the lack of a local minimum at the shortest orbital periods.) However these objects are as large as late M-dwarf stars and it is unclear what type of object they represent. Determination of their masses with RV techniques is clearly warranted because the results would not only provide masses, but densities as well when combined with the transit results.

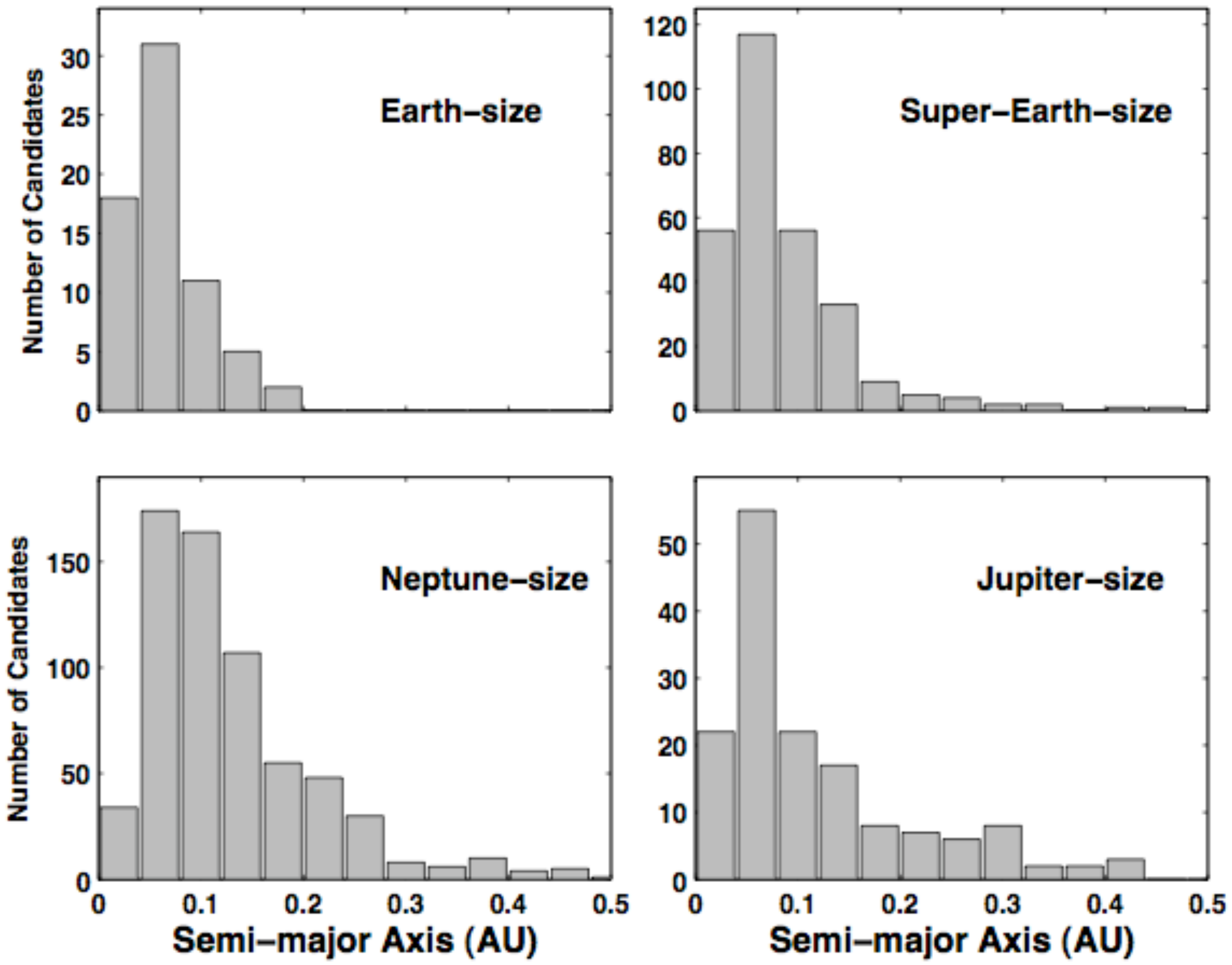

**Figure 7.** Number of observed candidates versus semi-major axis for four candidate size ranges. As defined in Table 6, Earth-size refers to $R_p < 1.25\ R_\oplus$, super-Earth-size to $1.25\ R_\oplus < R_p < 2\ R_\oplus$, Neptune-size to $2\ R_\oplus, < R_p < 6\ R_\oplus$, and Jupiter-size refers to $6\ R_\oplus < R_p < 15\ R_\oplus$. Bin size for the semi-major axis is 0.04 AU.

A breakout of the number of candidates versus semi-major axis is shown in Figure 7 using the definition for size in Table 6. "Earth-size" candidates and some of the "super-Earth-size" candidates are expected to be rocky type planets without a hydrogen-helium atmosphere. "Neptune-size" candidates could be similar to Neptune and the ice giants in composition. All size classes show a rise in the number of candidates for decreasing semi-major axis until a value of 0.04 AU and then a steep drop. The drop off in the number of Earth-size candidates for semi-major axes greater than 0.2AU is due at least in part to the decreasing probability of a favorable geometrical alignment and the difficulty of detecting small planets when only a few transits are available.

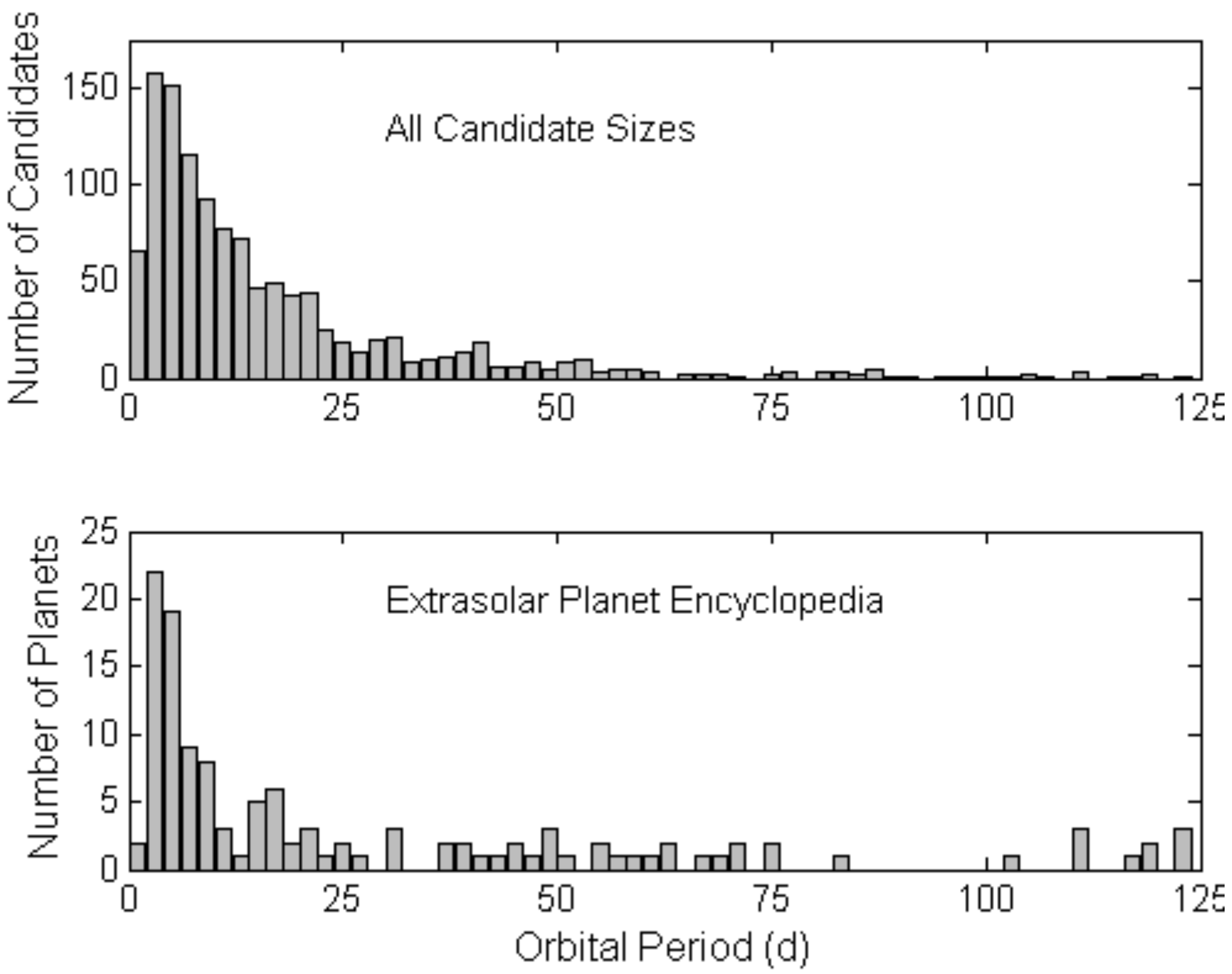

Figure 8. (Upper panel) Observed period distribution of Kepler planet candidates with orbital periods less than 125 days, uncorrected for observational selection effects. (Lower panel) Period distribution over the same range for RV-discovered planets listed in the Extrasolar Planet Encyclopedia (EPE) as of 7 Dec 2010 exclusive of *Kepler* planets. Bin size is 2 days.

Figure 8 compares the orbital period distribution of the Kepler planet candidates with the planets discovered by the RV method (as reported by the EPE.) Both detection methods show a prominent peak in the numbers for periods between two and four days and a large drop in the number for shorter periods. There are several references in the literature to the pile-up of giant planet orbital periods near 3 days (*e.g.* Santos and Mayor 2003). It is suggestive of a process that allows planets migrating inward to synchronize their orbital period with the rotation period of the star, raise tides of sufficient strength that enough momentum is transferred to the planet to halt its migration. Later, the star becomes sufficiently luminous that the dust and gas of the accretion disk are expelled leaving the planet in a stable, but short-period orbit. The cause of the much larger relative decrease seen in the RV-discovered planets compared to that seen in the Kepler results is not understood.

The planetary candidates observed at shorter distances could represent those that did not come into synchronism with the star, but stopped short of entering the star's atmosphere because a coincidence with the dissipation of the accretion disk. They could also represent a continued migration of the body into the star.

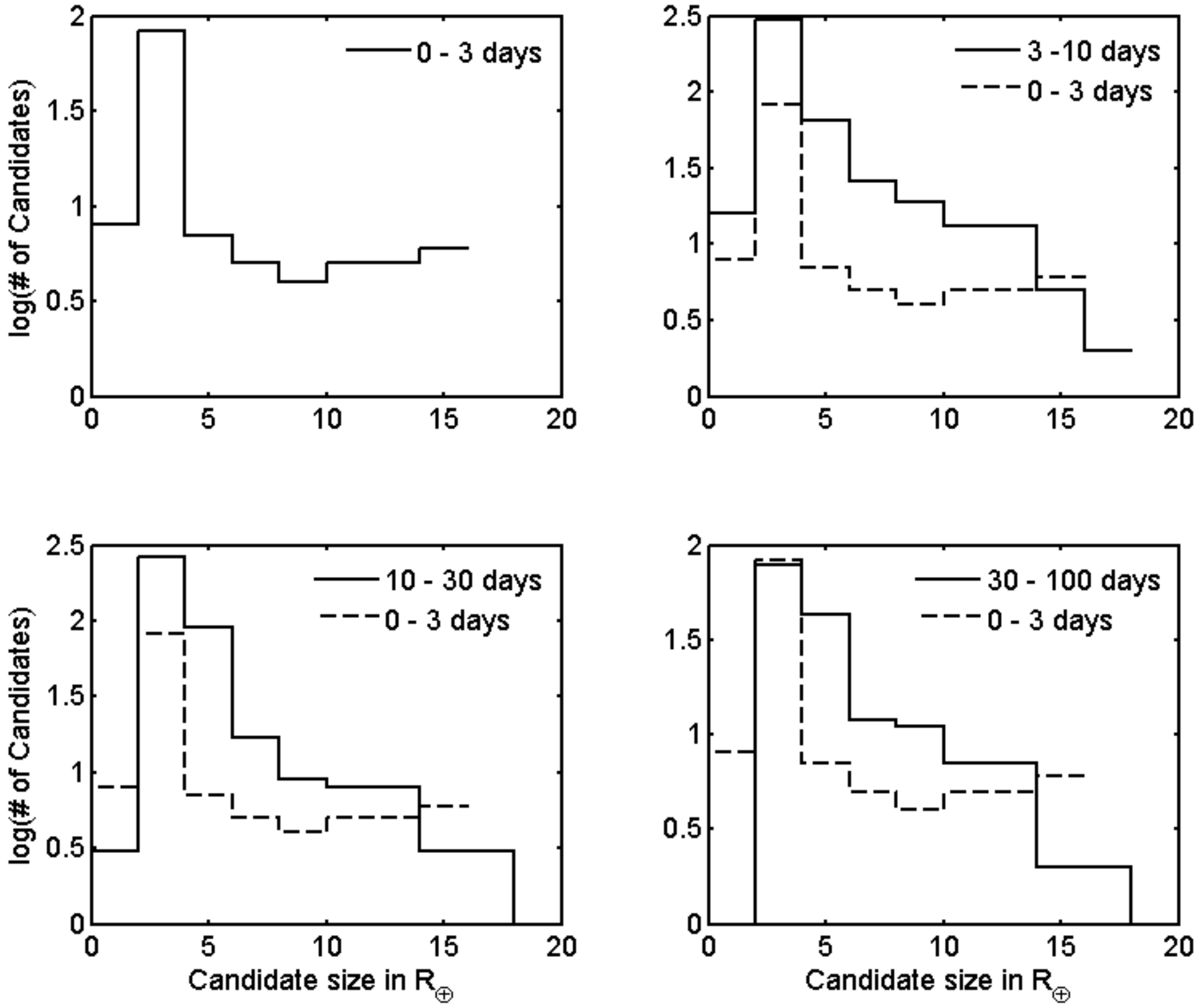

**Figure 9.** Observed distribution of candidate sizes for four ranges of orbital period, uncorrected for selection effects. Panels 2, 3, & 4 compare the distributions for longer periods with that of the shortest period range. Bin size is 2 $R_{\oplus}$.

Except for the peak between 2 to 4 $R_{\oplus}$, Figure 9 shows that the number of short-period (< 3 days) candidates is nearly independent of candidate size through 16 $R_{\oplus}$. However, small candidates are more numerous than large ones for longer orbital periods. This distribution suggests that short-period candidates might represent a different population than the populations at larger orbital periods and semi-major axes. In particular, they might represent rocky planets and the remnant cores of ice giants and gas giant planets that have lost their atmospheres. To confirm that this population is distinct from that of longer-period candidates will require a future investigation of the comparison of the mass-radius relationships of the populations.

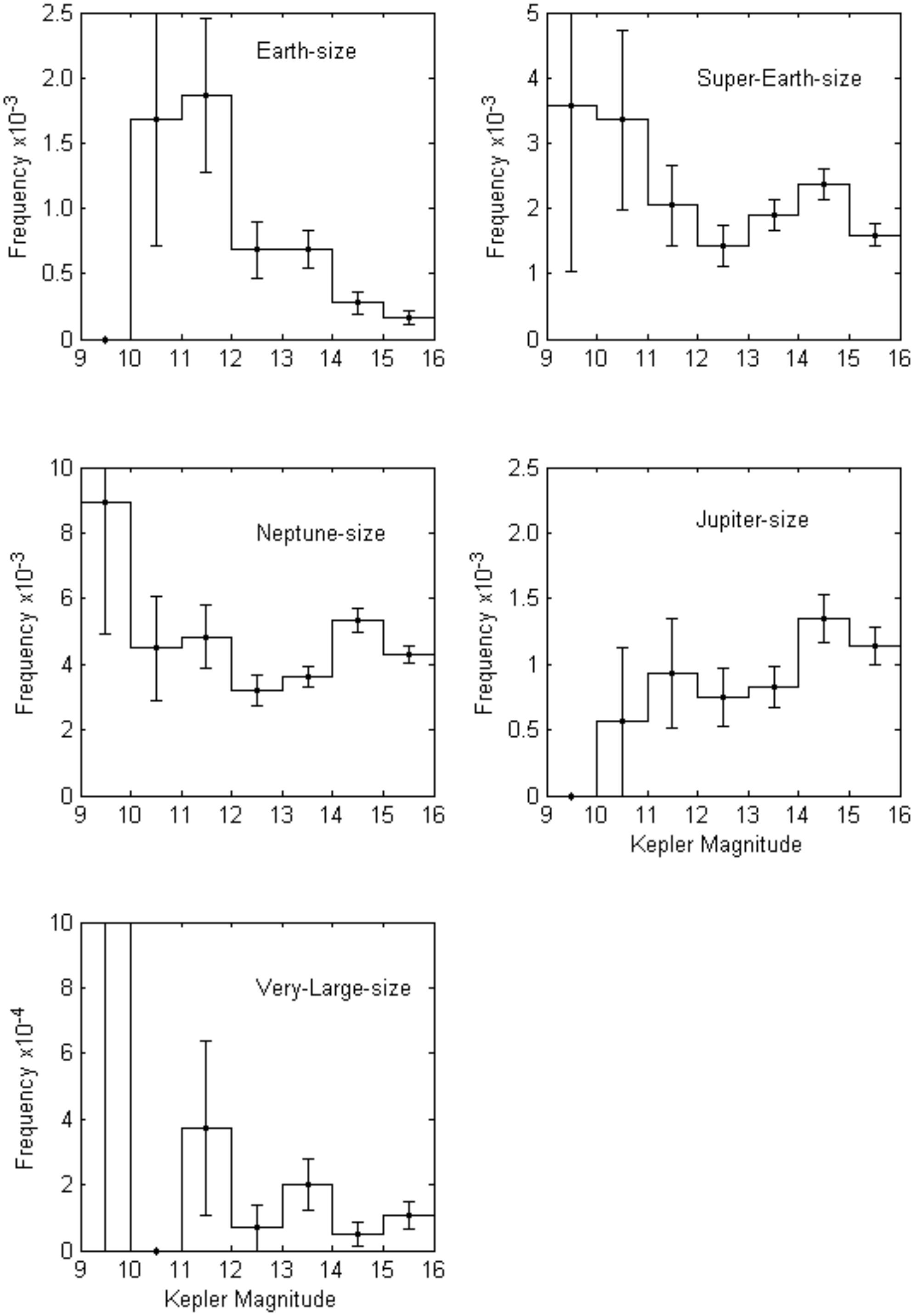

**Figure 10.** Observed frequencies, uncorrected for selection effects, of candidates for five size ranges defined in Table 6 as a function of *Kepler* magnitude. The error bars represent only the Poisson noise associated with the number of events in each bin, and the upper bar represents a single event if no events are observed.

In Figure 10, the observed frequency of candidates in each magnitude bin has been simply calculated from the number of candidates in each bin divided by the total number of stars monitored in each bin. The number of stars brighter than Kp = 9.0 or fainter than Kp = 16.0 in the current list is so small that the count is not shown.

The panels for Earth-size and super-Earth size candidates are consistent with a decrease in the observed frequency with increasing magnitude for magnitudes larger than Kp=11, and are indicative of difficulty in detecting small candidates around faint stars. Near-constant values of observed frequencies of the Neptune-size and larger candidates would be expected if the survey were mostly complete for the large candidates and for the orbital periods reported here and if the distribution of stellar types is independent of apparent magnitude. However, almost all M-dwarf stars in the Kepler FOV have Kp>14. Therefore if the frequency of large candidates around M-dwarfs is different than for other spectral types, then near-constant frequencies of Neptune- and larger-size candidates should not be expected. Perhaps the apparent decrease with increasing magnitude is due to this cause.

An examination of the upper left panel of Figure 10 indicates that several Earth-size candidates must be present in the 15$^{th}$ to 16$^{th}$ magnitude bin. The noise properties of the instrument are such that only the smallest stars or small stars with short-period candidates can appear in this bin. To get a measure of the variation of the observed frequency distributions with magnitude when the transit amplitude is held nearly constant, the distributions for five ranges of the ratio Rp /R* are displayed in Figure 11.

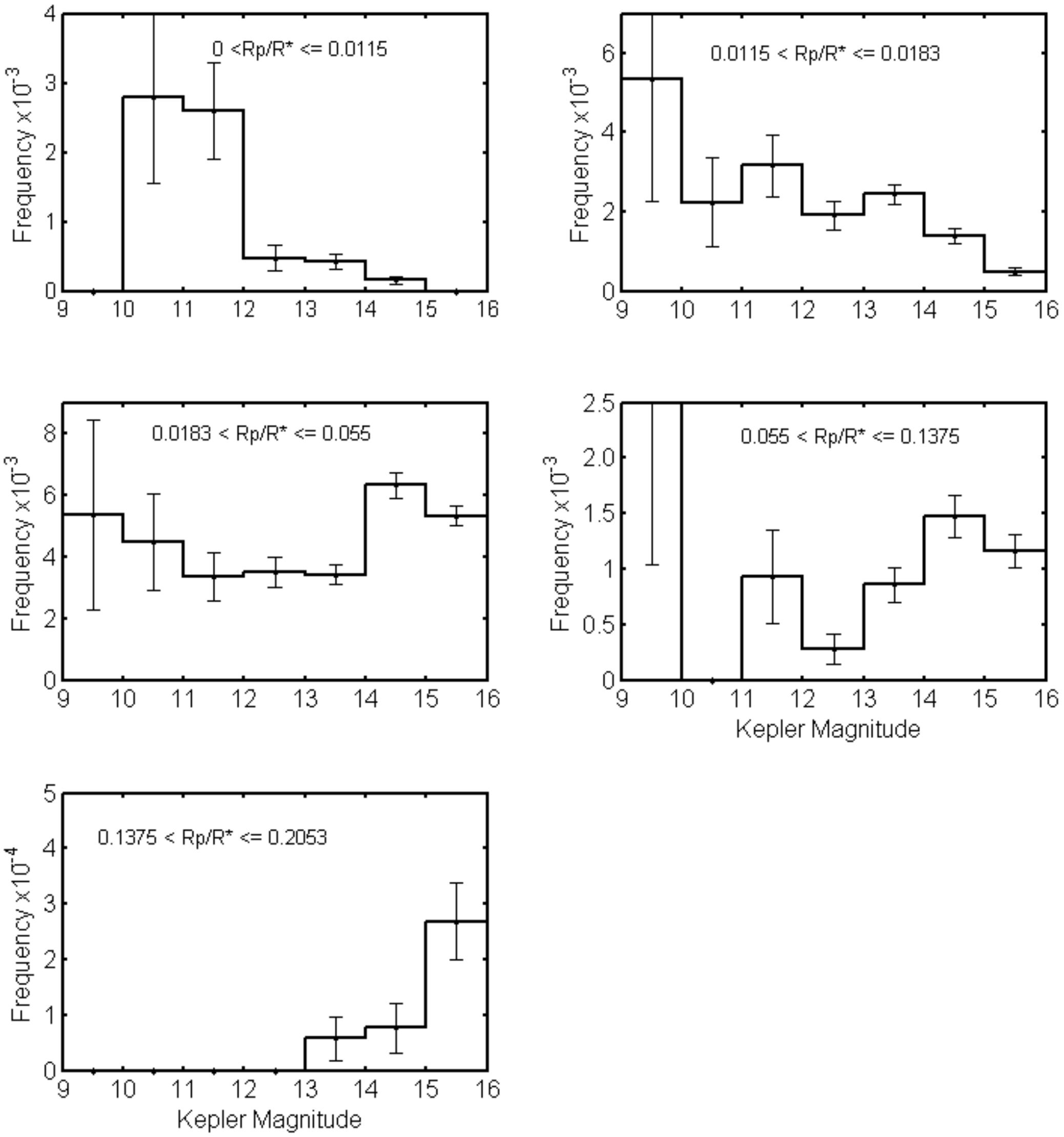

**Figure 11.** Frequency distribution (not corrected for selection effects) for 5 ranges of the ratio of the radius of the candidate to that of the host star versus magnitude.

The five ratios shown in Figure 11 are appropriate for Earth-size, super-Earth-size, Neptune-size, Jupiter-size, and very-Large-size candidates transiting stars of radius $R_\star = 1\ R_\odot$, where the subscript $\odot$ signifies solar values. An examination of the upper left hand panel shows no candidates are found for the 15 to 16 magnitude range. The Earth-size candidates around faint stars (Kp>15) shown in the upper left panel of Figure 10 orbit small stars and have a planet-star radius ratio greater than 0.0115. Thus they no longer appear in the upper left panel of Figure 11. The observed frequency distributions show a steeper decrease with increasing magnitude for the small Rp/R* shown in the two upper panels. The panels in the second row again show a nearly constant frequency with magnitude implying that such signal levels are readily detected over the magnitude range of interest. Contrary to what might be expected, a nearly constant frequency with magnitude is not seen for the largest ratio-range. This result is not understood.

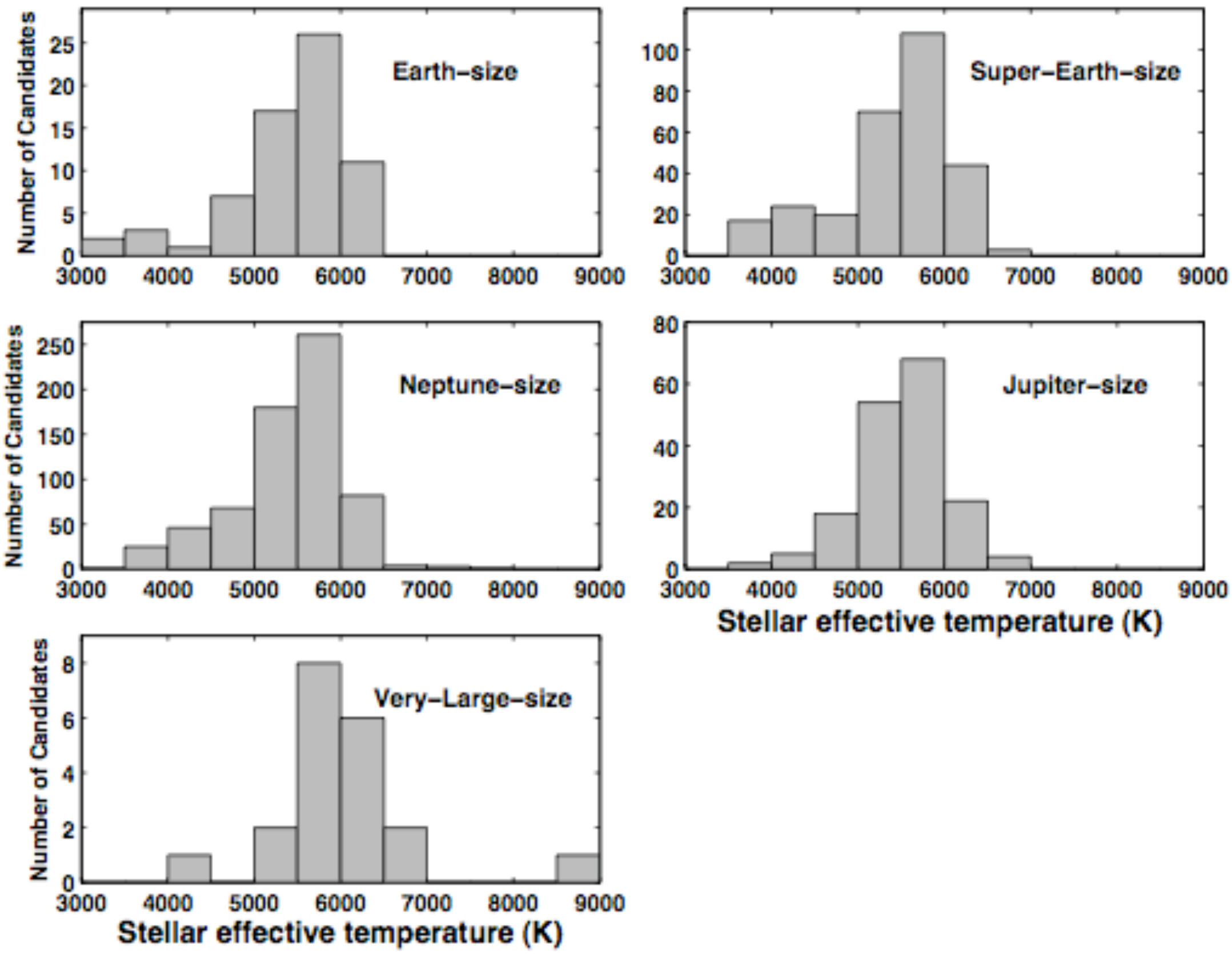

**Figure 12.** Observed number of candidates for various candidate sizes vs. stellar effective temperature, uncorrected for selection effects. Bin size is 500°K. Refer to Table 6 for the definition of each size category.

The number of candidates is a maximum for stars with temperatures between 5000 and 6000 K, i.e., G-type dwarfs (Figure 12). This result should be expected because the selection process explicitly emphasized these stars and because G-type stars are a large component of magnitude-limited surveys of dwarfs at the magnitudes of interest to the *Kepler* Mission.

To reduce the bias associated with the large fraction of K, G, and F type stars, the number of candidates in each bin was normalized to the number of star in the bin and frequencies calculated as a function of stellar temperature. However, because of the narrow-width temperature bins, many of the bins have a very small number of candidates which cause the frequencies to vary widely due to small-number statistics. To increase the number in each bin and reduce the large variations associated with small-number statistics, the bins in Figure 13 are twice as large as those in Figure 12.

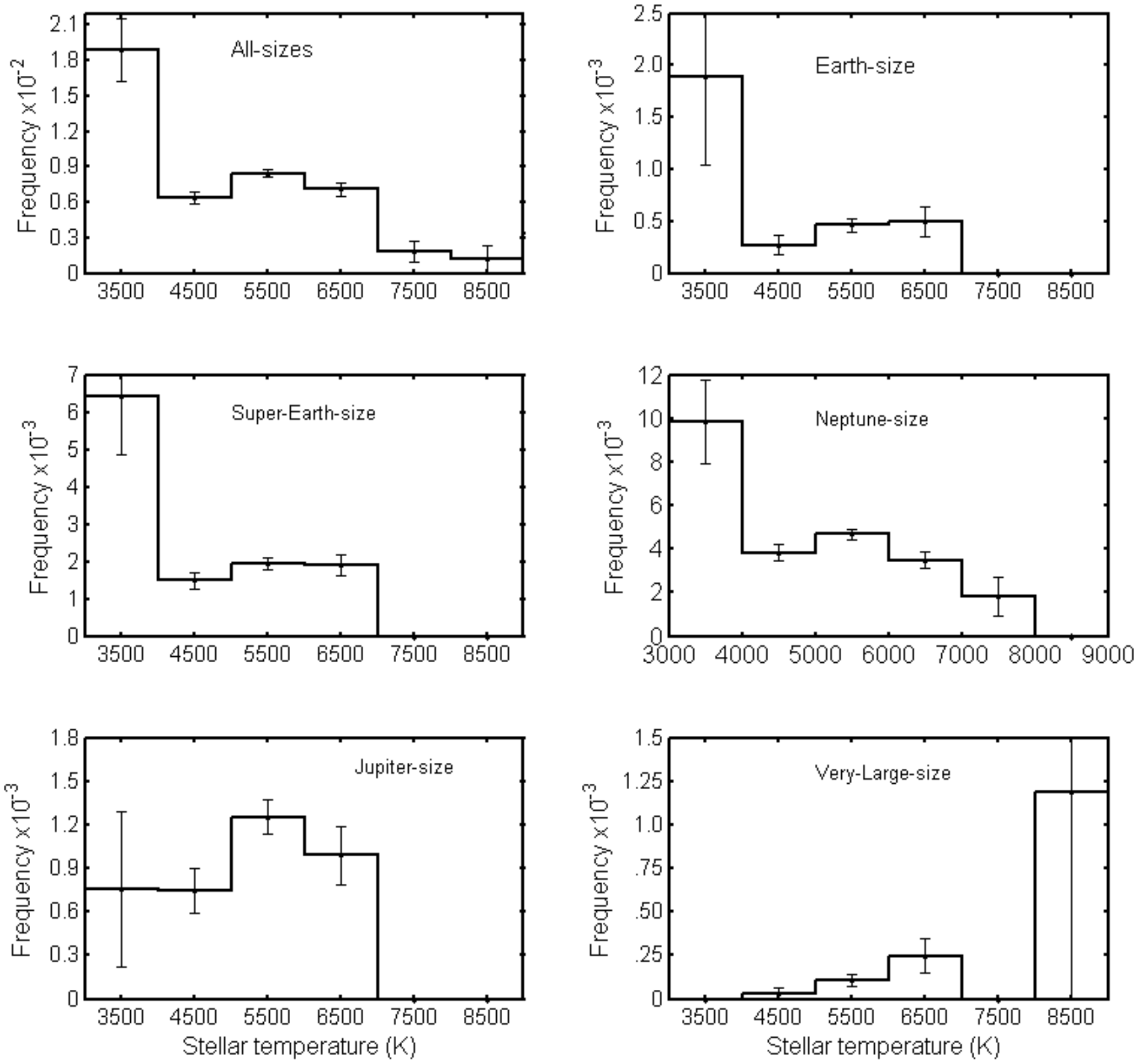

**Figure 13.** Measured frequency of candidates versus stellar temperature. The error bars shown with the distributions represents only that portion of the uncertainty due to Poisson noise. Bin size is 1000 K. Refer to Table 6 for the definition of each size category.

In Figure 13, a comparison of the frequencies of super-Earth-size and Neptune-size candidates shows an indication that candidates are preferentially found around stars cooler than 4000K. A similar distribution is also found for Earth-size candidates, but because of the very small number of candidates in that bin (i.e., 2), the maximum is not statistically significant. Main sequence stars with temperatures between 3000 K and 4000 K are classified as M-dwarfs. Giant and super-giant late K spectral-type stars are both more massive and larger than the M-dwarfs but have similar temperatures. A check of the KIC showed that none of the candidates were associated with log *g* less than 4.2; i.e., they are associated with dwarfs, not giants. Because M-dwarfs are much smaller than earlier spectral types, the amplitudes of the transits generated by small planets are substantially larger than those generated by hotter stars. This fact introduces a strong bias that will be considered in the next section.

## 4. Completeness Estimate

Although the primary purpose of the paper is to summarize the results of the observations and to act as a guide to content of the tables, a model was developed to provide a first estimate of the intrinsic frequency of planetary candidates. The "intrinsic frequency" of planetary candidates is used here to mean the observed number of candidates per number of target stars that must be observed to produce the observed number of candidates in the specified bins of semi-major axis $a$ and candidate size $R$ when all selection effects are applied. The bin limits used for $a$ are evenly spaced from 0.0 to 0.5 AU with a spacing of 0.02 AU. The bin limits for the planetary candidate size-classes are: Earth-size ($0.5 \leq R < 1.25$ $R_\oplus$), super-Earth-size ($1.25 \leq R < 2.0$ $R_\oplus$), Neptune-size ($2.0 \leq R < 6.0$ $R_\oplus$), Jupiter-size ($6.0 \leq R < 15.0 R_\oplus$), and very-Large-size ($15.0 \leq R < 22.4$ $R_\oplus$). It should be noted that the calculation of the intrinsic frequency is equivalent to ratio of the measured number of candidates divided by the expected number of candidates based on the ensemble of stars that are observed.

For every candidate in a $\Delta a$ $\Delta R$ bin, each of the 156,000 target stars was examined to determine if a planet orbiting it with the same size as the candidate and having the same $a$ could be detected during the Q0 through Q2 observation period. The number of target stars needed to produce a minimum of two transits in the period of interest with a signal $\geq 7$ σ was tabulated for each bin. (There is no need for three transits because confirmation as a planet is not considered here.) The actual period simulated is longer than the 138 days of the Q0 through Q2 period because the search for planetary candidates used data obtained during later periods to obtain accurate values of the epoch and period, as discussed earlier.

Inputs to the model include the observed noise for 3, 6, and 12-hour bins averaged over one quarter of data (Q3) for each target star and the target star's size, mass, and magnitude, as well as the values of the size and semi-major axis of each candidate in the $\Delta a$ $\Delta R$ bin. We also undertook an independent analysis that used the observed noise for 3-hour bins averaged over the Q3 data. Since the properties of the noises are not Gaussian, this serves as a check on our results.

The model computes the duration of the transits from the size and mass of the star at the specified value of the semi-major axis. The value of the noise for each target star is interpolated to the computed transit duration based on the values of the noise measured for 3, 6, and 12 hour samples. This a very important correction because for 80%of the stars, the variation of CDPP with the duration of the transit does not vary with the reciprocal of the square root of the time, but is less than that expected from a Poisson-distribution . The signal level is computed from the square of the ratio of the candidate size to the size of the target star. This value is then divided by the interpolated noise value to get the estimated single-transit SNR. The total SNR is based on the single-transit SNR multiplied by the square root of number of transits that occur during the observation period. A correction is made for the loss of transits (and consequently, the reduction in the total SNR) due to the monthly and quarterly interruptions of observations. The probability of a recognized detection event is then computed from the value of the total SNR and a threshold level of 7σ. In particular, if the total SNR is 7.0, then the transit pattern will be recognized 50% of the time while if the total SNR was estimated to be 8.0, then the transit pattern would be recognized 84% of the time. The value of this probability $p_1$ is tabulated and then an adjustment is made for the probability that the planet's orbit is correctly aligned to the line-of-sight $p_2$. The value of $p_2$ is based on the size of the target star and the semi-major axis specified for the candidate. The product of these probabilities $p_{nc}$ is the probability that the target star $n$ could have produced the observed candidate $c$.

The probability $p_{nc}$ is computed for each of the 156,000 stars and then summed to yield the estimated number of target stars $n^*_{c,a,R}$ that could have produced a detectable signal consistent with candidate's semi-major axis $a$ and size $R$. (Subscripts designate candidate "c", semi-major axis value "$a$", candidate size "$R$".) This procedure is repeated for each candidate in the $\Delta a$ $\Delta R$ bin.

The sum of the number of candidates of size-class "k" in a bin ($a$, $\Delta a$, $R$, $\Delta R$) is designated $S_{a,R,k}$. The size-class "k" (k=1 to 5) represents Earth-size, super-Earth-size, Neptune-size, Jupiter-size, and very-large size planetary candidates, respectively.

After a value of $n^*_{c,a,R,k}$ has been computed for each candidate in the bin, the median value $N^*_{a,R,k}$ of $n^*_{c,a,R,k}$ is computed and used to estimate the frequencies:

$$\text{Freq}(k, r_i, \Delta r_i, a_i, \Delta a_i) = \frac{S_{a,r,k}}{N^*_{a,R,k}} \quad \text{Eq. 1}$$

For each size-class, the sum of the frequencies over $a$ and $R$ is the estimate of the frequency for that size-class:

$$Freq(k) = \sum_{R=min}^{R=max} \sum_{a=0.01}^{a=0.5} \frac{S_{a,R,k}}{N^*_{a,R,k}}$$

Eq. 2

The summation for each size-class is done only for those bins that have at least 2 planetary candidates and a minimum of 10 target stars. These choices help to reduce the impact of outlier values.

The uncertainties in the results are quite large because the calculated number of stars $n^*_{c,a,R,k}$ for the observed number of candidates $S_{a,R,k}$ is a sensitive function of the position of each planetary candidate inside of the $\Delta a$ $\Delta Rr$ bin and because the number of candidates in each bin is often small. In particular, estimated frequencies based on the sum of the individual frequencies in each bin are very different than the estimates obtained by dividing the number of observed candidates by the average number of expected planets. Therefore medians are used instead of averages to reduce the effects of outliers.

To provide an estimate of the dispersion $D_{a,R,k}$ of the estimated frequencies for each bin, the relative error associated with the number of candidates used in the estimate of the frequency is added in quadrature to the variance due to the dispersion of the values of $n^*_{c,a,R,k}$.

$$D_{a,R,k} = \sqrt{\frac{1}{S_{a,R,k}} + \frac{Var(n^*_{c,a,R,k})}{(n^*_{c,a,R,k})^2}} \quad , \quad \text{Eq. 3}$$

where $n^*_{c,a,R,k} = \sum_{c=1}^{c=max} n^*_{c,a,R,k}$. Eq. 4

It is important to note that the estimated frequencies calculated by the model are based upon the number of candidates found in the data. In turn, the number and size distributions depend on both the results from the analysis pipeline and a manual inspection of the results of the pipeline product. The current version of the analysis pipeline provides "threshold crossing events" and checks that those data are consistent with an astrophysical process. However, it does not yet have the capability to stitch together quarterly records. Thus the number of candidates discussed here is based on a combination of pipeline results, manual inspection, and an *ad hoc* program that does not use the more comprehensive detrending that is done in the pipeline, but does allow a longer period of data to be examined. In some cases, the candidates in the Q0-Q2 data were not discovered until the Q3 and Q5 data were examined. As discussed later, the procedure is designed to quickly find candidates that can be followed up, but is not well controlled for the purpose of the model calculations. Consequently, the results must be considered very preliminary.

Table 7 presents an example of the calculated intrinsic frequencies, number of planetary candidates, mean value of the number of target stars, and dispersion values for the range of *a* from 0.01 to 0.50 AU for Earth-size candidates. The results for the all class-sizes are plotted in Figure 14.

Table 7. Intrinsic Frequency of Earth-size Candidates (Simulation of 1.0 year of observations)

Results for Earth-size Candidates

| ----- *a* (AU) | ----- | $S_{c,k,l=1}$ | $N_{a,R,k=1}$ | Freq(1) | Relative Dispersion |
|---|---|---|---|---|---|
| 0.001 | 0.02 | 7 | 22594.1 | 0.00031 | 0.558983 |
| 0.02 | 0.04 | 12 | 5815.89 | 0.002063 | 0.546926 |
| 0.04 | 0.06 | 17 | 3400.43 | 0.004999 | 0.504917 |
| 0.06 | 0.08 | 14 | 1541.07 | 0.009085 | 0.697659 |
| 0.08 | 0.1 | 6 | 744.935 | 0.008054 | 0.654883 |
| 0.1 | 0.12 | 5 | 722.045 | 0.006925 | 0.679988 |
| 0.12 | 0.14 | 4 | 667.149 | 0.005996 | 0.549324 |
| 0.14 | 0.16 | 1 | 0 | 0 | 1 |
| 0.16 | 0.18 | 0 | 0 | 0 | 0 |
| 0.18 | 0.2 | 2 | 117.749 | 0.016985 | 0.79465 |
| 0.2 | 0.22 | 0 | 0 | 0 | 0 |
| 0.22 | 0.24 | 0 | 0 | 0 | 0 |
| 0.24 | 0.26 | 0 | 0 | 0 | 0 |
| 0.26 | 0.28 | 0 | 0 | 0 | 0 |
| 0.28 | 0.3 | 0 | 0 | 0 | 0 |
| 0.3 | 0.32 | 0 | 0 | 0 | 0 |
| 0.32 | 0.34 | 0 | 0 | 0 | 0 |
| 0.34 | 0.36 | 0 | 0 | 0 | 0 |
| 0.36 | 0.38 | 0 | 0 | 0 | 0 |
| 0.38 | 0.4 | 0 | 0 | 0 | 0 |
| 0.4 | 0.42 | 0 | 0 | 0 | 0 |
| 0.42 | 0.44 | 0 | 0 | 0 | 0 |
| 0.44 | 0.46 | 0 | 0 | 0 | 0 |
| 0.46 | 0.48 | 0 | 0 | 0 | 0 |
| 0.48 | 0.5 | 0 | 0 | 0 | 0 |
| 0.5 | 0.52 | 0 | 0 | 0 | 0 |

The estimated intrinsic frequencies summed over semi-major axis are 0.054, 0.068, 0.193, 0.024, and 0.0015 for Earth-, super-Earth-, Neptune-, Jupiter- and very-Large-size planetary candidates, respectively. The sum over all values of the semi-major axis is 0.341. This value is interpreted to mean that the average number of candidates per star with semi-major axes less than 0.5 AU is 0.341 with a very large uncertainty.

When the model is run to simulate a six-month period, the results are very similar for candidates Neptune-size and larger, but the frequencies of super-Earth and Earth-size candidates are increased by 3 for Earth-size candidates and 2 for super-Earth size candidates. The uncertainty in the predictions will decrease as the mission duration increases and the number of transits and resulting SNR increase.

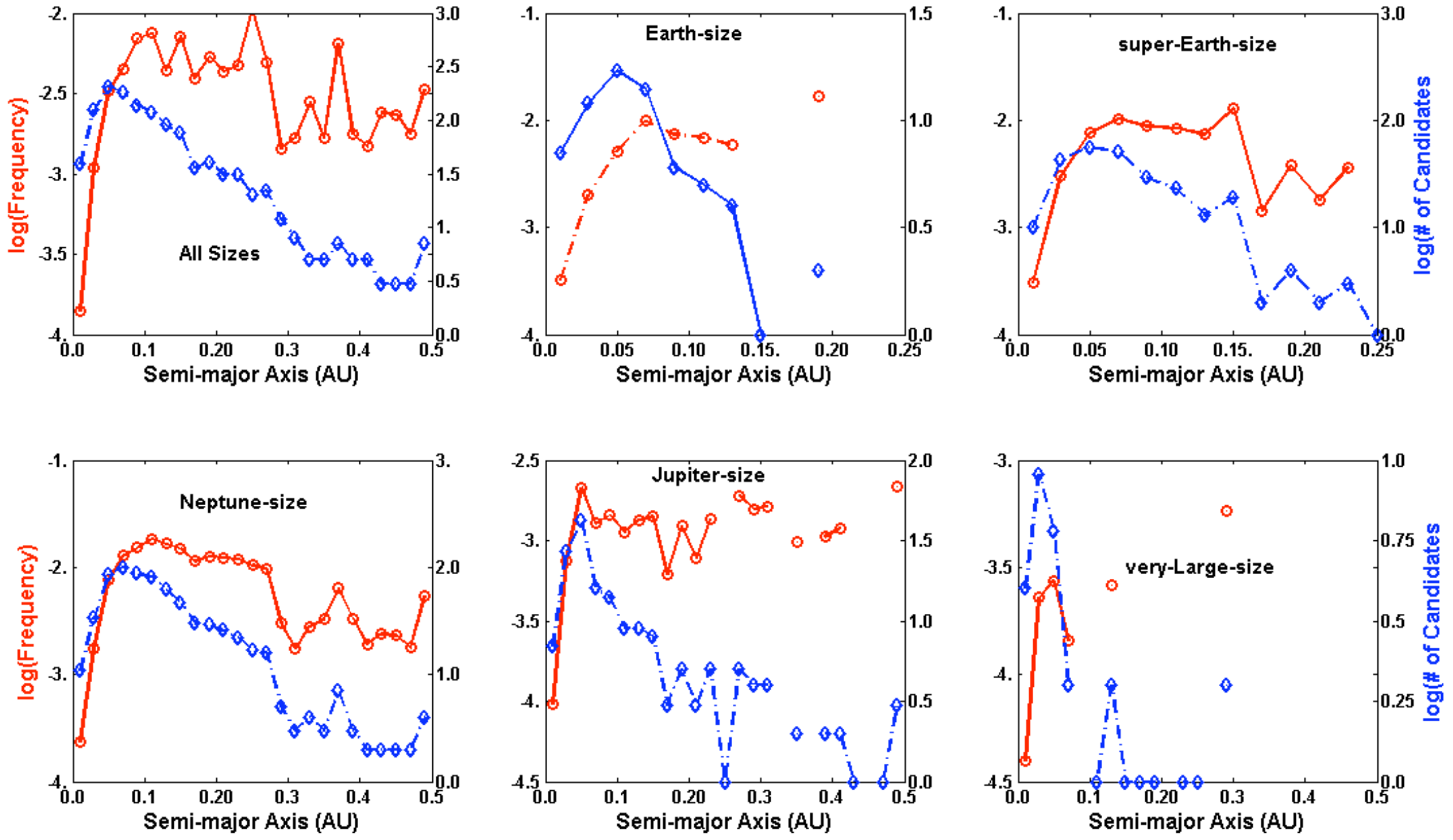

Figure 14. Comparisons of the logarithms of intrinsic frequencies ‘log(Frequency)’ to observations ‘log(# of Observations)’ as a function of semi-major axis for five size-classes. Red symbols (circles) denote intrinsic frequencies and use the scales on the left vertical axes. Blue symbols (diamonds) denote the number of observations and use the scales on the right vertical axes. Values of the observations are shown if at least one event is found in a bin. To reduce the effect of outliers, values for the intrinsic frequencies are shown only when at least two candidates are found in the bin. Frequencies are based on 0.02 AU bins.

All the panels in Figure 14 show a large increase in intrinsic frequency with semi-major axis from the 0.00 to approximately 0.07 AU and then show a negative or near-zero slope at larger values of the semi-major axis. (The variation of intrinsic frequency for the very-Large candidates is too noisy to characterize.) The result for the Jupiter-size candidates shows a nearly constant value with semi-major axis. The peak in the intrinsic frequencies for the three smallest class-sizes is located in the bin to the immediate right of the peak in the observations.

In Figure 15, the dependence of the intrinsic frequencies on the stellar temperature is examined. Note that these results subsume the entire range of semi-major axis just discussed.

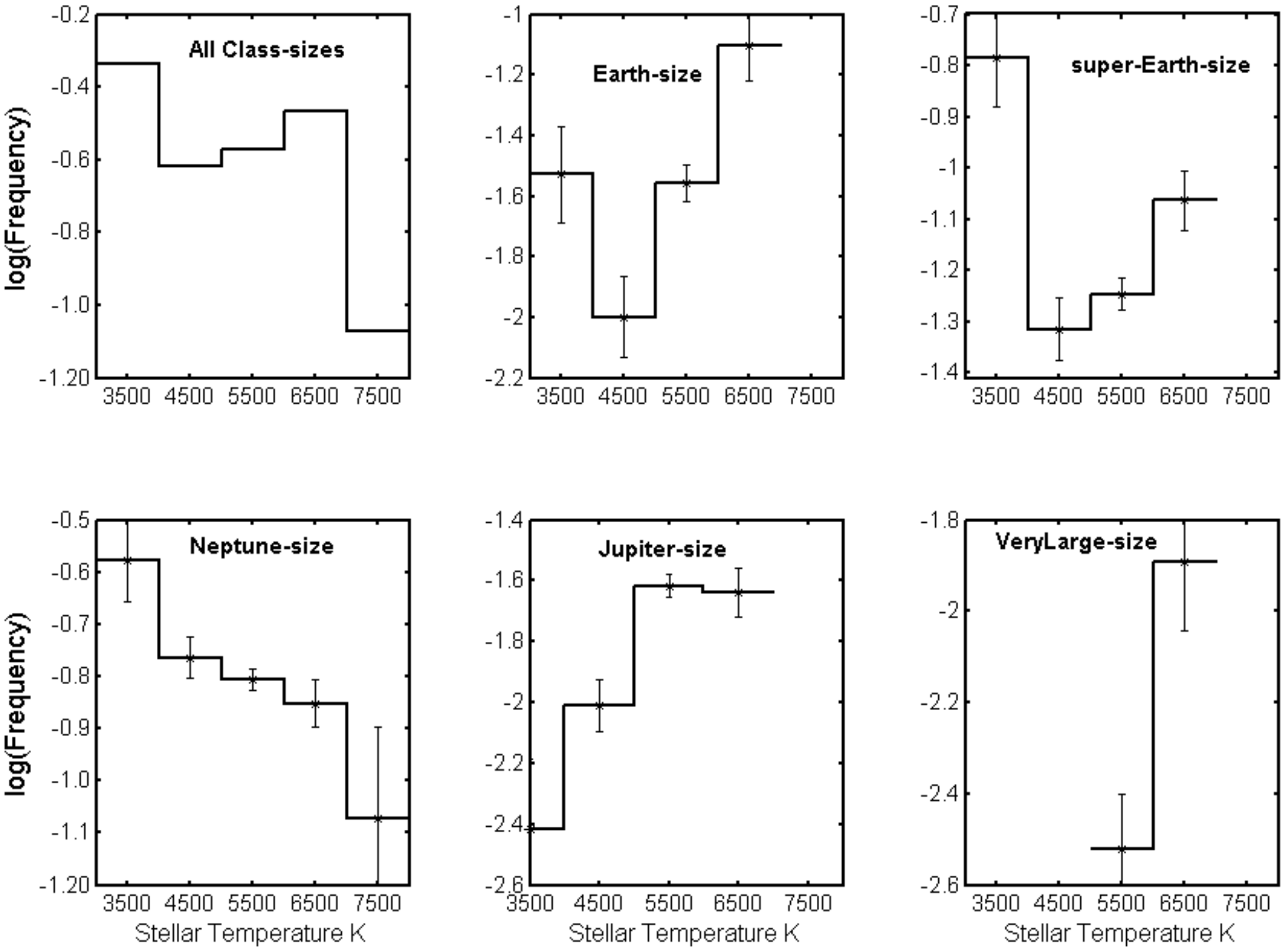

Figure 15. Logarithm of the mean number of candidates per star, as a function of stellar effective temperature, after implementing the sensitivity correction described in Section 4. The bins along the x-axis span 3000-4000K, 4000-5000K, 5000-6000K, 6000-7000K, with each bin labeled by the central value for each bin.

The results shown in Figure 15 indicate that once adjustments are made for the increased sensitivity to small planets orbiting small stars as opposed to Sun-like stars, the higher frequency of Earth-size candidates orbiting the coolest stars seen in Figure 13 disappears. However, the peak for super-Earth-size and Neptune size is still prominent and it is also clear that the Jupiter-size and very-Large candidates are much more frequent around hotter than they are for the cooler M- and K-type stars.

An examination of the panel in Figure 15 for the frequency dependence of Neptune-sized candidates, suggests a negative correlation with temperature. The linear correlation coefficient has a value of -0.95 with 95% confidence limits for the coefficient between -0.995 and -0.57. Although the intrinsic frequencies of Jupiter-sized and very-large-sized candidates also suggest a correlation with stellar effective temperature, because of the small number of data points, no formal estimation can be obtained for their correlation coefficients nor those for the Earth-size and super-Earth size candidates.

One of the surprising results shown in Figure 15 is the dip in the intrinsic frequency of Earth-size and super-Earth-size candidates orbiting stars with temperatures near 4500K, i.e., K-type stars. A careful inspection of the lower-left panel of Figure 3also shows a paucity of candidates for temperatures between 4000 and 5000K. The large values of the dispersion shown in Figure 15 indicate that the result should be interpreted with caution.

It should be noted that the values for the intrinsic frequencies in Table 7 and in Figures 14 and 15 must be considered preliminary estimates. These values will be lowered when more false positive events are recognized and removed, but they could also increase; the precision of the data is assumed to improve as the square root of the number of measurements in transit. If, however, the performance of the data does not achieve this ideal case, then fewer stars are being searched than assumed here. Thus, the inherent frequency would be higher than shown in Table 7 and associated figures. Furthermore, throughout the mission we will continue to make improvements to the data analysis pipeline. As the capability of the system to recognize small candidates improves, and more candidates in the data discussed here will be discovered. A significant improvement is expected in mid-year when the capability to stitch together quarters of observations becomes operational.

It is interesting to compare these results with those of Howard et al. 2010 for planets with periods $\leq$ 50 days discovered by RV. For planet masses 3 - 10 $M_{\oplus}$ (super-Earth-mass), they get approximately 10.7% to 11.8% while the present calculation for candidates with comparable periods days and super-Earth size gives 7.0%. For 10 - 30 $M_{\oplus}$, Howard et al. obtain 5.8 - 6.5% while the *Kepler* results for Neptune-size candidates predict 19.0%. The agreement is satisfactory given the many uncertainties involved in the estimates.

## 5. Overview of Multi-planet Systems

A total 170 target stars with multiple planet candidates have been detected among the 997 host stars in *Kepler* data. There are 115 stars with exactly two candidates, 45 with exactly three candidates, 8 stars with exactly 4 candidates, 1 star with 5, and 1 with 6 candidates. For these figures all candidates are included, whether they are validated planets or not. The fraction of host stars that have multi-candidate systems is 0.17 and the fraction of the candidates that are part of multi-candidate systems is 0.339, i.e., 408 among 1202 candidates. Because all the candidates discussed here show two or more transits, accurate orbital periods and epochs are available in Table 2.

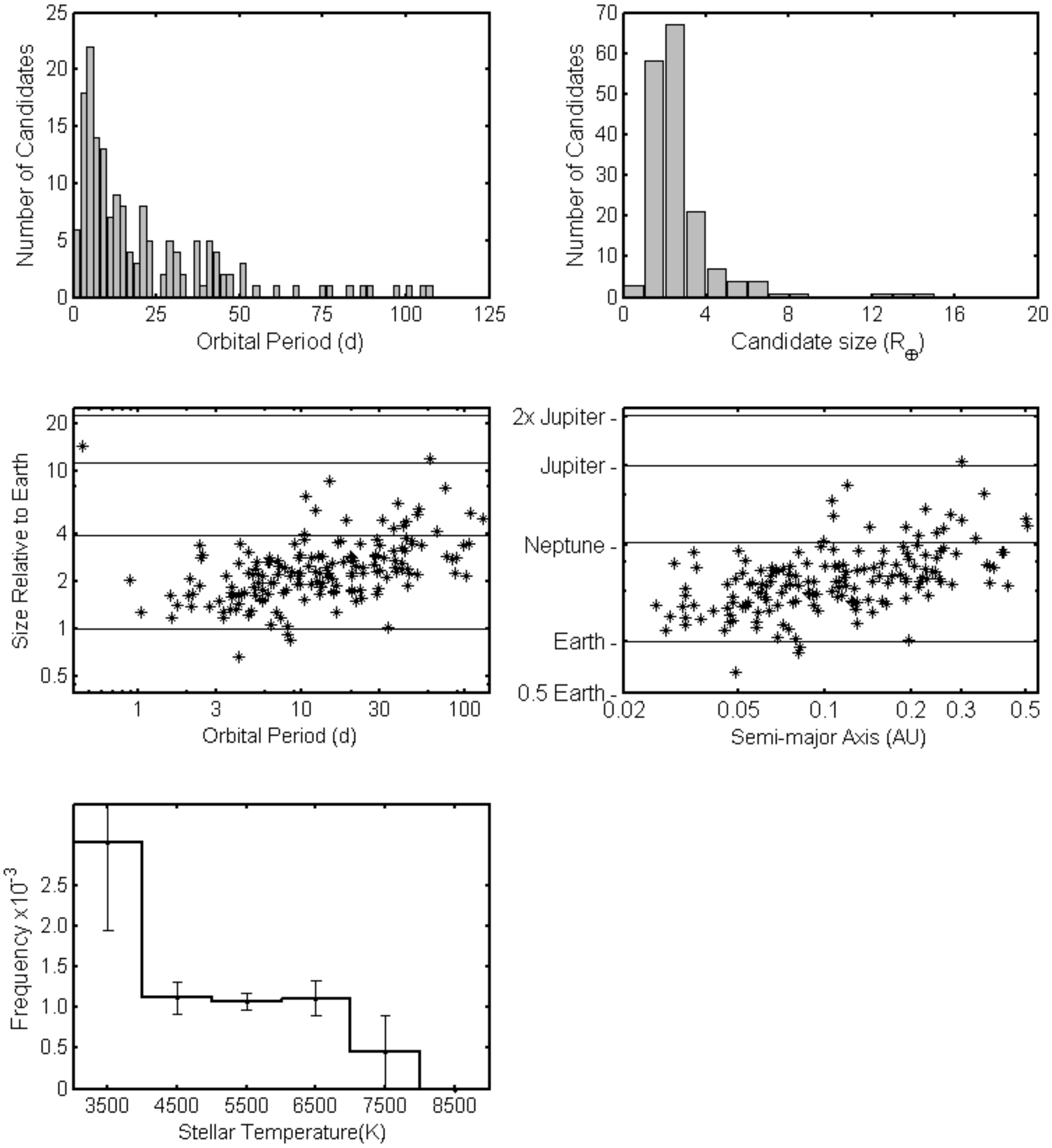

**Figure 16.** Observed distributions of planetary candidates in multi-planet candidate systems. Bin sizes for the upper two panels and the lower panel are 2 days, 1 $R_\oplus$ , and 1000 K, respectively. Refer to Table 6 for the definition of each size category.

Comparisons of the distributions presented in Figure 16 with previous figures show that they are similar to those for the ensemble of all candidates. The number versus orbital period is very much like that seen in Figure 6; a lack of candidates with orbital periods less than 2 days, a maxima near 4 days, and a gradual reduction in the number with orbital period. The number versus candidate size in Figure 16 is quite similar to that in Figure 2. The peak in the frequency with stellar temperature for cool stars is also repeated. However, the distributions displayed in the two scatter plots in the middle panel of Figure 16 show that the size versus orbital period and semi-major axis are different from those in Figure 3. In particular, both of the distributions shown in Figure 16 display a lack of giant planets for close-in/short-period orbits compared to the distributions in Figure 3. There is a clear paucity of giant planets in the observed multi-candidate and multi-planet systems (see Latham et al. 2011 for details). This result is consistent with radial velocity surveys which indicate short-period giant planets are significantly less common in multiple planet systems (Wright et al. 2009).

An unusual candidate KOI# 961.02, shows up in second row, left hand panel of Figure 16. It has a period of 0.45 days, a semi-major axis of 0.01 AU, and a size 28% larger than Jupiter. So far it has passed all vetting tests and will be on the list to get an RV confirmation.

Multiple planet candidate systems, as well as the single-planet candidate systems, could harbor additional planets that do not transit, or have not yet been recognized as such, and therefore are not seen in these data. Such planets might be detectable via transit timing variations (TTVs) of the transiting planets after several years of Kepler photometry (Agol et al. 2005, Holman and Murray 2005, Holman et al. 2010). A preliminary analysis of transit times of planetary candidates based on data up to and including quarter 2 provides hints that ~65 KOIs may already exhibit transit timing variations. A statistical analysis of these and many other marginal TTV signals has been submitted (Ford et al. 2011). Papers with TTV confirmation of three systems are already published (Holman et al. 2010; Lissauer et al. 2011a) or in preparation (Cochran et al. 2011). Ford et al. (2011) predicts that Kepler will confirm (or reject) at least ~12 systems with multiple transiting planet candidates via TTVs.

It is important to note that it is possible, though unlikely, for light from more than one background eclipsing binary star system to be within the photometric aperture, producing an apparent multi-planet transit signal in the light curve. While Latham et al. 2011 and Lissauer et al. 2011b present several arguments showing that candidates in multiples are more likely to be true planets, a thorough analysis of each system and a check of background binaries are required before any discovery can be claimed. Approximately 34% of Kepler candidates are part of multi-candidate systems. The corresponding fraction of RV planets in multi-planet systems is 30% based on the Extrasolar Planets Encyclopedia. The fraction of stars with multiple known planets or candidates is 17% for the Kepler sample and about 12% for the RV sample. Given the various limitations of these two observing techniques, these numbers are consistent. While an exhaustive study remains to be done, Lissauer et al. (2011b) investigated the dynamical attributes of Kepler multi-candidate systems and also suggest that nearly coplanar planetary systems might be common.

## 6. Summary and Conclusions

Distributions of the characteristics of 1202 planetary candidates have been given. These include number and frequency distributions with orbital size and period, stellar temperature and magnitude. These distributions are separated into five class-sizes; 68 candidates of approximately Earth-size ($R_p < 1.25\ R_\oplus$), 288 super-Earth size ($1.25\ R_\oplus < R_p < 2\ R_\oplus$), 662 Neptune-size ($2\ R_\oplus, < R_p < 6\ R_\oplus$), 165 Jupiter-size ($6\ R_\oplus < R_p < 15\ R_\oplus$), and 19 up to twice the size of Jupiter ($15\ R_\oplus < R_p < 22\ R_\oplus$). Over the temperature range appropriate for the habitable zone, 54 candidates are found with sizes ranging from Earth-size to larger than that of Jupiter. Six planetary candidates in the habitable zone are less than twice the size of the Earth.

Over 74% of the planetary candidates are smaller than Neptune. The observed number versus size distribution of planetary candidates increases to a peak at two to three times Earth-size and then declines inversely proportional to area of the candidate. For candidate sizes greater than $2\ R_\oplus$, t he dependence of the number of candidates on the candidate radius is proportional to the reciprocal of the square of the inverse radius on candidate radius.

However, there is a prominent decrease in the number of candidates with size in all class-sizes for semi-major axes smaller than 0.07 AU and for orbital periods less than 3 days. A group of candidates with orbital periods less 3 days is identified that appears distinctly different from those with longer periods in that the size distribution of candidates with short orbital periods is nearly constant with candidate size.

The intrinsic frequencies of super-Earth-size and Neptune-size candidates show maxima for the coolest stars. Both Earth-size and super-Earth-size candidates show minima for stars with temperatures near 4500K. Jupiter-size and very-Large-size candidates show much higher frequencies for hotter stars than for those cooler than 5500K.

The analysis of the first four months of Kepler observations is the first to estimate the frequency of small candidates (Earth-size, super-Earth-size, and Neptune-size) based on a uniform set of observations with the capability of detecting small candidates. After correcting for geometric and sensitivity biases, we find intrinsic frequencies of 5.4% for Earth-size candidates, 6.8% for super-Earth size candidates, 19.3% for Neptune-size candidates, and 2.4% for Jupiter-size candidates.

Multi-candidate, transiting systems are frequent; 17% of the host stars have multi-candidate systems, and 33.9% of all the candidates are part of multi-candidate systems.

There is also evidence for 34 candidates with sizes between 1.3 and 4.5 times that of Jupiter. The nature of these candidates is unclear. Those that are between 1.3 and 2.0 times the size of Jupiter are included in tables and figures presented in this paper because of the possibility that they are very inflated planetary objects, but the 15 larger than twice the size of Jupiter were omitted from the discussion because it is more likely that they are stellar objects or that the estimated size of the host star is much smaller than listed in the KIC.

In the coming years, many of these candidates are expected to be reclassified as exoplanets as the validation effort proceeds. The number of candidates is so large that the *Kepler* team must be selective in its follow up program and will devote the majority of its efforts to the detection and validation of the smallest candidates and to those with orbital periods appropriate for the habitable zone and those amenable to follow up. Many candidates will be left to future work or for follow up by the community. The release of the Q0 through Q1 data and the early release of the Q2 data and the descriptions of the candidates with accurate positions, magnitudes, epochs, and periods should help the community to confirm and validate many of these candidates.

The data released here should also provide to the community a more comprehensive source of data and distributions needed for further developments of the theories of planet structure and planetary systems. These results have concentrated upon discovery of candidates, and initial levels of validations sufficient to cull out many false positives. Future studies by the *Kepler* science team will include efforts to robustly quantify the completeness of these candidate lists through simulation studies, and provide more refined confidence levels on probabilities of candidates being planets. Discovery of additional candidates will of course continue and reduce incompleteness for weak signals whether those follow from small planets, long orbital periods, or faint stars.

The *Kepler* Mission was designed to determine the frequency of extrasolar planets, the distributions of their characteristics, and their association with host star characteristics. The present results are an important milestone toward the accomplishment of *Kepler*'s goals.

**Acknowledgements**

*Kepler* was competitively selected as the tenth Discovery mission. Funding for this mission is provided by NASA's Science Mission Directorate. Some of the data presented herein were obtained at the W. M. Keck Observatory, which is operated as a scientific partnership among the California Institute of Technology, the University of California, and the National Aeronautics and Space Administration. The Keck Observatory was made possible by the generous financial support of the W. M. Keck Foundation. We sincerely thank Andrew Gould for his timely, thorough, and very helpful review of this paper. The authors would like to thank the many people who gave so generously of their time to make this Mission a success.

**Table 1.**
Host Star Characteristics

All parameters are from the Kepler Input Catalog (KIC) except where $T_{eff}$ Flag = 1 indicates that no parameters were available in the KIC. In which case $T_{eff}$, log(g) and R are derived as noted.

Key:

| | |
|---|---|
| KOI | Kepler Object of Interest number |
| KIC | Kepler Input Catalogue Identifier |
| Kp | Kepler magnitude |
| CDPP | 6 hr Combined Differential Photometric Precision from Quarter 3 |
| RA | Right Ascension (J2000) |
| Dec | Declination (J2000) |
| $T_{eff}$ | Effective Temperature of host star as reported in the KIC. If $T_{eff}$ Flag = 1, then $T_{eff}$ , log(g), R are derived using KIC J-K colour and linear interpolation of luminosity class V stellar properties of Schmidt-Kaler (1982). |
| log(g) | Surface gravity reported by KIC. If $T_{eff}$ Flag = 1, then log(g) is based on J-K interpolation. |
| R | Stellar radius reported by KIC. If $T_{eff}$ Flag = 1, then R is based on J-K interpolation. |
| M | Stellar mass derived from log(g) and stellar radius. |

| **KOI** | **KIC** | **Kp** | **CDPP** | **RA** | **DEC** | **$T_{eff}$** | **log(g)** | **R** | **M** | **$T_{eff}$ Flag** |
|---|---|---|---|---|---|---|---|---|---|---|
| | | [mag] | [ppm] | [Hr] | [Deg] | [K] | [cgs] | [$R_{sun}$] | [$M_{sun}$] | |
| 1 | 11446443 | 11.338 | 14 | 19.12056 | 49.3164 | 5713 | 4.14 | 1.50 | 1.14 | |
| 2 | 10666592 | 10.463 | 21.9 | 19.48315 | 47.9695 | 6577 | 4.32 | 1.34 | 1.36 | 1 |
| 3 | 10748390 | 9.147 | 97.8 | 19.84729 | 48.0809 | 4628 | 4.53 | 0.76 | 0.71 | 1 |
| 4 | 3861595 | 11.432 | 126 | 19.62377 | 38.9474 | 6054 | 4.41 | 1.08 | 1.11 | |
| 5 | 8554498 | 11.665 | 20.2 | 19.31598 | 44.6474 | 5766 | 4.04 | 1.73 | 1.18 | |
| 7 | 11853905 | 12.211 | 71.2 | 19.04102 | 50.1358 | 5701 | 4.35 | 1.16 | 1.08 | |
| 10 | 6922244 | 13.563 | 58.6 | 18.75254 | 42.4511 | 6164 | 4.44 | 1.05 | 1.12 | |
| 12 | 5812701 | 11.353 | 82 | 19.83025 | 41.0110 | 6419 | 4.26 | 1.32 | 1.17 | |
| 13 | 9941662 | 9.958 | 10.4 | 19.13141 | 46.8684 | 8848 | 3.93 | 2.44 | 1.83 | |
| 17 | 10874614 | 13.000 | 38.6 | 19.78915 | 48.2399 | 5724 | 4.47 | 0.91 | 0.91 | 1 |
| 18 | 8191672 | 13.369 | 63.9 | 19.96047 | 44.0351 | 5816 | 4.46 | 0.95 | 0.95 | 1 |
| 20 | 11804465 | 13.438 | 46.8 | 19.08290 | 50.0404 | 6012 | 4.47 | 1.01 | 1.09 | |
| 22 | 9631995 | 13.435 | 63.5 | 18.84198 | 46.3234 | 5859 | 4.53 | 0.94 | 1.07 | |
| 41 | 6521045 | 11.000 | 32.7 | 19.42573 | 41.9903 | 5692 | 4.51 | 0.95 | 1.06 | |
| 42 | 8866102 | 9.364 | 41.6 | 18.87671 | 45.1398 | 6035 | 4.22 | 1.37 | 1.14 | |
| 44 | 8845026 | 13.483 | 324 | 20.01012 | 45.0896 | 5490 | 4.48 | 0.88 | 0.85 | 1 |
| 46 | 10905239 | 13.770 | 52.1 | 18.88370 | 48.3552 | 5562 | 4.48 | 0.89 | 0.87 | 1 |
| 49 | 9527334 | 13.704 | 142 | 19.48327 | 46.1648 | 5848 | 4.45 | 0.97 | 0.97 | 1 |
| 51 | 6056992 | 13.761 | 461 | 19.72792 | 41.3324 | 3240 | 4.90 | 0.27 | 0.21 | 1 |
| 63 | 11554435 | 11.582 | 171 | 19.28175 | 49.5482 | 5533 | 4.40 | 1.07 | 1.05 | |
| 64 | 7051180 | 13.143 | 119 | 19.76737 | 42.5474 | 5128 | 3.94 | 1.94 | 1.19 | |
| 69 | 3544595 | 9.931 | 11.1 | 19.42789 | 38.6724 | 5480 | 4.43 | 1.03 | 1.04 | |
| 70 | 6850504 | 12.498 | 73.7 | 19.17987 | 42.3387 | 5342 | 4.72 | 0.70 | 0.95 | |
| 72 | 11904151 | 10.961 | 44.2 | 19.04529 | 50.2413 | 5491 | 4.47 | 0.98 | 1.03 | |
| 75 | 7199397 | 10.775 | 27.6 | 19.43315 | 42.7285 | 5718 | 4.40 | 1.08 | 1.08 | |
| 82 | 10187017 | 11.492 | 59.6 | 18.76552 | 47.2080 | 4727 | 3.96 | 1.86 | 1.14 | |
| 84 | 2571238 | 11.898 | 50.9 | 19.36139 | 37.8518 | 5347 | 4.58 | 0.84 | 0.98 | |

| | | | | | | | | | |
|---|---|---|---|---|---|---|---|---|---|
| 85 | 5866724 | 11.018 | 41.6 | 19.24591 | 41.1512 | 6006 | 4.07 | 1.66 | 1.19 |
| 87 | 10593626 | 11.664 | 32.5 | 19.28117 | 47.8845 | 5606 | 4.36 | 1.14 | 1.07 |
| 89 | 8056665 | 11.642 | 28.4 | 19.98808 | 43.8143 | 7490 | 3.90 | 2.24 | 1.45 |
| 92 | 7941200 | 11.667 | 34 | 18.89165 | 43.7882 | 5850 | 4.28 | 1.26 | 1.11 |
| 94 | 6462863 | 12.205 | - | 19.82220 | 41.8911 | 6090 | 4.08 | 1.66 | 1.20 |
| 97 | 5780885 | 12.885 | 35.6 | 19.23877 | 41.0898 | 5944 | 4.27 | 1.29 | 1.12 |
| 98 | 10264660 | 12.128 | 46.7 | 19.18059 | 47.3331 | 6659 | 3.92 | 2.08 | 1.31 |
| 99 | 8505215 | 12.960 | 43.8 | 19.69562 | 44.5311 | 4951 | 4.33 | 1.11 | 0.97 |
| 100 | 4055765 | 12.598 | 115 | 19.41186 | 39.1995 | 6440 | 3.69 | 2.79 | 1.39 |
| 102 | 8456679 | 12.566 | 31.2 | 19.98032 | 44.4358 | 5919 | 3.90 | 2.08 | 1.24 |
| 103 | 2444412 | 12.593 | 78.7 | 19.44556 | 37.7516 | 5493 | 4.63 | 0.80 | 1.00 |
| 104 | 10318874 | 12.895 | 42.4 | 18.74632 | 47.4971 | 4411 | 4.56 | 0.73 | 0.71 |
| 105 | 8711794 | 12.870 | 33.6 | 19.92914 | 44.8579 | 5450 | 3.96 | 1.88 | 1.19 |
| 107 | 11250587 | 12.702 | 58.4 | 19.65568 | 48.9824 | 5816 | 4.46 | 1.01 | 1.08 |
| 108 | 4914423 | 12.287 | 30.5 | 19.26564 | 40.0645 | 5872 | 4.36 | 1.15 | 1.10 |
| 110 | 9450647 | 12.663 | 43.7 | 18.97038 | 46.0638 | 6344 | 4.35 | 1.18 | 1.14 |
| 111 | 6678383 | 12.596 | 32.5 | 19.17364 | 42.1668 | 5853 | 4.46 | 1.02 | 1.08 |
| 112 | 10984090 | 12.772 | 44.7 | 19.70991 | 48.4956 | 5839 | 4.31 | 1.22 | 1.11 |
| 113 | 2306756 | 12.394 | 36.7 | 19.48492 | 37.6716 | 5362 | 4.34 | 1.15 | 1.04 |
| 115 | 9579641 | 12.791 | 59.4 | 19.19249 | 46.2762 | 6202 | 4.25 | 1.34 | 1.15 |
| 116 | 8395660 | 12.882 | 71.9 | 20.05760 | 44.3376 | 5980 | 3.96 | 1.91 | 1.22 |
| 117 | 10875245 | 12.487 | 55.1 | 19.80188 | 48.2086 | 5725 | 4.47 | 1.00 | 1.07 |
| 118 | 3531558 | 12.377 | 41.6 | 19.15752 | 38.6496 | 5605 | 4.49 | 0.97 | 1.05 |
| 119 | 9471974 | 12.654 | 31.7 | 19.63728 | 46.0623 | 5380 | 4.44 | 1.00 | 1.02 |
| 122 | 8349582 | 12.346 | 46.5 | 18.96550 | 44.3980 | 5569 | 4.58 | 0.86 | 1.03 |
| 123 | 5094751 | 12.365 | 48.6 | 19.35951 | 40.2849 | 5897 | 4.29 | 1.25 | 1.11 |
| 124 | 11086270 | 12.935 | 41.1 | 19.52861 | 48.6028 | 6076 | 4.25 | 1.32 | 1.14 |
| 127 | 8359498 | 13.938 | 261 | 19.30720 | 44.3454 | 5570 | 4.53 | 0.92 | 1.04 |
| 128 | 11359879 | 13.758 | 70.7 | 19.74671 | 49.1401 | 5718 | 4.18 | 1.43 | 1.13 |
| 131 | 7778437 | 13.797 | 239 | 19.93984 | 43.4976 | 6244 | 4.40 | 1.10 | 1.13 |
| 135 | 9818381 | 13.958 | 266 | 19.01606 | 46.6683 | 5953 | 4.51 | 0.95 | 1.08 |
| 137 | 8644288 | 13.549 | 54.7 | 19.87196 | 44.7463 | 5289 | 4.25 | 1.27 | 1.05 |
| 138 | 8506766 | 13.960 | 55 | 19.72939 | 44.5784 | 6772 | 4.12 | 1.62 | 1.26 |
| 139 | 8559644 | 13.492 | 68.4 | 19.44355 | 44.6883 | 5921 | 4.56 | 0.90 | 1.07 |
| 141 | 12105051 | 13.687 | 74.6 | 19.20255 | 50.6516 | 5277 | 4.60 | 0.81 | 0.97 |
| 142 | 5446285 | 13.113 | 87.2 | 19.40987 | 40.6694 | 5361 | 4.68 | 0.74 | 0.96 |
| 144 | 4180280 | 13.698 | 122 | 19.76829 | 39.2498 | 4724 | 4.00 | 1.75 | 1.11 |
| 148 | 5735762 | 13.040 | 95.1 | 19.94261 | 40.9490 | 5063 | 4.51 | 0.89 | 0.94 |
| 149 | 3835670 | 13.397 | 86.8 | 19.10867 | 38.9456 | 6059 | 4.23 | 1.35 | 1.14 |
| 150 | 7626506 | 13.771 | 75.8 | 19.79873 | 43.2098 | 5538 | 4.29 | 1.23 | 1.08 |
| 151 | 2307199 | 14.000 | 104 | 19.49165 | 37.6310 | 6028 | 4.40 | 1.09 | 1.11 |
| 152 | 8394721 | 13.914 | 108 | 20.03448 | 44.3816 | 6187 | 4.54 | 0.94 | 1.10 |
| 153 | 12252424 | 13.461 | 89.6 | 19.19986 | 50.9443 | 4647 | 4.41 | 0.95 | 0.85 |
| 155 | 8030148 | 13.494 | 76.3 | 19.48810 | 43.8812 | 5651 | 4.18 | 1.42 | 1.12 |
| 156 | 10925104 | 13.738 | 142 | 19.60809 | 48.3495 | 4450 | 4.54 | 0.76 | 0.73 |
| 157 | 6541920 | 13.709 | 111 | 19.80767 | 41.9091 | 5675 | 4.47 | 1.00 | 1.06 |
| 159 | 8972058 | 13.431 | 78.6 | 19.84746 | 45.2619 | 5823 | 4.31 | 1.22 | 1.10 |
| 161 | 5084942 | 13.341 | 81.1 | 19.13678 | 40.2116 | 4768 | 4.04 | 1.66 | 1.09 |
| 162 | 8107380 | 13.837 | 66.9 | 19.67759 | 43.9630 | 5632 | 4.45 | 1.01 | 1.06 |
| 163 | 6851425 | 13.536 | 75.8 | 19.20451 | 42.3554 | 5151 | 4.37 | 1.08 | 1.00 |
| 165 | 9527915 | 13.938 | 82.9 | 19.49913 | 46.1962 | 4956 | 4.76 | 0.64 | 0.85 |
| 166 | 2441495 | 13.575 | 78.4 | 19.40146 | 37.7698 | 5216 | 4.24 | 1.28 | 1.04 |
| 167 | 11666881 | 13.273 | 66.5 | 19.63093 | 49.7650 | 6285 | 4.60 | 0.87 | 1.10 |
| 168 | 11512246 | 13.438 | 73.8 | 19.61460 | 49.4792 | 5877 | 3.97 | 1.88 | 1.21 |
| 171 | 7831264 | 13.717 | 79.7 | 19.64947 | 43.5368 | 6287 | 4.41 | 1.10 | 1.13 |
| 172 | 8692861 | 13.749 | 79.6 | 19.55073 | 44.8689 | 5603 | 4.80 | 0.66 | 0.98 |
| 173 | 11402995 | 13.844 | 57 | 19.46047 | 49.2621 | 5752 | 4.52 | 0.94 | 1.06 |
| 174 | 10810838 | 13.779 | 64.5 | 19.78819 | 48.1076 | 4654 | 4.54 | 0.80 | 0.80 |
| 176 | 6442377 | 13.432 | 56.9 | 19.44312 | 41.8847 | 6340 | 4.49 | 1.00 | 1.12 |
| 177 | 6803202 | 13.182 | 52.8 | 19.87848 | 42.2370 | 5620 | 4.39 | 1.10 | 1.07 |
| 179 | 9663113 | 13.955 | 57.8 | 19.80303 | 46.3287 | 5827 | 4.42 | 1.07 | 1.08 |
| 180 | 9573539 | 13.024 | 70.7 | 18.95962 | 46.2491 | 5549 | 4.62 | 0.82 | 1.01 |
| 183 | 9651668 | 14.290 | 343 | 19.52372 | 46.3912 | 5722 | 4.71 | 0.74 | 1.02 |
| 186 | 12019440 | 14.952 | 128 | 19.66642 | 50.4701 | 5826 | 4.56 | 0.89 | 1.06 |
| 187 | 7023960 | 14.857 | 103 | 19.24863 | 42.5509 | 5768 | 4.70 | 0.75 | 1.03 |
| 188 | 5357901 | 14.741 | 82.9 | 19.35720 | 40.5677 | 5087 | 4.73 | 0.67 | 0.89 |
| 189 | 11391018 | 14.388 | 88.8 | 18.99200 | 49.2670 | 4787 | 4.50 | 0.86 | 0.86 |
| 190 | 5771719 | 14.137 | 416 | 18.97659 | 41.0150 | 5425 | 4.23 | 1.33 | 1.08 |
| 191 | 5972334 | 14.991 | 134 | 19.68582 | 41.2220 | 5495 | 4.52 | 0.92 | 1.02 |
| 192 | 7950644 | 14.221 | 82.7 | 19.21697 | 43.7049 | 5936 | 4.46 | 1.01 | 1.09 |
| 193 | 10799735 | 14.904 | 96.7 | 19.52508 | 48.1953 | 5883 | 4.47 | 1.01 | 1.08 |
| 194 | 10904857 | 14.804 | 561 | 18.86441 | 48.3451 | 5883 | 4.63 | 0.82 | 1.05 |
| 195 | 11502867 | 14.835 | 284 | 19.29564 | 49.4734 | 5604 | 4.50 | 0.96 | 1.05 |
| 196 | 9410930 | 14.465 | 84.4 | 19.63422 | 45.9816 | 5585 | 4.51 | 0.94 | 1.04 |
| 197 | 2987027 | 14.018 | 267 | 19.38888 | 38.1844 | 4907 | 4.38 | 1.03 | 0.94 |
| 199 | 10019708 | 14.879 | 94.4 | 19.66838 | 46.9560 | 6214 | 4.60 | 0.87 | 1.09 |
| 200 | 6046540 | 14.412 | 82.1 | 19.53950 | 41.3555 | 5774 | 4.69 | 0.76 | 1.03 |
| 201 | 6849046 | 14.014 | 70.3 | 19.14204 | 42.3502 | 5491 | 4.45 | 1.00 | 1.04 |
| 202 | 7877496 | 14.309 | 141 | 19.07402 | 43.6810 | 5912 | 4.44 | 1.04 | 1.09 |
| 203 | 10619192 | 14.141 | 760 | 19.89302 | 47.8150 | 5634 | 4.49 | 0.97 | 1.05 |
| 204 | 9305831 | 14.678 | 202 | 20.00682 | 45.7621 | 5287 | 4.48 | 0.95 | 0.99 |
| 205 | 7046804 | 14.518 | 104 | 19.69978 | 42.5379 | 5060 | 4.57 | 0.83 | 0.93 |
| 206 | 5728139 | 14.463 | 134 | 19.83958 | 40.9773 | 5771 | 4.35 | 1.16 | 1.09 |
| 208 | 3762468 | 14.996 | 962 | 19.68717 | 38.8816 | 6094 | 4.59 | 0.88 | 1.08 |
| 209 | 10723750 | 14.274 | 74.9 | 19.25287 | 48.0402 | 6221 | 4.48 | 1.01 | 1.11 |
| 211 | 10656508 | 14.989 | 105 | 19.19802 | 47.9721 | 6072 | 4.41 | 1.09 | 1.11 |
| 212 | 6300348 | 14.858 | - | 19.74265 | 41.6032 | 5843 | 4.54 | 0.92 | 1.07 |
| 214 | 11046458 | 14.256 | 114 | 19.90833 | 48.5775 | 5322 | 4.44 | 1.00 | 1.01 |
| 216 | 6152974 | 14.711 | 164 | 19.94754 | 41.4248 | 5086 | 4.31 | 1.16 | 1.00 |
| 217 | 9595827 | 15.127 | 707 | 19.65770 | 46.2859 | 5504 | 4.72 | 0.71 | 0.98 |
| 219 | 6305192 | 14.153 | 235 | 19.81427 | 41.6640 | 5347 | 4.73 | 0.70 | 0.95 |
| 220 | 7132798 | 14.236 | 57.9 | 19.72976 | 42.6589 | 5388 | 4.87 | 0.59 | 0.92 |
| 221 | 3937519 | 14.622 | 98.9 | 19.06205 | 39.0981 | 5176 | 4.69 | 0.72 | 0.92 |
| 222 | 4249725 | 14.735 | 181 | 19.19278 | 39.3391 | 4353 | 4.71 | 0.58 | 0.64 |
| 223 | 4545187 | 14.708 | 139 | 19.07747 | 39.6780 | 5128 | 4.66 | 0.74 | 0.92 |
| 225 | 5801571 | 14.784 | 270 | 19.66089 | 41.0747 | 6037 | 4.55 | 0.92 | 1.08 |

| | | | | | | | | | | |
|---|---|---|---|---|---|---|---|---|---|---|
| 226 | 5959753 | 14.817 | 149 | 19.44273 | 41.2410 | 5043 | 4.89 | 0.54 | 0.84 | |
| 227 | 6185476 | 14.267 | 156 | 18.95680 | 41.5192 | 4043 | 4.54 | 0.67 | 0.57 | |
| 229 | 3847907 | 14.720 | 114 | 19.38439 | 38.9280 | 5608 | 4.37 | 1.12 | 1.07 | |
| 232 | 4833421 | 14.247 | 64.8 | 19.40746 | 39.9491 | 5868 | 4.68 | 0.77 | 1.04 | |
| 234 | 8491277 | 14.283 | 112 | 19.35721 | 44.5188 | 5735 | 4.36 | 1.15 | 1.09 | |
| 235 | 8107225 | 14.353 | 100 | 19.67400 | 43.9152 | 5041 | 4.65 | 0.74 | 0.90 | |
| 237 | 8041216 | 14.176 | 97.5 | 19.73009 | 43.8521 | 5679 | 4.53 | 0.92 | 1.05 | |
| 238 | 7219825 | 14.061 | 58.9 | 19.79991 | 42.7820 | 6032 | 4.44 | 1.05 | 1.10 | |
| 239 | 6383785 | 14.762 | 171 | 19.81344 | 41.7302 | 5983 | 4.54 | 0.92 | 1.08 | |
| 240 | 8026752 | 14.982 | 215 | 19.40534 | 43.8602 | 5996 | 4.60 | 0.86 | 1.07 | |
| 241 | 11288051 | 14.139 | 88.3 | 19.11685 | 49.0649 | 5055 | 4.85 | 0.57 | 0.85 | |
| 242 | 3642741 | 14.747 | 404 | 19.37577 | 38.7077 | 5437 | 4.51 | 0.93 | 1.02 | |
| 244 | 4349452 | 10.734 | 15.8 | 19.10923 | 39.4879 | 6104 | 4.37 | 1.14 | 1.10 | 1 |
| 245 | 8478994 | 9.710 | 22.6 | 18.93730 | 44.5182 | 5419 | 4.48 | 0.87 | 0.83 | 1 |
| 246 | 11295426 | 10.000 | 36.3 | 19.40215 | 49.0403 | 5658 | 4.41 | 1.06 | 1.07 | |
| 247 | 11852982 | 14.216 | 130 | 18.99991 | 50.1468 | 3804 | 4.56 | 0.60 | 0.48 | |
| 248 | 5364071 | 15.264 | 232 | 19.48631 | 40.5918 | 3974 | 4.53 | 0.67 | 0.55 | |
| 249 | 9390653 | 14.486 | 123 | 18.99479 | 45.9724 | 3654 | 4.42 | 0.73 | 0.51 | |
| 250 | 9757613 | 15.473 | 296 | 18.99607 | 46.5665 | 3933 | 4.51 | 0.68 | 0.55 | |
| 251 | 10489206 | 14.752 | 322 | 19.88388 | 47.6049 | 3846 | 4.58 | 0.59 | 0.49 | |
| 252 | 11187837 | 15.613 | 238 | 19.36010 | 48.8226 | 3897 | 4.60 | 0.58 | 0.49 | |
| 253 | 11752906 | 15.254 | 320 | 19.03829 | 49.9623 | 3951 | 4.57 | 0.62 | 0.52 | |
| 254 | 5794240 | 15.979 | 357 | 19.52486 | 41.0643 | 3948 | 4.54 | 0.65 | 0.53 | |
| 255 | 7021681 | 15.108 | 216 | 19.19054 | 42.5426 | 3989 | 4.48 | 0.73 | 0.58 | |
| 256 | 11548140 | 15.373 | 1996 | 19.01234 | 49.5654 | 3639 | 4.17 | 1.10 | 0.65 | |
| 257 | 5514383 | 10.868 | 43 | 18.97568 | 40.7198 | 6023 | 4.10 | 1.61 | 1.18 | |
| 258 | 11231334 | 9.890 | 115 | 18.96946 | 48.9613 | 6278 | 4.17 | 1.48 | 1.18 | |
| 260 | 8292840 | 10.500 | 24.5 | 19.28982 | 44.2085 | 6096 | 4.37 | 1.13 | 1.09 | 1 |
| 261 | 5383248 | 10.297 | 39 | 19.80464 | 40.5251 | 5588 | 3.96 | 1.89 | 1.20 | |
| 262 | 11807274 | 10.421 | 30.6 | 19.20672 | 50.0337 | 6143 | 4.24 | 1.35 | 1.15 | |
| 263 | 10514430 | 10.821 | 39.4 | 18.75159 | 47.7744 | 5550 | 4.33 | 1.17 | 1.07 | |
| 265 | 12024120 | 11.994 | 36 | 19.80126 | 50.4090 | 6032 | 4.38 | 1.13 | 1.11 | |
| 268 | 3425851 | 10.560 | 23.4 | 19.04859 | 38.5070 | 4808 | 4.51 | 0.79 | 0.73 | 1 |
| 269 | 7670943 | 10.927 | 22.6 | 19.15638 | 43.3784 | 6258 | 4.12 | 1.58 | 1.20 | |
| 270 | 6528464 | 11.411 | 26.7 | 19.58219 | 41.9008 | 5594 | 4.48 | 0.90 | 0.88 | 1 |
| 271 | 9451706 | 11.485 | 35.5 | 19.01267 | 46.0280 | 6089 | 4.47 | 1.01 | 1.10 | |
| 273 | 3102384 | 11.457 | 23.1 | 19.16523 | 38.2288 | 5503 | 4.76 | 0.68 | 0.97 | |
| 274 | 8077137 | 11.390 | 26.9 | 18.83282 | 43.9802 | 6013 | 4.37 | 1.14 | 1.11 | |
| 275 | 10586004 | 11.696 | 35.2 | 19.02075 | 47.8486 | 5809 | 4.46 | 0.95 | 0.94 | 1 |
| 276 | 11133306 | 11.854 | 22.8 | 19.31096 | 48.7062 | 5949 | 4.36 | 1.15 | 1.11 | |
| 277 | 11401755 | 11.866 | 42.1 | 19.41668 | 49.2318 | 5848 | 4.45 | 0.97 | 0.97 | 1 |
| 279 | 12314973 | 11.684 | 70.9 | 19.69910 | 51.0135 | 6152 | 4.27 | 1.30 | 1.14 | |
| 280 | 4141376 | 11.072 | 23.1 | 19.11263 | 39.2119 | 6435 | 4.22 | 1.39 | 1.18 | |
| 281 | 4143755 | 11.947 | 36.1 | 19.17700 | 39.2443 | 5699 | 3.89 | 2.08 | 1.23 | |
| 282 | 5088536 | 11.529 | 42.6 | 19.23004 | 40.2453 | 5900 | 4.43 | 1.01 | 0.99 | 1 |
| 283 | 5695396 | 11.525 | 37.1 | 19.23539 | 40.9423 | 5679 | 3.99 | 1.82 | 1.19 | |
| 284 | 6021275 | 11.818 | 25.5 | 18.88239 | 41.3430 | 5898 | 4.13 | 1.54 | 1.16 | |
| 285 | 6196457 | 11.565 | 48.1 | 19.27240 | 41.5630 | 5822 | 4.45 | 0.96 | 0.95 | 1 |
| 288 | 9592705 | 11.020 | 27.8 | 19.58110 | 46.2267 | 5946 | 4.41 | 1.04 | 1.01 | 1 |
| 289 | 10386922 | 12.747 | 46.2 | 18.86304 | 47.5749 | 5812 | 4.46 | 0.95 | 0.94 | 1 |
| 291 | 10933561 | 12.848 | 54.8 | 19.81853 | 48.3203 | 5491 | 4.61 | 0.82 | 1.01 | |
| 292 | 11075737 | 12.872 | 51 | 19.15511 | 48.6734 | 5743 | 4.26 | 1.29 | 1.11 | |
| 294 | 11259686 | 12.674 | 61.5 | 19.88460 | 48.9167 | 5861 | 4.46 | 1.02 | 1.08 | |
| 295 | 11547513 | 12.324 | 56.1 | 18.98260 | 49.5984 | 5936 | 4.41 | 1.04 | 1.01 | 1 |
| 296 | 11802615 | 12.935 | 43.7 | 19.00278 | 50.0754 | 5811 | 4.39 | 1.10 | 1.09 | |
| 297 | 11905011 | 12.182 | 41.9 | 19.08315 | 50.2424 | 6050 | 4.27 | 1.30 | 1.13 | |
| 298 | 12785320 | 12.713 | 47.1 | 19.36628 | 52.0555 | 5445 | 4.48 | 0.88 | 0.84 | 1 |
| 299 | 2692377 | 12.899 | 95.6 | 19.04411 | 37.9645 | 5544 | 3.97 | 1.88 | 1.19 | |
| 301 | 3642289 | 12.730 | 53.3 | 19.36634 | 38.7955 | 6131 | 4.34 | 1.19 | 1.13 | |
| 302 | 3662838 | 12.059 | 63.5 | 19.70725 | 38.7357 | 6711 | 4.15 | 1.56 | 1.24 | |
| 303 | 5966322 | 12.193 | 35.5 | 19.57835 | 41.2954 | 5497 | 4.50 | 0.95 | 1.03 | |
| 304 | 6029239 | 12.549 | 52.6 | 19.13933 | 41.3739 | 5885 | 3.98 | 1.86 | 1.21 | |
| 305 | 6063220 | 12.970 | 59.4 | 19.82360 | 41.3001 | 4653 | 4.27 | 1.17 | 0.92 | |
| 306 | 6071903 | 12.630 | 74.7 | 19.95464 | 41.3846 | 5120 | 4.16 | 1.42 | 1.07 | |
| 307 | 6289257 | 12.797 | 59.1 | 19.54536 | 41.6178 | 6081 | 4.39 | 1.11 | 1.11 | |
| 308 | 6291837 | 12.351 | 58.1 | 19.59531 | 41.6029 | 6054 | 4.14 | 1.53 | 1.17 | |
| 312 | 7050989 | 12.459 | 40.8 | 19.76449 | 42.5988 | 6014 | 4.38 | 1.09 | 1.04 | 1 |
| 313 | 7419318 | 12.990 | 50.9 | 18.80904 | 43.0391 | 5205 | 4.35 | 1.12 | 1.01 | |
| 314 | 7603200 | 12.925 | 201 | 19.35877 | 43.2930 | 3900 | 4.59 | 0.61 | 0.53 | 1 |
| 315 | 7700622 | 12.968 | 58.6 | 19.81813 | 43.3333 | 4711 | 4.17 | 1.35 | 0.99 | |
| 316 | 8008067 | 12.701 | 60.5 | 18.82613 | 43.8894 | 5561 | 4.32 | 1.18 | 1.07 | |
| 317 | 8121310 | 12.885 | 64.2 | 19.92109 | 43.9980 | 6464 | 4.06 | 1.72 | 1.24 | |
| 318 | 8156120 | 12.211 | 80.1 | 19.21027 | 44.0688 | 6366 | 4.25 | 1.34 | 1.17 | |
| 319 | 8684730 | 12.711 | 32.2 | 19.34117 | 44.8729 | 5835 | 4.45 | 0.97 | 0.96 | 1 |
| 321 | 8753657 | 12.520 | 34.7 | 19.45654 | 44.9682 | 5433 | 4.74 | 0.70 | 0.96 | |
| 323 | 9139084 | 12.465 | 89 | 18.93741 | 45.5069 | 5403 | 4.26 | 1.27 | 1.06 | |
| 326 | 9880467 | 12.960 | 189 | 19.11040 | 46.7835 | 3240 | 4.90 | 0.27 | 0.21 | 1 |
| 327 | 9881662 | 12.996 | 45.3 | 19.15797 | 46.7682 | 6144 | 4.47 | 1.02 | 1.11 | |
| 330 | 11361646 | 13.928 | 85.9 | 19.79061 | 49.1621 | 5749 | 4.25 | 1.31 | 1.11 | |
| 331 | 10285631 | 13.497 | 64.2 | 19.74556 | 47.3588 | 5335 | 4.86 | 0.59 | 0.91 | |
| 332 | 10290666 | 13.046 | 45.6 | 19.84340 | 47.3963 | 5552 | 4.73 | 0.71 | 0.99 | |
| 333 | 10337258 | 13.390 | 219 | 19.39719 | 47.4063 | 6310 | 4.45 | 1.05 | 1.13 | |
| 335 | 10470206 | 13.893 | 65.7 | 19.45457 | 47.6753 | 6380 | 4.14 | 1.54 | 1.20 | |
| 337 | 10545066 | 13.936 | 83.5 | 19.71232 | 47.7481 | 5778 | 4.55 | 0.91 | 1.06 | |
| 338 | 10552611 | 13.448 | 108 | 19.86473 | 47.7317 | 4910 | 4.18 | 1.36 | 1.02 | |
| 339 | 10587105 | 13.763 | 88.4 | 19.05922 | 47.8804 | 6013 | 4.69 | 0.77 | 1.05 | |
| 340 | 10616571 | 13.057 | 372 | 19.84431 | 47.8014 | 5544 | 3.86 | 2.15 | 1.23 | |
| 341 | 10878263 | 13.338 | 114 | 19.86410 | 48.2444 | 5425 | 4.33 | 1.16 | 1.05 | |
| 343 | 10982872 | 13.203 | 79.1 | 19.67459 | 48.4813 | 5703 | 4.45 | 1.02 | 1.07 | |
| 344 | 11015108 | 13.400 | 68.9 | 18.88935 | 48.5490 | 5715 | 4.34 | 1.16 | 1.08 | |
| 345 | 11074541 | 13.340 | 82.8 | 19.10165 | 48.6836 | 4794 | 4.10 | 1.51 | 1.05 | |
| 346 | 11100383 | 13.524 | 105 | 19.91073 | 48.6064 | 4962 | 4.39 | 1.04 | 0.96 | |
| 348 | 11194032 | 13.933 | 115 | 19.57905 | 48.8251 | 4667 | 4.20 | 1.28 | 0.96 | |
| 349 | 11394027 | 13.586 | 52.3 | 19.12351 | 49.2617 | 5652 | 4.36 | 1.14 | 1.08 | |
| 350 | 11395587 | 13.387 | 66 | 19.19076 | 49.2645 | 5775 | 4.37 | 1.12 | 1.09 | |
| 351 | 11442793 | 13.804 | 88.2 | 18.96223 | 49.3052 | 6103 | 4.53 | 0.94 | 1.09 | |
| 352 | 11521793 | 13.770 | 77.6 | 19.87118 | 49.4126 | 5714 | 4.39 | 1.10 | 1.08 | |
| 353 | 11566064 | 13.374 | 92.7 | 19.68085 | 49.5622 | 6679 | 4.37 | 1.18 | 1.18 | |
| 354 | 11568987 | 13.235 | 91.9 | 19.76124 | 49.5401 | 5941 | 3.90 | 2.09 | 1.25 | |
| 355 | 11621223 | 13.174 | 65.7 | 19.77111 | 49.6963 | 6062 | 4.47 | 1.01 | 1.10 | |
| 356 | 11624249 | 13.807 | 124 | 19.84909 | 49.6372 | 5124 | 4.07 | 1.61 | 1.12 | |
| 360 | 12107021 | 13.021 | - | 19.28099 | 50.6509 | 5994 | 4.39 | 1.08 | 1.04 | 1 |
| 361 | 12404954 | 13.100 | 90.5 | 19.32538 | 51.2771 | 5706 | 4.56 | 0.89 | 1.05 | |
| 364 | 7296438 | 10.087 | 17.6 | 19.72482 | 42.8812 | 5551 | 4.45 | 1.01 | 1.05 | |
| 365 | 11623629 | 11.195 | 17.8 | 19.83246 | 49.6235 | 5389 | 4.57 | 0.86 | 0.99 | |
| 366 | 3545478 | 11.714 | 32.6 | 19.44428 | 38.6193 | 6987 | 4.30 | 1.44 | 1.52 | 1 |
| 367 | 4815520 | 11.105 | 36.7 | 18.96481 | 39.9118 | 6046 | 4.33 | 1.19 | 1.12 | |
| 368 | 6603043 | 11.375 | 14.5 | 19.39031 | 42.0869 | 9034 | 4.13 | 1.89 | 1.77 | |
| 369 | 7175184 | 11.992 | 35.2 | 18.79438 | 42.7755 | 6153 | 4.46 | 1.03 | 1.11 | |
| 370 | 8494142 | 11.931 | 38.2 | 19.42585 | 44.5291 | 6207 | 3.83 | 2.29 | 1.30 | |
| 371 | 5652983 | 12.193 | 49.9 | 19.97841 | 40.8565 | 4997 | 3.60 | 3.01 | 1.33 | |
| 372 | 6471021 | 12.391 | 72.2 | 19.94150 | 41.8668 | 5638 | 4.50 | 0.95 | 1.05 | |
| 373 | 7364176 | 12.765 | 37.2 | 19.48233 | 42.9095 | 5816 | 4.26 | 1.30 | 1.11 | |
| 374 | 8686097 | 12.209 | 25.4 | 19.37502 | 44.8740 | 5829 | 4.28 | 1.26 | 1.11 | |
| 375 | 12356617 | 13.293 | 51.7 | 19.41341 | 51.1443 | 5692 | 4.43 | 1.04 | 1.07 | |
| 377 | 3323887 | 13.803 | 146 | 19.03826 | 38.4009 | 5722 | 4.78 | 0.68 | 1.00 | |
| 379 | 2446113 | 13.319 | 62 | 19.47045 | 37.7762 | 6203 | 4.11 | 1.59 | 1.19 | |
| 384 | 3353050 | 13.281 | 72.6 | 19.60998 | 38.4583 | 5744 | 4.30 | 1.22 | 1.09 | |
| 385 | 3446746 | 13.435 | 66.1 | 19.48101 | 38.5486 | 5466 | 4.42 | 1.04 | 1.04 | |
| 386 | 3656121 | 13.838 | 102 | 19.60738 | 38.7102 | 5969 | 4.38 | 1.12 | 1.11 | |
| 387 | 3733628 | 13.577 | 116 | 19.14791 | 38.8625 | 4460 | 4.54 | 0.74 | 0.69 | 1 |
| 388 | 3831053 | 13.644 | 59.2 | 18.98046 | 38.9368 | 5569 | 4.48 | 0.89 | 0.87 | 1 |
| 392 | 3942670 | 13.954 | 72.3 | 19.19838 | 39.0872 | 5684 | 4.28 | 1.25 | 1.09 | |
| 393 | 3964109 | 13.542 | 68.8 | 19.60194 | 39.0519 | 6084 | 4.75 | 0.72 | 1.05 | |
| 398 | 9946525 | 15.342 | 236 | 19.31908 | 46.8588 | 5101 | 4.55 | 0.86 | 0.94 | |
| 401 | 3217264 | 14.001 | 69.9 | 19.05691 | 38.3841 | 5264 | 4.18 | 1.40 | 1.08 | |
| 403 | 4247092 | 14.169 | 90.9 | 19.12531 | 39.3784 | 5565 | 4.44 | 1.02 | 1.05 | |
| 408 | 5351250 | 14.985 | 199 | 19.21561 | 40.5209 | 5631 | 4.49 | 0.96 | 1.05 | |
| 409 | 5444548 | 14.150 | 89.7 | 19.37109 | 40.6920 | 5709 | 5.01 | 0.51 | 0.95 | |
| 410 | 5449777 | 14.454 | 82.9 | 19.48320 | 40.6961 | 5968 | 4.38 | 1.12 | 1.10 | |
| 412 | 5683743 | 14.288 | 80.6 | 18.88384 | 40.9905 | 5584 | 4.28 | 1.26 | 1.09 | |
| 413 | 5791986 | 14.769 | 167 | 19.47752 | 41.0232 | 5236 | 4.56 | 0.86 | 0.97 | |
| 415 | 6289650 | 14.110 | 65.1 | 19.55374 | 41.6064 | 5823 | 4.36 | 1.15 | 1.09 | |
| 416 | 6508221 | 14.290 | 94 | 19.12437 | 41.9891 | 5083 | 4.65 | 0.75 | 0.91 | |
| 417 | 6879865 | 14.847 | 222 | 19.74997 | 42.3355 | 5635 | 4.59 | 0.85 | 1.04 | |
| 418 | 7975727 | 14.479 | 68.8 | 19.79329 | 43.7072 | 5153 | 4.42 | 1.01 | 0.98 | |
| 419 | 8219673 | 14.519 | 142 | 19.05942 | 44.1817 | 5723 | 4.70 | 0.75 | 1.02 | |
| 420 | 8352537 | 14.247 | 79.4 | 19.07458 | 44.3453 | 4687 | 4.51 | 0.83 | 0.82 | |
| 421 | 9115800 | 14.995 | 179 | 19.99373 | 45.4397 | 5181 | 4.32 | 1.16 | 1.02 | |
| 422 | 9214713 | 14.740 | 132 | 19.35932 | 45.6653 | 6002 | 4.40 | 1.09 | 1.10 | |
| 423 | 9478990 | 14.327 | 268 | 19.79735 | 46.0343 | 5992 | 4.45 | 1.03 | 1.10 | |
| 425 | 9967884 | 14.694 | 236 | 19.86495 | 46.8046 | 5689 | 4.54 | 0.91 | 1.05 | |
| 426 | 10016874 | 14.733 | 122 | 19.60153 | 46.9996 | 5796 | 4.33 | 1.19 | 1.10 | |
| 427 | 10189546 | 14.621 | 170 | 18.86819 | 47.2611 | 5293 | 4.50 | 0.93 | 0.99 | |

| | | | | | | | | | |
|---|---|---|---|---|---|---|---|---|---|
| 428 | 10418224 | 14.588 | 88.9 | 19.78758 | 47.5266 | 6127 | 4.55 | 0.92 | 1.09 |
| 429 | 10616679 | 14.486 | 292 | 19.84668 | 47.8638 | 5093 | 4.49 | 0.93 | 0.96 |
| 430 | 10717241 | 14.897 | 163 | 19.03234 | 48.0543 | 4124 | 4.58 | 0.64 | 0.57 |
| 431 | 10843590 | 14.262 | 117 | 18.83070 | 48.2571 | 5249 | 4.43 | 1.00 | 1.00 |
| 432 | 10858832 | 14.279 | 109 | 19.38318 | 48.2420 | 5830 | 4.46 | 1.02 | 1.08 |
| 433 | 10937029 | 14.924 | 289 | 19.90339 | 48.3325 | 5237 | 4.37 | 1.08 | 1.01 |
| 435 | 11709124 | 14.534 | 138 | 19.31870 | 49.8965 | 5709 | 4.66 | 0.78 | 1.03 |
| 438 | 12302530 | 14.258 | 174 | 19.23306 | 51.0819 | 4351 | 4.60 | 0.68 | 0.66 |
| 439 | 12470954 | 14.313 | 199 | 19.76046 | 51.3582 | 5267 | 4.90 | 0.55 | 0.89 |
| 440 | 2438264 | 14.172 | 161 | 19.35310 | 37.7495 | 4980 | 4.56 | 0.82 | 0.91 |
| 442 | 3745690 | 14.002 | 94.5 | 19.40648 | 38.8756 | 5750 | 4.54 | 0.92 | 1.06 |
| 443 | 3833007 | 14.200 | 93.5 | 19.03446 | 38.9324 | 5614 | 4.62 | 0.83 | 1.03 |
| 444 | 3847138 | 14.112 | 130 | 19.36863 | 38.9427 | 5732 | 4.52 | 0.94 | 1.06 |
| 446 | 4633570 | 14.427 | 137 | 18.93548 | 39.7813 | 4492 | 4.60 | 0.70 | 0.72 |
| 448 | 5640085 | 14.902 | 206 | 19.80468 | 40.8688 | 4264 | 4.55 | 0.71 | 0.65 |
| 452 | 6291033 | 14.641 | 110 | 19.58056 | 41.6151 | 5935 | 4.41 | 1.08 | 1.09 |
| 454 | 7098355 | 14.805 | 119 | 18.97719 | 42.6527 | 5138 | 4.57 | 0.84 | 0.94 |
| 456 | 7269974 | 14.619 | 122 | 19.18490 | 42.8693 | 5644 | 4.52 | 0.94 | 1.05 |
| 457 | 7440748 | 14.196 | 154 | 19.36777 | 43.0838 | 4931 | 4.65 | 0.73 | 0.87 |
| 458 | 7504328 | 14.708 | 132 | 18.85885 | 43.1920 | 5593 | 4.28 | 1.25 | 1.08 |
| 459 | 7977197 | 14.248 | 67 | 19.81859 | 43.7240 | 5601 | 4.43 | 1.04 | 1.06 |
| 460 | 8043638 | 14.743 | 125 | 19.77579 | 43.8028 | 5387 | 4.33 | 1.15 | 1.04 |
| 463 | 8845205 | 14.708 | 196 | 20.01374 | 45.0181 | 3576 | 4.44 | 0.68 | 0.47 |
| 464 | 8890783 | 14.361 | 165 | 19.58314 | 45.1072 | 5362 | 4.46 | 0.97 | 1.01 |
| 465 | 8891318 | 14.188 | 61.7 | 19.59523 | 45.1425 | 6029 | 4.48 | 1.00 | 1.10 |
| 466 | 9008220 | 14.663 | 191 | 19.07680 | 45.3326 | 5907 | 4.90 | 0.59 | 1.00 |
| 467 | 9583881 | 14.794 | 91.3 | 19.34131 | 46.2738 | 5583 | 4.54 | 0.91 | 1.04 |
| 468 | 9589524 | 14.767 | 126 | 19.49642 | 46.2898 | 4999 | 4.50 | 0.90 | 0.93 |
| 469 | 9703198 | 14.711 | 84.4 | 19.24252 | 46.4215 | 6005 | 4.63 | 0.83 | 1.07 |
| 470 | 9844088 | 14.749 | 253 | 19.78922 | 46.6264 | 5542 | 4.65 | 0.78 | 1.00 |
| 471 | 10019643 | 14.415 | 94.8 | 19.66685 | 46.9873 | 5548 | 4.67 | 0.77 | 1.00 |
| 472 | 10123064 | 15.000 | 218 | 18.90786 | 47.1967 | 5682 | 4.58 | 0.87 | 1.04 |
| 473 | 10155434 | 14.673 | 107 | 19.78724 | 47.1719 | 5379 | 4.69 | 0.74 | 0.96 |
| 474 | 10460984 | 14.282 | 102 | 19.18539 | 47.6299 | 6143 | 4.47 | 1.02 | 1.11 |
| 475 | 10577994 | 14.802 | 149 | 18.71273 | 47.8097 | 5056 | 4.54 | 0.86 | 0.93 |
| 476 | 10599206 | 14.958 | 138 | 19.43697 | 47.8145 | 4993 | 4.51 | 0.88 | 0.93 |
| 477 | 10934674 | 14.687 | 179 | 19.84484 | 48.3023 | 5039 | 4.51 | 0.89 | 0.94 |
| 478 | 10990886 | 14.273 | 185 | 19.87371 | 48.4012 | 3727 | 4.38 | 0.80 | 0.55 |
| 479 | 11015323 | 14.106 | 93.1 | 18.90048 | 48.5526 | 5602 | 4.53 | 0.92 | 1.04 |
| 480 | 11134879 | 14.332 | 108 | 19.36251 | 48.7919 | 5324 | 4.51 | 0.92 | 0.99 |
| 481 | 11192998 | 14.701 | 145 | 19.54401 | 48.8812 | 5227 | 4.64 | 0.77 | 0.94 |
| 483 | 11497977 | 14.675 | 140 | 19.10340 | 49.4187 | 5410 | 4.70 | 0.72 | 0.97 |
| 484 | 12061222 | 14.472 | 128 | 19.41474 | 50.5813 | 5065 | 4.76 | 0.65 | 0.88 |
| 486 | 12404305 | 14.118 | 79.6 | 19.30141 | 51.2373 | 5625 | 5.00 | 0.51 | 0.94 |
| 487 | 12834874 | 14.528 | 112 | 19.34980 | 52.1491 | 5463 | 4.51 | 0.93 | 1.02 |
| 488 | 2557816 | 14.720 | 141 | 19.13093 | 37.8297 | 5488 | 4.49 | 0.96 | 1.03 |
| 490 | 3239945 | 14.023 | 129 | 19.51056 | 38.3454 | 4781 | 4.40 | 1.00 | 0.90 |
| 492 | 3559935 | 14.424 | 145 | 19.67886 | 38.6542 | 5373 | 4.26 | 1.26 | 1.06 |
| 494 | 3966801 | 14.885 | 176 | 19.64746 | 39.0738 | 4854 | 4.90 | 0.52 | 0.78 |
| 496 | 4454752 | 14.411 | 113 | 19.25033 | 39.5637 | 5237 | 4.32 | 1.16 | 1.02 |
| 497 | 4757437 | 14.606 | 100 | 19.64734 | 39.8251 | 6045 | 4.50 | 0.98 | 1.09 |
| 499 | 4847534 | 14.272 | 85.6 | 19.66680 | 39.9529 | 5362 | 4.53 | 0.90 | 1.00 |
| 500 | 4852528 | 14.804 | 257 | 19.74084 | 39.9788 | 4250 | 4.52 | 0.74 | 0.66 |
| 501 | 4951877 | 14.612 | 153 | 19.89821 | 40.0759 | 5556 | 4.50 | 0.95 | 1.04 |
| 503 | 5340644 | 15.000 | 226 | 18.89999 | 40.5528 | 4110 | 4.55 | 0.67 | 0.59 |
| 504 | 5461440 | 14.560 | 106 | 19.69552 | 40.6482 | 5403 | 4.75 | 0.68 | 0.95 |
| 505 | 5689351 | 14.194 | 122 | 19.06666 | 40.9193 | 4985 | 4.24 | 1.26 | 1.01 |
| 506 | 5780715 | 14.731 | 116 | 19.23392 | 41.0959 | 5777 | 4.56 | 0.90 | 1.06 |
| 507 | 5812960 | 14.915 | 193 | 19.83358 | 41.0570 | 5117 | 4.41 | 1.02 | 0.98 |
| 508 | 6266741 | 14.387 | 109 | 18.96808 | 41.6296 | 5497 | 4.26 | 1.27 | 1.08 |
| 509 | 6381846 | 14.883 | 246 | 19.78479 | 41.7555 | 5437 | 4.57 | 0.87 | 1.00 |
| 510 | 6422155 | 14.532 | 161 | 18.89123 | 41.8219 | 5355 | 4.39 | 1.07 | 1.03 |
| 511 | 6451936 | 14.209 | 110 | 19.64311 | 41.8841 | 5802 | 4.40 | 1.08 | 1.09 |
| 512 | 6838050 | 14.825 | 157 | 18.80742 | 42.3545 | 5406 | 4.32 | 1.18 | 1.05 |
| 513 | 6937692 | 14.856 | 135 | 19.21537 | 42.4136 | 6288 | 4.58 | 0.90 | 1.10 |
| 517 | 8015907 | 14.034 | 122 | 19.10190 | 43.8734 | 5510 | 4.31 | 1.19 | 1.07 |
| 518 | 8017703 | 14.287 | 92.4 | 19.16261 | 43.8321 | 4822 | 4.60 | 0.76 | 0.84 |
| 519 | 8022244 | 14.939 | 157 | 19.30355 | 43.8762 | 5807 | 4.52 | 0.94 | 1.06 |
| 520 | 8037145 | 14.550 | 149 | 19.64453 | 43.8533 | 5048 | 4.47 | 0.95 | 0.95 |
| 521 | 8162789 | 14.633 | 165 | 19.38234 | 44.0913 | 5767 | 4.39 | 1.09 | 1.08 |
| 522 | 8265218 | 14.406 | 172 | 20.07653 | 44.1045 | 5663 | 4.91 | 0.57 | 0.97 |
| 523 | 8806123 | 15.000 | 98.3 | 19.06981 | 45.0532 | 5942 | 4.42 | 1.07 | 1.09 |
| 524 | 8934495 | 14.868 | 137 | 18.90294 | 45.2256 | 5187 | 4.70 | 0.71 | 0.92 |
| 525 | 9119458 | 14.539 | 150 | 20.06052 | 45.4579 | 5524 | 4.28 | 1.24 | 1.07 |
| 526 | 9157634 | 14.427 | 124 | 19.53715 | 45.5512 | 5467 | 4.63 | 0.80 | 0.99 |
| 528 | 9941859 | 14.598 | 109 | 19.14007 | 46.8965 | 5448 | 4.35 | 1.14 | 1.05 |
| 530 | 10266615 | 14.909 | 152 | 19.24471 | 47.3991 | 5517 | 4.84 | 0.61 | 0.96 |
| 531 | 10395543 | 14.418 | 129 | 19.17941 | 47.5469 | 3946 | 4.49 | 0.70 | 0.56 |
| 532 | 10454313 | 14.708 | 142 | 18.95048 | 47.6890 | 5874 | 4.54 | 0.92 | 1.07 |
| 533 | 10513530 | 14.680 | 133 | 18.70942 | 47.7519 | 5198 | 4.44 | 0.99 | 0.99 |
| 534 | 10554999 | 14.613 | 142 | 19.91092 | 47.7620 | 5145 | 4.79 | 0.63 | 0.89 |
| 535 | 10873260 | 14.434 | 140 | 19.75904 | 48.2335 | 5782 | 4.45 | 1.02 | 1.07 |
| 536 | 10965008 | 14.499 | 93.3 | 19.06827 | 48.4318 | 5614 | 4.60 | 0.84 | 1.03 |
| 537 | 11073351 | 14.665 | 112 | 19.05271 | 48.6833 | 5889 | 4.91 | 0.58 | 1.00 |
| 538 | 11090765 | 14.560 | 114 | 19.66343 | 48.6512 | 5923 | 4.43 | 1.06 | 1.09 |
| 541 | 11656721 | 14.748 | 81.4 | 19.26144 | 49.7620 | 5369 | 4.71 | 0.71 | 0.96 |
| 542 | 11669239 | 14.350 | 107 | 19.70476 | 49.7746 | 5509 | 4.36 | 1.13 | 1.06 |
| 543 | 11823054 | 14.707 | 158 | 19.74343 | 50.0958 | 5166 | 4.72 | 0.69 | 0.91 |
| 546 | 12058931 | 14.896 | 143 | 19.32201 | 50.5862 | 5989 | 4.49 | 0.99 | 1.09 |
| 547 | 12116489 | 14.773 | 156 | 19.63803 | 50.6730 | 5086 | 4.62 | 0.78 | 0.92 |
| 548 | 12600735 | 14.020 | 81.4 | 19.30005 | 51.6857 | 6154 | 4.57 | 0.90 | 1.09 |
| 550 | 4165473 | 14.070 | 125 | 19.56467 | 39.2528 | 5635 | 4.68 | 0.77 | 1.02 |
| 551 | 4270253 | 14.943 | 126 | 19.56872 | 39.3159 | 5627 | 4.67 | 0.78 | 1.02 |
| 552 | 5122112 | 14.741 | 237 | 19.82655 | 40.2292 | 6018 | 4.43 | 1.06 | 1.10 |
| 554 | 5443837 | 14.545 | 333 | 19.35674 | 40.6870 | 5835 | 4.64 | 0.81 | 1.05 |
| 555 | 5709725 | 14.759 | 114 | 19.54156 | 40.9348 | 5218 | 4.63 | 0.78 | 0.95 |
| 557 | 5774349 | 14.970 | 138 | 19.06077 | 41.0694 | 5002 | 4.42 | 1.01 | 0.96 |
| 558 | 5978361 | 14.874 | 135 | 19.77752 | 41.2612 | 5281 | 4.58 | 0.84 | 0.97 |
| 559 | 6422367 | 14.791 | 94.7 | 18.89807 | 41.8732 | 5187 | 4.47 | 0.96 | 0.98 |
| 560 | 6501635 | 14.721 | 146 | 18.92220 | 41.9787 | 5142 | 4.83 | 0.59 | 0.88 |
| 561 | 6665695 | 14.005 | 89.5 | 18.80031 | 42.1765 | 5059 | 4.60 | 0.80 | 0.92 |
| 563 | 6707833 | 14.519 | 79.4 | 19.73547 | 42.1420 | 5879 | 4.48 | 0.99 | 1.08 |
| 564 | 6786037 | 14.854 | 115 | 19.61873 | 42.2910 | 5686 | 4.53 | 0.93 | 1.05 |
| 566 | 7119481 | 14.718 | 96.8 | 19.48773 | 42.6263 | 5865 | 4.56 | 0.90 | 1.07 |
| 567 | 7445445 | 14.338 | 116 | 19.46346 | 43.0747 | 5536 | 4.52 | 0.92 | 1.03 |
| 568 | 7595157 | 14.140 | 70.4 | 19.15303 | 43.2799 | 5265 | 4.86 | 0.58 | 0.90 |
| 569 | 8008206 | 14.458 | 107 | 18.83121 | 43.8899 | 5039 | 4.55 | 0.85 | 0.93 |
| 571 | 8120608 | 14.625 | 187 | 19.91018 | 43.9550 | 3881 | 4.54 | 0.64 | 0.51 |
| 572 | 8193178 | 14.173 | 102 | 19.98643 | 44.0893 | 5666 | 4.31 | 1.21 | 1.09 |
| 573 | 8344004 | 14.674 | 124 | 18.75198 | 44.3155 | 5729 | 4.35 | 1.15 | 1.08 |
| 574 | 8355239 | 14.859 | 169 | 19.17098 | 44.3050 | 5047 | 4.67 | 0.73 | 0.90 |
| 575 | 8367113 | 14.686 | 89.8 | 19.49516 | 44.3813 | 5979 | 4.48 | 0.99 | 1.09 |
| 577 | 8558011 | 14.405 | - | 19.40381 | 44.6324 | 5043 | 4.31 | 1.15 | 1.00 |
| 578 | 8565266 | 14.684 | 159 | 19.59220 | 44.6381 | 5777 | 4.36 | 1.14 | 1.09 |
| 579 | 8616637 | 14.137 | 107 | 19.23894 | 44.7338 | 5074 | 4.60 | 0.80 | 0.92 |
| 580 | 8625925 | 14.856 | 143 | 19.48135 | 44.7000 | 5603 | 4.92 | 0.56 | 0.96 |
| 581 | 8822216 | 14.807 | 158 | 19.54476 | 45.0656 | 5514 | 4.86 | 0.60 | 0.95 |
| 582 | 9020160 | 14.808 | 159 | 19.41223 | 45.3232 | 5103 | 4.65 | 0.75 | 0.92 |
| 583 | 9076513 | 14.573 | 96.6 | 19.06700 | 45.4804 | 5735 | 4.55 | 0.90 | 1.05 |
| 584 | 9146018 | 14.129 | 102 | 19.19453 | 45.5929 | 5350 | 4.80 | 0.63 | 0.93 |
| 585 | 9279669 | 14.911 | 148 | 19.42364 | 45.7479 | 5437 | 4.74 | 0.70 | 0.96 |
| 586 | 9570741 | 14.608 | 122 | 18.85245 | 46.2450 | 5707 | 4.67 | 0.78 | 1.03 |
| 587 | 9607164 | 14.574 | 123 | 19.89232 | 46.2760 | 5112 | 4.42 | 1.01 | 0.98 |
| 588 | 9631762 | 14.337 | 114 | 18.83004 | 46.3214 | 4431 | 4.46 | 0.85 | 0.76 |
| 589 | 9763754 | 14.547 | 85.6 | 19.24331 | 46.5978 | 5880 | 4.64 | 0.81 | 1.05 |
| 590 | 9782691 | 14.615 | 134 | 19.76657 | 46.5772 | 6106 | 4.55 | 0.92 | 1.09 |
| 592 | 9957627 | 14.292 | 93.6 | 19.63084 | 46.8215 | 5810 | 4.41 | 1.08 | 1.08 |
| 593 | 9958962 | 14.957 | 214 | 19.66594 | 46.8383 | 5737 | 4.62 | 0.83 | 1.04 |
| 596 | 10388286 | 14.818 | 147 | 18.91605 | 47.5163 | 3740 | 4.55 | 0.60 | 0.47 |
| 597 | 10600261 | 14.915 | 142 | 19.46476 | 47.8642 | 5833 | 4.42 | 1.07 | 1.09 |
| 598 | 10656823 | 14.813 | 207 | 19.20806 | 47.9667 | 5171 | 4.81 | 0.61 | 0.89 |
| 599 | 10676824 | 14.854 | 131 | 19.74138 | 47.9215 | 5820 | 4.54 | 0.92 | 1.06 |
| 600 | 10718726 | 14.827 | 119 | 19.08332 | 48.0609 | 5869 | 4.45 | 1.03 | 1.08 |
| 601 | 10973664 | 14.697 | 144 | 19.38994 | 48.4160 | 5862 | 4.58 | 0.88 | 1.06 |
| 602 | 12459913 | 14.647 | 113 | 19.41063 | 51.3349 | 6007 | 4.41 | 1.09 | 1.10 |
| 605 | 4832837 | 14.915 | 161 | 19.39515 | 39.9142 | 4270 | 4.76 | 0.53 | 0.60 |
| 607 | 5441980 | 14.377 | - | 19.32061 | 40.6158 | 5497 | 4.61 | 0.83 | 1.01 |
| 609 | 5608566 | 14.491 | 179 | 19.21071 | 40.8789 | 5696 | 4.30 | 1.23 | 1.09 |
| 610 | 5686174 | 14.672 | 150 | 18.96523 | 40.9331 | 4072 | 4.53 | 0.69 | 0.58 |
| 611 | 6309763 | 14.022 | 106 | 19.88627 | 41.6838 | 6122 | 4.55 | 0.92 | 1.09 |
| 612 | 6587002 | 14.157 | 78.5 | 18.99790 | 42.0792 | 5105 | 4.22 | 1.32 | 1.04 |
| 614 | 7368664 | 14.517 | 89.3 | 19.57242 | 42.9289 | 5675 | 4.89 | 0.59 | 0.98 |
| 617 | 9846086 | 14.608 | 76.5 | 19.82791 | 46.6443 | 5594 | 4.53 | 0.92 | 1.04 |
| 618 | 10353968 | 14.959 | 227 | 19.79911 | 47.4774 | 5471 | 4.52 | 0.92 | 1.02 |

| | | | | | | | | | | |
|---|---|---|---|---|---|---|---|---|---|---|
| 620 | 11773022 | 14.669 | 385 | 19.76532 | 49.9377 | 5803 | 4.54 | 0.91 | 1.06 | |
| 622 | 12417486 | 14.932 | 153 | 19.72535 | 51.2638 | 5171 | 4.31 | 1.17 | 1.01 | |
| 623 | 12068975 | 11.811 | 33.3 | 19.68176 | 50.5590 | 6191 | 4.07 | 1.67 | 1.21 | |
| 624 | 3541946 | 13.597 | 141 | 19.37821 | 38.6910 | 5537 | 4.73 | 0.71 | 0.99 | |
| 625 | 4449034 | 13.592 | 104 | 19.10425 | 39.5345 | 6199 | 3.86 | 2.21 | 1.29 | |
| 626 | 4478168 | 13.490 | 62.2 | 19.67956 | 39.5397 | 6134 | 4.40 | 1.10 | 1.12 | |
| 627 | 4563268 | 13.307 | 86 | 19.47670 | 39.6376 | 5851 | 4.22 | 1.36 | 1.13 | |
| 628 | 4644604 | 13.946 | 82.7 | 19.24658 | 39.7083 | 5668 | 4.25 | 1.31 | 1.10 | |
| 629 | 4656049 | 13.949 | 122 | 19.47362 | 39.7679 | 6203 | 4.18 | 1.46 | 1.17 | |
| 632 | 4827723 | 13.359 | 65.5 | 19.29452 | 39.9450 | 5273 | 4.64 | 0.77 | 0.95 | |
| 633 | 4841374 | 13.871 | 63.5 | 19.56180 | 39.9424 | 5759 | 4.03 | 1.74 | 1.18 | |
| 635 | 5020319 | 13.034 | 170 | 19.62490 | 40.1946 | 6065 | 4.42 | 1.07 | 1.11 | |
| 638 | 5113822 | 13.595 | 121 | 19.70396 | 40.2363 | 5722 | 4.31 | 1.21 | 1.09 | |
| 639 | 5120087 | 13.500 | 87.7 | 19.79238 | 40.2282 | 6166 | 4.44 | 1.05 | 1.11 | |
| 640 | 5121511 | 13.332 | 69.3 | 19.81684 | 40.2886 | 5131 | 4.37 | 1.07 | 0.99 | |
| 641 | 5131180 | 13.583 | 146 | 19.95330 | 40.2351 | 4054 | 4.34 | 0.93 | 0.68 | |
| 644 | 5356593 | 13.725 | 69.9 | 19.33112 | 40.5327 | 5395 | 3.80 | 2.33 | 1.25 | |
| 645 | 5374854 | 13.716 | 60.6 | 19.68116 | 40.5923 | 5890 | 4.09 | 1.61 | 1.17 | |
| 647 | 5531694 | 13.550 | 48.1 | 19.41300 | 40.7027 | 6154 | 4.38 | 1.13 | 1.12 | |
| 649 | 5613330 | 13.310 | 57.8 | 19.31822 | 40.8007 | 6000 | 4.29 | 1.26 | 1.12 | |
| 650 | 5786676 | 13.594 | 83.8 | 19.35975 | 41.0401 | 4928 | 4.32 | 1.13 | 0.97 | |
| 652 | 5796675 | 13.653 | 166 | 19.57279 | 41.0952 | 4628 | 4.79 | 0.57 | 0.72 | |
| 654 | 5941160 | 13.984 | 101 | 18.96066 | 41.2375 | 5799 | 4.24 | 1.33 | 1.12 | |
| 655 | 5966154 | 13.004 | 48.3 | 19.57507 | 41.2728 | 6249 | 4.44 | 1.06 | 1.12 | |
| 657 | 6020753 | 13.872 | 109 | 18.86489 | 41.3220 | 4632 | 4.63 | 0.70 | 0.76 | |
| 658 | 6062088 | 13.989 | 119 | 19.80600 | 41.3880 | 5676 | 4.53 | 0.93 | 1.05 | |
| 659 | 6125481 | 13.413 | 55.3 | 19.49448 | 41.4169 | 6463 | 4.23 | 1.38 | 1.18 | |
| 660 | 6267535 | 13.532 | 59.9 | 18.99463 | 41.6172 | 5250 | 4.15 | 1.45 | 1.09 | |
| 661 | 6347299 | 13.909 | 75.4 | 19.02315 | 41.7619 | 5825 | 4.40 | 1.09 | 1.09 | |
| 662 | 6365156 | 13.336 | 57.3 | 19.47913 | 41.7271 | 5889 | 4.41 | 1.08 | 1.09 | |
| 663 | 6425957 | 13.506 | 74.8 | 19.01914 | 41.8612 | 4156 | 4.53 | 0.70 | 0.61 | |
| 664 | 6442340 | 13.484 | 55.3 | 19.44231 | 41.8339 | 5725 | 4.24 | 1.32 | 1.11 | |
| 665 | 6685609 | 13.182 | 103 | 19.33537 | 42.1661 | 5864 | 4.38 | 1.12 | 1.09 | |
| 666 | 6707835 | 13.721 | 65.9 | 19.73548 | 42.1317 | 5553 | 4.59 | 0.85 | 1.02 | |
| 667 | 6752502 | 13.826 | 847 | 18.81033 | 42.2346 | 4135 | 4.57 | 0.67 | 0.60 | 1 |
| 670 | 7033671 | 13.774 | 66.9 | 19.45490 | 42.5162 | 5608 | 4.35 | 1.15 | 1.07 | |
| 671 | 7040629 | 13.749 | 49.5 | 19.59268 | 42.5280 | 5845 | 4.45 | 0.97 | 0.96 | 1 |
| 672 | 7115785 | 13.998 | 110 | 19.41130 | 42.6408 | 5565 | 4.16 | 1.45 | 1.12 | |
| 673 | 7124613 | 13.343 | 95.7 | 19.58897 | 42.6250 | 6338 | 4.34 | 1.19 | 1.15 | |
| 674 | 7277317 | 13.781 | - | 19.35515 | 42.8983 | 4864 | 3.67 | 2.74 | 1.29 | |
| 676 | 7447200 | 13.822 | 249 | 19.50023 | 43.0832 | 4218 | 4.55 | 0.69 | 0.63 | |
| 678 | 7509886 | 13.283 | 102 | 19.02928 | 43.1685 | 5073 | 4.17 | 1.39 | 1.06 | |
| 679 | 7515212 | 13.178 | 51.1 | 19.17472 | 43.1417 | 5929 | 4.41 | 1.03 | 1.01 | 1 |
| 680 | 7529266 | 13.643 | 62.5 | 19.48582 | 43.1973 | 6060 | 4.35 | 1.16 | 1.12 | |
| 682 | 7619236 | 13.916 | 69.7 | 19.67987 | 43.2695 | 5504 | 4.50 | 0.95 | 1.03 | |
| 683 | 7630229 | 13.714 | 78.1 | 19.85495 | 43.2584 | 5624 | 4.67 | 0.78 | 1.02 | |
| 684 | 7730747 | 13.831 | 66.3 | 18.75269 | 43.4133 | 5331 | 3.96 | 1.88 | 1.18 | |
| 685 | 7764367 | 13.949 | 76.4 | 19.69838 | 43.4931 | 6187 | 4.16 | 1.49 | 1.18 | |
| 686 | 7906882 | 13.579 | 46.1 | 19.78938 | 43.6471 | 5360 | 4.47 | 0.96 | 1.01 | |
| 687 | 7976520 | 13.813 | 134 | 19.80781 | 43.7114 | 5606 | 4.50 | 0.96 | 1.05 | |
| 688 | 8161561 | 13.992 | 79.4 | 19.35485 | 44.0358 | 6157 | 4.26 | 1.31 | 1.15 | |
| 689 | 8361905 | 13.766 | - | 19.36476 | 44.3871 | 5438 | 4.60 | 0.83 | 1.00 | |
| 691 | 8480285 | 13.965 | 64.9 | 18.98589 | 44.5917 | 6037 | 4.35 | 1.17 | 1.12 | |
| 692 | 8557374 | 13.648 | 50.4 | 19.38828 | 44.6472 | 5608 | 4.90 | 0.58 | 0.96 | |
| 693 | 8738735 | 13.949 | 76.7 | 18.98366 | 44.9560 | 6121 | 4.50 | 0.97 | 1.10 | |
| 694 | 8802165 | 13.939 | 91.9 | 18.92664 | 45.0169 | 5596 | 4.87 | 0.60 | 0.97 | |
| 695 | 8805348 | 13.437 | 49.1 | 19.04373 | 45.0796 | 5980 | 4.36 | 1.16 | 1.11 | |
| 697 | 8878187 | 13.684 | - | 19.26673 | 45.1543 | 5779 | 4.05 | 1.70 | 1.18 | |
| 698 | 8891278 | 13.816 | 91.2 | 19.59424 | 45.1898 | 5705 | 4.47 | 0.91 | 0.90 | 1 |
| 700 | 8962094 | 13.580 | 54.7 | 19.66490 | 45.2137 | 5601 | 4.37 | 1.12 | 1.07 | |
| 701 | 9002278 | 13.725 | 83.4 | 18.88085 | 45.3499 | 4869 | 4.70 | 0.68 | 0.83 | |
| 703 | 9162741 | 13.361 | 60.2 | 19.66080 | 45.5667 | 6178 | 4.36 | 1.17 | 1.14 | 1 |
| 704 | 9266431 | 13.704 | 75.7 | 18.95908 | 45.7197 | 5276 | 4.47 | 0.96 | 0.99 | |
| 707 | 9458613 | 13.988 | 103 | 19.27184 | 46.0052 | 5933 | 4.27 | 1.29 | 1.12 | |
| 708 | 9530945 | 13.998 | 95 | 19.58187 | 46.1292 | 6036 | 4.53 | 0.94 | 1.09 | |
| 709 | 9578686 | 13.940 | 75 | 19.15552 | 46.2035 | 5468 | 4.53 | 0.90 | 1.02 | |
| 710 | 9590976 | 13.294 | 48.2 | 19.53585 | 46.2775 | 6653 | 4.31 | 1.36 | 1.38 | 1 |
| 711 | 9597345 | 13.967 | 63 | 19.69419 | 46.2665 | 5488 | 4.45 | 1.00 | 1.04 | |
| 712 | 9640976 | 13.720 | 61.1 | 19.19942 | 46.3569 | 5500 | 4.76 | 0.68 | 0.97 | |
| 714 | 9702072 | 13.393 | 93.4 | 19.20054 | 46.4736 | 5444 | 4.66 | 0.76 | 0.98 | |
| 716 | 9846348 | 13.754 | 53.3 | 19.83281 | 46.6945 | 5845 | 4.50 | 0.96 | 1.07 | |
| 717 | 9873254 | 13.387 | 51.6 | 18.81420 | 46.7178 | 5412 | 4.96 | 0.52 | 0.90 | |
| 718 | 9884104 | 13.764 | 93.4 | 19.24927 | 46.7626 | 5801 | 4.67 | 0.78 | 1.04 | |
| 719 | 9950612 | 13.177 | 55.6 | 19.43374 | 46.8957 | 4405 | 4.55 | 0.73 | 0.68 | 1 |
| 720 | 9963524 | 13.749 | 150 | 19.77698 | 46.8353 | 5123 | 4.54 | 0.86 | 0.95 | |
| 721 | 9964801 | 13.645 | 57.1 | 19.80456 | 46.8343 | 5812 | 4.10 | 1.59 | 1.17 | |
| 722 | 9965439 | 13.489 | 54.9 | 19.81727 | 46.8432 | 6133 | 4.63 | 0.83 | 1.08 | |
| 723 | 10002866 | 15.063 | 218 | 19.19374 | 46.9378 | 5244 | 4.66 | 0.76 | 0.94 | |
| 725 | 10068383 | 15.765 | 707 | 19.26705 | 47.0404 | 5046 | 4.65 | 0.74 | 0.90 | |
| 728 | 10221013 | 15.356 | 356 | 19.77615 | 47.2303 | 5976 | 4.54 | 0.92 | 1.08 | |
| 730 | 10227020 | 15.344 | 213 | 19.88789 | 47.2795 | 5599 | 4.39 | 1.10 | 1.07 | |
| 732 | 10265898 | 15.342 | 240 | 19.22186 | 47.3816 | 5360 | 4.59 | 0.83 | 0.98 | |
| 733 | 10271806 | 15.644 | 375 | 19.39562 | 47.3576 | 5038 | 4.85 | 0.58 | 0.85 | |
| 734 | 10272442 | 15.344 | 214 | 19.41315 | 47.3078 | 5719 | 4.70 | 0.75 | 1.02 | |
| 735 | 10287242 | 15.637 | 375 | 19.77956 | 47.3922 | 5080 | 4.47 | 0.94 | 0.96 | |
| 736 | 10340423 | 15.962 | 310 | 19.47960 | 47.4571 | 4157 | 4.55 | 0.68 | 0.60 | |
| 737 | 10345478 | 15.684 | 163 | 19.60807 | 47.4088 | 5117 | 4.60 | 0.80 | 0.93 | |
| 738 | 10358759 | 15.282 | 176 | 19.88989 | 47.4912 | 5711 | 4.54 | 0.91 | 1.05 | |
| 739 | 10386984 | 15.488 | 204 | 18.86559 | 47.5786 | 4050 | 4.54 | 0.67 | 0.57 | |
| 740 | 10395381 | 15.556 | 228 | 19.17388 | 47.5097 | 4711 | 4.64 | 0.70 | 0.79 | |
| 741 | 10418797 | 15.278 | 138 | 19.79715 | 47.5538 | 5556 | 4.73 | 0.71 | 0.99 | |
| 743 | 10464078 | 15.487 | 204 | 19.28388 | 47.6353 | 4877 | 4.30 | 1.14 | 0.96 | |
| 745 | 10485250 | 15.788 | 374 | 19.81030 | 47.6687 | 4957 | 4.43 | 0.97 | 0.94 | |
| 746 | 10526549 | 15.302 | 165 | 19.20922 | 47.7245 | 4681 | 4.55 | 0.79 | 0.81 | |
| 747 | 10583066 | 15.784 | 251 | 18.91406 | 47.8633 | 4357 | 4.68 | 0.61 | 0.65 | |
| 749 | 10601284 | 15.416 | 201 | 19.49174 | 47.8810 | 5374 | 4.78 | 0.65 | 0.94 | |

| | | | | | | | | | |
|---|---|---|---|---|---|---|---|---|---|
| 750 | 10662202 | 15.377 | 177 | 19.36434 | 47.9292 | 4619 | 4.62 | 0.70 | 0.76 |
| 751 | 10682541 | 15.861 | 486 | 19.85388 | 47.9467 | 5174 | 4.55 | 0.86 | 0.96 |
| 752 | 10797460 | 15.347 | 211 | 19.46228 | 48.1417 | 5584 | 4.41 | 1.07 | 1.06 |
| 753 | 10811496 | 15.436 | 263 | 19.80032 | 48.1341 | 5648 | 4.84 | 0.62 | 0.98 |
| 755 | 10854555 | 15.509 | 291 | 19.25033 | 48.2262 | 5781 | 4.44 | 1.04 | 1.08 |
| 756 | 10872983 | 15.714 | 249 | 19.75241 | 48.2247 | 5787 | 4.51 | 0.95 | 1.07 |
| 757 | 10910878 | 15.841 | 321 | 19.13330 | 48.3758 | 4956 | 4.69 | 0.70 | 0.87 |
| 758 | 10987985 | 15.390 | 270 | 19.80171 | 48.4786 | 4869 | 4.28 | 1.17 | 0.96 |
| 759 | 11018648 | 15.082 | 149 | 19.04790 | 48.5059 | 5401 | 4.56 | 0.86 | 1.00 |
| 760 | 11138155 | 15.263 | 118 | 19.47780 | 48.7276 | 5887 | 4.62 | 0.83 | 1.05 |
| 762 | 11153539 | 15.395 | 201 | 19.89712 | 48.7916 | 5779 | 4.60 | 0.85 | 1.05 |
| 763 | 11242721 | 15.536 | 142 | 19.40626 | 48.9338 | 5788 | 4.39 | 1.10 | 1.09 |
| 764 | 11304958 | 15.400 | 188 | 19.70367 | 49.0173 | 5263 | 4.37 | 1.09 | 1.02 |
| 765 | 11391957 | 15.317 | 287 | 19.03328 | 49.2774 | 5345 | 4.70 | 0.72 | 0.95 |
| 766 | 11403044 | 15.506 | 158 | 19.46241 | 49.2540 | 5913 | 4.47 | 1.00 | 1.09 |
| 767 | 11414511 | 15.052 | 301 | 19.80103 | 49.2253 | 5431 | 4.44 | 1.01 | 1.03 |
| 769 | 11460018 | 15.356 | 140 | 19.61772 | 49.3141 | 5461 | 4.64 | 0.79 | 0.99 |
| 771 | 11465813 | 15.207 | 126 | 19.77991 | 49.3165 | 5574 | 4.38 | 1.10 | 1.06 |
| 772 | 11493732 | 15.250 | 179 | 18.91506 | 49.4795 | 5885 | 4.41 | 1.08 | 1.09 |
| 773 | 11507101 | 15.172 | 109 | 19.44813 | 49.4809 | 5667 | 4.62 | 0.82 | 1.03 |
| 774 | 11656840 | 15.270 | 746 | 19.26579 | 49.7337 | 5873 | 4.46 | 1.01 | 1.08 |
| 775 | 11754553 | 15.095 | 239 | 19.11587 | 49.9758 | 4075 | 4.54 | 0.68 | 0.58 |
| 776 | 11812062 | 15.523 | 178 | 19.38572 | 50.0541 | 5309 | 4.83 | 0.61 | 0.91 |
| 777 | 11818800 | 15.487 | 547 | 19.62112 | 50.0802 | 5256 | 4.48 | 0.95 | 0.99 |
| 778 | 11853255 | 15.135 | 262 | 19.01175 | 50.1498 | 4082 | 4.61 | 0.61 | 0.55 |
| 779 | 11909839 | 15.562 | 177 | 19.28237 | 50.2410 | 5527 | 4.40 | 1.08 | 1.05 |
| 780 | 11918099 | 15.334 | 245 | 19.58889 | 50.2304 | 4833 | 4.67 | 0.70 | 0.82 |
| 781 | 11923270 | 15.937 | 405 | 19.74813 | 50.2872 | 3833 | 4.40 | 0.79 | 0.57 |
| 782 | 11960862 | 15.312 | 247 | 19.33987 | 50.3216 | 5733 | 4.41 | 1.07 | 1.08 |
| 783 | 12020329 | 15.080 | 328 | 19.69702 | 50.4946 | 5284 | 4.76 | 0.66 | 0.93 |
| 784 | 12066335 | 15.385 | 254 | 19.59822 | 50.5319 | 4112 | 4.57 | 0.65 | 0.58 |
| 785 | 12070811 | 15.505 | 214 | 19.73476 | 50.5678 | 5380 | 4.73 | 0.70 | 0.96 |
| 786 | 12110942 | 15.242 | 147 | 19.44149 | 50.6182 | 5638 | 4.72 | 0.73 | 1.01 |
| 787 | 12366084 | 15.367 | 300 | 19.72108 | 51.1217 | 5615 | 4.53 | 0.92 | 1.05 |
| 788 | 12404086 | 15.234 | 177 | 19.29331 | 51.2503 | 4950 | 4.63 | 0.75 | 0.88 |
| 790 | 12470844 | 15.339 | 312 | 19.75777 | 51.3195 | 5176 | 5.06 | 0.44 | 0.82 |
| 791 | 12644822 | 15.140 | 463 | 19.27415 | 51.7374 | 5564 | 4.53 | 0.92 | 1.03 |
| 794 | 2713049 | 15.026 | 203 | 19.42933 | 37.9057 | 5744 | 4.49 | 0.97 | 1.07 |
| 795 | 3114167 | 15.591 | 258 | 19.39137 | 38.2736 | 5455 | 4.80 | 0.64 | 0.95 |
| 797 | 3115833 | 15.657 | 477 | 19.41880 | 38.2417 | 5725 | 4.53 | 0.92 | 1.05 |
| 799 | 3246984 | 15.279 | 257 | 19.62455 | 38.3103 | 5491 | 4.41 | 1.05 | 1.04 |
| 800 | 3342970 | 15.541 | 353 | 19.44357 | 38.4947 | 5938 | 4.61 | 0.85 | 1.07 |
| 801 | 3351888 | 15.001 | 166 | 19.59146 | 38.4229 | 5472 | 4.39 | 1.08 | 1.04 |
| 802 | 3453214 | 15.562 | 391 | 19.58775 | 38.5636 | 5556 | 5.01 | 0.50 | 0.92 |
| 804 | 3641726 | 15.387 | 239 | 19.35558 | 38.7282 | 5136 | 4.53 | 0.87 | 0.95 |
| 805 | 3734868 | 15.646 | 462 | 19.17978 | 38.8865 | 5415 | 4.37 | 1.11 | 1.04 |
| 806 | 3832474 | 15.403 | 652 | 19.01891 | 38.9473 | 5206 | 4.53 | 0.88 | 0.97 |
| 809 | 3935914 | 15.530 | 279 | 19.01425 | 39.0275 | 5690 | 4.48 | 0.98 | 1.06 |
| 810 | 3940418 | 15.119 | 224 | 19.13960 | 39.0607 | 4997 | 4.57 | 0.82 | 0.91 |
| 811 | 4049131 | 15.398 | 322 | 19.28770 | 39.1533 | 4764 | 4.43 | 0.94 | 0.88 |
| 812 | 4139816 | 15.954 | 341 | 19.07194 | 39.2783 | 4097 | 4.66 | 0.57 | 0.55 |
| 813 | 4275191 | 15.725 | 239 | 19.64368 | 39.3084 | 5357 | 4.73 | 0.70 | 0.95 |
| 814 | 4476123 | 15.583 | 251 | 19.64737 | 39.5217 | 5236 | 4.86 | 0.58 | 0.89 |
| 815 | 4544670 | 15.684 | 255 | 19.06133 | 39.6248 | 5344 | 4.49 | 0.95 | 1.00 |
| 816 | 4664847 | 15.670 | 261 | 19.63266 | 39.7714 | 5699 | 4.50 | 0.96 | 1.06 |
| 817 | 4725681 | 15.414 | 253 | 18.92443 | 39.8981 | 3905 | 4.59 | 0.59 | 0.50 |
| 818 | 4913852 | 15.877 | 315 | 19.25413 | 40.0334 | 3785 | 4.36 | 0.83 | 0.58 |
| 821 | 5021899 | 15.540 | 263 | 19.65179 | 40.1537 | 5408 | 4.41 | 1.05 | 1.03 |
| 822 | 5077629 | 15.805 | 475 | 18.90256 | 40.2180 | 5458 | 4.61 | 0.82 | 1.00 |
| 823 | 5115978 | 15.202 | 249 | 19.73380 | 40.2954 | 5976 | 4.43 | 1.06 | 1.10 |
| 824 | 5164255 | 16.422 | - | 18.88357 | 40.3582 | 4829 | 4.44 | 0.94 | 0.90 |
| 825 | 5252423 | 15.289 | 203 | 18.88143 | 40.4218 | 4735 | 4.58 | 0.76 | 0.81 |
| 826 | 5272878 | 15.090 | 137 | 19.40997 | 40.4205 | 5557 | 4.84 | 0.62 | 0.97 |
| 827 | 5283542 | 15.546 | 197 | 19.61493 | 40.4179 | 5837 | 4.54 | 0.92 | 1.06 |
| 829 | 5358241 | 15.386 | 137 | 19.36412 | 40.5625 | 5858 | 4.57 | 0.89 | 1.06 |
| 830 | 5358624 | 15.224 | 111 | 19.37212 | 40.5774 | 4915 | 4.90 | 0.53 | 0.80 |
| 833 | 5376067 | 15.446 | 193 | 19.70029 | 40.5055 | 5781 | 4.66 | 0.79 | 1.04 |
| 834 | 5436502 | 15.084 | 123 | 19.19314 | 40.6378 | 5614 | 4.60 | 0.85 | 1.03 |
| 835 | 5456651 | 15.208 | 181 | 19.61348 | 40.6634 | 4817 | 4.95 | 0.48 | 0.75 |
| 837 | 5531576 | 15.660 | 237 | 19.41058 | 40.7503 | 4817 | 4.75 | 0.62 | 0.80 |
| 838 | 5534814 | 15.311 | 339 | 19.48096 | 40.7598 | 5794 | 4.48 | 0.99 | 1.07 |
| 840 | 5651104 | 15.028 | 194 | 19.95441 | 40.8224 | 4916 | 4.39 | 1.03 | 0.94 |
| 841 | 5792202 | 15.855 | 467 | 19.48245 | 41.0859 | 5226 | 4.66 | 0.76 | 0.94 |
| 842 | 5794379 | 15.389 | 254 | 19.52754 | 41.0609 | 4497 | 4.52 | 0.79 | 0.76 |
| 843 | 5881688 | 15.270 | 250 | 19.56644 | 41.1376 | 5784 | 4.40 | 1.09 | 1.08 |
| 844 | 6022556 | 15.581 | 304 | 18.92429 | 41.3465 | 5381 | 4.81 | 0.63 | 0.94 |
| 845 | 6032497 | 15.447 | 198 | 19.23400 | 41.3018 | 5646 | 4.44 | 1.02 | 1.06 |
| 846 | 6061119 | 15.482 | 175 | 19.79255 | 41.3961 | 5612 | 4.60 | 0.85 | 1.03 |
| 847 | 6191521 | 15.201 | 105 | 19.14362 | 41.5658 | 5469 | 4.56 | 0.88 | 1.01 |
| 849 | 6276477 | 15.018 | 199 | 19.26628 | 41.6333 | 5303 | 4.48 | 0.96 | 1.00 |
| 850 | 6291653 | 15.305 | 130 | 19.59187 | 41.6618 | 5236 | 4.55 | 0.87 | 0.97 |
| 851 | 6392727 | 15.287 | 486 | 19.94579 | 41.7427 | 5570 | 4.55 | 0.89 | 1.03 |
| 852 | 6422070 | 15.257 | 170 | 18.88776 | 41.8371 | 5448 | 4.47 | 0.98 | 1.03 |
| 853 | 6428700 | 15.376 | 246 | 19.10438 | 41.8084 | 4842 | 4.47 | 0.91 | 0.89 |
| 854 | 6435936 | 15.849 | 493 | 19.30057 | 41.8121 | 3743 | 4.69 | 0.49 | 0.43 |
| 855 | 6522242 | 15.196 | 122 | 19.45069 | 41.9441 | 5316 | 4.59 | 0.83 | 0.97 |
| 856 | 6526710 | 15.344 | 127 | 19.54529 | 41.9692 | 5858 | 4.59 | 0.86 | 1.06 |
| 857 | 6587280 | 15.086 | 162 | 19.00737 | 42.0340 | 5033 | 4.63 | 0.76 | 0.91 |
| 858 | 6599919 | 15.060 | 261 | 19.32652 | 42.0092 | 5440 | 4.45 | 1.00 | 1.03 |
| 861 | 6685526 | 15.001 | 247 | 19.33384 | 42.1165 | 5066 | 4.63 | 0.76 | 0.91 |
| 863 | 6784235 | 15.533 | 295 | 19.58776 | 42.2125 | 5651 | 4.59 | 0.85 | 1.04 |
| 864 | 6849310 | 15.604 | 214 | 19.14951 | 42.3014 | 5337 | 4.77 | 0.66 | 0.93 |
| 865 | 6862328 | 15.085 | 139 | 19.43892 | 42.3682 | 5560 | 4.70 | 0.73 | 1.00 |
| 867 | 6863998 | 15.219 | 231 | 19.47368 | 42.3803 | 5059 | 4.52 | 0.88 | 0.94 |
| 868 | 6867155 | 15.172 | 254 | 19.53386 | 42.3072 | 4118 | 4.52 | 0.71 | 0.61 |
| 869 | 6948054 | 15.599 | 277 | 19.44260 | 42.4363 | 5085 | 4.46 | 0.96 | 0.96 |
| 870 | 6949607 | 15.036 | 224 | 19.47580 | 42.4294 | 4590 | 4.29 | 1.12 | 0.89 |
| 871 | 7031517 | 15.215 | 1141 | 19.40798 | 42.5071 | 5650 | 5.05 | 0.48 | 0.93 |
| 872 | 7109675 | 15.262 | 349 | 19.28458 | 42.6042 | 5127 | 4.59 | 0.81 | 0.94 |
| 873 | 7118364 | 15.024 | 101 | 19.46428 | 42.6961 | 5470 | 4.78 | 0.66 | 0.96 |
| 874 | 7134976 | 15.024 | 154 | 19.76556 | 42.6634 | 5037 | 4.56 | 0.84 | 0.93 |
| 875 | 7135852 | 15.692 | 331 | 19.77960 | 42.6261 | 4198 | 4.87 | 0.45 | 0.54 |
| 876 | 7270230 | 15.877 | 727 | 19.19227 | 42.8324 | 5417 | 4.87 | 0.59 | 0.93 |
| 877 | 7287995 | 15.019 | 272 | 19.57580 | 42.8250 | 4211 | 4.57 | 0.68 | 0.62 |
| 878 | 7303253 | 15.316 | 164 | 19.83067 | 42.8399 | 4749 | 4.28 | 1.16 | 0.94 |
| 880 | 7366258 | 15.158 | 213 | 19.52489 | 42.9661 | 5512 | 4.49 | 0.96 | 1.03 |
| 881 | 7373451 | 15.859 | 243 | 19.66065 | 42.9353 | 5053 | 4.82 | 0.60 | 0.86 |
| 882 | 7377033 | 15.533 | 645 | 19.72236 | 42.9431 | 5081 | 4.57 | 0.83 | 0.93 |
| 883 | 7380537 | 15.766 | 342 | 19.77887 | 42.9679 | 4674 | 4.82 | 0.55 | 0.73 |
| 884 | 7434875 | 15.067 | 324 | 19.24283 | 43.0393 | 4931 | 4.87 | 0.55 | 0.81 |
| 886 | 7455287 | 15.847 | 288 | 19.65160 | 43.0563 | 3705 | 4.63 | 0.53 | 0.44 |
| 887 | 7458762 | 15.031 | 116 | 19.71010 | 43.0298 | 5601 | 4.53 | 0.92 | 1.04 |
| 889 | 757450 | 15.264 | 371 | 19.40917 | 36.5774 | 5101 | 4.48 | 0.93 | 0.96 |
| 890 | 7585481 | 15.261 | 131 | 18.84663 | 43.2727 | 5976 | 4.56 | 0.90 | 1.07 |
| 891 | 7663691 | 15.063 | 161 | 18.90872 | 43.3794 | 5851 | 4.59 | 0.86 | 1.06 |
| 892 | 7678434 | 15.193 | 241 | 19.35955 | 43.3639 | 5010 | 4.60 | 0.79 | 0.91 |
| 893 | 7685981 | 15.662 | 177 | 19.53104 | 43.3464 | 5729 | 4.46 | 1.01 | 1.07 |
| 895 | 7767559 | 15.403 | 593 | 19.76068 | 43.4554 | 5436 | 4.37 | 1.10 | 1.04 |
| 896 | 7825899 | 15.258 | 268 | 19.53743 | 43.5814 | 5206 | 4.63 | 0.78 | 0.94 |
| 897 | 7849854 | 15.257 | 263 | 19.97079 | 43.5036 | 5734 | 4.46 | 1.01 | 1.07 |
| 898 | 7870390 | 15.777 | 350 | 18.81550 | 43.6656 | 4051 | 4.53 | 0.68 | 0.57 |
| 899 | 7907423 | 15.234 | 176 | 19.79900 | 43.6585 | 3653 | 4.59 | 0.55 | 0.43 |
| 900 | 7938496 | 15.425 | 191 | 18.79737 | 43.7031 | 5692 | 4.34 | 1.17 | 1.09 |
| 901 | 8013419 | 15.750 | 163 | 19.01117 | 43.8763 | 4213 | 4.72 | 0.55 | 0.58 |
| 902 | 8018547 | 15.754 | 390 | 19.19016 | 43.8980 | 4312 | 4.62 | 0.65 | 0.64 |
| 903 | 8039892 | 15.813 | 167 | 19.70241 | 43.8845 | 5620 | 4.78 | 0.68 | 0.99 |
| 904 | 8150320 | 15.791 | 221 | 19.01122 | 44.0265 | 4362 | 4.56 | 0.72 | 0.69 |
| 905 | 8180063 | 15.289 | 410 | 19.76717 | 44.0361 | 5668 | 5.04 | 0.48 | 0.94 |
| 906 | 8226994 | 15.460 | 197 | 19.30632 | 44.1420 | 5017 | 4.56 | 0.84 | 0.92 |
| 907 | 8247638 | 15.223 | 153 | 19.77108 | 44.1059 | 5634 | 4.35 | 1.15 | 1.08 |
| 908 | 8255887 | 15.113 | 158 | 19.90904 | 44.1709 | 5391 | 4.25 | 1.29 | 1.07 |
| 910 | 8414716 | 15.651 | 179 | 18.96203 | 44.4805 | 5017 | 4.86 | 0.56 | 0.84 |
| 911 | 8490993 | 15.399 | 226 | 19.35089 | 44.5565 | 5820 | 4.78 | 0.68 | 1.02 |
| 912 | 8505670 | 15.058 | 183 | 19.70529 | 44.5460 | 4214 | 4.61 | 0.64 | 0.60 |
| 913 | 8544996 | 15.198 | 164 | 18.98870 | 44.6581 | 5463 | 4.75 | 0.69 | 0.97 |
| 914 | 8552202 | 15.371 | 195 | 19.24844 | 44.6075 | 5479 | 4.97 | 0.52 | 0.92 |
| 916 | 8628973 | 15.142 | 154 | 19.56366 | 44.7928 | 5401 | 4.48 | 0.96 | 1.01 |
| 917 | 8655354 | 15.172 | 204 | 20.06997 | 44.7063 | 5681 | 4.48 | 0.98 | 1.06 |
| 918 | 8672910 | 15.011 | 235 | 18.93183 | 44.8116 | 5321 | 4.54 | 0.88 | 0.99 |
| 920 | 8689031 | 15.067 | 140 | 19.44782 | 44.8873 | 5330 | 4.86 | 0.59 | 0.91 |
| 921 | 8689373 | 15.532 | 250 | 19.45613 | 44.8581 | 5046 | 4.71 | 0.69 | 0.89 |
| 922 | 8826878 | 15.365 | 261 | 19.65646 | 45.0344 | 5253 | 4.46 | 0.98 | 0.99 |
| 923 | 8883593 | 15.543 | 223 | 19.40616 | 45.1155 | 5669 | 4.60 | 0.85 | 1.04 |
| 924 | 8951215 | 15.229 | 174 | 19.41387 | 45.2444 | 5951 | 4.53 | 0.94 | 1.08 |

| | | | | | | | | | | |
|---|---|---|---|---|---|---|---|---|---|---|
| 926 | 9077124 | 15.603 | 227 | 19.09067 | 45.4143 | 5741 | 4.56 | 0.89 | 1.05 | |
| 928 | 9140402 | 15.251 | 128 | 18.98396 | 45.5991 | 5450 | 4.53 | 0.91 | 1.01 | |
| 929 | 9141746 | 15.649 | 152 | 19.03699 | 45.5789 | 5820 | 4.42 | 1.06 | 1.08 | |
| 931 | 9166862 | 15.272 | 144 | 19.75954 | 45.5686 | 5714 | 4.78 | 0.68 | 1.01 | |
| 934 | 9334289 | 15.843 | 309 | 19.21098 | 45.8165 | 5733 | 4.66 | 0.79 | 1.03 | |
| 935 | 9347899 | 15.237 | 186 | 19.60153 | 45.8531 | 6345 | 4.70 | 0.77 | 1.09 | |
| 936 | 9388479 | 15.073 | 188 | 18.91547 | 45.9588 | 3684 | 4.44 | 0.72 | 0.51 | |
| 937 | 9406990 | 15.412 | 176 | 19.53455 | 45.9147 | 5349 | 4.69 | 0.74 | 0.96 | |
| 938 | 9415172 | 15.596 | 199 | 19.73762 | 45.9768 | 5342 | 4.58 | 0.84 | 0.98 | |
| 939 | 9466668 | 15.065 | 157 | 19.50536 | 46.0974 | 5649 | 4.57 | 0.88 | 1.04 | |
| 940 | 9479273 | 15.017 | 73.1 | 19.80304 | 46.0274 | 5284 | 4.63 | 0.79 | 0.96 | |
| 941 | 9480189 | 15.471 | 305 | 19.82107 | 46.0233 | 4998 | 4.30 | 1.16 | 0.99 | |
| 942 | 9512687 | 15.386 | 203 | 18.97926 | 46.1723 | 4997 | 4.73 | 0.66 | 0.87 | |
| 943 | 9513865 | 15.733 | 202 | 19.02609 | 46.1809 | 5178 | 4.73 | 0.68 | 0.91 | |
| 944 | 9595686 | 15.361 | 196 | 19.65332 | 46.2076 | 5166 | 4.50 | 0.92 | 0.97 | |
| 945 | 9605514 | 15.083 | 166 | 19.86344 | 46.2647 | 6059 | 4.59 | 0.87 | 1.08 | |
| 947 | 9710326 | 15.190 | 228 | 19.45752 | 46.4293 | 3829 | 4.53 | 0.64 | 0.50 | |
| 949 | 9766437 | 15.485 | 218 | 19.33787 | 46.5791 | 5733 | 4.70 | 0.75 | 1.02 | |
| 951 | 9775938 | 15.223 | 198 | 19.60457 | 46.5793 | 4767 | 4.26 | 1.21 | 0.95 | |
| 952 | 9787239 | 15.801 | 291 | 19.85616 | 46.5743 | 3911 | 4.64 | 0.56 | 0.49 | |
| 953 | 9820483 | 15.954 | 349 | 19.09882 | 46.6925 | 5491 | 4.58 | 0.86 | 1.01 | |
| 954 | 9823457 | 15.219 | 139 | 19.21413 | 46.6150 | 5677 | 4.64 | 0.80 | 1.03 | |
| 955 | 9825625 | 15.067 | 133 | 19.29545 | 46.6175 | 6121 | 4.51 | 0.96 | 1.10 | |
| 956 | 9875711 | 15.223 | 197 | 18.92440 | 46.7896 | 4580 | 4.33 | 1.05 | 0.87 | |
| 960 | 8176650 | 15.513 | 165 | 19.69805 | 44.0107 | 5213 | 4.72 | 0.70 | 0.92 | |
| 961 | 8561063 | 15.920 | 157 | 19.48127 | 44.6192 | 4188 | 4.56 | 0.68 | 0.62 | 1 |
| 972 | 11013201 | 9.275 | 30.8 | 18.80002 | 48.5422 | 7779 | 3.82 | 2.54 | 1.56 | |
| 974 | 9414417 | 9.582 | 26 | 19.72018 | 45.9881 | 6069 | 4.34 | 1.19 | 1.12 | |
| 975 | 3632418 | 8.224 | 25.5 | 19.15745 | 38.7140 | 6017 | 4.38 | 1.09 | 1.05 | 1 |
| 976 | 3441784 | 9.729 | 149 | 19.39629 | 38.5389 | 7822 | 4.28 | 1.62 | 1.84 | 1 |
| 977 | 11192141 | 10.523 | 125 | 19.51464 | 48.8520 | 3240 | 4.90 | 0.27 | 0.21 | 1 |
| 981 | 8607720 | 10.733 | 101 | 18.91576 | 44.7133 | 5064 | 3.36 | 4.10 | 1.41 | |
| 984 | 1161345 | 11.631 | 101 | 19.40325 | 36.8399 | 5836 | 4.15 | 1.50 | 1.15 | |
| 986 | 2854698 | 14.138 | 175 | 19.46173 | 38.0141 | 5348 | 4.92 | 0.54 | 0.90 | |
| 987 | 7295235 | 12.550 | 42.4 | 19.70494 | 42.8064 | 5244 | 4.56 | 0.85 | 0.97 | |
| 988 | 2302548 | 13.562 | 126 | 19.42432 | 37.6092 | 5052 | 4.01 | 1.75 | 1.14 | |
| 991 | 10154388 | 13.581 | 48.4 | 19.76641 | 47.1829 | 5681 | 4.07 | 1.66 | 1.16 | |
| 992 | 1432789 | 15.214 | 195 | 19.43330 | 37.0593 | 5648 | 4.65 | 0.79 | 1.02 | |
| 993 | 1718189 | 14.246 | 129 | 19.38191 | 37.2527 | 5616 | 4.79 | 0.67 | 0.99 | |
| 994 | 1431122 | 14.613 | 174 | 19.40950 | 37.0613 | 5398 | 4.47 | 0.98 | 1.02 | |
| 998 | 1432214 | 15.661 | 278 | 19.42551 | 37.0730 | 5814 | 4.60 | 0.86 | 1.06 | |
| 999 | 2165002 | 15.391 | 351 | 19.49798 | 37.5678 | 5118 | 4.57 | 0.83 | 0.94 | |
| 1001 | 1871056 | 13.038 | 59.6 | 19.46964 | 37.3762 | 5956 | 3.81 | 2.33 | 1.29 | |
| 1002 | 1865042 | 13.615 | 82.4 | 19.37882 | 37.3551 | 5112 | 4.82 | 0.60 | 0.87 | |
| 1003 | 2438502 | 16.209 | 1381 | 19.35519 | 37.7267 | 5126 | 4.50 | 0.92 | 0.96 | |
| 1005 | 5780460 | 15.703 | 337 | 19.22777 | 41.0298 | 4975 | 4.46 | 0.95 | 0.94 | |
| 1010 | 1027438 | 13.620 | 94.8 | 19.42360 | 36.7845 | 6452 | 4.16 | 1.52 | 1.21 | |
| 1013 | 6047498 | 15.348 | 183 | 19.55904 | 41.3498 | 5359 | 4.88 | 0.58 | 0.92 | |

| | | | | | | | | | |
|---|---|---|---|---|---|---|---|---|---|
| 1014 | 8125580 | 15.759 | 261 | 19.99315 | 43.9039 | 4630 | 4.60 | 0.72 | 0.77 |
| 1015 | 8158127 | 14.500 | 111 | 19.27313 | 44.0044 | 6243 | 4.47 | 1.02 | 1.12 |
| 1017 | 8174625 | 15.007 | 146 | 19.65626 | 44.0777 | 5350 | 4.41 | 1.04 | 1.02 |
| 1019 | 8179973 | 10.266 | 58.6 | 19.76546 | 44.0092 | 4788 | 4.09 | 1.54 | 1.06 |
| 1020 | 2309719 | 12.899 | 74 | 19.52872 | 37.6066 | 5786 | 4.14 | 1.52 | 1.15 |
| 1022 | 2716853 | 15.762 | 340 | 19.48454 | 37.9556 | 5469 | 4.54 | 0.89 | 1.01 |
| 1024 | 2715135 | 14.496 | 204 | 19.46022 | 37.9023 | 4162 | 4.61 | 0.62 | 0.58 |
| 1026 | 1996399 | 14.748 | 189 | 19.09389 | 37.4268 | 3802 | 4.49 | 0.68 | 0.52 |
| 1029 | 2164169 | 14.757 | 163 | 19.48607 | 37.5808 | 5739 | 4.48 | 0.99 | 1.07 |
| 1030 | 2574338 | 15.399 | 232 | 19.40923 | 37.8593 | 6205 | 4.72 | 0.75 | 1.07 |
| 1031 | 2584163 | 15.160 | 251 | 19.55027 | 37.8462 | 5688 | 4.46 | 1.00 | 1.07 |
| 1032 | 2162635 | 13.862 | 125 | 19.46517 | 37.5326 | 4787 | 3.57 | 3.15 | 1.33 |
| 1050 | 5809890 | 13.999 | 136 | 19.79073 | 41.0973 | 4894 | 4.40 | 1.01 | 0.93 |
| 1051 | 6131236 | 15.391 | 127 | 19.61088 | 41.4518 | 5825 | 4.69 | 0.76 | 1.04 |
| 1052 | 5956342 | 15.381 | 215 | 19.36736 | 41.2448 | 6066 | 4.51 | 0.96 | 1.09 |
| 1053 | 5956656 | 15.376 | 264 | 19.37386 | 41.2011 | 5403 | 4.59 | 0.83 | 0.99 |
| 1054 | 6032981 | 11.899 | 188 | 19.24551 | 41.3073 | 5118 | 4.39 | 1.05 | 0.99 |
| 1059 | 6060203 | 14.803 | 114 | 19.77870 | 41.3865 | 5364 | 4.77 | 0.67 | 0.94 |
| 1060 | 5880320 | 14.348 | 104 | 19.53897 | 41.1354 | 6400 | 4.50 | 0.98 | 1.13 |
| 1061 | 6037187 | 14.502 | 114 | 19.34210 | 41.3815 | 5720 | 4.45 | 1.02 | 1.07 |
| 1072 | 8229696 | 14.743 | 136 | 19.36691 | 44.1436 | 5876 | 4.41 | 1.09 | 1.09 |
| 1078 | 10166274 | 15.438 | 294 | 19.98869 | 47.1575 | 3931 | 4.73 | 0.49 | 0.48 |
| 1081 | 10149023 | 15.220 | 125 | 19.65124 | 47.1328 | 5909 | 4.73 | 0.73 | 1.04 |
| 1082 | 10141900 | 15.692 | 270 | 19.47630 | 47.1574 | 5087 | 4.63 | 0.76 | 0.92 |
| 1083 | 10157458 | 15.302 | 124 | 19.82621 | 47.1646 | 5575 | 4.58 | 0.86 | 1.03 |
| 1085 | 10118816 | 15.233 | 188 | 18.73667 | 47.1882 | 3978 | 4.52 | 0.67 | 0.55 |
| 1086 | 10122255 | 14.609 | 107 | 18.87897 | 47.1557 | 5791 | 4.40 | 1.09 | 1.09 |
| 1089 | 3247268 | 14.696 | 127 | 19.62890 | 38.3555 | 5915 | 4.42 | 1.07 | 1.09 |
| 1094 | 2721030 | 15.678 | 296 | 19.54564 | 37.9698 | 5704 | 4.62 | 0.83 | 1.04 |
| 1095 | 3329204 | 15.617 | 183 | 19.17446 | 38.4023 | 5470 | 4.60 | 0.83 | 1.00 |
| 1099 | 2853093 | 15.435 | 238 | 19.43861 | 38.0358 | 5665 | 4.95 | 0.55 | 0.96 |
| 1101 | 3245969 | 15.681 | 320 | 19.60771 | 38.3935 | 4825 | 4.79 | 0.59 | 0.79 |
| 1102 | 3231341 | 14.925 | 204 | 19.36088 | 38.3438 | 5800 | 4.34 | 1.17 | 1.10 |
| 1106 | 3240158 | 14.818 | 163 | 19.51421 | 38.3582 | 5954 | 4.56 | 0.90 | 1.07 |
| 1108 | 3218908 | 14.604 | 128 | 19.09936 | 38.3749 | 5350 | 4.73 | 0.70 | 0.95 |
| 1109 | 3235672 | 14.792 | 222 | 19.44046 | 38.3481 | 4977 | 4.45 | 0.95 | 0.94 |
| 1110 | 2837111 | 14.794 | 120 | 19.16294 | 38.0884 | 5689 | 4.51 | 0.95 | 1.06 |
| 1111 | 3120276 | 15.212 | 236 | 19.48527 | 38.2845 | 5473 | 4.68 | 0.75 | 0.99 |
| 1112 | 3109930 | 14.633 | 144 | 19.31845 | 38.2623 | 5835 | 4.45 | 1.02 | 1.08 |
| 1113 | 2854914 | 13.703 | 90.9 | 19.46508 | 38.0551 | 6132 | 4.39 | 1.11 | 1.12 |
| 1114 | 3337425 | 14.928 | 187 | 19.34233 | 38.4042 | 5568 | 4.37 | 1.12 | 1.06 |
| 1115 | 3116412 | 13.974 | 98.5 | 19.42824 | 38.2166 | 5487 | 4.23 | 1.33 | 1.09 |
| 1116 | 2849805 | 13.333 | 64.4 | 19.39231 | 38.0576 | 5776 | 4.49 | 0.97 | 1.07 |
| 1117 | 3114811 | 12.808 | 57.3 | 19.40213 | 38.2120 | 6270 | 4.54 | 0.94 | 1.11 |
| 1118 | 2853446 | 13.835 | 87.6 | 19.44340 | 38.0282 | 6054 | 4.34 | 1.19 | 1.12 |
| 1128 | 6362874 | 13.507 | 61.1 | 19.42889 | 41.7034 | 5281 | 4.76 | 0.67 | 0.93 |
| 1129 | 6272413 | 15.456 | 131 | 19.15151 | 41.6356 | 4904 | 4.33 | 1.10 | 0.96 |
| 1141 | 8346392 | 15.950 | 422 | 18.84827 | 44.3465 | 4021 | 4.54 | 0.67 | 0.56 |

| | | | | | | | | | | |
|---|---|---|---|---|---|---|---|---|---|---|
| 1142 | 8288947 | 15.764 | 201 | 19.16516 | 44.2375 | 5141 | 4.69 | 0.71 | 0.91 | |
| 1144 | 8302450 | 15.282 | 84.8 | 19.52398 | 44.2006 | 5730 | 4.57 | 0.88 | 1.05 | |
| 1145 | 8313667 | 14.143 | 106 | 19.76991 | 44.2647 | 5808 | 4.56 | 0.90 | 1.06 | |
| 1146 | 8351704 | 15.649 | 178 | 19.04190 | 44.3109 | 3868 | 4.70 | 0.51 | 0.46 | |
| 1148 | 8410727 | 13.908 | 71.7 | 18.81334 | 44.4223 | 6158 | 4.28 | 1.28 | 1.14 | |
| 1149 | 8349405 | 15.602 | 136 | 18.95904 | 44.3913 | 5195 | 4.57 | 0.84 | 0.96 | |
| 1150 | 8278371 | 13.326 | 53.7 | 18.78443 | 44.2926 | 5601 | 4.52 | 0.93 | 1.04 | |
| 1151 | 8280511 | 13.404 | 44.1 | 18.86686 | 44.2842 | 5431 | 4.44 | 1.01 | 1.03 | |
| 1152 | 10287248 | 13.987 | 621 | 19.77971 | 47.3273 | 4069 | 4.57 | 0.65 | 0.58 | 1 |
| 1159 | 10354039 | 15.332 | 159 | 19.80099 | 47.4864 | 4886 | 4.48 | 0.91 | 0.90 | |
| 1160 | 10330115 | 15.987 | 352 | 19.17691 | 47.4148 | 5164 | 4.88 | 0.56 | 0.87 | |
| 1161 | 10426656 | 14.678 | 113 | 19.94183 | 47.5938 | 5109 | 4.36 | 1.10 | 0.99 | |
| 1162 | 10528068 | 12.783 | 45.7 | 19.25788 | 47.7594 | 5833 | 4.28 | 1.26 | 1.11 | |
| 1163 | 10468940 | 14.968 | 163 | 19.42015 | 47.6978 | 5419 | 4.50 | 0.93 | 1.01 | |
| 1164 | 10341831 | 14.960 | 188 | 19.51593 | 47.4907 | 3851 | 4.42 | 0.76 | 0.56 | |
| 1165 | 10337517 | 13.916 | 85.7 | 19.40420 | 47.4848 | 5549 | 4.27 | 1.26 | 1.08 | |
| 1166 | 10351231 | 15.440 | 128 | 19.74209 | 47.4388 | 5833 | 4.55 | 0.91 | 1.07 | |
| 1168 | 10460629 | 13.997 | 70.3 | 19.17245 | 47.6000 | 6209 | 4.23 | 1.37 | 1.16 | |
| 1169 | 10319385 | 13.248 | 49.5 | 18.77008 | 47.4700 | 5693 | 4.59 | 0.85 | 1.04 | |
| 1170 | 10482160 | 14.657 | 118 | 19.74881 | 47.6790 | 5558 | 4.94 | 0.55 | 0.94 | |
| 1175 | 10350571 | 13.290 | 63.7 | 19.72804 | 47.4486 | 5394 | 4.59 | 0.84 | 0.99 | |
| 1176 | 3749365 | 15.715 | 362 | 19.47122 | 38.8423 | 4601 | 4.69 | 0.65 | 0.74 | |
| 1177 | 3547091 | 15.537 | 861 | 19.47190 | 38.6315 | 5360 | 4.74 | 0.69 | 0.95 | |
| 1187 | 3848972 | 14.489 | 193 | 19.40475 | 38.9990 | 5286 | 4.85 | 0.59 | 0.91 | |
| 1192 | 3644071 | 14.215 | 86.6 | 19.40214 | 38.7039 | 5399 | 4.35 | 1.13 | 1.04 | |
| 1193 | 3942446 | 15.289 | 146 | 19.19254 | 39.0787 | 5486 | 4.37 | 1.11 | 1.05 | |
| 1198 | 3447722 | 15.319 | 257 | 19.49865 | 38.5149 | 6297 | 4.71 | 0.76 | 1.08 | |
| 1199 | 3859079 | 14.887 | 183 | 19.58293 | 38.9393 | 4746 | 4.50 | 0.86 | 0.85 | |
| 1201 | 4061149 | 15.597 | 388 | 19.50514 | 39.1209 | 3852 | 4.72 | 0.49 | 0.46 | |
| 1202 | 3444588 | 15.854 | 434 | 19.44616 | 38.5729 | 4196 | 4.74 | 0.53 | 0.57 | |
| 1203 | 3962243 | 15.368 | 210 | 19.57121 | 39.0363 | 5812 | 4.55 | 0.90 | 1.06 | |
| 1204 | 3438507 | 15.291 | 243 | 19.33336 | 38.5399 | 6084 | 4.53 | 0.95 | 1.09 | |
| 1205 | 3869014 | 14.507 | 143 | 19.73168 | 38.9466 | 5951 | 4.50 | 0.97 | 1.08 | |
| 1207 | 3732821 | 15.187 | 172 | 19.12781 | 38.8722 | 5058 | 4.47 | 0.94 | 0.95 | |
| 1208 | 3962440 | 13.594 | 65.8 | 19.57458 | 39.0226 | 6293 | 4.43 | 1.07 | 1.13 | |
| 1210 | 3962357 | 14.377 | 122 | 19.57306 | 39.0295 | 6156 | 4.74 | 0.73 | 1.06 | |
| 1212 | 3749134 | 14.981 | 201 | 19.46712 | 38.8207 | 5702 | 4.45 | 1.02 | 1.07 | |
| 1214 | 3660924 | 14.621 | 107 | 19.68111 | 38.7743 | 5538 | 4.35 | 1.15 | 1.06 | |
| 1215 | 3939150 | 13.420 | 60.9 | 19.10554 | 39.0772 | 6306 | 4.24 | 1.35 | 1.16 | |
| 1216 | 3839488 | 13.459 | 67.9 | 19.20891 | 38.9969 | 5842 | 4.33 | 1.19 | 1.10 | |
| 1218 | 3442055 | 13.331 | 64 | 19.40127 | 38.5456 | 5617 | 4.38 | 1.10 | 1.07 | |
| 1219 | 3440861 | 14.463 | 126 | 19.37779 | 38.5683 | 4972 | 4.44 | 0.97 | 0.94 | |
| 1220 | 4043190 | 12.988 | 54 | 19.15386 | 39.1447 | 4874 | 4.49 | 0.89 | 0.89 | |
| 1221 | 3640905 | 11.584 | 56.9 | 19.34048 | 38.7022 | 4972 | 3.39 | 3.93 | 1.39 | |
| 1222 | 4060815 | 12.203 | 75.2 | 19.49979 | 39.1932 | 4722 | 4.56 | 0.78 | 0.82 | |
| 1226 | 6621116 | 15.324 | 139 | 19.72446 | 42.0627 | 5045 | 4.47 | 0.94 | 0.95 | |
| 1227 | 6629332 | 13.997 | 401 | 19.84714 | 42.0474 | 5452 | 4.82 | 0.63 | 0.95 | |
| 1230 | 6470149 | 12.263 | 127 | 19.92988 | 41.8122 | 4864 | 3.02 | 6.16 | 1.46 | |
| 1236 | 6677841 | 13.659 | 124 | 19.15941 | 42.1948 | 6562 | 4.46 | 1.04 | 1.15 | |
| 1238 | 6383821 | 14.556 | 113 | 19.81393 | 41.7817 | 5408 | 4.52 | 0.92 | 1.01 | |
| 1240 | 6690082 | 14.466 | 99.9 | 19.42603 | 42.1806 | 5543 | 4.55 | 0.89 | 1.03 | |
| 1241 | 6448890 | 12.440 | 104 | 19.58389 | 41.8719 | 4857 | 3.23 | 4.83 | 1.43 | |
| 1242 | 6607447 | 13.750 | 48.4 | 19.48189 | 42.0354 | 6234 | 4.50 | 0.98 | 1.11 | |
| 1244 | 6692833 | 14.503 | 91.1 | 19.48248 | 42.1543 | 5871 | 4.65 | 0.80 | 1.05 | |
| 1245 | 6693640 | 14.200 | 69.6 | 19.49855 | 42.1423 | 6119 | 4.52 | 0.95 | 1.10 | |
| 1246 | 6441738 | 14.898 | 104 | 19.42850 | 41.8782 | 6032 | 4.43 | 1.06 | 1.10 | |
| 1257 | 8751933 | 14.651 | 104 | 19.41501 | 44.9274 | 5142 | 4.30 | 1.19 | 1.02 | |
| 1258 | 8630788 | 15.773 | 256 | 19.60721 | 44.7707 | 5517 | 4.72 | 0.72 | 0.99 | |
| 1261 | 8678594 | 15.120 | 118 | 19.14499 | 44.8786 | 5760 | 4.55 | 0.90 | 1.06 | |
| 1264 | 8612847 | 15.762 | 216 | 19.10499 | 44.7952 | 5418 | 4.62 | 0.81 | 0.99 | |
| 1266 | 8547140 | 15.314 | 189 | 19.07168 | 44.6646 | 4342 | 4.57 | 0.70 | 0.67 | |
| 1268 | 8813698 | 14.814 | 94.4 | 19.32598 | 45.0057 | 6064 | 4.43 | 1.06 | 1.11 | |
| 1270 | 8564587 | 14.809 | 118 | 19.57609 | 44.6570 | 5145 | 4.87 | 0.57 | 0.87 | |
| 1273 | 8806072 | 14.862 | 118 | 19.06772 | 45.0078 | 5468 | 4.60 | 0.84 | 1.00 | |
| 1275 | 8583696 | 13.672 | 78.4 | 19.94811 | 44.6939 | 5427 | 4.43 | 1.02 | 1.03 | |
| 1276 | 8804283 | 14.766 | 105 | 19.00241 | 45.0711 | 5440 | 4.85 | 0.60 | 0.94 | |
| 1278 | 8609450 | 15.204 | 129 | 18.97849 | 44.7977 | 5876 | 4.58 | 0.88 | 1.07 | |
| 1279 | 8628758 | 13.749 | 71.2 | 19.55815 | 44.7851 | 5582 | 4.81 | 0.65 | 0.98 | |
| 1281 | 8742590 | 14.427 | 93.6 | 19.13112 | 44.9925 | 5546 | 4.62 | 0.82 | 1.02 | |
| 1282 | 8822366 | 12.547 | 42 | 19.54866 | 45.0533 | 6060 | 4.25 | 1.33 | 1.14 | |
| 1283 | 8700771 | 11.731 | 23 | 19.73381 | 44.8535 | 5776 | 4.47 | 0.93 | 0.92 | 1 |
| 1285 | 10599397 | 14.546 | 147 | 19.44247 | 47.8323 | 5278 | 4.52 | 0.90 | 0.98 | |
| 1288 | 10790387 | 15.128 | 129 | 19.26801 | 48.1198 | 6130 | 4.43 | 1.07 | 1.11 | |
| 1298 | 10604335 | 15.847 | 338 | 19.57425 | 47.8390 | 4157 | 4.48 | 0.77 | 0.64 | |
| 1299 | 10864656 | 12.183 | 94.7 | 19.55215 | 48.2859 | 4950 | 3.47 | 3.58 | 1.37 | |
| 1300 | 10975146 | 14.285 | 161 | 19.43961 | 48.4456 | 4369 | 4.51 | 0.77 | 0.71 | |
| 1301 | 10538176 | 15.824 | - | 19.54310 | 47.7297 | 5414 | 4.81 | 0.63 | 0.94 | |
| 1302 | 10724369 | 14.755 | 118 | 19.27224 | 48.0181 | 5659 | 4.60 | 0.84 | 1.04 | |
| 1303 | 10867062 | 14.965 | 114 | 19.61534 | 48.2045 | 5239 | 4.22 | 1.32 | 1.06 | |
| 1304 | 10744335 | 15.803 | 203 | 19.77432 | 48.0826 | 5733 | 4.65 | 0.80 | 1.03 | |
| 1305 | 10730034 | 15.173 | 193 | 19.42606 | 48.0656 | 5235 | 4.71 | 0.70 | 0.93 | |
| 1306 | 10858691 | 15.587 | 198 | 19.37841 | 48.2943 | 5714 | 4.52 | 0.94 | 1.06 | |
| 1307 | 10973814 | 14.775 | 97.7 | 19.39516 | 48.4436 | 5559 | 4.40 | 1.07 | 1.06 | |
| 1308 | 10586208 | 13.971 | 75.2 | 19.02829 | 47.8346 | 5709 | 4.51 | 0.95 | 1.06 | |
| 1309 | 10854768 | 13.832 | 77.5 | 19.25702 | 48.2085 | 6869 | 4.27 | 1.33 | 1.22 | |
| 1310 | 10964440 | 14.569 | 94.3 | 19.04418 | 48.4354 | 5800 | 4.52 | 0.94 | 1.07 | |
| 1311 | 10713616 | 13.498 | 51.1 | 18.90220 | 48.0943 | 5920 | 4.20 | 1.40 | 1.14 | |
| 1312 | 10963242 | 14.706 | 101 | 18.99403 | 48.4316 | 6157 | 4.64 | 0.82 | 1.08 | |
| 1314 | 10585852 | 13.242 | 68.9 | 19.01474 | 47.8814 | 5033 | 3.70 | 2.66 | 1.29 | |
| 1315 | 10928043 | 13.137 | 57 | 19.68626 | 48.3748 | 6160 | 4.32 | 1.23 | 1.13 | |
| 1316 | 10794087 | 11.926 | 32.8 | 19.37016 | 48.1263 | 5861 | 4.44 | 0.98 | 0.97 | 1 |
| 1325 | 4282872 | 15.062 | 222 | 19.74901 | 39.3703 | 5589 | 4.83 | 0.63 | 0.98 | |
| 1328 | 4074736 | 15.671 | 206 | 19.72039 | 39.1780 | 5425 | 4.70 | 0.72 | 0.97 | |
| 1329 | 4072526 | 15.030 | 190 | 19.69139 | 39.1772 | 5157 | 4.61 | 0.80 | 0.94 | |
| 1335 | 4155328 | 13.968 | 135 | 19.40134 | 39.2207 | 5970 | 4.03 | 1.76 | 1.20 | |
| 1336 | 4077526 | 14.820 | 184 | 19.75724 | 39.1152 | 5843 | 4.38 | 1.12 | 1.09 | |
| 1337 | 4243911 | 14.829 | 169 | 19.04116 | 39.3795 | 5067 | 4.57 | 0.83 | 0.93 | |
| 1338 | 4466677 | 14.609 | 173 | 19.48507 | 39.5502 | 5597 | 4.44 | 1.03 | 1.06 | |
| 1339 | 4135665 | 14.801 | 125 | 18.97217 | 39.2202 | 5533 | 4.34 | 1.16 | 1.06 | |
| 1341 | 4650674 | 14.946 | 119 | 19.36516 | 39.7320 | 5933 | 4.40 | 1.09 | 1.10 | |
| 1342 | 4275721 | 14.207 | 110 | 19.65133 | 39.3985 | 5972 | 4.41 | 1.09 | 1.10 | |
| 1344 | 4136466 | 13.446 | 57.1 | 18.99068 | 39.2406 | 5731 | 4.57 | 0.88 | 1.05 | |
| 1353 | 7303287 | 13.956 | 113 | 19.83102 | 42.8828 | 6031 | 4.10 | 1.62 | 1.19 | |
| 1355 | 7211141 | 15.897 | 210 | 19.66440 | 42.7086 | 5529 | 4.97 | 0.52 | 0.93 | |
| 1360 | 7102227 | 15.596 | 287 | 19.09400 | 42.6816 | 4953 | 4.66 | 0.72 | 0.87 | |
| 1361 | 6960913 | 14.995 | 149 | 19.68697 | 42.4753 | 4050 | 4.62 | 0.59 | 0.53 | |
| 1363 | 6936909 | 15.934 | 271 | 19.19442 | 42.4373 | 5702 | 4.64 | 0.80 | 1.03 | |
| 1364 | 6962977 | 15.956 | 219 | 19.71774 | 42.4243 | 5311 | 4.48 | 0.96 | 1.00 | |
| 1366 | 6932987 | 15.368 | 179 | 19.09054 | 42.4065 | 5559 | 4.64 | 0.79 | 1.01 | |
| 1367 | 6934291 | 15.055 | 159 | 19.12582 | 42.4713 | 4927 | 4.64 | 0.74 | 0.87 | |
| 1369 | 7287415 | 14.878 | 124 | 19.56460 | 42.8224 | 5778 | 4.47 | 0.99 | 1.07 | |
| 1370 | 6924203 | 14.931 | 154 | 18.82219 | 42.4638 | 5489 | 4.67 | 0.77 | 0.99 | |
| 1372 | 6880531 | 15.384 | 120 | 19.75999 | 42.3871 | 5760 | 4.55 | 0.91 | 1.06 | |
| 1375 | 6766634 | 13.709 | 52.5 | 19.22136 | 42.2614 | 6169 | 4.36 | 1.17 | 1.14 | 1 |
| 1376 | 6774826 | 13.997 | 75.4 | 19.40091 | 42.2831 | 7211 | 4.29 | 1.50 | 1.61 | 1 |
| 1377 | 7211469 | 14.788 | 85.8 | 19.67029 | 42.7568 | 6027 | 4.41 | 1.09 | 1.11 | |
| 1378 | 7375795 | 13.601 | 51.1 | 19.70071 | 42.9306 | 5234 | 4.48 | 0.85 | 0.79 | 1 |
| 1379 | 7211221 | 13.687 | 51.9 | 19.66570 | 42.7277 | 5635 | 4.84 | 0.63 | 0.98 | |
| 1382 | 9446824 | 15.653 | 576 | 18.81229 | 46.0069 | 5921 | 4.39 | 1.10 | 1.10 | |
| 1385 | 9278553 | 15.840 | 215 | 19.39224 | 45.7659 | 5848 | 4.76 | 0.70 | 1.02 | |
| 1387 | 8949247 | 14.669 | 92 | 19.36620 | 45.2059 | 5696 | 4.47 | 0.99 | 1.06 | |
| 1391 | 8958035 | 14.360 | 152 | 19.58060 | 45.2864 | 6012 | 4.51 | 0.96 | 1.09 | |
| 1395 | 8935810 | 15.977 | 301 | 18.94731 | 45.2581 | 5279 | 4.83 | 0.61 | 0.91 | |
| 1396 | 9455556 | 15.843 | 305 | 19.15947 | 46.0593 | 5667 | 4.57 | 0.88 | 1.04 | |
| 1401 | 9030447 | 13.440 | 114 | 19.65267 | 45.3916 | 6693 | 4.30 | 1.28 | 1.19 | |
| 1402 | 9034103 | 15.906 | 184 | 19.72841 | 45.3839 | 5738 | 4.74 | 0.71 | 1.02 | |
| 1403 | 9214942 | 15.473 | 214 | 19.36533 | 45.6007 | 4209 | 4.57 | 0.67 | 0.61 | |
| 1404 | 8874090 | 15.931 | 280 | 19.13697 | 45.1804 | 4400 | 4.73 | 0.58 | 0.65 | |
| 1405 | 9264949 | 15.965 | 178 | 18.90495 | 45.7348 | 6009 | 4.61 | 0.85 | 1.07 | |
| 1406 | 9271752 | 14.630 | 103 | 19.16352 | 45.7044 | 6027 | 4.49 | 0.99 | 1.09 | |
| 1407 | 9007866 | 15.755 | 179 | 19.06591 | 45.3778 | 5487 | 4.81 | 0.64 | 0.96 | |
| 1408 | 9150827 | 14.688 | 166 | 19.35297 | 45.5646 | 4036 | 4.47 | 0.75 | 0.60 | |
| 1409 | 9391208 | 15.198 | 119 | 19.01645 | 45.9502 | 5593 | 4.38 | 1.10 | 1.07 | |
| 1410 | 9391506 | 15.300 | 137 | 19.02903 | 45.9760 | 5929 | 4.54 | 0.93 | 1.08 | |
| 1412 | 8950853 | 13.601 | 67.7 | 19.40498 | 45.2426 | 5997 | 4.40 | 1.10 | 1.10 | |
| 1413 | 9006449 | 14.447 | 99.9 | 19.02306 | 45.3678 | 5394 | 4.44 | 1.01 | 1.02 | |
| 1419 | 11125936 | 15.507 | 299 | 19.02746 | 48.7252 | 5848 | 4.46 | 1.01 | 1.08 | |
| 1422 | 11497958 | 15.921 | 300 | 19.10267 | 49.4373 | 3712 | 4.41 | 0.75 | 0.53 | |
| 1423 | 11177707 | 15.740 | 235 | 18.95009 | 48.8096 | 5288 | 4.77 | 0.66 | 0.93 | |
| 1424 | 11611600 | 15.127 | 137 | 19.48857 | 49.6535 | 4667 | 4.72 | 0.63 | 0.75 | |
| 1425 | 11254382 | 15.269 | 198 | 19.75480 | 48.9715 | 5639 | 4.85 | 0.61 | 0.98 | |
| 1426 | 11122894 | 14.232 | 76.1 | 18.88061 | 48.7776 | 5854 | 4.43 | 1.05 | 1.09 | |
| 1427 | 11129738 | 15.840 | 229 | 19.19052 | 48.7741 | 4027 | 4.53 | 0.68 | 0.56 | |
| 1428 | 11401182 | 14.631 | 105 | 19.39717 | 49.2038 | 4757 | 4.51 | 0.85 | 0.85 | |
| 1429 | 11030711 | 15.531 | 162 | 19.49010 | 48.5111 | 5595 | 4.58 | 0.87 | 1.03 | |
| 1430 | 11176127 | 15.415 | 212 | 18.86932 | 48.8254 | 4502 | 4.60 | 0.71 | 0.73 | |
| 1432 | 11014932 | 15.017 | 170 | 18.88046 | 48.5805 | 5675 | 4.65 | 0.80 | 1.03 | |
| 1433 | 11288505 | 15.650 | 203 | 19.13713 | 49.0460 | 5560 | 4.83 | 0.63 | 0.97 | |
| 1434 | 11493431 | 14.782 | 167 | 18.90032 | 49.4510 | 4731 | 4.47 | 0.89 | 0.86 | |
| 1435 | 11037335 | 14.201 | 77.7 | 19.68591 | 48.5997 | 5744 | 4.70 | 0.75 | 1.03 | |
| 1436 | 11389771 | 14.271 | 152 | 18.93220 | 49.2330 | 5616 | 4.52 | 0.93 | 1.05 | |
| 1437 | 11599038 | 15.280 | 179 | 18.98978 | 49.6924 | 5791 | 4.46 | 1.01 | 1.07 | |
| 1438 | 11193263 | 14.056 | 74.6 | 19.55255 | 48.8116 | 5596 | 4.43 | 1.04 | 1.06 | |
| 1439 | 11027624 | 12.849 | 44.6 | 19.39012 | 48.5213 | 5967 | 4.31 | 1.22 | 1.12 | |
| 1440 | 11032227 | 15.451 | 131 | 19.53896 | 48.5754 | 5970 | 4.68 | 0.77 | 1.05 | |
| 1441 | 11356260 | 15.135 | 132 | 19.64535 | 49.1402 | 5575 | 4.43 | 1.04 | 1.06 | |
| 1442 | 11600889 | 12.521 | 52.7 | 19.06909 | 49.6145 | 5469 | 4.24 | 1.30 | 1.08 | |
| 1444 | 11043167 | 13.949 | 92.2 | 19.82611 | 48.5607 | 6101 | 4.25 | 1.33 | 1.14 | |
| 1445 | 11336883 | 12.320 | 44.2 | 18.92912 | 49.1103 | 6336 | 4.36 | 1.18 | 1.14 | |
| 1448 | 9705459 | 15.418 | 1365 | 19.32035 | 46.4917 | 5658 | 4.36 | 1.14 | 1.08 | |
| 1452 | 7449844 | 13.630 | 440 | 19.55210 | 43.0558 | 6834 | 4.10 | 1.66 | 1.27 | |
| 1459 | 9761199 | 15.692 | 484 | 19.14288 | 46.5081 | 4060 | 4.40 | 0.84 | 0.65 | |
| 1463 | 7672940 | 12.328 | 48.2 | 19.21724 | 43.3765 | 6020 | 4.38 | 1.09 | 1.05 | 1 |
| 1465 | 11702948 | 14.245 | 179 | 19.07663 | 49.8675 | 5619 | 4.85 | 0.62 | 0.98 | |
| 1468 | 9851226 | 15.262 | 276 | 19.92150 | 46.6503 | 5635 | 4.44 | 1.02 | 1.06 | |
| 1472 | 7761545 | 15.061 | 103 | 19.64216 | 43.4005 | 5455 | 4.92 | 0.56 | 0.93 | |
| 1474 | 12365184 | 13.005 | 70.4 | 19.69453 | 51.1848 | 6498 | 4.08 | 1.68 | 1.23 | |
| 1475 | 4770365 | 15.937 | - | 19.82869 | 39.8478 | 4097 | 4.59 | 0.63 | 0.56 | |
| 1476 | 12406749 | 15.792 | 292 | 19.39028 | 51.2228 | 5275 | 4.52 | 0.90 | 0.98 | |
| 1477 | 7811397 | 15.917 | 299 | 19.17693 | 43.5057 | 5346 | 4.71 | 0.71 | 0.95 | |
| 1478 | 12403119 | 12.450 | 36.3 | 19.25658 | 51.2091 | 5441 | 4.73 | 0.70 | 0.97 | |
| 1480 | 7512982 | 15.887 | 310 | 19.11612 | 43.1901 | 4948 | 4.74 | 0.65 | 0.85 | |
| 1486 | 7898352 | 15.505 | 170 | 19.62113 | 43.6293 | 5688 | 4.62 | 0.83 | 1.04 | |
| 1488 | 9589323 | 15.623 | 212 | 19.49026 | 46.2263 | 4984 | 4.51 | 0.88 | 0.92 | |
| 1489 | 9823487 | 15.554 | 205 | 19.21578 | 46.6059 | 5014 | 4.61 | 0.78 | 0.91 | |
| 1494 | 11821363 | 15.858 | - | 19.69714 | 50.0772 | 4559 | 4.54 | 0.78 | 0.77 | |
| 1495 | 7629518 | 15.430 | 139 | 19.84430 | 43.2490 | 5661 | 4.50 | 0.95 | 1.05 | |
| 1498 | 9636135 | 15.776 | 188 | 19.00381 | 46.3207 | 5941 | 4.67 | 0.78 | 1.05 | |
| 1499 | 7841925 | 14.480 | 92.1 | 19.84829 | 43.5274 | 5264 | 4.38 | 1.07 | 1.01 | |
| 1501 | 7439316 | 15.835 | 269 | 19.34149 | 43.0856 | 4659 | 4.51 | 0.83 | 0.82 | |
| 1502 | 12061238 | 15.202 | 140 | 19.41541 | 50.5670 | 5034 | 4.64 | 0.75 | 0.90 | |
| 1503 | 12400538 | 14.827 | 134 | 19.15235 | 51.2500 | 5356 | 4.90 | 0.56 | 0.91 | |
| 1505 | 9813499 | 15.695 | 213 | 18.82423 | 46.6752 | 5701 | 4.54 | 0.92 | 1.05 | |
| 1506 | 12254792 | 14.982 | 177 | 19.29497 | 50.9941 | 5582 | 4.62 | 0.82 | 1.02 | |
| 1507 | 12020218 | 15.259 | 198 | 19.69350 | 50.4786 | 5881 | 4.54 | 0.92 | 1.07 | |
| 1508 | 7690844 | 15.689 | 208 | 19.63514 | 43.3682 | 5695 | 4.93 | 0.56 | 0.97 | |
| 1510 | 11870545 | 15.929 | 275 | 19.67228 | 50.1146 | 4772 | 4.79 | 0.59 | 0.77 | |
| 1511 | 7901948 | 15.106 | 128 | 19.69450 | 43.6056 | 5520 | 4.61 | 0.83 | 1.01 | |
| 1512 | 11955499 | 14.880 | 155 | 19.12583 | 50.3771 | 5068 | 4.70 | 0.70 | 0.90 | |
| 1515 | 7871954 | 14.390 | 121 | 18.87570 | 43.6571 | 4103 | 4.57 | 0.66 | 0.58 | |
| 1516 | 12418724 | 14.829 | 155 | 19.76021 | 51.2713 | 6092 | 4.56 | 0.90 | 1.09 | |
| 1517 | 7456001 | 14.683 | 117 | 19.66379 | 43.0143 | 5815 | 4.43 | 1.06 | 1.08 | |
| 1518 | 7549209 | 15.219 | 117 | 19.82009 | 43.1419 | 5563 | 4.57 | 0.88 | 1.03 | |
| 1519 | 7663405 | 15.369 | 154 | 18.89855 | 43.3838 | 4994 | 4.68 | 0.71 | 0.88 | |
| 1520 | 9765975 | 14.516 | 85.8 | 19.32394 | 46.5227 | 5235 | 4.51 | 0.91 | 0.98 | |
| 1521 | 9818462 | 14.817 | 128 | 19.01975 | 46.6034 | 4962 | 4.42 | 0.99 | 0.95 | |
| 1522 | 12266636 | 14.264 | 106 | 19.72044 | 50.9098 | 5609 | 4.41 | 1.07 | 1.06 | |
| 1523 | 9850893 | 14.673 | 134 | 19.91534 | 46.6887 | 5207 | 4.37 | 1.08 | 1.01 | |
| 1525 | 7869917 | 12.082 | 77.1 | 18.79808 | 43.6728 | 6680 | 4.24 | 1.38 | 1.20 | |
| 1526 | 9824805 | 15.216 | 135 | 19.26617 | 46.6708 | 6011 | 4.69 | 0.77 | 1.05 | |
| 1527 | 7768451 | 14.879 | 94.1 | 19.77809 | 43.4984 | 5470 | 4.24 | 1.31 | 1.08 | |
| 1528 | 7691260 | 14.083 | 73.3 | 19.64333 | 43.3356 | 5087 | 4.95 | 0.51 | 0.84 | |
| 1529 | 9821454 | 14.307 | 77.3 | 19.13597 | 46.6401 | 6074 | 4.53 | 0.94 | 1.09 | |
| 1530 | 11954842 | 13.029 | 43.9 | 19.09558 | 50.3003 | 5973 | 4.41 | 1.08 | 1.10 | |
| 1531 | 11764462 | 13.069 | 54.7 | 19.50563 | 49.9232 | 5811 | 4.41 | 1.07 | 1.08 | |
| 1532 | 11656246 | 12.841 | 48.9 | 19.24410 | 49.7372 | 6225 | 4.31 | 1.24 | 1.14 | |
| 1533 | 7808587 | 13.939 | 60.1 | 19.08478 | 43.5952 | 6122 | 4.43 | 1.07 | 1.11 | |
| 1534 | 4741126 | 13.470 | 51.5 | 19.33939 | 39.8170 | 6193 | 4.42 | 1.09 | 1.12 | |
| 1535 | 11669125 | 13.046 | 41.8 | 19.70141 | 49.7386 | 5924 | 4.33 | 1.19 | 1.11 | |
| 1536 | 12159249 | 12.710 | 47.6 | 19.43115 | 50.7595 | 5848 | 4.41 | 1.08 | 1.09 | |
| 1537 | 9872292 | 11.740 | 31.3 | 18.76411 | 46.7899 | 6063 | 4.37 | 1.12 | 1.07 | 1 |
| 1540 | 5649956 | 15.559 | 759 | 19.94022 | 40.8493 | 5390 | 4.53 | 0.90 | 1.00 | |
| 1541 | 4840513 | 15.189 | 705 | 19.54634 | 39.9073 | 6164 | 4.53 | 0.95 | 1.10 | |
| 1543 | 5270698 | 14.985 | 953 | 19.36316 | 40.4429 | 5821 | 4.54 | 0.92 | 1.06 | |
| 1546 | 5475431 | 14.456 | 177 | 19.90091 | 40.6396 | 5505 | 4.97 | 0.52 | 0.93 | |
| 1549 | 8053552 | 15.135 | 155 | 19.93706 | 43.8972 | 5401 | 4.36 | 1.12 | 1.04 | |
| 1553 | 7951018 | 15.182 | 122 | 19.22905 | 43.7635 | 5942 | 4.52 | 0.95 | 1.08 | |
| 1557 | 5371776 | 14.840 | 222 | 19.62945 | 40.5576 | 4783 | 4.25 | 1.22 | 0.96 | |
| 1560 | 8046659 | 15.042 | 111 | 19.82798 | 43.8766 | 5620 | 4.39 | 1.09 | 1.07 | |
| 1561 | 4940438 | 15.549 | 252 | 19.73556 | 40.0196 | 5659 | 4.48 | 0.98 | 1.06 | |
| 1564 | 5184584 | 15.287 | 127 | 19.40559 | 40.3553 | 5709 | 4.92 | 0.56 | 0.97 | |
| 1569 | 8009350 | 15.587 | 240 | 18.87531 | 43.8882 | 4639 | 4.61 | 0.72 | 0.77 | |
| 1573 | 5031857 | 14.373 | 115 | 19.78974 | 40.1386 | 5838 | 4.57 | 0.89 | 1.06 | |
| 1574 | 10028792 | 14.600 | 85.3 | 19.86113 | 46.9651 | 5537 | 4.58 | 0.85 | 1.02 | |
| 1576 | 5299459 | 14.072 | 152 | 19.85301 | 40.4177 | 5445 | 4.35 | 1.13 | 1.05 | |
| 1577 | 12506770 | 15.988 | 270 | 19.28604 | 51.4089 | 4165 | 4.61 | 0.62 | 0.58 | |
| 1581 | 7939330 | 15.481 | 200 | 18.82928 | 43.7345 | 5301 | 4.57 | 0.85 | 0.98 | |
| 1582 | 4918309 | 15.402 | 132 | 19.34191 | 40.0218 | 5384 | 4.80 | 0.64 | 0.94 | |
| 1583 | 12602568 | 15.068 | 135 | 19.37246 | 51.6958 | 5619 | 4.63 | 0.82 | 1.03 | |
| 1584 | 9941066 | 15.961 | 321 | 19.10569 | 46.8709 | 4147 | 4.52 | 0.72 | 0.62 | |
| 1585 | 5470739 | 15.401 | 217 | 19.83217 | 40.6808 | 5514 | 4.66 | 0.78 | 1.00 | |
| 1586 | 10022908 | 14.940 | 169 | 19.74515 | 46.9990 | 4692 | 4.70 | 0.65 | 0.77 | |
| 1587 | 9932970 | 15.700 | 254 | 18.77459 | 46.8142 | 5007 | 4.54 | 0.85 | 0.93 | |
| 1588 | 5617854 | 14.699 | 145 | 19.41020 | 40.8875 | 4106 | 4.58 | 0.64 | 0.57 | |
| 1589 | 5301750 | 14.764 | 160 | 19.88347 | 40.4961 | 5755 | 4.42 | 1.06 | 1.08 | |
| 1590 | 5542466 | 15.674 | 255 | 19.62432 | 40.7209 | 4830 | 4.50 | 0.88 | 0.88 | |
| 1591 | 10028140 | 15.372 | 173 | 19.84932 | 46.9419 | 5130 | 4.97 | 0.49 | 0.84 | |
| 1593 | 5289854 | 15.809 | 182 | 19.71901 | 40.4396 | 5676 | 4.54 | 0.91 | 1.05 | |
| 1595 | 10006581 | 14.904 | 128 | 19.31545 | 46.9538 | 5669 | 4.52 | 0.94 | 1.05 | |
| 1596 | 10027323 | 15.157 | 162 | 19.83399 | 46.9613 | 4656 | 4.39 | 0.99 | 0.87 | |
| 1597 | 5039228 | 12.681 | 94.7 | 19.89126 | 40.1734 | 6178 | 4.36 | 1.17 | 1.14 | 1 |
| 1598 | 10004738 | 14.279 | 107 | 19.25596 | 46.9867 | 5565 | 4.52 | 0.93 | 1.04 | |
| 1599 | 5474613 | 14.802 | 119 | 19.89159 | 40.6184 | 5627 | 4.48 | 0.98 | 1.05 | |
| 1601 | 5438757 | 14.659 | 108 | 19.25140 | 40.6642 | 5502 | 4.60 | 0.83 | 1.01 | |
| 1602 | 4860678 | 14.943 | 173 | 19.85313 | 39.9214 | 5596 | 4.58 | 0.87 | 1.03 | |
| 1603 | 5177104 | 14.429 | 96.4 | 19.25486 | 40.3891 | 5995 | 4.67 | 0.79 | 1.06 | |
| 1605 | 5009189 | 14.832 | 123 | 19.41592 | 40.1635 | 5680 | 4.33 | 1.17 | 1.08 | |
| 1606 | 9886661 | 13.984 | 76 | 19.33674 | 46.7134 | 5377 | 4.57 | 0.85 | 0.99 | |
| 1608 | 10055126 | 13.797 | 74.4 | 18.80409 | 47.0855 | 6030 | 4.39 | 1.12 | 1.11 | |
| 1609 | 5009743 | 13.956 | 62.5 | 19.42704 | 40.1255 | 6063 | 4.45 | 1.03 | 1.10 | |

**Table 2**
List of Planetary Candidates and their Characteristics

Key:

| | |
|---|---|
| KOI | Kepler Object of Interest number † indicates that this KOI was detected on the basis of a single transit with the period derived from the transit duration and stellar radius. |
| Dur | Transit duration, first contact to last contact |
| Depth | Transit depth at center of transit |
| SNR | Total SNR of all transits detected. SNR=Depth/(Std*sqrt(N)) where Std is the standard deviation of all data outside of transits (Q0 through Q5) and N is the total number of measurements inside of all transits. |
| $t_0$, $t_0$_unc | Time of a transit center based on a linear fit to all observed transits and its uncertainty |
| Period, P_unc | Average interval between transits based on a linear fit to all observed transits and uncertainty |
| a/R*, a/R*_unc | Ratioofsemi-majoraxistostellarradiusassumingzeroeccentricity,aparameterderivedfromthelightcurve,anduncertainty<br>Note: For planets in non-circular orbits, a/R* is the scaled planet-star separation at the time of transit. |
| r/R*, r/R*_unc | Ratio of planet radius to stellar radius and uncertainty |
| b, b_unc | Impact parameter of the transit and uncertainty. Note, there is a strong co-variance between b and a/R* |
| $R_p$ | Radius of planet in units of $R_{Earth}$=6378km |
| a | Semi-major axis of orbit based on Newton's generalization of Kepler's third law and the stellar mass in Appendix1. |
| $T_{eq}$ | Equilibrium temperature of the planet (see main text and Appendix 5 for discussion) |
| EB prob | Probability of background eclipsing binary confused with planet's host star (see text for discussion) |
| V | Vetting flag<br>1 Confirmed and published planet<br>2 Strong probability candidate, cleanly passes tests that were applied<br>3 Moderate probability candidate, not all tests cleanly passed but no definite test failures<br>4 Insufficient follow-up to perform full suite of vetting tests |
| FOP | Follow-up observation description (to be revised)<br>1 Reconnaissance spectra taken<br>2 Adaptive optics observations taken<br>3 Speckle observations taken<br>4 10m/s RV spectra taken<br>5 2m/s RV spectra taken<br>NoObs No observations yet taken |
| N | Notes flag. A "1" indicates a note on this KOI or its host star in Appendix3. |

| KOI | Dur | Depth | SNR | $t_0$ | $t_0$_unc | Period | P_unc | a/R* | a/R*_unc | r/R* | r/R*_unc | b | b_unc | $R_p$ | a | $T_{eq}$ | EBprob | V | FOP | N |
|---|---|---|---|---|---|---|---|---|---|---|---|---|---|---|---|---|---|---|---|---|
| | [h] | [ppm] | | [BJD-2454900] | | [days] | | | | | | | | [$R_{Earth}$] | [AU] | [K] | | | | |
| 1.01 | 1.7952 | 14174 | 2062 | 55.76258 | 0.00004 | 2.4706131 | 0.0000004 | 8.519 | 0.082 | 0.12429 | 0.00029 | 0.816 | 0.067 | 20.3 | 0.037 | 1603 | 1.4E-06 | 1 | | 1 |
| 2.01 | 3.9107 | 6716 | 2413 | 54.35781 | 0.00005 | 2.2047355 | 0.0000004 | 4.152 | 0.041 | 0.07931 | 0.00012 | 0.51 | 0.1 | 11.6 | 0.037 | 1743 | 2.4E-06 | 1 | | 1 |
| 3.01 | 2.3607 | 4197 | 328 | 57.81227 | 0.00033 | 4.8878177 | 0.0000089 | 16.1 | 9.1 | 0.0577 | 0.0073 | 0.29 | 0.86 | 4.8 | 0.05 | 796 | 2.0E-06 | 1 | | 1 |
| 4.01 | 2.3866 | 1193 | 136 | 90.5261 | 0.00055 | 3.84937 | 0.000014 | 10 | 24 | 0.034 | 0.015 | 0.7 | 1.4 | 4.0 | 0.05 | 1242 | - | 3 | | 1 |
| 5.01 | 2.0326 | 951 | 263 | 65.9735 | 0.00025 | 4.7803247 | 0.0000058 | 7.3 | 2.2 | 0.03707 | 0.0002 | 0.91 | 0.27 | 7.0 | 0.059 | 1376 | - | 3 | | 1 |
| 7.01 | 3.6234 | 741 | 231 | 56.61126 | 0.00041 | 3.213682 | 0.000011 | 3.94 | 0.56 | 0.02911 | 0.00069 | 0.86 | 0.23 | 3.7 | 0.044 | 1290 | 9.5E-06 | 1 | | 1 |
| 10.01 | 3.2860 | 9390 | 237 | 54.11809 | 0.00062 | 3.522297 | 0.00008 | 8.15 | 0.34 | 0.09138 | 0.00071 | 0.53 | 0.21 | 10.5 | 0.047 | 1287 | 6.5E-06 | 1 | | 1 |
| 12.01 | 7.4343 | 9253 | 604 | 79.59772 | 0.00038 | 17.855038 | 0.000038 | 19.9 | 0.025 | 0.0874 | 0.0001 | 0.0003 | - | 12.6 | 0.141 | 868 | 9.4E-06 | 3 | NoObs | 1 |
| 13.01 | 3.2029 | 4644 | 1147 | 53.56498 | 0.00012 | 1.7635892 | 0.0000014 | 4.51 | 0.2 | 0.07695 | 0.00043 | 0.26 | 0.24 | 20.5 | 0.035 | 3257 | 1.8E-06 | 2 | 2,3 | 1 |
| 17.01 | 3.9011 | 10738 | 724 | 54.48575 | 0.00007 | 3.2347003 | 0.0000012 | 6.9639 | 0.0036 | 0.09467 | 0.00004 | - | - | 9.4 | 0.041 | 1192 | 1.6E-05 | 1 | | 1 |
| 18.01 | 4.6271 | 7239 | 496 | 55.90127 | 0.00022 | 3.548461 | 0.000033 | 6.7257 | 0.0047 | 0.0788 | 0.00005 | 0.0001 | - | 8.2 | 0.045 | 1180 | 3.3E-05 | 1 | | 1 |
| 20.01 | 4.7062 | 16726 | 2001 | 104.00835 | 0.00006 | 4.4379643 | 0.0000013 | 8.0762 | 0.0038 | 0.11678 | 0.00004 | 0.0001 | - | 12.8 | 0.054 | 1145 | 4.8E-06 | 2 | 1,2,3 | |
| 22.01 | 4.3233 | 10570 | 1098 | 110.24939 | 0.00011 | 7.8914455 | 0.0000059 | 15.471 | 0.013 | 0.09222 | 0.00006 | 0.0271 | - | 9.4 | 0.079 | 890 | 5.3E-06 | 2 | 1,2,3 | |
| 41.01 | 6.3192 | 224 | 44 | 55.9589 | 0.0042 | 12.81521 | 0.00029 | 15.67 | 0.24 | 0.01353 | 0.00019 | 0.001 | 0.01 | 1.4 | 0.109 | 741 | 3.6E-05 | 2 | 1,2,3,4 | |
| 42.01 | 4.4845 | 251 | 34 | 114.235 | 0.018 | 17.8328 | 0.0022 | 21.8612 | 0.0027 | 0.01721 | - | 0.6978 | - | 2.6 | 0.14 | 834 | 6.1E-06 | 2 | 1,3 | |
| 44.01 | 14.0968 | 2758 | 67 | 93.395 | 0.013 | 66.5126 | 0.007 | 17.3 | 1.2 | 0.0782 | 0.0013 | 0.83 | 0.17 | 7.5 | 0.304 | 412 | - | 2 | | 1 |
| 46.01 | 3.8313 | 1347 | 145 | 103.931 | 0.00083 | 3.487714 | 0.000028 | 7.241 | 0.044 | 0.03279 | 0.00016 | 0.0021 | - | 3.2 | 0.043 | 1117 | 1.1E-05 | 2 | 1 | |
| 49.01 | 3.6652 | 1113 | 17 | 108.99 | 0.0053 | 8.31393 | 0.00041 | 19 | 1.1 | 0.0266 | 0.0014 | 0.03 | 0.072 | 2.8 | 0.079 | 906 | 3.4E-05 | 2 | 1 | |
| 51.01 | 3.3759 | 25812 | 275 | 66.93528 | 0.00052 | 10.431147 | 0.00002 | 21.1 | 6.3 | 0.16271 | 0.00076 | 0.56 | 0.17 | 4.8 | 0.056 | 314 | - | 3 | | 1 |
| 63.01 | 2.9910 | 4100 | 97 | 110.8421 | 0.0015 | 9.434152 | 0.000098 | 27.57 | 0.63 | 0.0566 | 0.001 | 0.025 | 0.036 | 6.6 | 0.089 | 844 | 5.3E-06 | 2 | 3 | 1 |
| 64.01 | 1.6811 | 1223 | 132 | 90.54051 | 0.00046 | 1.9510939 | 0.0000057 | 4.4 | 1.3 | 0.04 | 0.00034 | 0.84 | 0.25 | 8.5 | 0.032 | 1760 | 4.3E-05 | 3 | 1,2,3,4 | 1 |
| 69.01 | 2.8948 | 269 | 138 | 67.92512 | 0.00069 | 4.726745 | 0.000018 | 12.706 | 0.084 | 0.01465 | 0.00008 | 0 | 0.014 | 1.6 | 0.056 | 1036 | - | 3 | | 1 |
| 70.01 | 3.7978 | 1027 | 135 | 71.60857 | 0.00081 | 10.854042 | 0.000039 | 20 | 18 | 0.0297 | 0.0054 | 0.49 | 1 | 2.3 | 0.094 | 643 | 1.7E-05 | 2 | 1,2,3 | |
| 70.02 | 2.4785 | 375 | 58 | 67.5005 | 0.0011 | 3.696125 | 0.000017 | 6.2 | 9.6 | 0.0209 | 0.0055 | 0.85 | 0.78 | 1.6 | 0.046 | 919 | 2.3E-05 | 2 | 1,3 | |
| 70.03 | 7.2380 | 793 | 34 | 97.729 | 0.01 | 77.609 | 0.0063 | 83.39 | 0.99 | 0.02575 | 0.00033 | 0.198 | 0.035 | 2.0 | 0.35 | 333 | 1.9E-05 | 2 | 1,2,3 | |
| 70.04 | 2.7697 | 74 | 15 | 68.93 | 0.0055 | 6.09852 | 0.00014 | 16 | 95 | 0.0079 | 0.0087 | 0.4 | 2.7 | 0.6 | 0.064 | 779 | 3.4E-05 | 2 | | |
| 72.01 | 1.8200 | 184 | 101 | 64.57364 | 0.0007 | 0.8374958 | 0.0000042 | 3.609 | 0.035 | 0.01211 | 0.00009 | 0.029 | 0.027 | 1.3 | 0.018 | 1790 | 1.0E-05 | 1 | 1,2,3 | 1 |
| 72.02 | 6.8565 | 461 | 74 | 71.676 | 0.0021 | 45.29491 | 0.00069 | 35 | 38 | 0.0214 | 0.0042 | 0.73 | 0.84 | 2.3 | 0.252 | 478 | 8.1E-06 | 2 | 1,2 | |
| 75.01 | 17.4137 | 1275 | 165 | 89.9691 | 0.0019 | 105.8885 | 0.0014 | 47.1 | 2.3 | 0.0362 | 0.0012 | 0.01 | 0.23 | 4.3 | 0.449 | 391 | - | 2 | | |
| 82.01 | 4.1390 | 981 | 102 | 67.7519 | 0.0014 | 16.14583 | 0.00012 | 18.3 | 6.7 | 0.0337 | 0.0025 | 0.85 | 0.38 | 6.8 | 0.131 | 786 | 5.6E-06 | 2 | 1,2,3,4,5 | |
| 82.02 | 3.2778 | 248 | 31 | 67.0745 | 0.0044 | 10.312 | 0.00026 | 13 | 30 | 0.0182 | 0.0083 | 0.86 | 0.92 | 3.7 | 0.097 | 914 | 8.8E-06 | 2 | 1,2,3,4,5 | |
| 84.01 | 3.4186 | 692 | 141 | 68.99092 | 0.00073 | 9.287047 | 0.00003 | 21.25 | 0.15 | 0.02349 | 0.00013 | 0.0009 | - | 2.2 | 0.086 | 737 | 4.2E-05 | 2 | 1,2,3 | |
| 85.01 | 4.0746 | 326 | 117 | 65.0392 | 0.0012 | 5.859965 | 0.000039 | 7.7 | 5.7 | 0.0179 | 0.0022 | 0.73 | 0.69 | 3.2 | 0.067 | 1318 | 2.4E-05 | 2 | 1,2,3,4 | |
| 85.02 | 3.2040 | 99 | 53 | 66.5008 | 0.0019 | 2.15488 | 0.000018 | 4.5 | 1.3 | 0.00923 | 0.00013 | 0.277 | 0.083 | 1.7 | 0.035 | 1823 | - | 4 | | |
| 85.03 | 4.2219 | 103 | 34 | 70.9924 | 0.0035 | 8.13119 | 0.00012 | 7 | 15 | 0.0111 | 0.0039 | 0.88 | 0.81 | 2.0 | 0.084 | 1177 | - | 4 | | |
| 87.01 | 7.5747 | 481 | 38 | 66.6987 | 0.0033 | 289.8605 | 0.0047 | 311.1 | 5 | 0.01956 | 0.00033 | 0.02 | 0.066 | 2.4 | 0.877 | 282 | 1.2E-05 | 2 | 1,2 | |
| 89.01 | 10.4048 | 372 | 35 | 83.587 | 0.012 | 84.6763 | 0.0075 | 56 | 156 | 0.018 | 0.01 | 0.5 | 1.7 | 4.4 | 0.427 | 756 | 6.2E-05 | 2 | 1,2,3 | |
| 89.02 | 7.4240 | 377 | 22 | 222.855 | 0.005 | 107.5 | 0.0097 | 48 | 14 | 0.02241 | 0.00061 | 0.86 | 0.26 | 5.5 | 0.501 | 698 | 5.4E-05 | 2 | 1 | |
| 92.01 | 3.7259 | 703 | 39 | 70.4508 | 0.0042 | 65.7008 | 0.0019 | 56 | 17 | 0.03169 | 0.00074 | 0.89 | 0.27 | 4.4 | 0.33 | 505 | 8.0E-06 | 2 | 1,2,3 | |
| 94.01 | 7.0052 | 5674 | 382 | 65.74223 | 0.00047 | 22.34094 | 0.00065 | 24.3 | 4 | 0.07 | 0.0024 | 0.39 | 0.45 | 12.6 | 0.165 | 851 | 2.2E-05 | 2 | 1,2,3 | |
| 94.02 | 5.3435 | 749 | 34 | 71.0084 | 0.0051 | 10.42361 | 0.00022 | 13.73 | 0.76 | 0.022 | 0.051 | 0.201 | 0.081 | 4.0 | 0.099 | 1099 | 8.8E-05 | 2 | 1,3 | |
| 94.03 | 9.0070 | 1929 | 53 | 94.2397 | 0.0044 | 90.5323 | 0.0017 | 84.7 | 4.5 | 0.0382 | 0.0016 | 0.006 | 0.076 | 6.9 | 0.419 | 534 | 4.5E-05 | 2 | 1,3 | |
| 97.01 | 5.5128 | 7396 | 1275 | 67.27602 | 0.00013 | 4.8854906 | 0.0000024 | 7.8838 | 0.0076 | 0.07784 | 0.00007 | - | - | 11.0 | 0.059 | 1226 | 1.3E-05 | 1 | | 1 |
| 98.01 | 6.8561 | 2299 | 478 | 71.08749 | 0.00048 | 6.790119 | 0.000019 | 8.004903 | 0.000022 | 0.0534 | 0.0094 | 0.0001 | - | 12.1 | 0.077 | 1528 | 9.0E-06 | 2 | 1,2,3 | |
| †99.01 | 19.7865 | 1803 | 72 | 73.0486 | 0.0019 | 817 | 12 | 327.4 | 4.9 | 0.03768 | - | - | - | 4.6 | 1.693 | 177 | 4.2E-05 | 3 | 1 | 1 |
| 100.01 | 4.5162 | 1506 | 95 | 74.1515 | 0.0016 | 9.966512 | 0.000092 | 7.8 | 2.3 | 0.04475 | 0.00052 | 0.88 | 0.26 | 13.6 | 0.101 | 1493 | 3.1E-05 | 3 | 1,3 | 1 |
| 102.01 | 2.4232 | 933 | 240 | 68.05961 | 0.00042 | 1.7351339 | 0.0000041 | 4 | 1.8 | 0.0303 | 0.0024 | 0.76 | 0.52 | 6.9 | 0.03 | 2175 | 5.8E-05 | 2 | 1,2,3 | 1 |
| 103.01 | 3.4297 | 800 | 72 | 74.3331 | 0.0015 | 14.91155 | 0.00013 | 36.15 | 0.44 | 0.02642 | 0.00024 | 0.025 | 0.027 | 2.3 | 0.119 | 628 | 4.8E-05 | 2 | 1,2,3 | |
| 104.01 | 1.2727 | 1174 | 68 | 67.99798 | 0.0007 | 2.508091 | 0.000013 | 16 | 11 | 0.035 | 0.0048 | 0.45 | 0.86 | 2.8 | 0.032 | 927 | 8.4E-06 | 2 | 1,2,3 | |
| 105.01 | 2.3693 | 1082 | 114 | 69.6502 | 0.0014 | 8.98092 | 0.000096 | 12.2 | 3.6 | 0.0394 | 0.00059 | 0.91 | 0.27 | 8.1 | 0.09 | 1100 | 4.4E-05 | 2 | 1 | |
| 107.01 | 4.8694 | 463 | 91 | 67.0215 | 0.0015 | 7.257003 | 0.000062 | 11.608 | 0.094 | 0.01932 | 0.00014 | 0.009 | 0.017 | 2.1 | 0.075 | 943 | 2.6E-05 | 2 | 1 | |
| 108.01 | 4.5210 | 505 | 79 | 75.1762 | 0.0016 | 15.96534 | 0.00016 | 22 | 28 | 0.0211 | 0.0047 | 0.6 | 1.1 | 2.7 | 0.128 | 777 | 3.0E-05 | 2 | 1,2,3 | |
| 110.01 | 3.9364 | 522 | 98 | 68.2136 | 0.0013 | 9.94075 | 0.000079 | 13 | 12 | 0.0229 | 0.0034 | 0.74 | 0.76 | 2.9 | 0.095 | 985 | 1.5E-05 | 2 | 1 | |
| 111.01 | 4.5716 | 506 | 84 | 70.6135 | 0.0015 | 11.427514 | 0.000074 | 14 | 14 | 0.0222 | 0.0038 | 0.72 | 0.82 | 2.5 | 0.102 | 814 | 2.1E-05 | 2 | 1,2,3 | |
| 111.02 | 5.7168 | 456 | 65 | 65.7105 | 0.0022 | 23.6686 | 0.00022 | 28 | 40 | 0.0203 | 0.0052 | 0.5 | 1.2 | 2.3 | 0.166 | 638 | 2.1E-05 | 2 | 1 | |
| 111.04 | 7.5697 | 636 | 38 | 359.3572 | 0.0047 | 103.5112 | 0.0066 | 108 | 2 | 0.02271 | 0.00038 | 0.018 | 0.071 | 2.5 | 0.443 | 391 | - | 2 | | |
| 112.01 | 6.3832 | 756 | 67 | 118.1798 | 0.0018 | 51.07943 | 0.00063 | 40 | 29 | 0.0276 | 0.0033 | 0.78 | 0.63 | 3.7 | 0.279 | 539 | 2.2E-05 | 3 | 2,3 | 1 |
| 112.02 | 2.5592 | 118 | 24 | 66.9837 | 0.0036 | 3.709209 | 0.000057 | 6 | 20 | 0.0128 | 0.0079 | 0.9 | 1 | 1.7 | 0.048 | 1300 | - | 4 | | |
| †113.01 | 4.6696 | 1038 | 21 | 87.3183 | 0.0041 | 300 | - | 167 | 50 | 0.066 | 0.021 | 1.02 | 0.31 | 8.3 | 0.888 | 269 | 1.9E-05 | 2 | 1,2 | |
| 115.01 | 2.9014 | 595 | 96 | 66.1414 | 0.0011 | 5.412245 | 0.000032 | 13 | 19 | 0.0231 | 0.006 | 0.5 | 1.2 | 3.4 | 0.063 | 1260 | 1.6E-05 | 2 | 1,2 | |
| 115.02 | 2.7875 | 192 | 28 | 72.0079 | 0.0042 | 7.12575 | 0.00017 | 11 | 41 | 0.0151 | 0.0088 | 0.9 | 1.2 | 2.2 | 0.076 | 1147 | 2.1E-05 | 2 | 1 | |
| 116.01 | 3.5494 | 471 | 43 | 69.2763 | 0.003 | 13.57096 | 0.00023 | 21 | 42 | 0.0227 | 0.0078 | 0.8 | 1.1 | 4.7 | 0.119 | 1057 | 8.1E-05 | 2 | 1,2,3 | |
| 116.02 | 6.7062 | 601 | 44 | 84.9344 | 0.0044 | 43.8448 | 0.0013 | 54.1 | 0.84 | 0.02306 | 0.00033 | 0.021 | 0.025 | 4.8 | 0.26 | 715 | 7.9E-05 | 2 | 1 | |
| 117.01 | 6.1280 | 397 | 68 | 71.7751 | 0.003 | 14.74942 | 0.00025 | 8.8 | 3 | 0.0219 | 0.0012 | 0.89 | 0.32 | 2.4 | 0.12 | 729 | 4.4E-05 | 2 | 1,2,3 | |
| 117.02 | 4.0855 | 117 | 29 | 71.6097 | 0.0052 | 4.90134 | 0.00015 | 5 | 13 | 0.0121 | 0.0052 | 0.87 | 0.98 | 1.3 | 0.058 | 1048 | 6.1E-05 | 3 | 1,2 | 1 |
| 117.03 | 3.3490 | 113 | 33 | 66.5106 | 0.0039 | 3.17985 | 0.000052 | 3.9 | 9.8 | 0.0118 | 0.0048 | 0.8 | 1 | 1.3 | 0.043 | 1217 | 5.8E-05 | 3 | 1,2 | 1 |
| 117.04 | 4.1670 | 43 | 8.8 | 70.838 | 0.014 | 7.9572 | 0.00051 | 11.11552 | 0.00071 | 0.0062 | - | 0.6594 | - | 0.7 | 0.08 | 892 | - | 4 | | |
| 118.01 | 5.7166 | 259 | 36 | 71.6798 | 0.0044 | 24.99348 | 0.0006 | 17 | 22 | 0.0172 | 0.0036 | 0.86 | 0.68 | 1.8 | 0.17 | 590 | 3.0E-05 | 2 | 2 | |
| 119.01 | 11.2962 | 1579 | 144 | 74.9125 | 0.0016 | 49.18381 | 0.00044 | 34.44916 | 0.00031 | 0.035 | 0.018 | 0.0009 | - | 3.9 | 0.264 | 462 | 2.3E-05 | 3 | 1 | 1 |
| 122.01 | 3.9263 | 509 | 89 | 64.9714 | 0.0013 | 11.522904 | 0.000062 | 22.89 | 0.21 | 0.02023 | 0.00016 | 0.0038 | - | 1.9 | 0.101 | 716 | 1.3E-05 | 2 | 1,2,3 | |
| 123.01 | 3.6301 | 293 | 66 | 55.9766 | 0.0017 | 6.481654 | 0.000047 | 11 | 14 | 0.0166 | 0.0039 | 0.6 | 1.1 | 2.3 | 0.071 | 1091 | 3.6E-05 | 2 | 1,2,3,4 | |
| 123.02 | 6.4510 | 365 | 65 | 70.5763 | 0.0024 | 21.22222 | 0.00023 | 19 | 25 | 0.0183 | 0.0042 | 0.7 | 1 | 2.5 | 0.155 | 739 | 3.4E-05 | 2 | 1,4 | |
| 124.01 | 3.8682 | 223 | 32 | 70.1234 | 0.0039 | 12.69116 | 0.00029 | 14 | 32 | 0.0161 | 0.0057 | 0.85 | 0.95 | 2.3 | 0.111 | 926 | 2.5E-05 | 2 | 1,2 | |
| 124.02 | 5.0544 | 352 | 39 | 75.822 | 0.0038 | 31.71954 | 0.00071 | 28 | 46 | 0.0196 | 0.005 | 0.84 | 0.84 | 2.8 | 0.205 | 681 | 2.2E-05 | 2 | 1 | |
| 127.01 | 2.9074 | 11637 | 1078 | 67.02957 | 0.0001 | 3.5787781 | 0.000002 | 10.33 | 0.01 | 0.09665 | 0.00007 | 0.0038 | - | 9.7 | 0.046 | 1098 | 1.4E-05 | 2 | 1,2 | |
| 128.01 | 3.5022 | 11241 | 1191 | 69.32863 | 0.00024 | 4.9427813 | 0.000002 | 12.11 | 0.93 | 0.10066 | 0.00052 | 0.58 | 0.16 | 15.7 | 0.059 | 1240 | 1.1E-05 | 2 | 1 | |
| 131.01 | 4.6736 | 6925 | 458 | 66.17565 | 0.00071 | 5.014177 | 0.000021 | 8.885 | 0.017 | 0.07492 | 0.00012 | 0.0008 | - | 9.0 | 0.06 | 1181 | - | 3 | | 1 |
| 135.01 | 2.7842 | 7916 | 766 | 65.41537 | 0.00014 | 3.0241018 | 0.0000017 | 8.998787 | 0.000005 | 0.0795 | 0.0074 | 0.0089 | - | 8.3 | 0.042 | 1250 | 9.3E-06 | 2 | 1 | 1 |
| 137.01 | 3.5437 | 2354 | 155 | 68.40567 | 0.00078 | 7.641577 | 0.000023 | 17.82 | 0.11 | 0.04345 | 0.00022 | 0.027 | 0.02 | 6.0 | 0.077 | 949 | 6.2E-05 | 2 | 1,2,3,4 | |
| 137.02 | 3.7766 | 3279 | 297 | 61.15248 | 0.00045 | 14.859006 | 0.000025 | 19.3 | 2.1 | 0.0616 | 0.0013 | 0.84 | 0.21 | 8.6 | 0.12 | 760 | 4.8E-05 | 2 | 1,4 | |
| 137.03 | 2.0761 | 269 | 38 | 66.5071 | 0.0022 | 3.50472 | 0.000029 | 8 | 26 | 0.0167 | 0.0093 | 0.8 | 1.3 | 2.3 | 0.046 | 1228 | - | 4 | | |
| 138.01 | 6.1103 | 7085 | 245 | 73.7648 | 0.0019 | 48.93807 | 0.00058 | 37 | 11 | 0.09401 | 0.00057 | 0.82 | 0.25 | 16.6 | 0.282 | 716 | - | 2 | | |
| 139.01 | 10.7262 | 3542 | 102 | 75.0881 | 0.0028 | 224.7937 | 0.0032 | 126 | 22 | 0.0576 | 0.0022 | 0.67 | 0.37 | 5.7 | 0.741 | 288 | 1.9E-05 | 2 | 1 | |
| 139.02 | 2.6756 | 116 | 13 | 74.354 | 0.0066 | 3.3418 | 0.00012 | 7 | 36 | 0.012 | 0.011 | 0.7 | 2 | 1.2 | 0.045 | 1169 | 4.4E-05 | 2 | 1 | |
| 141.01 | 1.4408 | 2421 | 167 | 65.30523 | 0.00055 | 2.624219 | 0.000011 | 10.7 | 8.4 | 0.0489 | 0.0077 | 0.73 | 0.72 | 4.3 | 0.037 | 1090 | 1.5E-05 | 2 | 1,2 | |
| 142.01 | 3.7354 | 1188 | 105 | 66.0242 | 0.0013 | 10.914785 | 0.000076 | 23 | 6.9 | 0.03073 | 0.0003 | - | - | 2.5 | 0.095 | 661 | 2.9E-05 | 2 | 1,3 | |
| 144.01 | 3.7045 | 1396 | 166 | 66.08938 | 0.00084 | 4.176263 | 0.00002 | 7.9 | 4.7 | 0.0345 | 0.0047 | 0.53 | 0.78 | 6.6 | 0.053 | 1198 | 7.5E-05 | 2 | 1 | 1 |
| 148.01 | 2.6944 | 453 | 54 | 57.0628 | 0.0021 | 4.777978 | 0.000053 | 9 | 16 | 0.0218 | 0.0076 | 0.8 | 1 | 2.1 | 0.054 | 908 | 1.5E-04 | 2 | 1,2,3 | |
| 148.02 | 3.2778 | 960 | 107 | 58.3427 | 0.0013 | 9.67374 | 0.000068 | 16 | 16 | 0.0305 | 0.006 | 0.72 | 0.82 | 3.0 | 0.087 | 716 | 1.1E-04 | 2 | 1 | |
| 148.03 | 5.6283 | 519 | 37 | 79.0652 | 0.0037 | 42.89554 | 0.0007 | 53 | 16 | 0.02093 | 0.00041 | 0.216 | 0.065 | 2.0 | 0.235 | 435 | 1.3E-04 | 2 | 1 | |
| 149.01 | 7.8599 | 961 | 114 | 78.0897 | 0.0017 | 14.55792 | 0.00015 | 14.722 | 0.083 | 0.02799 | 0.00016 | 0.0005 | - | 4.1 | 0.122 | 890 | 2.9E-05 | 2 | 1 | |
| 150.01 | 3.5591 | 783 | 75 | 67.0052 | 0.0016 | 8.408848 | 0.000076 | 18.63 | 0.21 | 0.02552 | 0.00024 | 0.006 | - | 3.4 | 0.083 | 939 | 6.7E-05 | 2 | 1 | |
| 150.02 | 5.0711 | 796 | 53 | 76.8337 | 0.003 | 28.57378 | 0.0005 | 33 | 55 | 0.0276 | 0.0081 | 0.7 | 1.1 | 3.7 | 0.187 | 625 | 6.5E-05 | 2 | 1 | |
| 151.01 | 2.6612 | 1302 | 58 | 65.8279 | 0.0017 | 13.44739 | 0.00013 | 22.5 | 6.8 | 0.03831 | 0.00062 | 0.74 | 0.22 | 4.6 | 0.114 | 823 | 5.9E-05 | 3 | 2 | 1 |
| 152.01 | 8.5892 | 2845 | 118 | 91.7438 | 0.0019 | 52.09119 | 0.00063 | 48.52 | 0.27 | 0.04816 | 0.00026 | 0.0002 | - | 4.9 | 0.282 | 497 | 5.6E-05 | 2 | | |
| 152.02 | 6.7876 | 753 | 39 | 66.6192 | 0.0059 | 27.40415 | 0.00089 | 31.74 | 0.67 | 0.02485 | 0.00051 | 0.024 | 0.032 | 2.5 | 0.184 | 616 | 9.1E-05 | 2 | | |
| 152.03 | 5.0646 | 636 | 40 | 69.6257 | 0.005 | 13.484 | 0.00039 | 18 | 66 | 0.024 | 0.015 | 0.5 | 1.9 | 2.4 | 0.114 | 782 | 9.4E-05 | 2 | | |
| 153.01 | 2.7092 | 990 | 57 | 72.7136 | 0.0017 | 8.925031 | 0.000085 | 19 | 34 | 0.03 | 0.012 | 0.7 | 1.1 | 3.2 | 0.08 | 708 | 1.8E-05 | 2 | 1,2 | |
| 153.02 | 2.5456 | 797 | 67 | 61.5475 | 0.0015 | 4.753978 | 0.000039 | 8 | 12 | 0.0295 | 0.0086 | 0.83 | 0.8 | 3.1 | 0.053 | 870 | 2.0E-05 | 2 | 1,2 | |
| 155.01 | 4.5785 | 733 | 84 | 70.4159 | 0.0017 | 5.660629 | 0.000057 | 9.64 | 0.09 | 0.02428 | 0.0002 | 0.0011 | - | 3.8 | 0.065 | 1164 | 2.9E-05 | 2 | 1 | 1 |
| 156.01 | 2.4887 | 554 | 35 | 76.0363 | 0.0028 | 8.04144 | 0.00013 | 18 | 46 | 0.023 | 0.013 | 0.7 | 1.4 | 1.9 | 0.071 | 642 | 2.4E-05 | 2 | 1 | |
| 156.02 | 2.4671 | 330 | 26 | 78.3602 | 0.0039 | 5.18856 | 0.00012 | 10 | 36 | 0.02 | 0.014 | 0.8 | 1.3 | 1.6 | 0.053 | 743 | 2.8E-05 | 2 | 1 | |
| 156.03 | 2.8987 | 1423 | 96 | 75.7045 | 0.001 | 11.776179 | 0.000054 | 32.64 | 0.37 | 0.03337 | 0.00028 | 0.027 | 0.027 | 2.8 | 0.091 | 567 | 1.9E-05 | 2 | 1 | |
| 157.01 | 4.6051 | 757 | 83 | 71.1768 | 0.002 | 13.02495 | 0.000093 | 15 | 11 | 0.0278 | 0.0038 | 0.76 | 0.69 | 3.0 | 0.111 | 751 | 1.5E-04 | 1 | 1 | 1 |
| 157.02 | 5.5677 | 975 | 80 | 81.4549 | 0.002 | 22.68696 | 0.00015 | 17.8 | 5 | 0.0322 | 0.0015 | 0.84 | 0.34 | 3.5 | 0.16 | 625 | 1.4E-04 | 1 | 1 | 1 |
| 157.03 | 4.3105 | 1352 | 86 | 87.1604 | 0.0015 | 31.99541 | 0.00018 | 33.5 | 9.7 | 0.0387 | 0.002 | 0.84 | 0.35 | 4.2 | 0.201 | 558 | 1.1E-04 | 1 | 1 | 1 |
| 157.04 | 6.3817 | 524 | 36 | 158.0328 | 0.0034 | 46.68872 | 0.00081 | 57.4 | 1.1 | 0.02041 | 0.00036 | 0.014 | 0.025 | 2.2 | 0.259 | 491 | 1.7E-04 | 1 | 1 | 1 |
| 157.05 | 9.7575 | 1119 | 54 | 220.288 | 0.004 | 118.3772 | 0.0031 | 95.9 | 1.2 | 0.02976 | 0.00034 | 0.03 | 0.032 | 3.2 | 0.481 | 361 | 1.3E-04 | 1 | 1 | 1 |
| 157.06 | 4.1041 | 297 | 36 | 71.5062 | 0.0038 | 10.3039 | 0.00013 | 18.2 | 5.5 | 0.01559 | 0.00033 | 0.128 | 0.038 | 1.7 | 0.095 | 811 | 2.0E-05 | 1 | | 1 |
| 159.01 | 4.1073 | 477 | 46 | 69.7359 | 0.0033 | 8.99093 | 0.00017 | 8.9 | 9.1 | 0.0235 | 0.0038 | 0.86 | 0.62 | 3.1 | 0.087 | 963 | 1.1E-04 | 2 | 1 | |
| 161.01 | 1.9363 | 898 | 137 | 66.21002 | 0.00063 | 3.105501 | 0.000011 | 11 | 12 | 0.0293 | 0.0068 | 0.69 | 0.91 | 5.3 | 0.043 | 1306 | 2.5E-05 | 2 | 1 | |
| 162.01 | 4.6255 | 706 | 62 | 75.2079 | 0.0024 | 14.00656 | 0.00019 | 23.73 | 0.3 | 0.02379 | 0.00026 | 0.0008 | - | 2.6 | 0.116 | 733 | 7.2E-05 | 2 | 1 | |
| 163.01 | 3.3145 | 654 | 58 | 72.7543 | 0.002 | 11.11978 | 0.00013 | 22 | 37 | 0.0242 | 0.0082 | 0.6 | 1.3 | 2.8 | 0.097 | 756 | 2.4E-05 | 2 | 1 | |
| 165.01 | 2.7694 | 916 | 50 | 72.5644 | 0.002 | 13.22216 | 0.00016 | 29 | 69 | 0.028 | 0.014 | 0.7 | 1.4 | 1.9 | 0.104 | 541 | 3.8E-05 | 2 | 1 | |
| 166.01 | 2.4622 | 704 | 45 | 71.4619 | 0.002 | 12.49334 | 0.00015 | 34 | 84 | 0.026 | 0.013 | 0.7 | 1.4 | 3.7 | 0.107 | 796 | 5.9E-05 | 2 | 1 | |
| 167.01 | 4.2389 | 415 | 70 | 67.602 | 0.002 | 4.919592 | 0.000055 | 8 | 13 | 0.0195 | 0.0055 | 0.6 | 1.3 | 1.8 | 0.058 | 1072 | 3.0E-05 | 2 | 1 | |
| 168.01 | 6.1290 | 358 | 42 | 66.2761 | 0.0041 | 10.74356 | 0.00025 | 12 | 22 | 0.018 | 0.006 | 0.5 | 1.4 | 3.7 | 0.102 | 1112 | 3.2E-05 | 2 | 1 | |
| 168.02 | 5.7377 | 80 | 13 | 70.389 | 0.011 | 5.0918 | 0.00025 | 6.93 | 0.29 | 0.00908 | 0.00035 | 0.029 | 0.01 | 1.9 | 0.062 | 1427 | - | 4 | | |
| 168.03 | 4.7708 | 98 | 13 | 71.317 | 0.011 | 7.10664 | 0.00032 | 11.4 | 0.53 | 0.00978 | 0.00042 | 0.015 | 0.01 | 2.0 | 0.077 | 1280 | - | 4 | | |
| 171.01 | 3.6443 | 489 | 62 | 70.1549 | 0.002 | 5.968839 | 0.000069 | 12.89 | 0.17 | 0.02039 | 0.00022 | 0.0172 | - | 2.5 | 0.067 | 1125 | 7.6E-05 | 2 | 1 | |
| 172.01 | 4.9815 | 586 | 46 | 70.8361 | 0.0034 | 13.72288 | 0.00026 | 21.44 | 0.35 | 0.02184 | 0.00031 | 0.0102 | - | 1.6 | 0.112 | 598 | 4.0E-05 | 2 | 1 | |
| 173.01 | 4.4592 | 421 | 44 | 71.9655 | 0.0033 | 10.06092 | 0.00019 | 17.75 | 0.31 | 0.01863 | 0.00028 | 0 | 0.025 | 1.9 | 0.093 | 808 | 2.8E-05 | 2 | 1 | |
| 174.01 | 3.3658 | 1060 | 27 | 77.8417 | 0.0054 | 56.3509 | 0.0025 | 142.2 | 5.3 | 0.02897 | 0.00083 | 0.027 | 0.05 | 2.5 | 0.267 | 355 | 4.3E-05 | 2 | 1 | |
| 176.01 | 6.5324 | 397 | 36 | 67.5141 | 0.0049 | 30.2303 | 0.0008 | 24 | 41 | 0.0202 | 0.0056 | 0.8 | 1 | 2.2 | 0.197 | 630 | 5.4E-05 | 2 | 1 | |
| 177.01 | 4.8343 | 281 | 26 | 76.6066 | 0.0059 | 21.05996 | 0.00071 | 28 | 86 | 0.0162 | 0.0092 | 0.6 | 1.7 | 1.9 | 0.153 | 664 | 1.7E-04 | 2 | 1 | |
| 179.01 | 10.2799 | 972 | 94 | 75.8002 | 0.0027 | 20.74007 | 0.00033 | 15.93 | 0.1 | 0.02832 | 0.00019 | 0.0005 | - | 3.3 | 0.152 | 681 | - | 2 | | 1 |
| 180.01 | 3.2069 | 626 | 20 | 62.0919 | 0.0038 | 10.0456 | 0.0002 | 25.5 | 1.4 | 0.02066 | 0.00093 | 0.028 | 0.062 | 1.8 | 0.092 | 731 | 1.6E-05 | 2 | 1,2,3 | 1 |
| 183.01 | 2.6997 | 18301 | 1464 | 66.35411 | 0.0001 | 2.684327 | 0.0000015 | 8.581648 | 0.000005 | 0.122 | 0.019 | 0.0072 | - | 9.9 | 0.038 | 1113 | 1.5E-05 | 2 | 1 | |
| 186.01 | 3.0609 | 17321 | 1071 | 66.66821 | 0.00012 | 3.2432615 | 0.0000021 | 9.1033 | 0.0094 | 0.11805 | 0.00009 | - | - | 11.5 | 0.044 | 1156 | 1.6E-05 | 2 | NoObs | |
| 187.01 | 5.3569 | 24933 | 739 | 84.52867 | 0.00025 | 30.88252 | 0.000048 | 50.291303 | 0.000079 | 0.142 | 0.013 | 0.0007 | - | 11.6 | 0.194 | 499 | 1.2E-05 | 2 | | |
| 188.01 | 2.2281 | 14813 | 1052 | 66.5079 | 0.00009 | 3.7970199 | 0.0000018 | 14.451 | 0.017 | 0.10841 | 0.00009 | 0.0004 | - | 8.0 | 0.046 | 859 | 2.1E-05 | 2 | | |
| 189.01 | 6.0251 | 22443 | 778 | 81.09127 | 0.00024 | 30.360407 | 0.000057 | 42.5 | 1.1 | 0.13323 | 0.00083 | 0.29 | 0.18 | 12.6 | 0.181 | 461 | 6.0E-06 | 2 | NoObs | |
| 190.01 | 4.4351 | 11465 | 431 | 72.30197 | 0.00032 | 12.265011 | 0.000024 | 15.27 | 0.32 | 0.11222 | 0.0006 | 0.814 | 0.1 | 16.3 | 0.107 | 843 | 1.0E-05 | 2 | NoObs | |
| 191.01 | 4.1332 | 15314 | 523 | 65.38412 | 0.00025 | 15.358776 | 0.000016 | 28.4 | 1.3 | 0.115 | 0.0011 | 0.49 | 0.22 | 11.6 | 0.122 | 666 | 4.6E-05 | 2 | | 1 |
| 191.02 | 2.1206 | 680 | 41 | 65.5052 | 0.0019 | 2.418402 | 0.00002 | 8.45 | 0.5 | 0.028 | 0.0013 | 0.0176 | - | 2.8 | 0.036 | 1226 | 1.3E-04 | 3 | | 1 |
| 191.03 | 1.4863 | 210 | 21 | 66.3662 | 0.0029 | 0.7086217 | 0.000009 | 3.035907 | 0.000039 | 0.01386 | - | 0.615 | - | 1.4 | 0.016 | 1839 | - | 2 | | |
| 191.04 | 5.3874 | 265 | 9.6 | 77.73 | 0.015 | 19.3245 | 0.0013 | 29.5 | 7.9 | 0.0148 | 0.0037 | 0.006 | 0.027 | 1.5 | 0.142 | 617 | - | 2 | | |
| 192.01 | 4.2999 | 10023 | 511 | 70.02102 | 0.0003 | 10.291006 | 0.000018 | 19.951591 | 0.000035 | 0.09 | 0.021 | 0.0001 | - | 9.9 | 0.095 | 855 | 1.4E-05 | 2 | 1 | |
| 193.01 | 5.0147 | 20885 | 667 | 90.34941 | 0.00029 | 37.590346 | 0.00007 | 74.98 | 0.12 | 0.12875 | 0.00016 | 0.0017 | - | 14.2 | 0.226 | 548 | 1.1E-05 | 2 | | 1 |
| 194.01 | 2.0031 | 16271 | 118 | 72.46531 | 0.00061 | 3.120831 | 0.000011 | 8.9 | 1 | 0.1338 | 0.0037 | 0.79 | 0.24 | 12.0 | 0.043 | 1133 | 6.7E-06 | 2 | | |
| 195.01 | 2.1873 | 14694 | 494 | 66.631 | 0.00019 | 3.2175236 | 0.0000035 | 11.28 | 0.91 | 0.1144 | 0.0019 | 0.57 | 0.28 | 11.9 | 0.043 | 1165 | 1.2E-05 | 2 | | |
| 196.01 | 2.2855 | 10797 | 1009 | 70.18042 | 0.00009 | 1.8555565 | 0.000001 | 5.87 | 0.27 | 0.09728 | 0.00088 | 0.51 | 0.22 | 10.0 | 0.03 | 1377 | 2.1E-05 | 2 | NoObs | |
| 197.01 | 4.2447 | 10776 | 593 | 66.83869 | 0.00027 | 17.276276 | 0.000027 | 36.447 | 0.065 | 0.09147 | 0.00012 | 0.0002 | - | 10.3 | 0.128 | 615 | 2.6E-05 | 2 | NoObs | |
| 199.01 | 3.4738 | 10575 | 847 | 70.48111 | 0.00015 | 3.2686931 | 0.0000029 | 7.899543 | 0.000007 | 0.093 | 0.017 | 0.0001 | - | 8.7 | 0.044 | 1216 | 2.2E-05 | 2 | NoObs | |
| 200.01 | 2.8848 | 8465 | 542 | 67.34424 | 0.0002 | 7.3407361 | 0.0000086 | 21.103 | 0.04 | 0.08279 | 0.00012 | - | - | 6.9 | 0.075 | 810 | 2.8E-05 | 2 | | |
| 201.01 | 2.8358 | 6043 | 777 | 70.5598 | 0.0003 | 4.2253865 | 0.0000072 | 12.387 | 0.032 | 0.07155 | 0.00014 | 0.0103 | - | 7.8 | 0.052 | 1064 | - | 3 | | |
| 202.01 | 1.9597 | 10265 | 1056 | 66.02029 | 0.00008 | 1.7208618 | 0.0000008 | 5.27 | 0.12 | 0.10151 | 0.00047 | 0.74 | 0.12 | 11.5 | 0.029 | 1559 | 1.3E-05 | 2 | 1 | |
| 203.01 | 2.2791 | 20902 | 500 | 65.79282 | 0.00014 | 1.4857106 | 0.0000012 | 5.682 | 0.014 | 0.13028 | 0.00024 | - | - | 13.8 | 0.026 | 1518 | 1.6E-05 | 2 | NoObs | |
| 204.01 | 3.0172 | 7176 | 430 | 66.37805 | 0.0003 | 3.2467374 | 0.0000053 | 8.855 | 0.021 | 0.07562 | 0.00014 | - | 0.01 | 7.9 | 0.043 | 1098 | 6.4E-05 | 2 | | 1 |
| 205.01 | 3.0463 | 10043 | 554 | 75.17334 | 0.00022 | 11.720115 | 0.000015 | 30.4 | 3 | 0.0918 | 0.002 | 0.46 | 0.33 | 8.3 | 0.099 | 647 | 3.4E-05 | 2 | NoObs | |
| 206.01 | 6.2693 | 5001 | 464 | 64.98094 | 0.00042 | 5.334076 | 0.000013 | 6.941 | 0.011 | 0.06342 | 0.00009 | 0.0016 | - | 8.0 | 0.062 | 1102 | - | 2 | | |
| 208.01 | 3.2881 | 9253 | 99 | 67.711 | 0.0011 | 3.00385 | 0.000019 | 7.443 | 0.071 | 0.08887 | 0.00066 | 0.0003 | - | 8.5 | 0.042 | 1229 | 5.6E-05 | 3 | | 1 |
| 209.01 | 10.8828 | 5841 | 209 | 68.6335 | 0.0011 | 50.78974 | 0.00031 | 38.12 | 0.12 | 0.06873 | 0.00021 | 0.031 | 0.02 | 7.6 | 0.278 | 522 | 1.2E-05 | 2 | | |
| 209.02 | 7.1448 | 2349 | 147 | 78.8225 | 0.0031 | 18.79567 | 0.00034 | 21.0514 | 0.003 | 0.044 | 0.04 | 0 | 0.014 | 4.9 | 0.143 | 728 | 1.6E-05 | 2 | | |
| 211.01 | 4.8294 | 7800 | 91 | 69.0141 | 0.001 | 372.1084 | 0.0015 | 639.12 | 0.68 | 0.081 | 0.096 | 0.023 | 0.039 | 9.6 | 1.048 | 273 | - | 3 | | 1 |
| 212.01 | 3.5920 | 5153 | 250 | 72.23136 | 0.00055 | 5.69584 | 0.000048 | 12.88171 | 0.00011 | 0.064 | 0.014 | 0.0002 | - | 6.5 | 0.064 | 977 | 5.3E-05 | 2 | | |
| 214.01 | 1.4852 | 5803 | 324 | 64.74221 | 0.00022 | 3.3118618 | 0.0000042 | 9.4 | 2.8 | 0.1106 | 0.0035 | 0.93 | 0.28 | 12.1 | 0.044 | 1119 | 2.2E-05 | 2 | | |
| 216.01 | 3.6988 | 5400 | 146 | 74.2081 | 0.00097 | 20.17213 | 0.00013 | 52.37 | 0.4 | 0.06487 | 0.00038 | 0.0005 | - | 8.2 | 0.145 | 635 | 8.3E-05 | 2 | NoObs | |
| 217.01 | 2.8377 | 22161 | 818 | 66.41389 | 0.00019 | 3.9050889 | 0.0000051 | 11.892 | 0.017 | 0.13352 | 0.00014 | 0.0004 | - | 10.4 | 0.048 | 935 | 2.0E-05 | 2 | | 1 |
| 219.01 | 5.3499 | 3196 | 228 | 65.46923 | 0.00078 | 8.025085 | 0.000037 | 12.067 | 0.042 | 0.0505 | 0.00015 | 0.0012 | - | 3.8 | 0.077 | 710 | 1.0E-04 | 2 | | |
| 220.01 | 2.5693 | 1912 | 265 | 65.93893 | 0.00038 | 2.4220912 | 0.0000053 | 6.3 | 3.3 | 0.041 | 0.0042 | 0.55 | 0.72 | 2.6 | 0.034 | 987 | 6.1E-05 | 2 | | |
| 220.02 | 2.8695 | 109 | 12 | 66.6388 | 0.0072 | 4.12515 | 0.00012 | 11.37 | 0.64 | 0.01042 | 0.00052 | 0.0198 | - | 0.7 | 0.049 | 822 | - | 4 | | |
| 221.01 | 2.6291 | 3998 | 283 | 65.44215 | 0.00037 | 3.4130416 | 0.0000073 | 10.544 | 0.04 | 0.05615 | 0.00016 | - | - | 4.4 | 0.043 | 935 | 2.7E-05 | 2 | | |
| 222.01 | 2.8034 | 1305 | 65 | 65.6572 | 0.0017 | 6.312382 | 0.000058 | 18.28 | 0.28 | 0.03254 | 0.00039 | 0.0298 | - | 2.1 | 0.057 | 613 | 3.7E-05 | 2 | 1 | |
| 222.02 | 3.2313 | 891 | 32 | 63.7728 | 0.0044 | 12.79397 | 0.0003 | 30.47 | 0.95 | 0.02647 | 0.00065 | 0.0253 | - | 1.7 | 0.092 | 482 | 3.9E-05 | 2 | 1 | |
| 223.01 | 1.5125 | 1128 | 67 | 67.477 | 0.0011 | 3.177431 | 0.000019 | 12 | 24 | 0.034 | 0.013 | 0.7 | 1.2 | 2.7 | 0.041 | 964 | 3.6E-05 | 2 | | 1 |
| 223.02 | 4.0259 | 991 | 25 | 80.0334 | 0.004 | 41.0084 | 0.00084 | 68 | 381 | 0.03 | 0.034 | 0.6 | 2.4 | 2.4 | 0.226 | 410 | - | 4 | | |
| 225.01 | 1.2452 | 2571 | 200 | 74.537 | 0.00032 | 0.838598 | 0.0000022 | 4.2 | 1.3 | 0.04932 | 0.00032 | 0.47 | 0.14 | 4.9 | 0.018 | 1903 | - | 3 | | 1 |
| 226.01 | 3.0258 | 773 | 39 | 71.1091 | 0.0029 | 8.3089 | 0.00014 | 15 | 39 | 0.028 | 0.014 | 0.7 | 1.3 | 1.6 | 0.076 | 595 | 6.1E-05 | 3 | | 1 |
| 227.01 | 4.6968 | 1304 | 55 | 69.5662 | 0.0024 | 17.66076 | 0.00025 | 42.02 | 0.57 | 0.03955 | 0.00039 | 0.031 | 0.033 | 2.9 | 0.11 | 440 | 1.9E-05 | 2 | 3 | |
| 229.01 | 2.9170 | 3054 | 228 | 67.93353 | 0.0005 | 3.5732 | 0.00001 | 9.976 | 0.044 | 0.04922 | 0.00017 | 0.0003 | - | 6.0 | 0.047 | 1207 | 5.9E-05 | 2 | | 1 |
| 232.01 | 4.9765 | 2256 | 158 | 67.00465 | 0.00095 | 12.465891 | 0.000067 | 19.996 | 0.1 | 0.04269 | 0.00018 | 0.028 | 0.017 | 3.6 | 0.107 | 694 | 3.4E-05 | 2 | NoObs | |
| 232.02 | 3.7903 | 354 | 32 | 67.0179 | 0.0056 | 5.76607 | 0.00018 | 11.83 | 0.36 | 0.0185 | 0.00048 | 0.0102 | - | 1.6 | 0.064 | 897 | 5.6E-05 | 2 | NoObs | |
| 234.01 | 4.5866 | 778 | 53 | 65.1832 | 0.0026 | 9.61391 | 0.00014 | 11 | 15 | 0.0278 | 0.0067 | 0.73 | 0.95 | 3.5 | 0.091 | 897 | 3.4E-05 | 2 | | |
| 235.01 | 2.0142 | 661 | 46 | 66.8175 | 0.0019 | 5.632479 | 0.000061 | 21.29 | 0.51 | 0.02282 | 0.00042 | 0.0027 | - | 1.8 | 0.06 | 781 | 7.9E-05 | 2 | | 1 |
| 237.01 | 3.4487 | 601 | 50 | 67.7859 | 0.0024 | 8.50827 | 0.00011 | 19.7 | 0.33 | 0.02262 | 0.00031 | 0.021 | - | 2.3 | 0.083 | 834 | 8.0E-05 | 2 | | |
| 238.01 | 4.4245 | 472 | 32 | 68.0935 | 0.0047 | 17.23217 | 0.00045 | 23 | 61 | 0.0214 | 0.0098 | 0.7 | 1.4 | 2.5 | 0.135 | 742 | 8.5E-05 | 2 | NoObs | |
| 239.01 | 2.8293 | 1378 | 76 | 71.5556 | 0.0013 | 5.640649 | 0.000043 | 9.3 | 5.9 | 0.0382 | 0.0041 | 0.81 | 0.56 | 3.9 | 0.064 | 1003 | 1.6E-04 | 2 | | 1 |
| 240.01 | 4.2286 | 1322 | 84 | 71.6146 | 0.0016 | 4.286837 | 0.000041 | 8.04 | 0.076 | 0.03302 | 0.00027 | 0.0098 | - | 3.1 | 0.053 | 1063 | 3.5E-05 | 2 | | 1 |
| 241.01 | 3.5173 | 825 | 46 | 64.7933 | 0.0027 | 13.82145 | 0.00027 | 22 | 49 | 0.028 | 0.012 | 0.7 | 1.3 | 1.7 | 0.107 | 516 | 1.4E-05 | 2 | | |
| 242.01 | 5.7720 | 3941 | 193 | 71.34318 | 0.00098 | 7.258477 | 0.000041 | 10.183 | 0.042 | 0.05567 | 0.0002 | 0.0013 | - | 5.7 | 0.074 | 850 | 5.6E-05 | 2 | | |
| 244.01 | 2.8876 | 1176 | 211 | 111.52718 | 0.00043 | 12.720359 | 0.000038 | 19.1 | 2.7 | 0.03659 | 0.00081 | 0.87 | 0.22 | 4.5 | 0.11 | 865 | - | 2 | | |
| 244.02 | 3.5837 | 395 | 157 | 104.7062 | 0.0013 | 6.23855 | 0.000058 | 7.5 | 6.3 | 0.0214 | 0.0027 | 0.84 | 0.59 | 2.6 | 0.068 | 1101 | - | 2 | | |
| 245.01 | 4.7135 | 575 | 96 | 108.239 | 0.0013 | 39.79454 | 0.00038 | 67.02 | 0.57 | 0.02158 | 0.00015 | 0.0171 | - | 2.1 | 0.215 | 482 | 5.3E-06 | 2 | 1,2,3 | |
| 246.01 | 3.4381 | 283 | 127 | 106.85727 | 0.0007 | 5.398753 | 0.000027 | 9.5 | 6.7 | 0.0167 | 0.0022 | 0.63 | 0.78 | 1.9 | 0.062 | 1032 | 1.2E-05 | 2 | 1,2,3 | |
| 247.01 | 2.0575 | 999 | 22 | 114.1234 | 0.0023 | 13.81524 | 0.00032 | 52.5 | 2.3 | 0.0312 | 0.0011 | 0.0309 | - | 2.1 | 0.089 | 437 | 1.3E-05 | 2 | 1,2,3 | |
| 248.01 | 2.5695 | 1803 | 54 | 103.2885 | 0.0013 | 7.203494 | 0.000065 | 22.73 | 0.4 | 0.03948 | 0.00055 | 0.025 | 0.033 | 2.9 | 0.06 | 584 | 5.5E-05 | 2 | 3 | |
| 248.02 | 2.0752 | 1348 | 28 | 102.8387 | 0.0022 | 10.91401 | 0.00018 | 43.1 | 1.8 | 0.0343 | 0.0011 | 0.0088 | - | 2.5 | 0.079 | 509 | 5.5E-05 | 2 | 3 | |
| 248.03 | 1.6378 | 764 | 31 | 105.1278 | 0.0019 | 2.576536 | 0.000033 | 13 | 38 | 0.027 | 0.012 | 0.1 | 2 | 2.0 | 0.03 | 825 | 6.7E-05 | 2 | 3 | |
| 249.01 | 1.8415 | 1775 | 74 | 108.75705 | 0.00086 | 9.549259 | 0.000058 | 43 | 101 | 0.039 | 0.014 | 0.5 | 1.6 | 3.1 | 0.07 | 519 | 1.5E-05 | 2 | 1,2,3 | |
| 250.01 | 2.8200 | 2855 | 74 | 103.4024 | 0.0012 | 12.282356 | 0.000097 | 36.32 | 0.48 | 0.04892 | 0.00048 | 0.028 | 0.03 | 3.6 | 0.085 | 491 | 1.8E-05 | 2 | 3 | |
| 250.02 | 2.1205 | 2011 | 39 | 82.8815 | 0.0023 | 17.25204 | 0.00025 | 34 | 10 | 0.0486 | 0.0015 | 0.81 | 0.24 | 3.6 | 0.107 | 438 | 1.8E-05 | 2 | 3 | |
| 250.03 | 1.9814 | 343 | 15 | 69.2598 | 0.0046 | 3.543871 | 0.000071 | 14.5 | 1 | 0.0173 | 0.001 | 0.0412 | - | 1.3 | 0.037 | 744 | - | 4 | | |
| 251.01 | 1.8489 | 2342 | 82 | 104.08747 | 0.0008 | 4.164371 | 0.000023 | 18.37 | 0.28 | 0.04436 | 0.0005 | 0.02 | 0.027 | 2.9 | 0.04 | 653 | 4.6E-05 | 2 | 1,2,3 | |
| 252.01 | 3.5652 | 2052 | 45 | 103.5012 | 0.002 | 17.60439 | 0.00024 | 40 | 135 | 0.042 | 0.022 | 0.1 | 2.2 | 2.7 | 0.104 | 406 | 2.8E-05 | 2 | 3 | |
| 253.01 | 1.7694 | 2157 | 49 | 103.6019 | 0.0012 | 6.38324 | 0.000071 | 31.13 | 0.81 | 0.04291 | 0.00079 | 0.0427 | - | 2.9 | 0.054 | 592 | 1.4E-05 | 2 | 3 | |
| 254.01 | 1.9011 | 39093 | 795 | 103.82108 | 0.00009 | 2.4552389 | 0.0000016 | 11.95 | 0.42 | 0.1841 | 0.0012 | 0.44 | 0.2 | 13.0 | 0.029 | 824 | 2.5E-05 | 3 | NoObs | 1 |
| 255.01 | 4.1157 | 2457 | 65 | 122.8256 | 0.0015 | 27.52156 | 0.00031 | 54.25 | 0.7 | 0.04484 | 0.00047 | 0.018 | 0.01 | 3.6 | 0.149 | 388 | 2.5E-05 | 2 | NoObs | |
| 256.01 | 1.2273 | 16968 | 55 | 102.77735 | 0.00093 | 1.378681 | 0.000012 | 9 | 2.7 | 0.1235 | 0.0023 | 0.158 | 0.048 | 14.8 | 0.021 | 1160 | - | 3 | | 1 |
| 257.01 | 2.3916 | 514 | 33 | 105.6621 | 0.0021 | 6.883344 | 0.000098 | 18 | 65 | 0.023 | 0.014 | 0.7 | 1.6 | 4.0 | 0.075 | 1230 | - | 2 | | |
| 258.01 | 5.2610 | 998 | 39 | 105.4942 | 0.0018 | 4.157642 | 0.000069 | 5.2 | 7.6 | 0.0281 | 0.0069 | 0 | 1.4 | 4.5 | 0.054 | 1449 | - | 3 | | 1 |
| 260.01 | 4.5271 | 94 | 20 | 105.786 | 0.0063 | 10.49577 | 0.00046 | 17.32 | 0.52 | 0.0096 | 0.00026 | 0.012 | 0.01 | 1.2 | 0.097 | 919 | - | 2 | | |
| 260.02 | 10.7302 | 324 | 37 | 178.0423 | 0.0039 | 100.27937 | 0.00052 | 61 | 127 | 0.0178 | 0.006 | 0.6 | 1.4 | 2.2 | 0.435 | 434 | - | 2 | | |
| 261.01 | 3.8752 | 693 | 45 | 104.0206 | 0.0022 | 16.23844 | 0.00017 | 20 | 29 | 0.027 | 0.0067 | 0.8 | 0.85 | 5.6 | 0.133 | 929 | - | 2 | | |
| 262.01 | 4.1589 | 111 | 27 | 105.6257 | 0.0035 | 7.81279 | 0.00019 | 7 | 21 | 0.0111 | 0.0046 | 0.9 | 1 | 1.6 | 0.081 | 1106 | 1.7E-05 | 2 | 1,2 | |
| 263.01 | 4.1456 | 165 | 13 | 117.7331 | 0.0086 | 20.7183 | 0.0013 | 40.1 | 2.3 | 0.01143 | 0.00058 | 0.021 | 0.054 | 1.5 | 0.151 | 682 | 8.0E-06 | 3 | 1,2,3 | 1 |
| 265.01 | 3.1809 | 88 | 19 | 102.7477 | 0.0036 | 3.567971 | 0.000083 | 8.67 | 0.27 | 0.0087 | 0.00024 | 0.012 | 0.036 | 1.1 | 0.047 | 1303 | 3.2E-05 | 2 | 1,2,3 | |
| 268.01 | 12.0373 | 486 | 68 | 108.9277 | 0.0032 | 110.3742 | 0.003 | 61 | 73 | 0.0205 | 0.0041 | 0.5 | 1.1 | 1.8 | 0.406 | 295 | 1.6E-05 | 2 | | 1 |
| 269.01 | 6.1397 | 89 | 25 | 118.0993 | 0.0059 | 18.01136 | 0.00073 | 12 | 30 | 0.0099 | 0.0037 | 0.8 | 1 | 1.7 | 0.143 | 918 | 2.3E-05 | 2 | 1,2 | |
| 270.01 | 5.9033 | 105 | 32 | 108.033 | 0.0043 | 12.58084 | 0.00036 | 16.01 | 0.34 | 0.00925 | 0.00018 | 0.01 | 0.01 | 0.9 | 0.101 | 735 | 4.8E-05 | 2 | 1,2 | |
| 270.02 | 8.3939 | 139 | 31 | 95.044 | 0.0059 | 33.67205 | 0.00087 | 31.08 | 0.65 | 0.01051 | 0.00022 | 0.017 | 0.01 | 1.0 | 0.195 | 529 | - | 4 | | |
| 271.01 | 6.9832 | 312 | 43 | 105.5501 | 0.0041 | 48.6292 | 0.0014 | 33 | 48 | 0.0181 | 0.0042 | 0.79 | 0.87 | 2.0 | 0.269 | 520 | 1.2E-05 | 2 | 1,2,3 | |
| 271.02 | 6.9360 | 338 | 66 | 142.0721 | 0.0022 | 29.39292 | 0.00036 | 32.9 | 9.9 | 0.01655 | 0.00021 | - | - | 1.8 | 0.192 | 615 | 1.2E-05 | 3 | 1,2 | 1 |
| 273.01 | 1.8180 | 281 | 49 | 108.0667 | 0.0012 | 10.573667 | 0.000089 | 46.9 | 1 | 0.01548 | 0.00028 | 0.016 | 0.033 | 1.1 | 0.093 | 655 | 2.3E-05 | 2 | 1,2,3 | |
| 274.01 | 4.5111 | 61 | 14 | 108.9264 | 0.0072 | 15.09124 | 0.00055 | 12 | 63 | 0.0089 | 0.0067 | 0.9 | 1.2 | 1.1 | 0.124 | 805 | 1.7E-05 | 3 | 1,2,3 | 1 |
| 275.01 | 7.0060 | 156 | 32 | 109.8237 | 0.0048 | 15.79142 | 0.00077 | 14 | 30 | 0.0118 | 0.0045 | 0.6 | 1.4 | 1.2 | 0.121 | 717 | 1.3E-05 | 2 | 1,2,3 | |
| 276.01 | 4.6992 | 429 | 58 | 143.4006 | 0.0023 | 41.74523 | 0.00077 | 47 | 65 | 0.0207 | 0.0048 | 0.75 | 0.93 | 2.6 | 0.244 | 569 | 1.5E-05 | 2 | 1 | |
| 277.01 | 7.6827 | 421 | 93 | 104.6804 | 0.0018 | 16.23675 | 0.00021 | 16.85 | 0.1 | 0.01945 | 0.00012 | 0.029 | 0.022 | 2.1 | 0.124 | 723 | 1.7E-05 | 2 | 1 | |
| 279.01 | 8.0747 | 1369 | 176 | 109.70148 | 0.00091 | 28.45557 | 0.00019 | 24.1 | 8.6 | 0.0346 | 0.0022 | 0.5 | 0.62 | 4.9 | 0.191 | 708 | 1.4E-05 | 2 | 1 | |
| 279.02 | 6.4306 | 213 | 31 | 69.9418 | 0.0054 | 15.41304 | 0.00034 | 13 | 30 | 0.0147 | 0.0068 | 0.7 | 1.3 | 2.1 | 0.127 | 868 | - | 4 | | |
| 280.01 | 2.5779 | 341 | 80 | 113.89515 | 0.00098 | 11.87286 | 0.000081 | 16.6 | 5 | 0.02031 | 0.00023 | 0.82 | 0.25 | 3.1 | 0.108 | 1018 | 1.8E-05 | 2 | 1 | |
| 281.01 | 8.0749 | 284 | 57 | 122.0403 | 0.0029 | 19.55687 | 0.00048 | 13 | 14 | 0.0163 | 0.0031 | 0.71 | 0.87 | 3.7 | 0.152 | 930 | 2.7E-05 | 2 | 1,2 | |
| 282.01 | 5.8542 | 648 | 112 | 115.1037 | 0.0014 | 27.50882 | 0.00026 | 26 | 15 | 0.0251 | 0.0027 | 0.72 | 0.64 | 2.8 | 0.178 | 620 | 2.3E-05 | 2 | 1 | |
| 282.02 | 3.2310 | 73 | 18 | 72.0688 | 0.0048 | 8.45735 | 0.00017 | 20 | 124 | 0.008 | 0.01 | 0.2 | 2.9 | 0.9 | 0.081 | 919 | - | 4 | | |
| 283.01 | 3.3342 | 460 | 37 | 103.5999 | 0.0028 | 16.09173 | 0.00031 | 20 | 28 | 0.0237 | 0.0053 | 0.89 | 0.65 | 4.7 | 0.132 | 930 | 2.2E-05 | 2 | 1,2 | |
| 284.01 | 2.7486 | 173 | 20 | 112.423 | 0.0037 | 18.01109 | 0.00046 | 19.8 | 5.9 | 0.01509 | 0.00058 | 0.88 | 0.26 | 2.5 | 0.141 | 859 | 1.8E-05 | 2 | 1,2,3 | 1 |
| 284.02 | 3.4231 | 107 | 23 | 102.6342 | 0.0039 | 6.41508 | 0.00017 | 8 | 26 | 0.0117 | 0.0062 | 0.9 | 1.1 | 2.0 | 0.071 | 1211 | 2.2E-05 | 3 | 1,2 | |
| 284.03 | 3.2542 | 106 | 23 | 101.8617 | 0.0036 | 6.17824 | 0.00015 | 7 | 27 | 0.0112 | 0.0063 | 0.9 | 1.1 | 1.9 | 0.069 | 1228 | 2.3E-05 | 3 | 1,2 | |
| 285.01 | 5.8394 | 406 | 30 | 112.2811 | 0.0054 | 13.74872 | 0.00052 | 13 | 29 | 0.0201 | 0.008 | 0.7 | 1.2 | 2.1 | 0.11 | 757 | 2.7E-05 | 2 | 1,2 | |
| 288.01 | 6.2875 | 202 | 57 | 110.2711 | 0.0026 | 10.2754 | 0.00019 | 12.72 | 0.15 | 0.01287 | 0.00014 | 0.0046 | - | 1.5 | 0.093 | 878 | 2.8E-05 | 2 | 1 | |
| 289.01 | 7.8380 | 453 | 52 | 124.9949 | 0.0026 | 26.62907 | 0.00054 | 26.43 | 0.33 | 0.01911 | 0.00023 | 0.007 | 0.017 | 2.0 | 0.171 | 604 | 1.2E-05 | 2 | 1 | |
| 291.01 | 7.3578 | 339 | 38 | 118.1545 | 0.0041 | 31.51605 | 0.00086 | 33.08 | 0.58 | 0.01626 | 0.00027 | 0.016 | 0.014 | 1.5 | 0.196 | 495 | 5.1E-05 | 2 | 1 | |
| 291.02 | 2.1775 | 140 | 16 | 67.1034 | 0.0048 | 8.12993 | 0.00017 | 28 | 197 | 0.012 | 0.017 | 0.4 | 2.9 | 1.0 | 0.079 | 780 | - | 4 | | |
| 292.01 | 2.2814 | 219 | 48 | 104.8415 | 0.0015 | 2.58665 | 0.000037 | 9 | 20 | 0.0139 | 0.0057 | 0.2 | 1.7 | 2.0 | 0.038 | 1478 | 1.5E-05 | 2 | 1,2 | |
| 294.01 | 5.5281 | 422 | 32 | 126.0347 | 0.0037 | 34.4361 | 0.001 | 35 | 83 | 0.0195 | 0.008 | 0.7 | 1.4 | 2.2 | 0.213 | 565 | 4.5E-05 | 2 | 1 | |
| 295.01 | 2.8818 | 277 | 19 | 104.8769 | 0.0018 | 5.317406 | 0.000085 | 7 | 33 | 0.017 | 0.012 | 0.9 | 1.2 | 2.0 | 0.06 | 1088 | 1.3E-05 | 2 | 1 | |
| 296.01 | 5.2805 | 437 | 25 | 111.4863 | 0.0038 | 28.8605 | 0.0017 | 30 | 96 | 0.02 | 0.011 | 0.7 | 1.4 | 2.4 | 0.189 | 619 | 1.3E-05 | 2 | 1 | |
| 297.01 | 2.9110 | 123 | 21 | 105.2745 | 0.0035 | 5.65189 | 0.00019 | 8 | 27 | 0.0126 | 0.0067 | 0.8 | 1.2 | 1.8 | 0.065 | 1192 | 1.5E-05 | 2 | 1 | |
| 298.01 | 2.6557 | 259 | 17 | 111.308 | 0.0039 | 19.96343 | 0.00059 | 29 | 146 | 0.017 | 0.014 | 0.9 | 1.3 | 1.7 | 0.136 | 610 | 2.2E-05 | 3 | NoObs | |
| 299.01 | 1.9962 | 312 | 51 | 103.5428 | 0.0012 | 1.541677 | 0.000013 | 3.8 | 9.7 | 0.0176 | 0.0076 | 0.8 | 1.1 | 3.6 | 0.028 | 2002 | 3.1E-05 | 2 | 1 | |
| 301.01 | 3.8551 | 184 | 34 | 104.716 | 0.0029 | 6.00245 | 0.00011 | 11.91 | 0.27 | 0.01253 | 0.00024 | 0.029 | 0.044 | 1.6 | 0.067 | 1142 | 7.7E-05 | 2 | 1 | |
| 302.01 | 8.6779 | 540 | 52 | 106.9965 | 0.0031 | 24.85467 | 0.00051 | 22.35 | 0.23 | 0.02266 | 0.00024 | 0.02 | 0.02 | 3.8 | 0.179 | 873 | 5.8E-05 | 2 | 1 | |
| 303.01 | 6.3927 | 762 | 64 | 106.3643 | 0.0021 | 60.92983 | 0.00085 | 48 | 35 | 0.0277 | 0.0036 | 0.77 | 0.64 | 2.9 | 0.306 | 426 | 3.3E-05 | 2 | | |
| 304.01 | 2.6128 | 604 | 58 | 107.9095 | 0.0016 | 8.511982 | 0.000093 | 18 | 35 | 0.0249 | 0.0079 | 0.8 | 0.98 | 5.0 | 0.087 | 1199 | 1.7E-05 | 2 | 1 | |
| 305.01 | 2.3541 | 416 | 36 | 104.8397 | 0.0022 | 4.603576 | 0.000072 | 12 | 29 | 0.02 | 0.011 | 0.7 | 1.4 | 2.5 | 0.053 | 963 | 1.4E-04 | 2 | 1 | |
| 306.01 | 2.9825 | 550 | 12 | 111.3658 | 0.0068 | 24.3077 | 0.0013 | 71.9 | 5.5 | 0.0232 | 0.0014 | 0.011 | 0.022 | 3.6 | 0.168 | 657 | 8.5E-05 | 2 | 1,2 | |
| 307.01 | 3.6490 | 224 | 24 | 109.2686 | 0.004 | 19.67445 | 0.00055 | 42.2 | 1.4 | 0.01374 | 0.0004 | 0.008 | 0.01 | 1.7 | 0.148 | 736 | 5.2E-05 | 3 | 1 | |
| 308.01 | 6.3242 | 754 | 55 | 120.5441 | 0.0024 | 35.59061 | 0.00058 | 22 | 8 | 0.0281 | 0.0016 | 0.88 | 0.34 | 4.7 | 0.223 | 699 | 3.3E-05 | 2 | 1 | |
| 312.01 | 2.7215 | 197 | 24 | 108.586 | 0.0036 | 11.57898 | 0.00029 | 33.1 | 1.1 | 0.01322 | 0.00037 | 0.015 | 0.037 | 1.6 | 0.102 | 867 | 6.6E-05 | 2 | 1 | |
| 313.01 | 3.1424 | 547 | 36 | 110.6353 | 0.0022 | 18.73564 | 0.00031 | 25 | 37 | 0.0254 | 0.0068 | 0.87 | 0.72 | 3.1 | 0.139 | 651 | 1.3E-05 | 2 | 1,2,3 | |
| 313.02 | 3.2232 | 320 | 33 | 112.8879 | 0.0026 | 8.43628 | 0.00015 | 14 | 39 | 0.0178 | 0.0095 | 0.7 | 1.3 | 2.2 | 0.081 | 852 | 1.7E-05 | 2 | 1,2 | |
| 314.01 | 2.4517 | 747 | 42 | 110.852 | 0.0015 | 13.78105 | 0.00016 | 24.2 | 7.3 | 0.02918 | 0.00058 | 0.74 | 0.22 | 1.9 | 0.091 | 446 | 2.3E-05 | 2 | 1 | |
| 314.02 | 2.0157 | 558 | 21 | 103.9994 | 0.0028 | 23.0904 | 0.00034 | 73 | 680 | 0.023 | 0.039 | 0.7 | 2.5 | 1.6 | 0.128 | 376 | 2.3E-05 | 2 | 1 | |
| 315.01 | 4.3681 | 979 | 68 | 121.9796 | 0.0017 | 35.5917 | 0.00045 | 41 | 30 | 0.0325 | 0.0049 | 0.81 | 0.59 | 4.8 | 0.211 | 526 | 4.6E-05 | 2 | 1 | |
| 316.01 | 5.0479 | 524 | 56 | 117.9011 | 0.002 | 15.77135 | 0.00022 | 18 | 26 | 0.0221 | 0.0056 | 0.7 | 1.1 | 2.8 | 0.126 | 750 | 1.4E-05 | 3 | 1,2 | |
| 317.01 | 7.2681 | 427 | 44 | 139.3631 | 0.003 | 22.20767 | 0.0006 | 18 | 33 | 0.0199 | 0.0058 | 0.6 | 1.2 | 3.7 | 0.166 | 918 | 9.9E-05 | 3 | 1 | |
| 318.01 | 10.3373 | 1286 | 44 | 107.8303 | 0.0025 | 38.58439 | 0.00038 | 29.99 | 0.45 | 0.03035 | 0.00048 | 0.03 | 0.036 | 4.4 | 0.235 | 671 | 1.2E-05 | 3 | 1 | |
| 319.01 | 5.3951 | 1793 | 137 | 109.6254 | 0.001 | 46.15115 | 0.00034 | 54 | 16 | 0.04055 | 0.00022 | 0.39 | 0.12 | 4.3 | 0.248 | 508 | 1.7E-05 | 2 | 1 | |
| 321.01 | 2.5947 | 174 | 39 | 103.4551 | 0.002 | 2.426307 | 0.000032 | 7.27 | 0.16 | 0.01226 | 0.00022 | 0.0016 | - | 0.9 | 0.035 | 1068 | - | 2 | | |
| 323.01 | 3.3770 | 592 | 11 | 102.8509 | 0.0073 | 5.83674 | 0.00029 | 14.5 | 1.2 | 0.021 | 0.0014 | 0.03 | 0.085 | 2.9 | 0.065 | 1051 | 1.3E-05 | 2 | 1 | |
| 326.01 | 2.9995 | 890 | 26 | 104.0345 | 0.0031 | 8.97297 | 0.0002 | 19 | 80 | 0.029 | 0.023 | 0.6 | 1.9 | 0.9 | 0.05 | 332 | 1.4E-05 | 3 | NoObs | |
| 327.01 | 2.9306 | 147 | 30 | 105.6621 | 0.0025 | 3.254241 | 0.000056 | 8.974 | 0.082 | 0.01156 | 0.00024 | 0.0593 | - | 1.3 | 0.044 | 1304 | 2.5E-05 | 2 | 1 | |
| 330.01 | 5.2342 | 281 | 27 | 107.5328 | 0.0048 | 7.97398 | 0.00025 | 7 | 17 | 0.0174 | 0.0071 | 0.8 | 1.1 | 2.5 | 0.081 | 1021 | 3.3E-05 | 3 | NoObs | |
| 331.01 | 6.3739 | 355 | 32 | 103.8303 | 0.0047 | 18.68416 | 0.00059 | 22.47 | 0.48 | 0.01684 | 0.00034 | 0.012 | 0.01 | 1.1 | 0.134 | 494 | 4.0E-05 | 2 | 1 | |
| 332.01 | 3.8993 | 221 | 40 | 105.0885 | 0.0028 | 5.458491 | 0.0001 | 11.27 | 0.22 | 0.01399 | 0.00022 | 0.0377 | - | 1.1 | 0.06 | 841 | 4.7E-05 | 2 | 1 | |
| 333.01 | 6.0214 | 345 | 24 | 102.8642 | 0.0039 | 13.28468 | 0.00025 | 17.1828 | 0.0013 | 0.02 | 0.13 | 0.011 | 0.01 | 1.9 | 0.114 | 843 | 2.6E-05 | 2 | 1 | |
| 335.01 | 7.4741 | 763 | 53 | 129.3109 | 0.0024 | 46.56623 | 0.00083 | 45 | 95 | 0.0257 | 0.0093 | 0.4 | 1.6 | 4.3 | 0.269 | 674 | 2.3E-05 | 2 | NoObs | |
| 337.01 | 5.3936 | 324 | 25 | 110.7155 | 0.004 | 19.78404 | 0.00042 | 28.67 | 0.78 | 0.01613 | 0.00043 | 0.011 | 0.01 | 1.6 | 0.146 | 636 | 6.3E-05 | 3 | NoObs | |
| 338.01 | 2.9400 | 293 | 27 | 107.5777 | 0.0033 | 7.01048 | 0.00016 | 19.24 | 0.66 | 0.01587 | 0.00043 | 0.011 | 0.035 | 2.3 | 0.072 | 940 | 5.7E-05 | 3 | NoObs | |
| 339.01 | 2.4238 | 286 | 44 | 103.1393 | 0.0017 | 1.980349 | 0.000031 | 3.9 | 9.1 | 0.0175 | 0.0064 | 0.8 | 1.1 | 1.5 | 0.031 | 1319 | 1.8E-05 | 3 | NoObs | |
| 339.02 | 3.1373 | 167 | 17 | 71.3385 | 0.0052 | 6.41681 | 0.00015 | 16.6 | 0.19 | 0.01286 | 0.00043 | 0.0625 | - | 1.1 | 0.069 | 884 | - | 4 | | |
| 340.01 | 14.2340 | 21220 | 172 | 93.6222 | 0.0073 | 23.67378 | 0.00058 | 14.366 | 0.071 | 0.12942 | 0.00052 | 0.001 | 0.01 | 30.4 | 0.173 | 862 | - | 3 | | 1 |
| 341.01 | 2.9799 | 796 | 31 | 109.6648 | 0.0026 | 7.17068 | 0.00013 | 19.09 | 0.54 | 0.02618 | 0.0006 | 0.027 | 0.042 | 3.3 | 0.074 | 949 | 4.5E-05 | 2 | 1 | |
| 341.02 | 2.7109 | 300 | 14 | 110.6123 | 0.0054 | 4.69975 | 0.00012 | 14.3 | 0.8 | 0.01781 | 0.00077 | 0.011 | 0.046 | 2.3 | 0.056 | 1091 | 5.7E-05 | 2 | 1 | |
| 343.01 | 3.3150 | 474 | 68 | 103.3364 | 0.0013 | 4.76166 | 0.000044 | 11.21 | 0.14 | 0.01969 | 0.0002 | 0.0089 | - | 2.2 | 0.057 | 1065 | 2.9E-05 | 2 | NoObs | |
| 343.02 | 2.4768 | 234 | 47 | 103.4963 | 0.0018 | 2.024138 | 0.000025 | 6.593 | 0.042 | 0.0147 | 0.00021 | 0.041 | - | 1.6 | 0.032 | 1421 | 3.4E-05 | 2 | 1 | |
| 344.01 | 5.8847 | 1113 | 70 | 104.3354 | 0.0018 | 39.3095 | 0.00054 | 42 | 49 | 0.0318 | 0.0067 | 0.6 | 1 | 4.0 | 0.232 | 564 | 1.0E-05 | 2 | 1 | |
| 345.01 | 4.8165 | 1246 | 79 | 106.1889 | 0.0015 | 29.88569 | 0.00055 | 37 | 32 | 0.0351 | 0.0063 | 0.72 | 0.76 | 5.8 | 0.192 | 592 | 9.9E-06 | 2 | NoObs | |
| 346.01 | 2.7973 | 989 | 20 | 103.7797 | 0.0045 | 12.92463 | 0.00041 | 36 | 241 | 0.03 | 0.045 | 0.1 | 3.1 | 3.4 | 0.106 | 684 | 4.0E-05 | 2 | NoObs | |
| 348.01 | 4.6221 | 1888 | 108 | 120.3634 | 0.0012 | 28.51109 | 0.00027 | 51.65 | 0.43 | 0.03813 | 0.00026 | 0.018 | 0.017 | 5.3 | 0.18 | 549 | 1.0E-05 | 2 | 1 | |
| 349.01 | 2.3194 | 582 | 38 | 103.4557 | 0.0019 | 14.38666 | 0.00025 | 51.7 | 1.4 | 0.02231 | 0.00044 | 0.028 | 0.04 | 2.8 | 0.119 | 772 | 1.3E-05 | 2 | NoObs | |
| 350.01 | 2.5804 | 397 | 24 | 110.2199 | 0.0027 | 12.99192 | 0.00033 | 22 | 96 | 0.02 | 0.014 | 0.9 | 1.3 | 2.5 | 0.111 | 810 | 1.4E-05 | 2 | 1 | |
| 351.01 | 14.4207 | 8331 | 210 | 73.4753 | 0.0012 | 331.6457 | 0.0017 | 190.34 | 0.16 | 0.0827 | 0.0064 | 0.011 | 0.035 | 8.5 | 0.966 | 266 | 1.9E-05 | 3 | NoObs | |
| 351.02 | 11.9806 | 4236 | 60 | 80.098 | 0.0015 | 210.4526 | 0.0021 | 142.4 | 1.6 | 0.05855 | 0.00061 | 0.001 | 0.069 | 6.0 | 0.713 | 309 | - | 4 | | |
| 351.03 | 8.0060 | 460 | 21 | 91.9518 | 0.0078 | 59.7389 | 0.0022 | 53 | 713 | 0.019 | 0.044 | 0.4 | 4 | 1.9 | 0.308 | 471 | - | 4 | | |
| 352.01 | 4.4932 | 393 | 21 | 124.8039 | 0.0049 | 27.08268 | 0.00098 | 28 | 85 | 0.02 | 0.01 | 0.8 | 1.2 | 2.4 | 0.181 | 621 | 5.9E-05 | 2 | NoObs | |
| 353.01 | 7.0683 | 3685 | 93 | 109.5299 | 0.0019 | 152.1011 | 0.0027 | 95.2 | 8.3 | 0.064 | 0.0012 | 0.88 | 0.17 | 8.2 | 0.589 | 417 | 1.5E-05 | 2 | | |
| 354.01 | 4.5028 | 473 | 32 | 104.5198 | 0.0034 | 15.95999 | 0.00037 | 21 | 51 | 0.0217 | 0.0094 | 0.7 | 1.3 | 4.9 | 0.134 | 1034 | 2.6E-05 | 2 | 1 | |
| 355.01 | 2.8403 | 304 | 38 | 105.8966 | 0.0021 | 4.903454 | 0.000074 | 10 | 25 | 0.0179 | 0.0077 | 0.7 | 1.4 | 2.0 | 0.058 | 1115 | 3.2E-05 | 2 | 1 | |
| 356.01 | 2.1029 | 1195 | 143 | 103.52521 | 0.00051 | 1.8270789 | 0.0000064 | 6 | 6.3 | 0.0328 | 0.0071 | 0.59 | 0.99 | 5.8 | 0.03 | 1655 | 2.4E-05 | 2 | NoObs | |
| 360.01 | 4.5515 | 150 | 15 | 104.5813 | 0.0051 | 5.94042 | 0.00063 | 9.13 | 0.41 | 0.01108 | 0.00046 | 0.03 | 0.066 | 1.3 | 0.065 | 1075 | 2.8E-05 | 3 | 1 | |
| 361.01 | 2.5126 | 191 | 24 | 104.2741 | 0.0031 | 3.247565 | 0.000069 | 10.37 | 0.37 | 0.01302 | 0.00038 | 0.0439 | - | 1.3 | 0.044 | 1132 | 2.7E-05 | 2 | 1 | |
| †364.01 | 4.5649 | 451 | 28 | 156.9082 | 0.0022 | 173.9 | 1.6 | 132 | 39 | 0.02395 | 0.00058 | 0.84 | 0.25 | 2.6 | 0.619 | 313 | 1.7E-05 | 3 | | |
| 365.01 | 6.6552 | 638 | 83 | 144.6778 | 0.0019 | 81.7378 | 0.0015 | 64 | 44 | 0.0251 | 0.0032 | 0.75 | 0.66 | 2.3 | 0.368 | 363 | 1.5E-05 | 2 | 1,2 | |
| 366.01 | 4.9911 | 3808 | 271 | 140.7142 | 0.00036 | 75.1119 | 0.00021 | 77.4 | 3.8 | 0.06441 | 0.00061 | 0.83 | 0.15 | 10.2 | 0.4 | 586 | 1.3E-05 | 2 | NoObs | |
| 367.01 | 2.8404 | 2267 | 176 | 110.20526 | 0.00049 | 31.57867 | 0.00011 | 98 | 108 | 0.0438 | 0.0085 | 0.6 | 1 | 5.7 | 0.203 | 646 | 3.6E-06 | 2 | 1 | |
| 368.01 | 12.7602 | 7307 | 1409 | 130.36472 | 0.0002 | 110.32148 | 0.00015 | 51.53 | 0.31 | 0.08456 | 0.00008 | 0.715 | 0.064 | 17.4 | 0.544 | 742 | 6.1E-06 | 2 | | |
| 369.01 | 1.8748 | 143 | 19 | 107.4265 | 0.003 | 5.88521 | 0.00013 | 20 | 90 | 0.0116 | 0.0089 | 0.5 | 2.2 | 1.3 | 0.066 | 1073 | 1.5E-05 | 2 | 1 | |
| 370.01 | 9.7635 | 348 | 46 | 136.6501 | 0.0035 | 42.8821 | 0.0012 | 21 | 20 | 0.0194 | 0.003 | 0.8 | 0.69 | 4.9 | 0.262 | 810 | 2.4E-05 | 2 | 1 | |
| †371.01 | 10.1954 | 1202 | 58 | 177.0753 | 0.0048 | 278 | 1488 | 65 | 20 | 0.2 | 3.8 | 1.3 | 0.39 | 60.1 | 0.916 | 400 | 5.1E-06 | 3 | | |
| 372.01 | 9.1846 | 8267 | 46 | 186.349 | 0.0044 | 125.6125 | 0.0064 | 111 | 65 | 0.0813 | 0.0093 | 0.17 | 0.9 | 8.5 | 0.499 | 344 | 2.0E-05 | 2 | NoObs | |
| 373.01 | 8.7090 | 648 | 35 | 123.9263 | 0.005 | 135.1937 | 0.0072 | 92 | 142 | 0.025 | 0.0067 | 0.7 | 1.1 | 3.5 | 0.534 | 400 | 3.9E-05 | 2 | | |
| 374.01 | 11.2542 | 646 | 44 | 169.9597 | 0.0038 | 172.6735 | 0.0051 | 96 | 124 | 0.0243 | 0.0055 | 0.6 | 1.1 | 3.3 | 0.628 | 365 | 2.1E-05 | 2 | | |
| †375.01 | 7.0221 | 5087 | 121 | 172.22424 | 0.00095 | 220 | - | 150.80509 | - | 0.07725 | - | 0.8738 | - | 8.8 | 0.729 | 300 | 9.7E-06 | 2 | NoObs | |
| 377.01 | 4.5242 | 7507 | 129 | 115.66381 | 0.00078 | 19.25832 | 0.00017 | 32 | 6.4 | 0.0776 | 0.003 | 0.56 | 0.44 | 5.7 | 0.141 | 553 | 1.8E-05 | 1 | 1,2,3 | 1 |
| 377.02 | 5.0267 | 6754 | 83 | 108.4011 | 0.0014 | 38.9116 | 0.0006 | 36.2 | 2.9 | 0.0839 | 0.0017 | 0.88 | 0.16 | 6.2 | 0.225 | 438 | 2.1E-05 | 1 | 1,2 | 1 |
| 377.03 | 1.7745 | 244 | 22 | 115.0924 | 0.0059 | 1.592928 | 0.000068 | 7 | 41 | 0.014 | 0.016 | 0.2 | 2.9 | 1.0 | 0.027 | 1264 | 3.6E-05 | 1 | | 1 |
| 379.01 | 2.5183 | 273 | 24 | 103.9955 | 0.003 | 6.71743 | 0.00014 | 13 | 54 | 0.018 | 0.012 | 0.8 | 1.3 | 3.1 | 0.074 | 1267 | 6.8E-05 | 3 | 1 | |
| 384.01 | 5.0735 | 191 | 30 | 107.4333 | 0.0036 | 5.07977 | 0.00012 | 4.7 | 9.5 | 0.015 | 0.0051 | 0.8 | 1 | 2.0 | 0.06 | 1143 | 1.2E-04 | 2 | 1 | |
| 385.01 | 3.4446 | 292 | 19 | 107.8373 | 0.0042 | 13.14613 | 0.0004 | 30 | 151 | 0.016 | 0.016 | 0.2 | 2.6 | 1.8 | 0.11 | 743 | 7.7E-05 | 2 | NoObs | |
| 386.01 | 5.3012 | 917 | 46 | 106.905 | 0.0029 | 31.15847 | 0.00063 | 47.12 | 0.75 | 0.02782 | 0.00038 | 0.029 | 0.036 | 3.4 | 0.2 | 623 | 9.3E-05 | 2 | NoObs | |
| 386.02 | 5.9833 | 716 | 24 | 133.6702 | 0.0064 | 76.735 | 0.0034 | 101.1 | 3.2 | 0.02374 | 0.00071 | 0.011 | 0.04 | 2.9 | 0.366 | 461 | 9.7E-05 | 2 | NoObs | |
| 387.01 | 3.3365 | 1021 | 32 | 115.8646 | 0.003 | 13.89952 | 0.0003 | 28 | 98 | 0.031 | 0.02 | 0.5 | 1.9 | 2.5 | 0.1 | 534 | 2.8E-05 | 2 | NoObs | |
| 388.01 | 5.3710 | 290 | 34 | 102.5496 | 0.0035 | 6.14974 | 0.00015 | 10 | 158 | 0.006 | 0.017 | 0 | 4.8 | 0.6 | 0.063 | 925 | 4.1E-05 | 2 | NoObs | |
| 392.01 | 7.3677 | 235 | 14 | 104.316 | 0.01 | 33.4205 | 0.0024 | 35.4 | 1.6 | 0.01417 | 0.00061 | 0.013 | 0.042 | 1.9 | 0.209 | 614 | 4.6E-05 | 3 | NoObs | |
| 393.01 | 6.7617 | 249 | 21 | 109.2556 | 0.0066 | 21.41586 | 0.00096 | 19 | 72 | 0.0152 | 0.0097 | 0.6 | 1.8 | 1.2 | 0.154 | 578 | 1.2E-04 | 2 | NoObs | |
| 398.01 | 4.7975 | 9436 | 155 | 103.085 | 0.001 | 51.84581 | 0.00033 | 80 | 14 | 0.092 | 0.0036 | 0.57 | 0.41 | 8.6 | 0.267 | 403 | 1.4E-05 | 2 | 3 | |
| 398.02 | 2.4429 | 1704 | 74 | 106.7183 | 0.0015 | 4.180054 | 0.000043 | 13.67 | 0.27 | 0.03726 | 0.00056 | 0.0002 | - | 3.5 | 0.05 | 932 | 2.3E-05 | 2 | 3 | |
| 398.03 | 1.7407 | 492 | 28 | 66.8188 | 0.0024 | 1.729364 | 0.000018 | 6.9 | 2.1 | 0.02047 | 0.00065 | 0.218 | 0.065 | 1.9 | 0.028 | 1246 | - | 4 | | |
| 401.01 | 5.3596 | 2154 | 100 | 118.4415 | 0.0013 | 29.19859 | 0.00029 | 45.63 | 0.38 | 0.041 | 0.00028 | 0.0002 | - | 6.2 | 0.19 | 629 | 2.5E-05 | 2 | NoObs | |
| 401.02 | 6.2359 | 1531 | 30 | 184.2868 | 0.0049 | 160.0112 | 0.0069 | 114 | 43 | 0.0434 | 0.0034 | 0.9 | 0.32 | 6.6 | 0.591 | 357 | - | 4 | | |
| 403.01 | 1.5859 | 1309 | 41 | 104.1305 | 0.0016 | 21.0569 | 0.00024 | 31.9 | 9.6 | 0.35598 | 0.00096 | 1.74 | 0.52 | 39.7 | 0.152 | 637 | 1.0E-05 | 3 | | |
| 408.01 | 3.1757 | 1485 | 63 | 106.0728 | 0.0015 | 7.381987 | 0.000074 | 18.92 | 0.29 | 0.03466 | 0.00041 | 0.0189 | - | 3.6 | 0.075 | 889 | 4.4E-05 | 2 | 1 | |
| 408.02 | 3.7260 | 869 | 31 | 99.7951 | 0.0039 | 12.56093 | 0.00034 | 25.58 | 0.69 | 0.02735 | 0.00062 | 0.011 | 0.01 | 2.9 | 0.108 | 741 | 5.0E-05 | 2 | 1 | |
| 408.03 | 5.0363 | 734 | 20 | 85.9989 | 0.0061 | 30.82869 | 0.00088 | 49 | 315 | 0.024 | 0.032 | 0 | 3 | 2.6 | 0.196 | 550 | - | 4 | | |
| 409.01 | 4.3307 | 632 | 49 | 112.5253 | 0.0025 | 13.24874 | 0.00024 | 17 | 34 | 0.0247 | 0.0098 | 0.7 | 1.2 | 1.4 | 0.108 | 545 | 4.6E-05 | 2 | | |
| 410.01 | 1.8788 | 4057 | 211 | 109.28616 | 0.00033 | 7.216812 | 0.000017 | 14.5 | 4.4 | 0.1016 | 0.005 | 0.97 | 0.29 | 12.4 | 0.076 | 1009 | - | 2 | | |
| 412.01 | 3.0251 | 3615 | 226 | 103.32514 | 0.00037 | 4.147024 | 0.00001 | 11.416 | 0.052 | 0.05341 | 0.00019 | 0.0011 | - | 7.3 | 0.052 | 1211 | 1.6E-05 | 2 | | |

| | | | | | | | | | | | | | | | | | | | |
|---|---|---|---|---|---|---|---|---|---|---|---|---|---|---|---|---|---|---|---|
| 413.01 | 2.6270 | 1023 | 34 | 109.5582 | 0.0025 | 15.22926 | 0.00027 | 47.8 | 1.4 | 0.02991 | 0.00062 | 0.042 | 0.01 | 2.8 | 0.119 | 619 | 1.2E-04 | 2 | |
| 415.01 | 7.0021 | 4827 | 118 | 178.1412 | 0.0014 | 166.7879 | 0.0019 | 229.111 | 0.016 | 0.062 | 0.063 | 0.006 | 0.01 | 7.7 | 0.611 | 352 | 2.8E-05 | 2 | |
| 416.01 | 3.7319 | 1645 | 83 | 118.8413 | 0.0013 | 18.20811 | 0.00017 | 38.65 | 0.11 | 0.03601 | 0.00028 | 0.0038 | - | 2.9 | 0.131 | 536 | 2.4E-05 | 2 | |
| 416.02 | 4.0059 | 1143 | 32 | 86.78 | 0.0033 | 88.2547 | 0.0013 | 115 | 200 | 0.035 | 0.012 | 0.78 | 0.97 | 2.8 | 0.376 | 317 | - | 4 | |
| 417.01 | 2.4571 | 6728 | 185 | 109.96607 | 0.00044 | 19.193112 | 0.000062 | 36 | 11 | 0.0972 | 0.0017 | 0.84 | 0.25 | 9.0 | 0.142 | 608 | 3.9E-05 | 2 | |
| 418.01 | 4.8665 | 12155 | 610 | 105.79613 | 0.00023 | 22.418338 | 0.000036 | 24.82 | 0.36 | 0.11484 | 0.00044 | 0.803 | 0.085 | 12.7 | 0.155 | 580 | 3.1E-05 | 2 | |
| 419.01 | 2.7205 | 7678 | 292 | 122.38996 | 0.0003 | 20.13146 | 0.000044 | 42 | 13 | 0.09084 | 0.00037 | 0.65 | 0.19 | 7.5 | 0.146 | 573 | 1.1E-05 | 2 | |
| 420.01 | 2.2582 | 2700 | 195 | 107.08404 | 0.0004 | 6.010401 | 0.000017 | 21 | 17 | 0.0474 | 0.0081 | 0.39 | 1 | 4.3 | 0.061 | 763 | 1.4E-05 | 2 | |
| 421.01 | 2.6993 | 17361 | 687 | 105.81931 | 0.00025 | 4.4542074 | 0.000009 | 16.852 | 0.036 | 0.11481 | 0.00018 | 0.001 | - | 14.5 | 0.053 | 1068 | 3.9E-05 | 2 | |
| †422.01 | 9.0034 | 17493 | 274 | 183.63055 | 0.00071 | 200 | 0.024 | 136.12 | 0.016 | 0.13831 | - | 0.8077 | - | 16.5 | 0.692 | 333 | 1.4E-05 | 2 | |
| 423.01 | 6.0128 | 9104 | 241 | 135.856 | 0.00051 | 21.087391 | 0.000085 | 29.034 | 0.097 | 0.08496 | 0.00025 | 0.0003 | - | 9.6 | 0.154 | 685 | 2.7E-05 | 2 | |
| 425.01 | 1.5268 | 12252 | 122 | 102.75274 | 0.00034 | 5.428352 | 0.000013 | 15 | 4.2 | 0.133 | 0.014 | 0.9 | 0.26 | 13.2 | 0.061 | 967 | - | 2 | |
| 426.01 | 3.4978 | 907 | 36 | 105.1505 | 0.0027 | 16.30089 | 0.00035 | 37.91 | 0.94 | 0.02709 | 0.00055 | 0.029 | 0.01 | 3.5 | 0.13 | 773 | 4.2E-05 | 2 | |
| 427.01 | 2.9072 | 1812 | 44 | 124.7364 | 0.002 | 24.6157 | 0.00034 | 38 | 11 | 0.0457 | 0.001 | 0.75 | 0.22 | 4.6 | 0.165 | 554 | 1.1E-05 | 2 | |
| 428.01 | 6.8663 | 3825 | 361 | 105.51811 | 0.00049 | 6.873163 | 0.000023 | 8.103 | 0.017 | 0.05591 | 0.0001 | 0 | 0.01 | 5.6 | 0.073 | 959 | 4.1E-05 | 2 | |
| 429.01 | 4.0370 | 2892 | 150 | 105.52804 | 0.00079 | 8.600087 | 0.000048 | 17.1 | 0.1 | 0.04762 | 0.00023 | 0.027 | 0.017 | 4.8 | 0.081 | 760 | 4.3E-05 | 3 | |
| 430.01 | 2.7448 | 1720 | 52 | 112.4041 | 0.0016 | 12.37645 | 0.00019 | 33 | 80 | 0.038 | 0.018 | 0.4 | 1.7 | 2.7 | 0.087 | 493 | 1.4E-05 | 2 | |
| 431.01 | 3.1580 | 1128 | 44 | 111.7122 | 0.002 | 18.86998 | 0.00031 | 36 | 84 | 0.033 | 0.015 | 0.7 | 1.3 | 3.6 | 0.139 | 622 | 1.2E-05 | 2 | |
| 431.02 | 3.8717 | 833 | 24 | 87.3073 | 0.0045 | 46.90198 | 0.00094 | 51 | 68 | 0.032 | 0.008 | 0.88 | 0.66 | 3.5 | 0.254 | 460 | - | 4 | |
| 432.01 | 2.1018 | 1041 | 72 | 107.35008 | 0.0009 | 5.263436 | 0.000033 | 14 | 24 | 0.0321 | 0.0096 | 0.7 | 1.1 | 3.6 | 0.061 | 1049 | 2.3E-05 | 2 | |
| 433.01 | 2.8751 | 2960 | 104 | 104.09249 | 0.00084 | 4.03042 | 0.000023 | 11 | 19 | 0.049 | 0.018 | 0.3 | 1.5 | 5.8 | 0.05 | 1076 | 4.7E-05 | 3 | |
| 433.02 | 11.7758 | 12968 | 195 | 132.2029 | 0.0013 | 328.2403 | 0.0019 | 176.2 | 8.4 | 0.1133 | 0.0013 | 0.73 | 0.18 | 13.4 | 0.935 | 249 | - | 4 | |
| 435.01 | 5.4169 | 1558 | 57 | 111.9483 | 0.0019 | 20.54902 | 0.00033 | 29.53 | 0.4 | 0.0352 | 0.00041 | 0.003 | 0.01 | 3.0 | 0.148 | 579 | 2.2E-05 | 2 | |
| 438.01 | 2.3764 | 940 | 42 | 107.7956 | 0.0017 | 5.931204 | 0.00007 | 16 | 39 | 0.03 | 0.016 | 0.6 | 1.4 | 2.2 | 0.056 | 668 | 2.3E-05 | 2 | |
| 439.01 | 2.1891 | 2240 | 201 | 103.44904 | 0.00038 | 1.9022064 | 0.000005 | 5.8 | 4.1 | 0.0447 | 0.0063 | 0.56 | 0.83 | 2.7 | 0.029 | 1014 | 2.5E-05 | 2 | NoObs |
| 440.01 | 4.0463 | 984 | 41 | 110.9313 | 0.0023 | 15.90655 | 0.00027 | 22 | 47 | 0.031 | 0.014 | 0.7 | 1.2 | 2.8 | 0.12 | 576 | 7.2E-05 | 2 | NoObs |
| 440.02 | 1.6038 | 748 | 36 | 103.8861 | 0.0015 | 4.973444 | 0.00005 | 23.7 | 0.77 | 0.02488 | 0.00048 | 0.031 | 0.048 | 2.2 | 0.055 | 850 | 9.6E-05 | 2 | NoObs |
| 442.01 | 4.4306 | 426 | 26 | 104.6832 | 0.0043 | 13.53981 | 0.00041 | 23.22 | 0.68 | 0.01862 | 0.00048 | 0.008 | 0.01 | 1.9 | 0.113 | 722 | 8.2E-05 | 2 | NoObs |
| 442.02 | 2.3333 | 188 | 23 | 67.5316 | 0.0036 | 1.732341 | 0.000026 | 6.29 | 0.2 | 0.01435 | 0.00037 | 0.06 | - | 1.4 | 0.029 | 1425 | - | 4 | |
| 443.01 | 4.7403 | 745 | 44 | 113.0459 | 0.0027 | 16.21718 | 0.0003 | 26.46 | 0.44 | 0.02481 | 0.00037 | 0.0056 | - | 2.2 | 0.127 | 631 | 3.7E-05 | 2 | |
| 444.01 | 4.0665 | 469 | 29 | 110.3174 | 0.0039 | 11.7228 | 0.00031 | 22 | 114 | 0.02 | 0.019 | 0 | 2.7 | 2.0 | 0.103 | 763 | 8.4E-05 | 2 | NoObs |
| 446.01 | 2.7087 | 857 | 23 | 107.7539 | 0.0032 | 16.70916 | 0.00036 | 32 | 129 | 0.03 | 0.025 | 0.8 | 1.5 | 2.3 | 0.115 | 490 | 2.4E-05 | 2 | NoObs |
| 446.02 | 3.7173 | 609 | 14 | 118.4722 | 0.0066 | 28.5532 | 0.0015 | 59 | 364 | 0.022 | 0.031 | 0.2 | 2.9 | 1.7 | 0.164 | 411 | 2.6E-05 | 2 | NoObs |
| 448.01 | 2.8620 | 1263 | 27 | 111.4491 | 0.0033 | 10.13961 | 0.00023 | 28.6 | 1.2 | 0.03002 | 0.00094 | 0.013 | 0.04 | 2.3 | 0.079 | 564 | 1.7E-04 | 2 | NoObs |
| 448.02 | 5.0051 | 2340 | 33 | 127.4625 | 0.0037 | 43.6205 | 0.0013 | 45 | 38 | 0.0488 | 0.0088 | 0.79 | 0.67 | 3.8 | 0.21 | 346 | 1.5E-04 | 2 | NoObs |
| 452.01 | 5.0013 | 458 | 50 | 102.9417 | 0.0025 | 3.705996 | 0.000064 | 5.758 | 0.08 | 0.01956 | 0.00025 | 0.0028 | - | 2.3 | 0.048 | 1242 | 6.7E-05 | 2 | |
| 454.01 | 4.6580 | 831 | 23 | 103.5546 | 0.0047 | 29.008 | 0.001 | 40 | 156 | 0.027 | 0.022 | 0.6 | 1.9 | 2.4 | 0.181 | 487 | 2.8E-05 | 2 | |
| 456.01 | 4.2628 | 1093 | 51 | 104.4739 | 0.0023 | 13.70035 | 0.00023 | 25.57 | 0.38 | 0.03051 | 0.00039 | 0.009 | 0.022 | 3.1 | 0.114 | 714 | 2.9E-05 | 2 | |
| 456.02 | 2.9208 | 241 | 18 | 67.0265 | 0.0054 | 4.30954 | 0.0001 | 13.06 | 0.56 | 0.01621 | 0.00054 | 0.0615 | - | 1.7 | 0.053 | 1047 | - | 4 | |
| 457.01 | 1.8924 | 742 | 40 | 107.2985 | 0.0017 | 4.921331 | 0.000057 | 13 | 40 | 0.028 | 0.017 | 0.8 | 1.3 | 2.2 | 0.054 | 799 | 3.2E-05 | 2 | |
| 458.01 | 4.0333 | 3343 | 60 | 141.0775 | 0.0018 | 53.71858 | 0.00079 | 49 | 15 | 0.0768 | 0.0062 | 0.93 | 0.28 | 10.5 | 0.286 | 515 | 1.1E-05 | 3 | |
| 459.01 | 3.5933 | 976 | 47 | 103.1027 | 0.0022 | 19.44639 | 0.00036 | 24 | 20 | 0.0325 | 0.0045 | 0.84 | 0.59 | 3.7 | 0.144 | 664 | 6.9E-05 | 2 | |
| 459.02 | 3.3100 | 157 | 13 | 67.1191 | 0.0079 | 6.91977 | 0.00023 | 16.04 | 0.25 | 0.0113 | 0.00058 | 0.12 | 0.01 | 1.3 | 0.072 | 939 | - | 4 | |
| 460.01 | 4.2947 | 1445 | 64 | 109.0751 | 0.0018 | 17.58782 | 0.00022 | 32.97 | 0.43 | 0.03389 | 0.00036 | 0.031 | 0.036 | 4.3 | 0.134 | 696 | 7.1E-05 | 2 | |
| 463.01 | 2.2233 | 2830 | 60 | 118.2658 | 0.0014 | 18.47817 | 0.00019 | 83.5 | 2.1 | 0.04736 | 0.00092 | 0.0236 | - | 3.5 | 0.107 | 398 | 7.9E-05 | 2 | NoObs |
| 464.01 | 6.3270 | 5551 | 162 | 129.55353 | 0.00077 | 58.3625 | 0.00033 | 76.39 | 0.36 | 0.06667 | 0.00028 | 0.0095 | - | 7.1 | 0.295 | 430 | 2.7E-05 | 2 | 1 |
| 464.02 | 2.2822 | 725 | 42 | 128.7576 | 0.0023 | 5.350243 | 0.000097 | 17 | 74 | 0.025 | 0.022 | 0.4 | 2.3 | 2.7 | 0.06 | 953 | 5.1E-05 | 2 | 1 |
| †465.01 | 7.8994 | 1609 | 25 | 137.0202 | 0.0047 | 350 | - | 177 | 53 | 0.0442 | 0.0015 | 0.82 | 0.24 | 4.8 | 1.002 | 266 | 3.2E-05 | 3 | |
| 466.01 | 2.1253 | 2724 | 71 | 103.53919 | 0.00086 | 9.391009 | 0.000057 | 31.7 | 9.5 | 0.04855 | 0.00057 | 0.216 | 0.065 | 3.1 | 0.087 | 679 | - | 2 | |
| 467.01 | 4.8878 | 3161 | 129 | 115.4428 | 0.0011 | 18.00891 | 0.00013 | 29.66 | 0.19 | 0.05027 | 0.00026 | 0.002 | - | 5.0 | 0.136 | 636 | 1.8E-05 | 2 | |
| 468.01 | 3.2229 | 1547 | 48 | 107.5957 | 0.0019 | 22.18452 | 0.0003 | 58.1 | 1.2 | 0.03537 | 0.00057 | 0.019 | - | 3.5 | 0.151 | 538 | 3.7E-05 | 2 | |
| 469.01 | 1.5075 | 2449 | 80 | 107.60632 | 0.00065 | 10.329116 | 0.00005 | 24.7 | 7.4 | 0.0612 | 0.0029 | 0.9 | 0.27 | 5.5 | 0.095 | 782 | 1.6E-05 | 2 | |
| 470.01 | 1.9936 | 2462 | 140 | 104.15067 | 0.00048 | 3.750839 | 0.000012 | 13 | 13 | 0.0464 | 0.0088 | 0.5 | 1 | 4.0 | 0.047 | 997 | 4.7E-05 | 2 | |
| 471.01 | 3.6749 | 505 | 24 | 104.7317 | 0.0048 | 21.34737 | 0.00069 | 25 | 58 | 0.0242 | 0.0095 | 0.84 | 0.98 | 2.0 | 0.151 | 551 | 4.6E-05 | 3 | |
| 472.01 | 3.4034 | 1464 | 85 | 106.5627 | 0.0012 | 4.243748 | 0.000035 | 9.95 | 0.11 | 0.03403 | 0.0003 | 0.0015 | - | 3.2 | 0.052 | 1023 | 1.5E-05 | 2 | |
| 473.01 | 2.3654 | 884 | 36 | 113.6359 | 0.0023 | 12.70512 | 0.00021 | 43.8 | 1.3 | 0.02715 | 0.0006 | 0.0281 | - | 2.2 | 0.105 | 629 | 5.6E-05 | 2 | |
| 474.01 | 3.1718 | 535 | 32 | 109.72 | 0.0027 | 10.94564 | 0.0002 | 27 | 132 | 0.021 | 0.018 | 0.2 | 2.6 | 2.3 | 0.1 | 863 | 2.4E-05 | 2 | |
| 474.02 | 3.3543 | 447 | 17 | 67.7739 | 0.0059 | 28.98843 | 0.00077 | 47 | 309 | 0.021 | 0.027 | 0.7 | 2.1 | 2.3 | 0.191 | 625 | - | 4 | |
| 475.01 | 2.5576 | 756 | 27 | 109.7 | 0.003 | 8.18066 | 0.00017 | 25.89 | 0.96 | 0.02514 | 0.00071 | 0.0033 | - | 2.4 | 0.078 | 740 | 1.6E-05 | 2 | 1 |
| 475.02 | 2.8820 | 888 | 26 | 104.7906 | 0.0035 | 15.31341 | 0.00037 | 33 | 154 | 0.028 | 0.027 | 0.6 | 2.1 | 2.6 | 0.118 | 602 | 1.5E-05 | 2 | 1 |
| 476.01 | 2.9423 | 743 | 22 | 111.4386 | 0.0038 | 18.42776 | 0.0005 | 40 | 188 | 0.025 | 0.025 | 0.6 | 2 | 2.4 | 0.133 | 567 | 3.0E-05 | 2 | |
| 477.01 | 3.7966 | 825 | 28 | 102.6428 | 0.0038 | 16.54318 | 0.00044 | 28 | 98 | 0.027 | 0.02 | 0.6 | 1.8 | 2.6 | 0.125 | 593 | 5.9E-05 | 2 | |
| 478.01 | 1.7873 | 1908 | 62 | 104.1148 | 0.0011 | 11.023452 | 0.000085 | 21.4 | 6.4 | 0.0515 | 0.0018 | 0.9 | 0.27 | 4.5 | 0.079 | 522 | 3.7E-05 | 3 | NoObs |
| 479.01 | 5.3442 | 1106 | 48 | 126.387 | 0.0025 | 34.18966 | 0.00069 | 41 | 85 | 0.031 | 0.012 | 0.6 | 1.4 | 3.2 | 0.209 | 518 | 1.3E-05 | 2 | NoObs |
| 480.01 | 2.0071 | 736 | 37 | 105.3089 | 0.0015 | 4.301702 | 0.000044 | 14 | 51 | 0.027 | 0.019 | 0.6 | 1.8 | 2.7 | 0.052 | 985 | 2.5E-05 | 2 | |
| 481.01 | 2.7127 | 988 | 56 | 104.969 | 0.0015 | 7.650377 | 0.000079 | 19 | 40 | 0.029 | 0.013 | 0.6 | 1.4 | 2.5 | 0.075 | 738 | 2.7E-05 | 2 | 1 |
| 481.02 | 1.6900 | 421 | 42 | 102.8291 | 0.0016 | 1.554014 | 0.000017 | 7 | 17 | 0.0197 | 0.0098 | 0.1 | 1.8 | 1.7 | 0.026 | 1253 | 3.3E-05 | 2 | 1 |
| 481.03 | 4.8919 | 1044 | 38 | 116.228 | 0.0039 | 34.2603 | 0.00092 | 28 | 14 | 0.0355 | 0.0034 | 0.87 | 0.41 | 3.0 | 0.202 | 450 | 2.8E-05 | 2 | 1 |
| 483.01 | 3.0292 | 827 | 50 | 106.2564 | 0.0017 | 4.798596 | 0.000077 | 9 | 18 | 0.028 | 0.011 | 0.7 | 1.2 | 2.3 | 0.055 | 866 | 1.7E-05 | 2 | |
| 484.01 | 3.6637 | 1050 | 38 | 108.059 | 0.0024 | 17.20516 | 0.0003 | 36.19 | 0.83 | 0.02881 | 0.00051 | 0.029 | 0.041 | 2.0 | 0.125 | 509 | 2.4E-05 | 2 | |
| 486.01 | 5.1182 | 678 | 31 | 102.4949 | 0.0041 | 22.1831 | 0.00058 | 26 | 84 | 0.025 | 0.015 | 0.6 | 1.6 | 1.4 | 0.152 | 454 | 2.5E-05 | 2 | |
| 487.01 | 3.0191 | 653 | 26 | 106.0424 | 0.0033 | 7.65867 | 0.00017 | 21.02 | 0.78 | 0.02338 | 0.00067 | 0.0423 | - | 2.4 | 0.077 | 837 | 3.1E-05 | 3 | |
| 488.01 | 3.3557 | 470 | 17 | 109.447 | 0.0048 | 9.37924 | 0.00032 | 14 | 61 | 0.021 | 0.015 | 0.8 | 1.5 | 2.2 | 0.088 | 797 | 4.7E-05 | 3 | |
| 490.01 | 2.3764 | 415 | 12 | 105.8695 | 0.007 | 4.39312 | 0.00021 | 10 | 76 | 0.021 | 0.033 | 0.7 | 2.2 | 2.3 | 0.051 | 932 | 7.9E-05 | 3 | NoObs |
| 490.03 | 2.8056 | 391 | 26 | 67.0685 | 0.0035 | 7.4063 | 0.00011 | 13 | 57 | 0.021 | 0.017 | 0.8 | 1.6 | 2.2 | 0.072 | 784 | - | 4 | |
| 492.01 | 6.3675 | 889 | 34 | 127.7091 | 0.0038 | 29.91151 | 0.00093 | 38.59 | 0.81 | 0.02692 | 0.00051 | 0.014 | 0.01 | 3.7 | 0.192 | 607 | 1.1E-04 | 2 | |
| 494.01 | 3.7520 | 1039 | 26 | 121.7839 | 0.0035 | 25.69719 | 0.00067 | 44 | 152 | 0.031 | 0.024 | 0.6 | 1.8 | 1.8 | 0.157 | 388 | 1.2E-04 | 2 | |
| 496.01 | 1.5390 | 408 | 8 | 102.8045 | 0.0067 | 1.616844 | 0.000073 | 5 | 1.5 | 0.021 | 0.0026 | 0.66 | 0.2 | 2.7 | 0.027 | 1514 | 5.0E-05 | 3 | NoObs |
| 497.01 | 4.9474 | 560 | 32 | 108.6103 | 0.0036 | 13.19281 | 0.00032 | 15 | 37 | 0.0233 | 0.0095 | 0.7 | 1.3 | 2.5 | 0.113 | 785 | 1.3E-04 | 2 | |
| 497.02 | 3.6013 | 161 | 13 | 67.5572 | 0.0082 | 4.4255 | 0.00015 | 6 | 24 | 0.016 | 0.013 | 0.8 | 1.3 | 1.7 | 0.054 | 1135 | - | 4 | |
| 499.01 | 2.4759 | 405 | 24 | 107.5364 | 0.0032 | 9.66856 | 0.00022 | 18 | 79 | 0.021 | 0.016 | 0.8 | 1.5 | 2.0 | 0.089 | 750 | 1.2E-04 | 2 | |
| 500.01 | 2.5052 | 1457 | 51 | 109.4766 | 0.0017 | 7.053478 | 0.000083 | 22.62 | 0.47 | 0.03412 | 0.00054 | 0.0244 | - | 2.7 | 0.063 | 642 | 1.1E-04 | 2 | 1 |

| | | | | | | | | | | | | | | | | | | | | |
|---|---|---|---|---|---|---|---|---|---|---|---|---|---|---|---|---|---|---|---|---|
| 500.02 | 2.4154 | 1417 | 44 | 110.4801 | 0.0021 | 9.5217 | 0.00014 | 29 | 83 | 0.035 | 0.023 | 0.4 | 1.8 | 2.8 | 0.076 | 584 | 1.0E-04 | 2 | 1 | |
| 500.03 | 2.0724 | 434 | 22 | 67.025 | 0.0035 | 3.072166 | 0.000046 | 11.53 | 0.6 | 0.01825 | 0.00074 | 0.0162 | - | 1.5 | 0.036 | 849 | - | 4 | | |
| 500.04 | 2.0486 | 532 | 22 | 67.5914 | 0.0034 | 4.645353 | 0.000067 | 11 | 51 | 0.026 | 0.023 | 0.8 | 1.5 | 2.1 | 0.047 | 743 | - | 4 | | |
| 500.05 | 1.4084 | 289 | 22 | 66.1708 | 0.0027 | 0.986779 | 0.000012 | 5 | 19 | 0.015 | 0.014 | 0.4 | 2.2 | 1.2 | 0.017 | 1235 | - | 4 | | |
| 501.01 | 7.9692 | 490 | 20 | 103.3382 | 0.0086 | 24.794 | 0.0015 | 23 | 123 | 0.02 | 0.02 | 0.2 | 2.7 | 2.1 | 0.169 | 580 | 2.0E-04 | 2 | | |
| 503.01 | 2.6971 | 1414 | 41 | 105.9557 | 0.0018 | 8.22236 | 0.0001 | 24.44 | 0.6 | 0.03383 | 0.00065 | 0 | 0.01 | 2.5 | 0.067 | 575 | 2.5E-05 | 2 | | |
| 504.01 | 5.3925 | 680 | 23 | 132.256 | 0.0043 | 40.6068 | 0.0015 | 58 | 1.8 | 0.02324 | 0.00066 | 0.017 | 0.028 | 1.7 | 0.228 | 411 | 1.2E-04 | 2 | | |
| 505.01 | 2.9555 | 645 | 29 | 107.809 | 0.0026 | 13.76725 | 0.00025 | 39.7 | 1.4 | 0.02282 | 0.0006 | 0.011 | 0.01 | 3.1 | 0.113 | 734 | 2.4E-05 | 2 | | |
| 506.01 | 1.1470 | 727 | 61 | 102.96485 | 0.00078 | 1.5831619 | 0.0000085 | 12.11 | 0.26 | 0.02594 | 0.00045 | 0.0211 | - | 2.5 | 0.027 | 1468 | 4.6E-05 | 3 | | |
| 507.01 | 3.5887 | 1578 | 26 | 106.4976 | 0.0026 | 18.49248 | 0.00032 | 36 | 150 | 0.039 | 0.033 | 0.5 | 2.1 | 4.3 | 0.136 | 619 | 1.5E-04 | 2 | | |
| 508.01 | 3.5907 | 744 | 46 | 102.5153 | 0.002 | 7.93059 | 0.00011 | 12 | 23 | 0.0272 | 0.0099 | 0.7 | 1.1 | 3.8 | 0.08 | 965 | 2.5E-05 | 2 | 1 | |
| 508.02 | 4.0833 | 700 | 33 | 113.2053 | 0.0031 | 16.66519 | 0.00038 | 27 | 84 | 0.025 | 0.015 | 0.6 | 1.7 | 3.5 | 0.131 | 754 | 2.6E-05 | 2 | 1 | |
| 509.01 | 2.7553 | 856 | 50 | 102.7129 | 0.0018 | 4.167068 | 0.000051 | 9 | 24 | 0.028 | 0.014 | 0.6 | 1.5 | 2.7 | 0.051 | 988 | 1.9E-04 | 2 | | |
| 509.02 | 2.7152 | 912 | 33 | 70.3827 | 0.0028 | 11.46349 | 0.00014 | 23 | 94 | 0.03 | 0.025 | 0.7 | 1.6 | 2.9 | 0.1 | 705 | - | 4 | | |
| 510.01 | 2.7078 | 445 | 29 | 102.8992 | 0.0026 | 2.940409 | 0.000053 | 5 | 16 | 0.023 | 0.012 | 0.8 | 1.2 | 2.7 | 0.041 | 1208 | 3.1E-05 | 2 | 1 | |
| 510.02 | 3.1918 | 534 | 25 | 108.4732 | 0.0033 | 6.38914 | 0.00015 | 12 | 46 | 0.023 | 0.017 | 0.7 | 1.8 | 2.7 | 0.068 | 938 | 2.9E-05 | 2 | 1 | |
| 511.01 | 3.2069 | 647 | 43 | 103.5031 | 0.0021 | 8.00573 | 0.00012 | 19.76 | 0.4 | 0.02329 | 0.00038 | 0.0106 | - | 2.8 | 0.081 | 936 | 5.6E-05 | 2 | | |
| 512.01 | 3.3326 | 514 | 20 | 105.921 | 0.0044 | 6.5098 | 0.0002 | 15 | 90 | 0.02 | 0.025 | 0.4 | 2.6 | 2.6 | 0.069 | 985 | 2.2E-05 | 3 | | |
| 513.01 | 7.0369 | 868 | 33 | 103.0988 | 0.0042 | 35.18059 | 0.00096 | 39.22 | 0.74 | 0.02732 | 0.0005 | 0.006 | 0.01 | 2.7 | 0.217 | 563 | 3.2E-05 | 2 | | |
| 517.01 | 2.0898 | 753 | 55 | 104.2588 | 0.0012 | 2.75236 | 0.000024 | 7 | 15 | 0.028 | 0.011 | 0.8 | 1.1 | 3.6 | 0.039 | 1343 | 2.6E-05 | 2 | NoObs | |
| 518.01 | 3.0039 | 1006 | 50 | 114.9745 | 0.002 | 13.98172 | 0.00019 | 36.78 | 0.72 | 0.02854 | 0.00041 | 0.027 | 0.035 | 2.4 | 0.107 | 567 | 2.7E-05 | 2 | 1 | |
| 518.02 | 5.0627 | 570 | 21 | 143.7633 | 0.0076 | 43.9985 | 0.0025 | 57 | 224 | 0.023 | 0.019 | 0.5 | 2 | 1.9 | 0.23 | 387 | 3.1E-05 | 2 | 1 | |
| 519.01 | 4.1728 | 592 | 23 | 111.3384 | 0.005 | 11.90376 | 0.00043 | 16 | 66 | 0.024 | 0.017 | 0.7 | 1.8 | 2.4 | 0.104 | 768 | 4.2E-05 | 2 | | |
| 520.01 | 3.4450 | 886 | 39 | 103.3046 | 0.0025 | 12.7599 | 0.00022 | 19 | 41 | 0.03 | 0.012 | 0.8 | 1.1 | 3.1 | 0.105 | 668 | 7.7E-05 | 2 | | |
| 520.02 | 2.3531 | 282 | 16 | 71.3679 | 0.0049 | 5.43316 | 0.00012 | 11 | 62 | 0.019 | 0.02 | 0.8 | 1.7 | 2.0 | 0.06 | 884 | - | 4 | | |
| 520.03 | 2.7291 | 725 | 22 | 69.4611 | 0.004 | 25.75125 | 0.00044 | 62 | 396 | 0.027 | 0.034 | 0.7 | 2.3 | 2.8 | 0.168 | 528 | - | 4 | | |
| 521.01 | 3.0740 | 1302 | 57 | 105.0012 | 0.0017 | 10.16104 | 0.00012 | 23 | 64 | 0.033 | 0.016 | 0.5 | 1.7 | 3.9 | 0.094 | 868 | 3.0E-05 | 3 | | |
| 522.01 | 2.8393 | 1131 | 39 | 102.9428 | 0.0022 | 12.83012 | 0.00019 | 35.55 | 0.15 | 0.03071 | 0.00052 | 0.0148 | - | 1.9 | 0.106 | 580 | 1.1E-04 | 2 | | |
| 523.01 | 4.8395 | 3388 | 81 | 131.2301 | 0.0014 | 49.41297 | 0.00062 | 43.4 | 4.8 | 0.0629 | 0.0015 | 0.88 | 0.19 | 7.3 | 0.272 | 519 | 1.5E-05 | 2 | | |
| 523.02 | 7.2861 | 636 | 21 | 71.9294 | 0.0081 | 36.8539 | 0.0014 | 39.6089 | 0.0058 | 0.02 | 0.19 | 0.073 | 0.014 | 2.7 | 0.223 | 573 | - | 4 | | |
| 524.01 | 2.3748 | 1002 | 56 | 105.0022 | 0.0014 | 4.592522 | 0.000042 | 15.85 | 0.3 | 0.02923 | 0.00041 | 0.0227 | - | 2.3 | 0.053 | 838 | 2.1E-05 | 2 | | |
| 525.01 | 2.1406 | 1205 | 34 | 106.6783 | 0.0023 | 11.53219 | 0.00019 | 52 | 2 | 0.03138 | 0.00085 | 0.026 | 0.045 | 4.3 | 0.102 | 850 | 9.7E-05 | 2 | | |
| 526.01 | 1.7639 | 926 | 73 | 104.04377 | 0.0008 | 2.104719 | 0.000012 | 6 | 10 | 0.0305 | 0.0088 | 0.7 | 1 | 2.6 | 0.032 | 1203 | 4.1E-05 | 2 | | |
| 528.01 | 3.3634 | 703 | 38 | 109.6802 | 0.0023 | 9.57676 | 0.00016 | 17 | 44 | 0.025 | 0.012 | 0.7 | 1.4 | 3.1 | 0.09 | 855 | 2.3E-05 | 2 | | |
| 528.02 | 6.0786 | 961 | 26 | 73.1959 | 0.006 | 96.6704 | 0.0024 | 130.6 | 3.9 | 0.02763 | 0.00072 | 0.031 | 0.06 | 3.4 | 0.419 | 396 | - | 4 | | |
| 528.03 | 2.4763 | 718 | 23 | 78.0287 | 0.0039 | 20.55273 | 0.00034 | 61 | 625 | 0.026 | 0.053 | 0.6 | 3.1 | 3.2 | 0.149 | 664 | - | 4 | | |
| 530.01 | 2.1729 | 567 | 18 | 103.3025 | 0.0036 | 10.94062 | 0.00026 | 40.2 | 3 | 0.0197 | 0.0011 | 0.001 | 0.01 | 1.3 | 0.095 | 619 | 2.7E-05 | 2 | NoObs | |
| 531.01 | 1.3319 | 2812 | 156 | 103.88016 | 0.00035 | 3.6874622 | 0.0000089 | 26 | 19 | 0.0554 | 0.0076 | 0.52 | 0.86 | 4.3 | 0.038 | 749 | 1.6E-05 | 3 | | 1 |
| 532.01 | 3.1402 | 629 | 42 | 106.6964 | 0.0022 | 4.221737 | 0.000066 | 11.06 | 0.23 | 0.02321 | 0.0004 | 0.0195 | - | 2.3 | 0.052 | 1089 | 1.7E-05 | 2 | | |
| 533.01 | 4.2317 | 661 | 23 | 104.7024 | 0.0052 | 16.54915 | 0.0006 | 32.7 | 1.1 | 0.02425 | 0.00065 | 0.031 | 0.057 | 2.6 | 0.126 | 641 | 1.7E-05 | 2 | | |
| 534.01 | 1.8349 | 752 | 31 | 107.0234 | 0.0021 | 6.400136 | 0.000094 | 17 | 67 | 0.03 | 0.022 | 0.8 | 1.4 | 2.0 | 0.065 | 706 | 5.8E-05 | 2 | NoObs | |

| | | | | | | | | | | | | | | | | | | | |
|---|---|---|---|---|---|---|---|---|---|---|---|---|---|---|---|---|---|---|---|
| 534.02 | 1.8465 | 417 | 27 | 104.0967 | 0.0025 | 2.735879 | 0.000047 | 12.24 | 0.46 | 0.02015 | 0.00059 | 0.0291 | - | 1.4 | 0.037 | 936 | 6.8E-05 | 2 | NoObs |
| 535.01 | 4.5283 | 1100 | 89 | 104.1811 | 0.0014 | 5.852997 | 0.000054 | 10.152 | 0.089 | 0.02989 | 0.00023 | 0.0075 | - | 3.3 | 0.065 | 1011 | 5.4E-05 | 2 | |
| 536.01 | 7.5537 | 1222 | 29 | 111.5934 | 0.0054 | 162.3361 | 0.0078 | 150.1292 | 0.0072 | 0.03241 | - | 0.4793 | - | 3.0 | 0.588 | 296 | 1.7E-05 | 2 | |
| 537.01 | 2.4773 | 448 | 34 | 103.7859 | 0.0022 | 2.820204 | 0.000057 | 6 | 18 | 0.022 | 0.011 | 0.7 | 1.4 | 1.4 | 0.039 | 1003 | 2.0E-05 | 2 | |
| 538.01 | 5.3723 | 659 | 30 | 104.6569 | 0.0044 | 21.2147 | 0.00074 | 20 | 38 | 0.0255 | 0.0082 | 0.8 | 1 | 2.9 | 0.155 | 683 | 3.6E-05 | 2 | |
| 541.01 | 3.4665 | 565 | 18 | 113.3505 | 0.0044 | 13.64591 | 0.00044 | 20 | 89 | 0.025 | 0.02 | 0.8 | 1.5 | 1.9 | 0.11 | 603 | 3.0E-05 | 2 | |
| 542.01 | 5.8419 | 589 | 21 | 111.6873 | 0.0058 | 41.8867 | 0.0017 | 55.7 | 1.8 | 0.02195 | 0.00066 | 0.023 | 0.028 | 2.7 | 0.241 | 526 | 3.5E-05 | 3 | |
| 543.01 | 1.9101 | 753 | 36 | 106.4363 | 0.0019 | 4.302187 | 0.000057 | 17.38 | 0.52 | 0.02529 | 0.00057 | 0.0378 | - | 1.9 | 0.05 | 844 | 3.5E-05 | 2 | |
| 543.02 | 1.8077 | 330 | 18 | 66.5566 | 0.0038 | 3.137846 | 0.000051 | 15.57 | 0.68 | 0.02041 | 0.00067 | 0.0847 | - | 1.5 | 0.041 | 932 | - | 4 | |
| 546.01 | 6.2656 | 842 | 28 | 103.1884 | 0.0048 | 20.68457 | 0.00068 | 26.2 | 0.66 | 0.02575 | 0.0006 | 0.003 | 0.01 | 2.8 | 0.152 | 673 | 2.8E-05 | 3 | |
| 547.01 | 4.4977 | 2224 | 68 | 121.059 | 0.0017 | 25.30298 | 0.0003 | 44.79 | 0.56 | 0.04137 | 0.00043 | 0.0096 | - | 3.5 | 0.164 | 489 | 2.8E-05 | 2 | |
| 548.01 | 4.0011 | 633 | 25 | 122.7062 | 0.0042 | 21.30056 | 0.00071 | 30 | 111 | 0.026 | 0.016 | 0.7 | 1.7 | 2.6 | 0.155 | 654 | 2.4E-05 | 3 | NoObs |
| 550.01 | 4.1038 | 585 | 33 | 111.5233 | 0.0031 | 13.02371 | 0.00027 | 25.5 | 0.64 | 0.02143 | 0.00047 | 0.032 | 0.01 | 1.8 | 0.109 | 660 | 1.1E-04 | 2 | NoObs |
| 551.01 | 3.2954 | 612 | 24 | 111.8449 | 0.0039 | 11.63684 | 0.00034 | 19 | 75 | 0.025 | 0.018 | 0.7 | 1.6 | 2.1 | 0.101 | 688 | 1.3E-04 | 3 | |
| 551.02 | 2.1328 | 441 | 20 | 66.9544 | 0.0039 | 5.688042 | 0.000092 | 23.6 | 1.1 | 0.02122 | 0.00074 | 0.0662 | - | 1.8 | 0.063 | 871 | - | 4 | |
| 552.01 | 1.7536 | 7643 | 133 | 104.09836 | 0.0005 | 3.055172 | 0.000011 | 8.6 | 2.6 | 0.0967 | 0.0013 | 0.79 | 0.24 | 11.2 | 0.043 | 1316 | 6.9E-05 | 2 | |
| 554.01 | 2.2625 | 3528 | 43 | 103.5436 | 0.0016 | 3.658495 | 0.00004 | 6 | 1.3 | 0.0687 | 0.0034 | 0.9 | 0.25 | 6.1 | 0.047 | 1068 | 3.4E-05 | 3 | |
| 555.01 | 2.4562 | 246 | 18 | 105.4452 | 0.0043 | 3.70178 | 0.00011 | 7 | 31 | 0.018 | 0.015 | 0.8 | 1.5 | 1.5 | 0.046 | 947 | 8.0E-05 | 2 | NoObs |
| 555.02 | 6.8853 | 876 | 21 | 114.886 | 0.01 | 86.4958 | 0.0038 | 100.4 | 3.3 | 0.02676 | 0.00079 | 0.041 | 0.039 | 2.3 | 0.376 | 331 | - | 4 | |
| 557.01 | 3.7725 | 897 | 30 | 103.7845 | 0.0032 | 15.65554 | 0.00034 | 28 | 92 | 0.028 | 0.019 | 0.5 | 1.8 | 3.1 | 0.121 | 636 | 2.8E-05 | 3 | |
| 558.01 | 2.1952 | 753 | 30 | 106.0859 | 0.0025 | 9.17892 | 0.00016 | 34 | 128 | 0.025 | 0.019 | 0.1 | 2.3 | 2.3 | 0.085 | 730 | 1.8E-04 | 2 | |
| 559.01 | 2.4524 | 214 | 9.7 | 106.705 | 0.0067 | 4.33065 | 0.0002 | 11.6 | 3.5 | 0.0137 | 0.0011 | 0.29 | 0.087 | 1.4 | 0.052 | 981 | 3.4E-05 | 3 | |
| 560.01 | 4.2117 | 882 | 14 | 112.2715 | 0.0067 | 23.6758 | 0.0011 | 35 | 234 | 0.028 | 0.039 | 0.6 | 2.4 | 1.8 | 0.154 | 445 | 3.0E-05 | 2 | |
| 561.01 | 2.5786 | 490 | 32 | 102.6083 | 0.0024 | 5.379017 | 0.000089 | 11 | 33 | 0.024 | 0.013 | 0.7 | 1.4 | 2.1 | 0.058 | 829 | 1.9E-05 | 2 | NoObs |
| 563.01 | 5.4260 | 270 | 17 | 108.6355 | 0.0069 | 15.28368 | 0.00064 | 16 | 65 | 0.017 | 0.012 | 0.7 | 1.7 | 1.8 | 0.124 | 734 | 1.0E-04 | 2 | |
| 564.01 | 7.4094 | 629 | 27 | 104.8916 | 0.0054 | 21.05821 | 0.00075 | 22.39 | 0.5 | 0.02326 | 0.00051 | 0.019 | 0.01 | 2.4 | 0.152 | 619 | 6.7E-05 | 2 | |
| 564.02 | 13.5324 | 3079 | 86 | 179.495 | 0.003 | 127.8872 | 0.0024 | 75.79 | 0.58 | 0.04929 | 0.00037 | 0.013 | 0.04 | 5.0 | 0.505 | 340 | - | 4 | |
| 566.01 | 4.0033 | 686 | 16 | 125.5736 | 0.0064 | 25.8548 | 0.002 | 50 | 682 | 0.024 | 0.058 | 0.2 | 4.3 | 2.3 | 0.175 | 586 | 6.1E-05 | 3 | |
| 567.01 | 3.3311 | 763 | 40 | 102.9293 | 0.0024 | 10.68782 | 0.00018 | 15 | 23 | 0.0286 | 0.0075 | 0.82 | 0.83 | 2.9 | 0.096 | 758 | 5.6E-05 | 2 | NoObs |
| 567.02 | 4.4409 | 503 | 23 | 109.8004 | 0.005 | 20.3032 | 0.00071 | 21 | 73 | 0.023 | 0.013 | 0.8 | 1.3 | 2.3 | 0.147 | 612 | 6.1E-05 | 2 | NoObs |
| 567.03 | 3.4536 | 615 | 20 | 131.3635 | 0.0042 | 29.02356 | 0.00067 | 66 | 682 | 0.022 | 0.043 | 0.1 | 3.8 | 2.2 | 0.187 | 543 | 5.6E-05 | 2 | NoObs |
| 568.01 | 1.3950 | 276 | 22 | 102.64 | 0.0023 | 3.383517 | 0.000051 | 16 | 66 | 0.016 | 0.012 | 0.6 | 1.9 | 1.0 | 0.043 | 856 | 3.6E-05 | 2 | NoObs |
| 569.01 | 2.7713 | 573 | 17 | 118.4404 | 0.0043 | 20.72804 | 0.00067 | 60.2 | 3.3 | 0.02221 | 0.00096 | 0.017 | 0.01 | 2.1 | 0.144 | 540 | 1.9E-05 | 3 | |
| 571.01 | 2.2027 | 702 | 28 | 107.316 | 0.0026 | 7.26733 | 0.00013 | 26 | 106 | 0.024 | 0.015 | 0.1 | 2.4 | 1.7 | 0.059 | 563 | 1.2E-04 | 2 | 1 |
| 571.02 | 2.6903 | 887 | 32 | 109.9 | 0.0023 | 13.34331 | 0.00015 | 40 | 1.2 | 0.02827 | 0.00067 | 0.0141 | - | 2.0 | 0.088 | 461 | 1.2E-04 | 2 | 1 |
| 571.03 | 1.8168 | 543 | 29 | 66.3315 | 0.0023 | 3.886758 | 0.000038 | 9 | 17 | 0.026 | 0.007 | 0.87 | 0.8 | 1.8 | 0.039 | 692 | - | 4 | |
| 572.01 | 5.2873 | 386 | 25 | 112.774 | 0.0054 | 10.6405 | 0.00042 | 16.33 | 0.39 | 0.01827 | 0.00045 | 0.024 | 0.01 | 2.4 | 0.097 | 882 | 1.3E-04 | 2 | |
| 573.01 | 3.1200 | 701 | 35 | 105.5032 | 0.0026 | 5.9966 | 0.00011 | 14 | 52 | 0.025 | 0.018 | 0.3 | 2.1 | 3.2 | 0.066 | 1054 | 2.0E-05 | 2 | |
| 573.02 | 2.0590 | 229 | 16 | 66.2019 | 0.0048 | 2.061872 | 0.000042 | 8.378 | 0.08 | 0.01674 | 0.00053 | 0.0437 | - | 2.1 | 0.033 | 1491 | - | 4 | |
| 574.01 | 3.4080 | 1118 | 36 | 104.3661 | 0.003 | 20.13504 | 0.0004 | 41 | 162 | 0.031 | 0.025 | 0.4 | 2.1 | 2.4 | 0.14 | 507 | 3.0E-05 | 2 | |
| 575.01 | 4.1934 | 563 | 16 | 116.4044 | 0.0064 | 24.3178 | 0.0012 | 32 | 165 | 0.024 | 0.022 | 0.7 | 1.9 | 2.6 | 0.169 | 640 | 3.8E-05 | 3 | |

| | | | | | | | | | | | | | | | | | | | | |
|---|---|---|---|---|---|---|---|---|---|---|---|---|---|---|---|---|---|---|---|---|
| 577.01 | 5.2888 | 500 | 9 | 111.55 | 0.011 | 39.6729 | 0.003 | 59 | 477 | 0.02 | 0.034 | 0.1 | 3.4 | 2.6 | 0.227 | 502 | 3.5E-05 | 3 | | |
| 578.01 | 5.1938 | 1173 | 90 | 102.8794 | 0.0015 | 6.412547 | 0.000063 | 9.811 | 0.082 | 0.03075 | 0.00023 | 0.004 | - | 3.8 | 0.069 | 1035 | 4.1E-05 | 2 | | |
| 579.01 | 1.8672 | 319 | 31 | 103.0698 | 0.0021 | 2.020003 | 0.000028 | 9.093 | 0.05 | 0.01747 | 0.00038 | 0.0295 | - | 1.5 | 0.03 | 1154 | 4.1E-05 | 2 | NoObs | |
| 580.01 | 2.7903 | 743 | 32 | 108.7093 | 0.0025 | 6.52125 | 0.00011 | 18 | 95 | 0.025 | 0.025 | 0 | 2.7 | 1.5 | 0.067 | 716 | 3.9E-05 | 2 | | |
| 581.01 | 2.6717 | 1206 | 49 | 108.9144 | 0.0016 | 6.996895 | 0.000074 | 20.47 | 0.39 | 0.03159 | 0.00048 | 0.001 | - | 2.1 | 0.07 | 714 | 4.2E-05 | 2 | | |
| 582.01 | 2.5505 | 844 | 34 | 103.4687 | 0.002 | 5.945053 | 0.000083 | 17 | 74 | 0.027 | 0.024 | 0.4 | 2.3 | 2.2 | 0.062 | 783 | 3.7E-05 | 3 | | |
| 583.01 | 3.1312 | 246 | 28 | 103.7392 | 0.0033 | 2.436893 | 0.000055 | 4 | 12 | 0.0164 | 0.0083 | 0.7 | 1.4 | 1.6 | 0.036 | 1266 | 2.8E-05 | 3 | | |
| 584.01 | 3.7109 | 720 | 47 | 108.6885 | 0.0021 | 9.9265 | 0.00015 | 21.68 | 0.46 | 0.02287 | 0.0004 | 0.0173 | - | 1.6 | 0.088 | 633 | 2.2E-05 | 2 | 1 | |
| 584.02 | 4.4730 | 553 | 27 | 103.3728 | 0.0044 | 21.22343 | 0.00069 | 37 | 1.1 | 0.02177 | 0.00053 | 0.025 | 0.01 | 1.5 | 0.146 | 492 | 2.3E-05 | 2 | 1 | |
| 585.01 | 1.9189 | 818 | 37 | 104.558 | 0.0017 | 3.722176 | 0.000042 | 15.33 | 0.43 | 0.02686 | 0.00053 | 0.0528 | - | 2.0 | 0.046 | 932 | 3.8E-05 | 2 | | |
| 586.01 | 3.7971 | 571 | 23 | 108.9754 | 0.0041 | 15.77916 | 0.00044 | 22 | 72 | 0.024 | 0.014 | 0.7 | 1.4 | 2.1 | 0.124 | 630 | 1.7E-05 | 3 | | |
| 587.01 | 3.4973 | 816 | 38 | 104.6019 | 0.0028 | 14.03513 | 0.00026 | 25 | 63 | 0.028 | 0.014 | 0.6 | 1.5 | 3.0 | 0.113 | 672 | 5.7E-05 | 2 | | |
| 588.01 | 2.6497 | 595 | 19 | 108.6871 | 0.0041 | 10.35547 | 0.0003 | 23 | 100 | 0.023 | 0.023 | 0.7 | 1.9 | 2.2 | 0.085 | 619 | 1.4E-05 | 2 | | |
| 589.01 | 4.2971 | 189 | 8.6 | 119.5356 | 0.0092 | 17.4808 | 0.00085 | 21 | 180 | 0.014 | 0.02 | 0.8 | 2.2 | 1.2 | 0.134 | 637 | 3.2E-05 | 3 | NoObs | |
| 590.01 | 3.7243 | 412 | 22 | 107.5461 | 0.0045 | 11.38933 | 0.00035 | 17 | 72 | 0.021 | 0.015 | 0.7 | 1.7 | 2.1 | 0.102 | 809 | 7.0E-05 | 2 | | |
| 590.02 | 5.8287 | 615 | 20 | 74.3251 | 0.0073 | 50.6962 | 0.0017 | 68.1 | 2.4 | 0.02208 | 0.00071 | 0.031 | 0.063 | 2.2 | 0.276 | 492 | - | 4 | | |
| 592.01 | 4.7864 | 478 | 19 | 108.4815 | 0.0061 | 39.7521 | 0.0018 | 35 | 85 | 0.0229 | 0.0088 | 0.85 | 0.98 | 2.7 | 0.234 | 550 | 4.4E-05 | 3 | | |
| 593.01 | 3.2745 | 524 | 16 | 104.7889 | 0.0051 | 9.99757 | 0.00034 | 18 | 124 | 0.023 | 0.028 | 0.7 | 2.4 | 2.1 | 0.092 | 760 | 5.2E-05 | 2 | | |
| 596.01 | 1.3419 | 689 | 42 | 103.4508 | 0.0012 | 1.682706 | 0.000014 | 10.097 | 0.022 | 0.02575 | 0.00042 | 0.0338 | - | 1.7 | 0.022 | 864 | 1.6E-05 | 2 | NoObs | |
| 597.01 | 4.8348 | 510 | 17 | 109.9401 | 0.0056 | 17.30819 | 0.00069 | 27.63 | 0.97 | 0.02229 | 0.00072 | 0.014 | 0.01 | 2.6 | 0.135 | 724 | 3.4E-05 | 2 | | |
| 597.02 | 2.6017 | 184 | 14 | 66.0722 | 0.0064 | 2.092181 | 0.000057 | 6.2 | 1.9 | 0.01204 | 0.00079 | - | - | 1.4 | 0.033 | 1464 | - | 4 | | |
| 598.01 | 2.9641 | 767 | 34 | 104.1666 | 0.0027 | 8.30811 | 0.00016 | 21.48 | 0.56 | 0.02527 | 0.00052 | 0.0223 | - | 1.7 | 0.077 | 644 | 2.6E-05 | 2 | | |
| 599.01 | 2.4207 | 581 | 29 | 106.2091 | 0.0027 | 6.45469 | 0.00012 | 17 | 93 | 0.023 | 0.022 | 0.6 | 2.2 | 2.3 | 0.069 | 935 | 6.6E-05 | 3 | | |
| 600.01 | 2.5733 | 388 | 26 | 103.3635 | 0.0032 | 3.59594 | 0.00011 | 10.27 | 0.33 | 0.01839 | 0.00049 | 0.0478 | - | 2.1 | 0.047 | 1213 | 2.2E-05 | 3 | | |
| 601.01 | 2.5206 | 825 | 19 | 105.1847 | 0.0038 | 5.40425 | 0.00014 | 17.1 | 1.6 | 0.0182 | 0.0015 | 0.11 | 0.01 | 1.7 | 0.062 | 974 | 2.8E-05 | 2 | NoObs | |
| 602.01 | 5.2372 | 445 | 19 | 110.2739 | 0.0063 | 12.91408 | 0.00054 | 19.55 | 0.77 | 0.01904 | 0.00068 | 0.002 | 0.01 | 2.3 | 0.111 | 831 | 3.2E-05 | 3 | | |
| 605.01 | 1.7663 | 980 | 54 | 102.7178 | 0.0011 | 2.628144 | 0.00002 | 11.29 | 0.26 | 0.02757 | 0.00044 | 0.023 | 0.033 | 1.6 | 0.031 | 782 | 4.8E-05 | 2 | | |
| 607.01 | 1.5864 | 6629 | 74 | 106.48563 | 0.00036 | 5.894028 | 0.000052 | 39.45 | 0.91 | 0.07544 | 0.00093 | 0.031 | 0.032 | 6.8 | 0.064 | 871 | - | 3 | | 1 |
| 609.01 | 1.8343 | 4272 | 56 | 105.028 | 0.0011 | 4.396913 | 0.000034 | 9.4 | 2.8 | 0.089 | 0.011 | 0.92 | 0.28 | 12.0 | 0.054 | 1200 | 2.1E-05 | 3 | | |
| 610.01 | 2.5145 | 886 | 26 | 113.8431 | 0.0026 | 14.28246 | 0.00026 | 45.2 | 1.8 | 0.02681 | 0.00084 | 0.022 | 0.044 | 2.0 | 0.096 | 481 | 2.6E-05 | 2 | | |
| 611.01 | 1.4497 | 4347 | 319 | 104.05987 | 0.00018 | 3.2516578 | 0.0000041 | 10.3 | 3.1 | 0.07259 | 0.00037 | 0.79 | 0.24 | 7.3 | 0.044 | 1235 | 6.6E-05 | 2 | NoObs | |
| 612.01 | 3.3581 | 540 | 26 | 106.2164 | 0.0031 | 20.74022 | 0.00046 | 32 | 102 | 0.024 | 0.015 | 0.8 | 1.3 | 3.5 | 0.15 | 668 | 2.4E-05 | 2 | NoObs | |
| 612.02 | 5.3336 | 799 | 28 | 149.5591 | 0.0046 | 47.4276 | 0.0019 | 72.4 | 2.1 | 0.02517 | 0.00061 | 0.004 | 0.01 | 3.6 | 0.26 | 507 | 2.2E-05 | 2 | NoObs | |
| 614.01 | 1.8723 | 3854 | 122 | 103.02216 | 0.00059 | 12.874706 | 0.00005 | 39 | 12 | 0.06268 | 0.0006 | 0.58 | 0.17 | 4.0 | 0.107 | 587 | - | 3 | | |
| 617.01 | 2.9128 | 7003 | 158 | 131.59768 | 0.00057 | 37.86537 | 0.00017 | 52 | 16 | 0.177 | 0.021 | 1.11 | 0.33 | 17.8 | 0.224 | 499 | - | 3 | | |
| 618.01 | 2.6232 | 1028 | 38 | 111.3474 | 0.0022 | 9.07071 | 0.00014 | 19 | 54 | 0.031 | 0.017 | 0.7 | 1.4 | 3.2 | 0.086 | 790 | 5.8E-05 | 2 | | |
| 620.01 | 5.8762 | 6402 | 53 | 92.1077 | 0.0026 | 45.15416 | 0.00078 | 63.25 | 0.93 | 0.07225 | 0.00091 | 0.016 | 0.022 | 7.2 | 0.253 | 486 | 2.6E-05 | 2 | | |
| 622.01 | 8.9133 | 4669 | 58 | 146.4969 | 0.0031 | 155.0467 | 0.0044 | 81 | 12 | 0.0732 | 0.0025 | 0.85 | 0.24 | 9.3 | 0.568 | 327 | 2.5E-05 | 3 | | |
| 623.01 | 4.2819 | 106 | 23 | 107.0644 | 0.0043 | 10.3496 | 0.00032 | 9 | 31 | 0.0112 | 0.0055 | 0.9 | 1 | 2.0 | 0.099 | 1121 | 2.9E-05 | 3 | 1 | |
| 623.02 | 5.5070 | 103 | 21 | 112.4629 | 0.0054 | 15.67781 | 0.00065 | 22.78 | 0.74 | 0.00954 | 0.00026 | 0.032 | 0.078 | 1.7 | 0.13 | 978 | 3.0E-05 | 3 | 1 | |
| 623.03 | 3.7922 | 74 | 21 | 104.4771 | 0.0049 | 5.5992 | 0.00018 | 6 | 19 | 0.0099 | 0.0048 | 0.9 | 1 | 1.8 | 0.066 | 1373 | 3.1E-05 | 3 | 1 | |

| | | | | | | | | | | | | | | | | | | | |
|---|---|---|---|---|---|---|---|---|---|---|---|---|---|---|---|---|---|---|---|
| 624.01 | 4.4288 | 921 | 23 | 115.4407 | 0.0046 | 17.78948 | 0.00056 | 31.6 | 1.1 | 0.02703 | 0.00078 | 0.031 | 0.066 | 2.1 | 0.133 | 565 | - | 2 | |
| 625.01 | 4.4579 | 1257 | 58 | 113.4422 | 0.0025 | 38.13719 | 0.00073 | 23.4 | 7 | 0.0623 | 0.0097 | 1 | 0.3 | 15.1 | 0.241 | 828 | 2.0E-05 | 2 | 1 |
| 626.01 | 3.9008 | 343 | 29 | 105.2217 | 0.0039 | 14.58635 | 0.00037 | 21 | 68 | 0.0186 | 0.0099 | 0.7 | 1.5 | 2.2 | 0.121 | 817 | 1.1E-04 | 2 | NoObs |
| 627.01 | 3.5631 | 400 | 38 | 109.1711 | 0.0028 | 7.75193 | 0.00015 | 12 | 30 | 0.0197 | 0.0082 | 0.7 | 1.3 | 2.9 | 0.08 | 1066 | 7.1E-05 | 2 | 1 |
| 628.01 | 3.0462 | 413 | 22 | 108.0042 | 0.0041 | 14.48612 | 0.00041 | 21 | 89 | 0.022 | 0.015 | 0.9 | 1.3 | 3.1 | 0.12 | 826 | 4.3E-05 | 2 | NoObs |
| 629.01 | 6.7051 | 383 | 18 | 105.5659 | 0.008 | 40.7013 | 0.0023 | 35 | 142 | 0.019 | 0.013 | 0.7 | 1.8 | 3.0 | 0.244 | 669 | 7.8E-05 | 2 | NoObs |
| 632.01 | 3.1532 | 267 | 21 | 104.22 | 0.0044 | 7.23848 | 0.00022 | 17.76 | 0.71 | 0.01483 | 0.00047 | 0.028 | 0.055 | 1.2 | 0.072 | 761 | 4.6E-05 | 2 | NoObs |
| 633.01 | 10.3584 | 713 | 31 | 103.6091 | 0.007 | 161.4682 | 0.0099 | 64 | 31 | 0.0283 | 0.0023 | 0.86 | 0.42 | 5.4 | 0.614 | 428 | 9.5E-05 | 2 | |
| 635.01 | 3.4572 | 602 | 20 | 104.4013 | 0.0042 | 16.71985 | 0.0005 | 28 | 163 | 0.023 | 0.022 | 0.7 | 2 | 2.7 | 0.132 | 762 | 8.4E-05 | 2 | 1 |
| 638.01 | 5.2480 | 1147 | 64 | 105.6573 | 0.002 | 23.63591 | 0.00031 | 19.5 | 6.3 | 0.036 | 0.002 | 0.84 | 0.36 | 4.8 | 0.166 | 682 | 8.0E-05 | 2 | NoObs |
| 638.02 | 7.1418 | 1245 | 47 | 79.5663 | 0.0034 | 67.0936 | 0.0011 | 75.5 | 1.2 | 0.03133 | 0.00043 | 0.017 | 0.022 | 4.1 | 0.333 | 482 | - | 4 | |
| 639.01 | 5.6758 | 422 | 33 | 115.2433 | 0.0039 | 17.97984 | 0.0005 | 25.46 | 0.56 | 0.01868 | 0.00037 | 0.026 | 0.035 | 2.1 | 0.139 | 747 | 1.7E-04 | 2 | 1 |
| 640.01 | 3.1182 | 679 | 39 | 124.7843 | 0.0023 | 30.99665 | 0.00051 | 72 | 225 | 0.025 | 0.016 | 0.6 | 1.8 | 2.9 | 0.193 | 533 | 1.2E-04 | 2 | 1 |
| 641.01 | 3.3757 | 1172 | 37 | 110.9992 | 0.0023 | 14.85198 | 0.00025 | 32 | 70 | 0.031 | 0.016 | 0.5 | 1.6 | 3.2 | 0.104 | 535 | 1.2E-04 | 3 | NoObs |
| 644.01 | 7.3023 | 24143 | 903 | 173.59859 | 0.00015 | 45.977503 | 0.000048 | 54.776631 | 0.000058 | 0.1387 | 0.0078 | 0.0008 | - | 35.3 | 0.271 | 698 | 2.1E-06 | 2 | NoObs |
| 645.01 | 2.8982 | 184 | 17 | 103.8637 | 0.0049 | 8.50365 | 0.00029 | 13 | 64 | 0.015 | 0.012 | 0.8 | 1.5 | 2.6 | 0.086 | 1124 | 1.3E-04 | 2 | 1 |
| 645.02 | 7.6570 | 209 | 21 | 112.7424 | 0.0075 | 23.7847 | 0.0013 | 21 | 71 | 0.0141 | 0.0085 | 0.5 | 1.9 | 2.5 | 0.171 | 797 | 1.3E-04 | 2 | NoObs |
| 647.01 | 4.6876 | 187 | 33 | 103.3185 | 0.0035 | 5.16923 | 0.00012 | 6 | 12 | 0.0145 | 0.005 | 0.8 | 1.1 | 1.8 | 0.061 | 1168 | 5.3E-05 | 2 | 1 |
| 649.01 | 8.1220 | 245 | 29 | 115.9258 | 0.0058 | 23.44942 | 0.00088 | 22.24 | 0.46 | 0.01421 | 0.0003 | 0.01 | 0.02 | 2.0 | 0.167 | 727 | 4.8E-05 | 2 | 1 |
| 650.01 | 2.3266 | 865 | 58 | 111.7215 | 0.0014 | 11.95458 | 0.00011 | 26 | 40 | 0.031 | 0.0093 | 0.83 | 0.83 | 3.8 | 0.101 | 728 | 3.2E-05 | 2 | 1 |
| 652.01 | 2.9306 | 3206 | 87 | 115.75856 | 0.00095 | 16.08075 | 0.0001 | 44 | 87 | 0.049 | 0.023 | 0.1 | 1.7 | 3.0 | 0.112 | 459 | 3.2E-05 | 2 | 1 |
| 654.01 | 2.9044 | 328 | 20 | 104.6352 | 0.0037 | 8.59449 | 0.00022 | 13 | 51 | 0.02 | 0.013 | 0.8 | 1.3 | 2.9 | 0.085 | 1013 | 2.7E-05 | 2 | NoObs |
| 655.01 | 5.6200 | 399 | 44 | 125.0972 | 0.003 | 25.67234 | 0.00061 | 35.25 | 0.57 | 0.01781 | 0.00027 | 0.0157 | - | 2.1 | 0.177 | 674 | 5.0E-05 | 2 | 1 |
| 657.01 | 2.0229 | 532 | 35 | 104.0183 | 0.0018 | 4.069378 | 0.000049 | 15.71 | 0.51 | 0.02084 | 0.00049 | 0.014 | 0.036 | 1.6 | 0.046 | 798 | 2.5E-05 | 2 | NoObs |
| 657.02 | 2.8468 | 799 | 34 | 113.7963 | 0.0022 | 16.28267 | 0.00025 | 45 | 1.4 | 0.02489 | 0.00057 | 0 | 0.01 | 1.9 | 0.115 | 504 | 2.3E-05 | 2 | NoObs |
| 658.01 | 1.9455 | 499 | 46 | 102.6422 | 0.0015 | 3.162668 | 0.000033 | 12.13 | 0.46 | 0.01513 | 0.0005 | 0.04 | - | 1.5 | 0.043 | 1162 | 1.7E-04 | 2 | NoObs |
| 658.02 | 2.0263 | 477 | 35 | 105.2367 | 0.0022 | 5.370662 | 0.000079 | 15 | 60 | 0.022 | 0.016 | 0.7 | 1.7 | 2.2 | 0.061 | 975 | 1.7E-04 | 2 | NoObs |
| 659.01 | 4.2939 | 291 | 20 | 113.7619 | 0.0048 | 23.2056 | 0.0008 | 31 | 144 | 0.016 | 0.013 | 0.7 | 1.9 | 2.5 | 0.168 | 818 | 5.4E-05 | 2 | NoObs |
| 660.01 | 6.7029 | 243 | 37 | 103.5844 | 0.0039 | 6.07977 | 0.00016 | 7.05 | 0.12 | 0.01413 | 0.00024 | 0.0093 | - | 2.2 | 0.067 | 1077 | 2.9E-05 | 2 | 1 |
| 661.01 | 3.4869 | 349 | 20 | 107.9834 | 0.004 | 14.40135 | 0.00042 | 27 | 143 | 0.018 | 0.017 | 0.5 | 2.3 | 2.1 | 0.119 | 776 | 2.7E-05 | 2 | NoObs |
| 662.01 | 5.5389 | 226 | 30 | 103.7094 | 0.0048 | 10.21362 | 0.00033 | 13 | 53 | 0.0127 | 0.0094 | 0 | 2.4 | 1.5 | 0.095 | 875 | 5.8E-05 | 2 | 1 |
| 663.01 | 1.8327 | 527 | 64 | 103.84688 | 0.00089 | 2.755602 | 0.000017 | 7 | 11 | 0.0245 | 0.0065 | 0.84 | 0.81 | 1.9 | 0.033 | 846 | 2.3E-05 | 2 | 1 |
| 663.02 | 2.8399 | 644 | 39 | 105.6543 | 0.0021 | 20.30708 | 0.00029 | 57 | 1.5 | 0.02282 | 0.00043 | 0.031 | 0.049 | 1.7 | 0.124 | 436 | 2.1E-05 | 2 | 1 |
| 664.01 | 4.6566 | 202 | 22 | 103.228 | 0.006 | 13.13755 | 0.00052 | 15 | 44 | 0.0149 | 0.0077 | 0.7 | 1.4 | 2.1 | 0.113 | 863 | 6.1E-05 | 3 | NoObs |
| 665.01 | 4.0178 | 423 | 59 | 103.3258 | 0.0018 | 5.867973 | 0.000072 | 11.55 | 0.16 | 0.01863 | 0.00022 | 0.0003 | - | 2.3 | 0.066 | 1066 | 4.0E-05 | 2 | 1 |
| 665.02 | 3.1579 | 89 | 21 | 66.5664 | 0.0048 | 1.611912 | 0.000033 | 1.9 | 7.9 | 0.0097 | 0.0068 | 0.9 | 1.1 | 1.2 | 0.028 | 1636 | - | 4 | |
| 665.03 | 3.7875 | 75 | 14 | 66.3005 | 0.0079 | 3.07154 | 0.0001 | 6.2 | 0.4 | 0.00685 | 0.00043 | 0.008 | 0.01 | 0.8 | 0.043 | 1320 | - | 4 | |
| 666.01 | 3.9017 | 615 | 44 | 107.1338 | 0.0026 | 22.24844 | 0.00039 | 45 | 183 | 0.022 | 0.017 | 0.1 | 2.4 | 2.0 | 0.156 | 570 | 7.2E-05 | 2 | 1 |
| 667.01 | 2.7220 | 10130 | 124 | 103.45173 | 0.00088 | 4.305252 | 0.000024 | 13.95 | 0.13 | 0.08972 | 0.00064 | 0.024 | 0.022 | 6.5 | 0.044 | 711 | 7.1E-06 | 3 | |
| 670.01 | 3.2407 | 252 | 20 | 104.926 | 0.0042 | 9.49006 | 0.00028 | 14 | 57 | 0.016 | 0.011 | 0.8 | 1.5 | 2.1 | 0.09 | 883 | 6.2E-05 | 3 | NoObs |
| 671.01 | 3.2752 | 152 | 21 | 103.7411 | 0.0045 | 4.22875 | 0.00012 | 6 | 22 | 0.0135 | 0.0085 | 0.8 | 1.3 | 1.4 | 0.051 | 1126 | 7.2E-05 | 2 | NoObs |

| | | | | | | | | | | | | | | | | | | | | |
|---|---|---|---|---|---|---|---|---|---|---|---|---|---|---|---|---|---|---|---|---|
| 672.01 | 3.0784 | 554 | 32 | 105.8116 | 0.0028 | 16.08822 | 0.00033 | 25 | 67 | 0.025 | 0.011 | 0.8 | 1.1 | 4.0 | 0.13 | 821 | 5.1E-05 | 2 | 1 | |
| 672.02 | 5.8817 | 966 | 50 | 86.8426 | 0.0037 | 41.749 | 0.0012 | 58.85 | 0.87 | 0.02823 | 0.00037 | 0.016 | 0.022 | 4.5 | 0.245 | 598 | 4.7E-05 | 2 | 1 | |
| 673.01 | 3.0279 | 247 | 20 | 103.7904 | 0.0041 | 4.41748 | 0.00013 | 11.98 | 0.55 | 0.01412 | 0.00056 | 0.0305 | - | 1.8 | 0.055 | 1303 | 5.6E-05 | 2 | 1 | |
| 674.01 | 9.4755 | 1610 | 97 | 110.9192 | 0.0022 | 16.33886 | 0.00021 | 11.3 | 5.8 | 0.0378 | 0.0042 | 0.59 | 0.69 | 11.3 | 0.137 | 959 | 2.4E-05 | 3 | | |
| 676.01 | 2.8769 | 3080 | 53 | 104.5826 | 0.0014 | 7.972513 | 0.000076 | 12 | 3.6 | 0.0593 | 0.0034 | 0.84 | 0.35 | 4.5 | 0.067 | 598 | 3.4E-05 | 2 | NoObs | |
| 676.02 | 1.7470 | 1693 | 41 | 103.8934 | 0.0014 | 2.453224 | 0.000023 | 9.5 | 2.8 | 0.03888 | 0.00086 | 0.308 | 0.092 | 2.9 | 0.03 | 894 | 3.9E-05 | 2 | NoObs | |
| 678.01 | 2.7421 | 113 | 21 | 105.59 | 0.013 | 6.04097 | 0.00054 | 10.2 | 3.1 | 0.011 | 0.0026 | 0.66 | 0.2 | 1.7 | 0.066 | 1028 | 3.3E-05 | 3 | 1 | |
| 679.01 | 8.1270 | 307 | 33 | 123.2486 | 0.0051 | 31.8049 | 0.0012 | 30.4 | 9.1 | 0.01568 | 0.00039 | - | - | 1.8 | 0.197 | 598 | 2.8E-05 | 2 | 1 | |
| 680.01 | 8.9336 | 4384 | 586 | 110.64238 | 0.00045 | 8.600116 | 0.000027 | 7.817 | 0.011 | 0.05966 | 0.00008 | 0.0002 | - | 7.6 | 0.085 | 989 | - | 2 | | |
| †682.01 | 9.9203 | 4927 | 163 | 118.99358 | 0.00099 | 163.7133 | 0.0083 | 156.9 | 1.4 | 0.04688 | 0.0004 | 0.002 | 0.032 | 4.9 | 0.591 | 307 | 5.1E-05 | 2 | | |
| 683.01 | 4.4673 | 2328 | 55 | 110.5186 | 0.0021 | 278.1232 | 0.003 | 385 | 340 | 0.0489 | 0.0078 | 0.8 | 0.67 | 4.2 | 0.839 | 239 | 3.8E-05 | 2 | | |
| 684.01 | 1.8215 | 794 | 50 | 105.2568 | 0.0012 | 4.034923 | 0.000033 | 6 | 1.8 | 0.0407 | 0.0042 | 0.96 | 0.29 | 8.3 | 0.052 | 1414 | 1.3E-05 | 2 | NoObs | |
| 685.01 | 3.4760 | 286 | 41 | 103.9261 | 0.0024 | 3.173885 | 0.000053 | 4.7 | 9.9 | 0.0175 | 0.006 | 0.8 | 1.1 | 2.8 | 0.045 | 1570 | 8.9E-05 | 2 | NoObs | |
| 686.01 | 3.0231 | 14530 | 620 | 104.67404 | 0.00014 | 52.513492 | 0.000039 | 147.001 | 0.00011 | 0.1077 | 0.0032 | 0.0028 | - | 11.3 | 0.275 | 442 | 2.7E-06 | 3 | | |
| 687.01 | 2.1209 | 285 | 13 | 104.983 | 0.0059 | 4.17853 | 0.00017 | 11.6 | 3.5 | 0.0164 | 0.0013 | 0.44 | 0.13 | 1.7 | 0.052 | 1060 | - | 3 | | 1 |
| 688.01 | 2.9239 | 270 | 31 | 103.2535 | 0.0027 | 3.275814 | 0.00006 | 5 | 16 | 0.0175 | 0.0083 | 0.8 | 1.2 | 2.5 | 0.045 | 1465 | 4.1E-05 | 2 | NoObs | |
| 689.01 | 3.5876 | 582 | 21 | 115.398 | 0.0047 | 15.87403 | 0.00046 | 34.3 | 1.3 | 0.02165 | 0.00074 | 0.021 | 0.01 | 2.0 | 0.123 | 622 | 6.5E-05 | 2 | | |
| 691.01 | 8.3024 | 614 | 47 | 122.3661 | 0.0038 | 29.66717 | 0.00092 | 27.7 | 0.36 | 0.02253 | 0.0003 | 0.018 | 0.01 | 2.9 | 0.195 | 653 | 2.0E-05 | 2 | NoObs | |
| 691.02 | 6.1582 | 108 | 9.7 | 77.039 | 0.015 | 16.2245 | 0.001 | 20.8 | 1.3 | 0.01006 | 0.00059 | 0.056 | 0.014 | 1.3 | 0.13 | 799 | - | 4 | | |
| 692.01 | 1.8674 | 173 | 20 | 104.8412 | 0.0028 | 2.462367 | 0.000047 | 10.33 | 0.41 | 0.01309 | 0.00042 | 0.0614 | - | 0.8 | 0.035 | 1004 | 4.2E-05 | 2 | 1 | |
| 693.01 | 7.3509 | 311 | 27 | 126.3107 | 0.0058 | 28.7784 | 0.0013 | 24 | 74 | 0.0169 | 0.0088 | 0.6 | 1.7 | 1.8 | 0.19 | 611 | 2.3E-05 | 2 | NoObs | |
| 693.02 | 7.0385 | 321 | 36 | 79.3547 | 0.0051 | 15.66002 | 0.00032 | 18.23 | 0.23 | 0.01624 | 0.00029 | 0.022 | 0.01 | 1.7 | 0.126 | 750 | - | 4 | | |
| 694.01 | 4.8698 | 827 | 56 | 117.2445 | 0.0023 | 17.42154 | 0.00028 | 27.97 | 0.39 | 0.02593 | 0.00028 | 0.031 | 0.055 | 1.7 | 0.13 | 530 | 1.9E-05 | 2 | NoObs | |
| 695.01 | 4.9238 | 566 | 45 | 108.293 | 0.0027 | 29.90653 | 0.00063 | 32 | 51 | 0.0241 | 0.0065 | 0.76 | 0.98 | 3.1 | 0.195 | 643 | 1.7E-05 | 2 | NoObs | |
| 697.01 | 3.6358 | 480 | 45 | 104.7324 | 0.0021 | 3.032186 | 0.00004 | 5 | 10 | 0.0215 | 0.008 | 0.7 | 1.3 | 4.0 | 0.043 | 1601 | 3.3E-05 | 2 | | |
| 698.01 | 2.4663 | 7776 | 386 | 105.99432 | 0.00028 | 12.718733 | 0.000023 | 23 | 6.9 | 0.1209 | 0.0025 | 0.92 | 0.28 | 12.0 | 0.103 | 748 | 1.2E-05 | 2 | | |
| 700.01 | 2.9694 | 564 | 28 | 105.9348 | 0.0031 | 30.86436 | 0.00069 | 45 | 113 | 0.025 | 0.01 | 0.86 | 0.97 | 3.1 | 0.197 | 588 | 3.5E-05 | 2 | 1 | |
| 700.02 | 3.2972 | 211 | 19 | 104.9588 | 0.0044 | 9.36127 | 0.00028 | 13 | 45 | 0.0159 | 0.0091 | 0.8 | 1.2 | 1.9 | 0.089 | 875 | 4.6E-05 | 2 | 1 | |
| 701.01 | 2.9877 | 919 | 48 | 113.8108 | 0.0017 | 18.16428 | 0.00025 | 35 | 74 | 0.03 | 0.013 | 0.7 | 1.2 | 2.2 | 0.127 | 496 | 1.7E-05 | 2 | 1 | |
| 701.02 | 2.3257 | 411 | 34 | 103.9187 | 0.0023 | 5.714973 | 0.000087 | 14 | 41 | 0.021 | 0.013 | 0.7 | 1.5 | 1.5 | 0.059 | 728 | 2.0E-05 | 2 | 1 | |
| 701.03 | 6.8714 | 697 | 25 | 83.398 | 0.0045 | 122.3894 | 0.0017 | 138.6 | 4 | 0.02341 | 0.00059 | 0.015 | 0.046 | 1.7 | 0.454 | 262 | - | 4 | | |
| 703.01 | 1.6971 | 135 | 24 | 102.9528 | 0.0025 | 1.3686 | 0.000024 | 3 | 15 | 0.013 | 0.0093 | 0.9 | 1.3 | 1.7 | 0.025 | 1866 | 5.1E-05 | 2 | NoObs | |
| 704.01 | 2.5400 | 498 | 14 | 118.13 | 0.0051 | 18.39714 | 0.00067 | 60.8 | 5.1 | 0.0167 | 0.0014 | 0.015 | 0.01 | 1.7 | 0.136 | 619 | 1.6E-05 | 2 | NoObs | |
| 707.01 | 7.8208 | 677 | 48 | 122.631 | 0.0036 | 21.77654 | 0.00055 | 21.9 | 0.29 | 0.02384 | 0.0003 | 0.0044 | - | 3.4 | 0.159 | 745 | 2.1E-05 | 2 | 1 | |
| 707.02 | 9.8121 | 396 | 23 | 105.5817 | 0.0091 | 41.0315 | 0.0025 | 32.14 | 0.77 | 0.01873 | 0.00047 | 0.041 | 0.01 | 2.6 | 0.242 | 604 | 2.5E-05 | 2 | 1 | |
| 707.03 | 8.5541 | 373 | 22 | 68.8687 | 0.0091 | 31.7845 | 0.0012 | 29.17 | 0.77 | 0.01761 | 0.00047 | 0.029 | 0.04 | 2.5 | 0.204 | 658 | - | 4 | | |
| 707.04 | 6.4886 | 246 | 21 | 76.6803 | 0.0079 | 13.17535 | 0.00045 | 11 | 38 | 0.0156 | 0.009 | 0.7 | 1.6 | 2.2 | 0.113 | 884 | - | 4 | | |
| 708.01 | 6.7713 | 525 | 42 | 104.0034 | 0.0035 | 17.40696 | 0.00044 | 20.02 | 0.3 | 0.02136 | 0.0003 | 0.027 | 0.028 | 2.2 | 0.135 | 703 | 4.4E-05 | 2 | 1 | |
| 708.02 | 4.9436 | 269 | 26 | 109.5127 | 0.0047 | 7.69315 | 0.00025 | 10 | 33 | 0.0167 | 0.0098 | 0.6 | 1.7 | 1.7 | 0.078 | 924 | 5.2E-05 | 2 | 1 | |
| 709.01 | 3.7885 | 615 | 31 | 111.791 | 0.0029 | 21.38418 | 0.00044 | 43 | 202 | 0.023 | 0.021 | 0.3 | 2.5 | 2.2 | 0.152 | 588 | 2.1E-05 | 2 | NoObs | |
| 710.01 | 3.9341 | 140 | 21 | 103.9326 | 0.0044 | 5.37503 | 0.00016 | 10.67 | 0.16 | 0.0113 | 0.00031 | 0.0171 | - | 1.7 | 0.067 | 1320 | 5.1E-05 | 2 | 1 | |
| 711.01 | 6.1018 | 786 | 34 | 107.8257 | 0.0048 | 44.6987 | 0.0019 | 56.3 | 1.2 | 0.02508 | 0.00047 | 0.004 | 0.025 | 2.7 | 0.249 | 486 | 4.6E-05 | 2 | NoObs | |
| 711.02 | 3.1710 | 184 | 22 | 68.4376 | 0.0044 | 3.619344 | 0.000069 | 4 | 23 | 0.012 | 0.011 | 0.9 | 1.3 | 1.3 | 0.047 | 1118 | - | 4 | | |
| 711.03 | 9.8334 | 685 | 25 | 187.1803 | 0.0076 | 124.5229 | 0.0058 | 87 | 382 | 0.024 | 0.021 | 0.5 | 2.2 | 2.6 | 0.494 | 345 | - | 4 | | |
| 712.01 | 1.9895 | 132 | 17 | 104.23 | 0.0031 | 2.178191 | 0.000046 | 9.893 | 0.073 | 0.01118 | 0.00044 | 0.0809 | - | 0.8 | 0.033 | 1101 | - | 3 | | |
| 714.01 | 2.2176 | 862 | 89 | 105.78778 | 0.00081 | 4.182017 | 0.000023 | 14.8 | 0.18 | 0.02607 | 0.00024 | 0.0001 | - | 2.2 | 0.051 | 929 | 1.8E-05 | 2 | 1 | |
| 716.01 | 2.2744 | 2182 | 103 | 112.95307 | 0.0008 | 26.89291 | 0.00016 | 40 | 12 | 0.0606 | 0.0027 | 0.93 | 0.28 | 6.3 | 0.18 | 595 | 3.0E-05 | 2 | NoObs | |
| 717.01 | 3.1618 | 285 | 22 | 108.7924 | 0.0038 | 14.70752 | 0.00037 | 37.7 | 1.5 | 0.01569 | 0.0005 | 0.044 | 0.01 | 0.9 | 0.114 | 510 | 1.5E-05 | 3 | 1 | |
| 718.01 | 3.2541 | 367 | 42 | 102.8579 | 0.0021 | 4.585494 | 0.000048 | 9 | 21 | 0.0186 | 0.0082 | 0.6 | 1.5 | 1.6 | 0.055 | 964 | 2.4E-05 | 2 | 1 | |
| 718.02 | 5.7207 | 499 | 33 | 80.2934 | 0.0041 | 22.71449 | 0.00042 | 31.07 | 0.65 | 0.02078 | 0.00038 | 0.031 | 0.054 | 1.8 | 0.159 | 567 | - | 4 | | |
| 718.03 | 5.5414 | 332 | 16 | 74.9787 | 0.0082 | 47.9042 | 0.0019 | 55 | 327 | 0.017 | 0.018 | 0.6 | 2.2 | 1.5 | 0.261 | 443 | - | 4 | | |
| 719.01 | 1.6354 | 538 | 41 | 104.013 | 0.0015 | 9.034227 | 0.000093 | 32 | 111 | 0.024 | 0.014 | 0.8 | 1.4 | 1.9 | 0.075 | 605 | 2.0E-05 | 2 | 1 | |
| 720.01 | 2.4833 | 1235 | 27 | 107.0488 | 0.0026 | 5.69057 | 0.0001 | 17 | 102 | 0.032 | 0.04 | 0.4 | 2.7 | 3.0 | 0.061 | 849 | 4.3E-05 | 2 | NoObs | |
| 721.01 | 6.8323 | 206 | 25 | 113.6486 | 0.0064 | 13.72423 | 0.0006 | 15.7 | 0.39 | 0.01352 | 0.00032 | 0.029 | 0.044 | 2.3 | 0.118 | 942 | 6.5E-05 | 3 | NoObs | |
| 722.01 | 6.9235 | 468 | 35 | 125.9894 | 0.0038 | 46.408 | 0.0013 | 51.86 | 0.95 | 0.01951 | 0.00036 | 0.001 | 0.025 | 1.8 | 0.259 | 485 | 6.0E-05 | 2 | 1 | |
| 723.01 | 1.8098 | 1328 | 24 | 102.6479 | 0.0026 | 3.936985 | 0.000069 | 17.4 | 0.84 | 0.034 | 0.0011 | 0.0378 | - | 2.8 | 0.048 | 918 | 2.3E-05 | 2 | NoObs | |
| 723.02 | 4.3932 | 1302 | 13 | 127.916 | 0.0049 | 28.08205 | 0.00043 | 50.5 | 2.5 | 0.0349 | 0.0015 | 0.025 | 0.049 | 2.9 | 0.177 | 478 | 2.3E-05 | 2 | NoObs | |
| 723.03 | 1.4752 | 1541 | 15 | 106.0831 | 0.0034 | 10.0888 | 0.00025 | 66.1 | 7.2 | 0.0384 | 0.0031 | 0.015 | - | 3.2 | 0.09 | 670 | 2.1E-05 | 2 | NoObs | |
| 725.01 | 3.3669 | 8800 | 31 | 102.6447 | 0.0023 | 7.305 | 0.00011 | 18 | 5.4 | 0.0833 | 0.0024 | - | - | 6.8 | 0.071 | 720 | - | 2 | | |
| 728.01 | 2.0043 | 7786 | 211 | 103.11774 | 0.00035 | 7.18937 | 0.000018 | 17.5 | 5.2 | 0.09887 | 0.00097 | 0.8 | 0.24 | 9.9 | 0.075 | 922 | - | 3 | | |
| 730.01 | 5.7701 | 746 | 20 | 109.796 | 0.0062 | 14.7845 | 0.00099 | 17 | 77 | 0.026 | 0.021 | 0.5 | 2.1 | 3.1 | 0.12 | 746 | 7.0E-05 | 2 | | |
| 730.02 | 5.4941 | 393 | 13 | 71.36 | 0.011 | 9.84978 | 0.00047 | 13.71 | 0.67 | 0.01915 | 0.00083 | 0.04 | 0.01 | 2.3 | 0.092 | 852 | - | 4 | | |
| 730.03 | 4.3590 | 343 | 9.7 | 68.131 | 0.011 | 9.85997 | 0.00051 | 12 | 103 | 0.021 | 0.033 | 0.7 | 2.4 | 2.5 | 0.092 | 852 | - | 4 | | |
| 730.04 | 5.2290 | 259 | 9.1 | 70.475 | 0.014 | 7.38469 | 0.00047 | 11.25 | 0.44 | 0.01532 | 0.00098 | 0.049 | 0.01 | 1.8 | 0.076 | 937 | - | 4 | | |
| 732.01 | 1.7619 | 1149 | 72 | 103.4084 | 0.00081 | 1.2602586 | 0.000007 | 5.834 | 0.018 | 0.0318 | 0.00031 | 0.0447 | - | 2.9 | 0.023 | 1424 | 2.6E-05 | 2 | | |
| 733.01 | 2.6528 | 1540 | 41 | 102.7156 | 0.002 | 5.924992 | 0.000081 | 17.95 | 0.43 | 0.03568 | 0.00066 | 0.0002 | - | 2.2 | 0.061 | 683 | 3.0E-05 | 2 | | |
| 733.02 | 2.9996 | 1152 | 23 | 67.3199 | 0.0042 | 11.34917 | 0.0002 | 30 | 1.2 | 0.0307 | 0.001 | 0.025 | 0.01 | 1.9 | 0.094 | 551 | - | 4 | | |
| 733.03 | 2.1611 | 460 | 16 | 68.6747 | 0.0047 | 3.132968 | 0.000064 | 9 | 51 | 0.024 | 0.027 | 0.6 | 2.2 | 1.5 | 0.04 | 844 | - | 4 | | |
| 734.01 | 7.1431 | 1075 | 34 | 120.9151 | 0.0043 | 24.54369 | 0.0008 | 26.84 | 0.52 | 0.02948 | 0.00055 | 0.012 | 0.01 | 2.4 | 0.166 | 535 | 3.1E-05 | 2 | | |
| 735.01 | 4.7686 | 2970 | 28 | 104.5711 | 0.0051 | 22.34101 | 0.00073 | 37.4 | 1.1 | 0.0483 | 0.0011 | 0.032 | 0.1 | 5.0 | 0.153 | 557 | 5.4E-05 | 2 | | |
| 736.01 | 3.0657 | 1422 | 19 | 110.7903 | 0.0042 | 18.79523 | 0.0006 | 43 | 244 | 0.036 | 0.038 | 0.5 | 2.5 | 2.6 | 0.117 | 442 | 3.1E-05 | 2 | | |
| 736.02 | 3.1139 | 607 | 14 | 68.2816 | 0.0074 | 6.7388 | 0.00021 | 20.4 | 1.1 | 0.0264 | 0.0011 | 0.064 | 0.01 | 2.0 | 0.059 | 623 | - | 4 | | |
| 737.01 | 3.1240 | 3249 | 73 | 115.6786 | 0.0014 | 14.49847 | 0.00015 | 18 | 2.5 | 0.0638 | 0.002 | 0.89 | 0.2 | 5.6 | 0.114 | 597 | 3.9E-05 | 2 | | |
| 738.01 | 3.1466 | 1219 | 34 | 103.4322 | 0.0028 | 10.33677 | 0.0002 | 22 | 83 | 0.033 | 0.024 | 0.6 | 1.9 | 3.3 | 0.095 | 781 | 6.0E-05 | 2 | NoObs | |
| 738.02 | 3.3240 | 933 | 23 | 105.0429 | 0.0041 | 13.29175 | 0.0004 | 32 | 248 | 0.029 | 0.042 | 0.3 | 3.1 | 2.9 | 0.112 | 719 | 6.5E-05 | 2 | NoObs | |
| 739.01 | 1.4949 | 779 | 42 | 102.8178 | 0.0013 | 1.287052 | 0.000012 | 7.23 | 0.2 | 0.02741 | 0.00051 | 0.0416 | - | 2.0 | 0.019 | 1058 | 1.9E-05 | 2 | NoObs | |
| 740.01 | 3.1145 | 864 | 18 | 119.3644 | 0.0045 | 17.67221 | 0.00061 | 25 | 94 | 0.03 | 0.023 | 0.8 | 1.4 | 2.3 | 0.123 | 497 | 3.0E-05 | 3 | | |
| 741.01 | 3.9182 | 37406 | 874 | 102.83287 | 0.00017 | 23.355367 | 0.000028 | 33.6 | 3 | 0.2416 | 0.0062 | 0.89 | 0.15 | 18.7 | 0.159 | 518 | - | 3 | | 1 |
| 743.01 | 10.5338 | 9931 | 189 | 105.4889 | 0.0014 | 19.40335 | 0.00021 | 15.3162 | 0.0011 | 0.087 | 0.021 | 0.006 | 0.01 | 10.9 | 0.139 | 617 | 1.6E-05 | 2 | | |
| 745.01 | 9.3456 | 10890 | 98 | 113.8723 | 0.0031 | 16.47063 | 0.00034 | 14.67 | 0.12 | 0.0917 | 0.00065 | 0.0082 | - | 9.7 | 0.124 | 613 | 4.2E-05 | 2 | | |
| 746.01 | 3.3082 | 1296 | 46 | 106.2476 | 0.0021 | 9.27391 | 0.00013 | 16 | 29 | 0.036 | 0.014 | 0.7 | 1.1 | 3.1 | 0.08 | 648 | 2.5E-05 | 2 | | |
| 747.01 | 1.5804 | 1906 | 37 | 104.6042 | 0.0016 | 6.029222 | 0.000065 | 24 | 82 | 0.043 | 0.032 | 0.7 | 1.6 | 2.8 | 0.056 | 633 | 1.5E-05 | 2 | | |
| 749.01 | 3.0377 | 848 | 33 | 104.8065 | 0.0028 | 5.349518 | 0.000099 | 14.486 | 0.099 | 0.02772 | 0.00052 | 0.0612 | - | 2.0 | 0.059 | 789 | 3.3E-05 | 2 | | |
| 749.02 | 2.3564 | 375 | 15 | 69.0904 | 0.0052 | 3.940973 | 0.000088 | 12.83 | 0.7 | 0.01895 | 0.00081 | 0.0316 | - | 1.4 | 0.048 | 875 | - | 4 | | |
| 750.01 | 3.3702 | 985 | 21 | 104.5295 | 0.0049 | 21.67821 | 0.00074 | 30 | 57 | 0.034 | 0.013 | 0.85 | 0.86 | 2.6 | 0.139 | 458 | 3.2E-05 | 2 | | |
| 751.01 | 2.0883 | 995 | 20 | 104.74 | 0.0045 | 4.99682 | 0.00014 | 13 | 61 | 0.034 | 0.031 | 0.7 | 1.7 | 3.2 | 0.056 | 896 | 7.0E-05 | 2 | | |
| 752.01 | 3.0592 | 593 | 19 | 103.5366 | 0.0044 | 9.48851 | 0.0003 | 22 | 130 | 0.023 | 0.026 | 0.4 | 2.6 | 2.7 | 0.089 | 853 | 3.4E-05 | 2 | | |
| 752.02 | 4.1260 | 885 | 14 | 95.5164 | 0.0081 | 54.4154 | 0.0022 | 84 | 693 | 0.029 | 0.048 | 0.7 | 2.6 | 3.4 | 0.286 | 476 | - | 4 | | |
| 753.01 | 1.8813 | 7451 | 48 | 108.8504 | 0.0014 | 19.89939 | 0.00019 | 49 | 15 | 0.1015 | 0.0072 | 0.84 | 0.25 | 6.9 | 0.143 | 519 | - | 3 | | |
| 755.01 | 1.6211 | 660 | 25 | 104.5925 | 0.0029 | 2.525605 | 0.000046 | 12.53 | 0.55 | 0.02434 | 0.00072 | 0.0327 | - | 2.8 | 0.037 | 1350 | 2.9E-05 | 2 | | |
| 756.01 | 4.6503 | 1502 | 39 | 104.2018 | 0.004 | 11.09431 | 0.00028 | 18.85 | 0.37 | 0.03516 | 0.00059 | 0.027 | 0.033 | 3.7 | 0.099 | 791 | 6.5E-05 | 2 | | |
| 756.02 | 3.0514 | 691 | 25 | 105.9703 | 0.0047 | 4.13463 | 0.00013 | 10.55 | 0.36 | 0.02509 | 0.0007 | 0.0209 | - | 2.6 | 0.052 | 1092 | 7.9E-05 | 2 | | |
| 756.03 | 2.4819 | 231 | 9.4 | 112.5496 | 0.0086 | 2.5667 | 0.00011 | 8.67 | 0.63 | 0.017 | 0.0011 | 0.2236 | - | 1.8 | 0.037 | 1294 | - | 4 | | |
| 757.01 | 3.5566 | 5214 | 99 | 106.622 | 0.001 | 16.0686 | 0.00017 | 37.1 | 0.37 | 0.06336 | 0.0005 | 0.0188 | - | 4.8 | 0.119 | 530 | 1.4E-05 | 2 | NoObs | |
| 757.02 | 4.6490 | 2395 | 35 | 98.035 | 0.0038 | 41.19249 | 0.00069 | 72 | 2.1 | 0.0431 | 0.0011 | 0.013 | 0.022 | 3.3 | 0.223 | 387 | 1.7E-05 | 2 | NoObs | |
| 757.03 | 2.4488 | 840 | 22 | 104.3013 | 0.0044 | 6.25288 | 0.00025 | 14 | 58 | 0.03 | 0.026 | 0.7 | 1.7 | 2.3 | 0.063 | 729 | 2.2E-05 | 2 | NoObs | |
| 758.01 | 3.4108 | 1282 | 22 | 109.354 | 0.0041 | 16.016 | 0.00074 | 39.8 | 2 | 0.03 | 0.0011 | 0.003 | 0.01 | 3.8 | 0.123 | 663 | 5.9E-05 | 2 | | |
| 759.01 | 4.8992 | 1713 | 40 | 127.1363 | 0.0031 | 32.6272 | 0.001 | 55 | 1.1 | 0.03774 | 0.00064 | 0.0003 | - | 3.6 | 0.2 | 495 | 1.5E-05 | 2 | | |
| 760.01 | 2.0728 | 9959 | 425 | 105.25698 | 0.00017 | 4.9593304 | 0.0000059 | 12.01 | 0.35 | 0.10675 | 0.00076 | 0.85 | 0.11 | 9.7 | 0.058 | 982 | 1.6E-05 | 2 | | |
| 762.01 | 3.7651 | 483 | 18 | 104.3455 | 0.0055 | 4.4987 | 0.00016 | 9.19 | 0.46 | 0.01832 | 0.0008 | 0.027 | 0.01 | 1.7 | 0.054 | 1013 | 7.8E-05 | 3 | | |
| 763.01 | 4.9957 | 12565 | 307 | 112.40123 | 0.00055 | 19.65119 | 0.00012 | 24.49 | 0.94 | 0.10979 | 0.0009 | 0.71 | 0.16 | 13.2 | 0.147 | 700 | 1.7E-05 | 2 | | |
| 764.01 | 10.3560 | 2681 | 71 | 141.9341 | 0.0031 | 41.43958 | 0.00098 | 31.83 | 0.27 | 0.04707 | 0.0004 | 0.003 | 0.02 | 5.6 | 0.236 | 500 | 3.1E-05 | 2 | | |
| 765.01 | 2.2835 | 990 | 24 | 104.6302 | 0.003 | 8.35404 | 0.00023 | 20 | 106 | 0.03 | 0.03 | 0.7 | 1.9 | 2.4 | 0.079 | 713 | 1.9E-05 | 2 | | |
| 766.01 | 3.1048 | 1447 | 63 | 102.7521 | 0.0016 | 4.125488 | 0.000042 | 10.66653 | 0.00011 | 0.035 | 0.073 | 0.0065 | - | 3.8 | 0.052 | 1144 | 3.0E-05 | 2 | | |
| 767.01 | 2.6063 | 17466 | 915 | 103.96676 | 0.00009 | 2.8165077 | 0.0000018 | 7.05 | 0.18 | 0.12849 | 0.00076 | 0.68 | 0.14 | 14.2 | 0.039 | 1221 | 1.6E-05 | 2 | NoObs | |
| 769.01 | 2.8675 | 688 | 31 | 104.8993 | 0.0027 | 4.280889 | 0.000079 | 11.636 | 0.1 | 0.02389 | 0.0005 | 0.019 | - | 2.0 | 0.051 | 946 | 4.3E-05 | 2 | | |
| †771.01 | 48.1018 | 19287 | 236 | 142.0388 | 0.0035 | 10389 | 133 | 1855 | 24 | 0.12445 | - | - | - | 15.0 | 9.514 | 84 | - | 2 | | |
| 772.01 | 5.5841 | 4359 | 30 | 106.8349 | 0.0042 | 61.2592 | 0.0032 | 53 | 17 | 0.0695 | 0.0044 | 0.84 | 0.36 | 8.2 | 0.313 | 482 | 1.2E-05 | 2 | | |
| 773.01 | 5.5792 | 692 | 20 | 105.817 | 0.005 | 38.37813 | 0.00087 | 51.4 | 1.9 | 0.02311 | 0.00078 | 0.015 | 0.01 | 2.1 | 0.225 | 477 | 3.2E-05 | 3 | | |
| 774.01 | 2.8950 | 26414 | 359 | 102.9705 | 0.00031 | 7.442665 | 0.000016 | 27.75 | 0.11 | 0.14349 | 0.00043 | 0.0002 | - | 15.8 | 0.077 | 939 | - | 3 | | 1 |
| 775.01 | 2.8272 | 955 | 13 | 105.724 | 0.0047 | 16.38523 | 0.0007 | 44.3 | 3.2 | 0.0285 | 0.0016 | 0.043 | 0.01 | 2.1 | 0.105 | 458 | 1.8E-05 | 2 | NoObs | |
| 775.02 | 2.4216 | 1320 | 22 | 109.3746 | 0.0023 | 7.87761 | 0.00018 | 26.4 | 1.2 | 0.0332 | 0.0012 | 0.027 | 0.052 | 2.5 | 0.065 | 582 | 1.6E-05 | 2 | NoObs | |
| 776.01 | 2.6154 | 5364 | 212 | 104.79264 | 0.00039 | 3.7287253 | 0.0000099 | 11.606493 | 0.000031 | 0.065 | 0.023 | 0.0007 | - | 4.3 | 0.046 | 852 | 2.0E-05 | 2 | | |
| 777.01 | 2.9867 | 6883 | 29 | 106.5648 | 0.0031 | 40.41887 | 0.00089 | 52 | 16 | 0.3477 | 0.0036 | 1.56 | 0.47 | 36.0 | 0.23 | 471 | 1.2E-05 | 3 | | |
| 778.01 | 1.1499 | 811 | 27 | 103.6798 | 0.0017 | 2.24334 | 0.000036 | 17.99 | 0.98 | 0.0279 | 0.001 | 0.01 | 0.042 | 1.9 | 0.027 | 857 | 1.8E-05 | 2 | | |
| 779.01 | 6.4631 | 14710 | 377 | 110.19884 | 0.0006 | 10.405935 | 0.000066 | 13.752186 | 0.000088 | 0.109 | 0.021 | - | - | 12.8 | 0.095 | 821 | 1.6E-05 | 2 | | |
| 780.01 | 1.9891 | 940 | 41 | 104.7599 | 0.0016 | 2.337466 | 0.000026 | 9.901 | 0.046 | 0.02856 | 0.00048 | 0.0496 | - | 2.2 | 0.032 | 995 | 3.8E-05 | 2 | NoObs | |
| 781.01 | 2.5888 | 3141 | 44 | 113.3936 | 0.0018 | 11.59823 | 0.00015 | 39 | 113 | 0.05 | 0.029 | 0.2 | 2 | 4.3 | 0.083 | 521 | 3.0E-05 | 2 | NoObs | |
| 782.01 | 4.3279 | 2926 | 107 | 106.63389 | 0.00093 | 6.57526 | 0.000045 | 12.201573 | 0.000083 | 0.048 | 0.093 | 0.0001 | - | 5.6 | 0.07 | 989 | 2.3E-05 | 2 | | |
| 783.01 | 7.3311 | 3022 | 176 | 102.9929 | 0.001 | 7.27509 | 0.000053 | 7.987 | 0.034 | 0.04895 | 0.00019 | 0.029 | 0.017 | 3.5 | 0.072 | 707 | 2.7E-05 | 2 | | |
| 784.01 | 2.9416 | 1246 | 20 | 119.7842 | 0.0044 | 19.2693 | 0.00059 | 47 | 256 | 0.032 | 0.033 | 0.5 | 2.5 | 2.3 | 0.117 | 429 | 3.3E-05 | 2 | | |
| 785.01 | 3.3041 | 937 | 19 | 111.7494 | 0.0046 | 12.39325 | 0.00037 | 29.4 | 1.4 | 0.027 | 0.0011 | 0.042 | 0.01 | 2.1 | 0.103 | 620 | 4.0E-05 | 2 | | |
| 786.01 | 2.3232 | 472 | 19 | 103.36 | 0.0035 | 3.689947 | 0.000087 | 13.82 | 0.62 | 0.02241 | 0.00076 | 0.0624 | - | 1.8 | 0.047 | 980 | 3.5E-05 | 3 | | |

| | | | | | | | | | | | | | | | | | | | |
|---|---|---|---|---|---|---|---|---|---|---|---|---|---|---|---|---|---|---|---|
| 787.01 | 3.1631 | 1031 | 39 | 104.0208 | 0.0025 | 4.431061 | 0.000075 | 11.1 | 0.27 | 0.02873 | 0.00055 | 0 | 0.03 | 2.9 | 0.054 | 1020 | 3.8E-05 | 2 | |
| 787.02 | 2.1124 | 435 | 12 | 66.8549 | 0.0064 | 5.68952 | 0.00015 | 21.8 | 1.5 | 0.0219 | 0.0012 | 0.0618 | - | 2.2 | 0.063 | 944 | - | 4 | |
| 788.01 | 3.3280 | 1708 | 29 | 109.0511 | 0.0032 | 26.3953 | 0.00057 | 47 | 158 | 0.04 | 0.028 | 0.7 | 1.6 | 3.2 | 0.166 | 463 | 2.4E-05 | 2 | |
| 790.01 | 2.6555 | 991 | 23 | 107.1626 | 0.0039 | 8.47239 | 0.00023 | 25.1 | 1.1 | 0.02876 | 0.00094 | 0.02 | 0.01 | 1.4 | 0.076 | 551 | 3.8E-05 | 2 | |
| 791.01 | 4.9811 | 6273 | 181 | 113.89115 | 0.00077 | 12.611939 | 0.000069 | 20.757 | 0.099 | 0.07053 | 0.00028 | 0.0007 | - | 7.1 | 0.107 | 718 | 1.8E-05 | 2 | |
| 794.01 | 2.4001 | 401 | 22 | 102.6744 | 0.0036 | 2.539147 | 0.000061 | 8.98 | 0.35 | 0.02019 | 0.00063 | 0.0296 | - | 2.1 | 0.037 | 1296 | 1.0E-04 | 2 | NoObs |
| 795.01 | 1.8917 | 1430 | 31 | 103.5759 | 0.0022 | 6.77034 | 0.0001 | 29.7 | 1.2 | 0.03458 | 0.00099 | 0.02 | 0.042 | 2.4 | 0.069 | 733 | 8.6E-05 | 2 | |
| 797.01 | 2.9281 | 6082 | 111 | 110.1418 | 0.0011 | 10.18151 | 0.000069 | 21.8 | 6.5 | 0.07652 | 0.00073 | 0.49 | 0.15 | 7.7 | 0.094 | 791 | 5.9E-05 | 3 | |
| 799.01 | 1.6033 | 1525 | 70 | 102.81727 | 0.00087 | 1.6266615 | 0.0000097 | 5.6 | 1.7 | 0.03934 | 0.00063 | 0.57 | 0.17 | 4.5 | 0.027 | 1511 | 1.1E-04 | 3 | |
| 800.01 | 3.0081 | 961 | 38 | 103.0385 | 0.0023 | 2.711437 | 0.000043 | 7.37 | 0.17 | 0.02941 | 0.00053 | 0.0402 | - | 2.7 | 0.039 | 1224 | 9.4E-05 | 2 | NoObs |
| 800.02 | 3.9728 | 924 | 26 | 105.8128 | 0.0044 | 7.21227 | 0.00022 | 14 | 99 | 0.027 | 0.034 | 0.2 | 3.1 | 2.5 | 0.075 | 882 | 9.5E-05 | 2 | NoObs |
| 801.01 | 2.4077 | 8385 | 533 | 103.82617 | 0.00016 | 1.6255204 | 0.0000017 | 5.639292 | 0.000006 | 0.081 | 0.013 | 0.001 | - | 9.6 | 0.027 | 1525 | 6.4E-05 | 2 | NoObs |
| 802.01 | 2.2149 | 22158 | 264 | 114.88094 | 0.00028 | 19.620402 | 0.00004 | 83 | 15 | 0.1345 | 0.0049 | 0.26 | 0.49 | 7.3 | 0.139 | 464 | 4.6E-05 | 2 | |
| 804.01 | 3.1943 | 987 | 29 | 110.2 | 0.0033 | 9.02971 | 0.00022 | 16 | 56 | 0.031 | 0.021 | 0.7 | 1.6 | 3.0 | 0.083 | 735 | 9.3E-05 | 2 | |
| 805.01 | 7.8222 | 18439 | 430 | 107.58168 | 0.00069 | 10.327948 | 0.000074 | 11.377346 | 0.000082 | 0.1195 | 0.0095 | 0.0002 | - | 14.4 | 0.094 | 820 | 2.2E-05 | 3 | |
| 806.01 | 8.9776 | 10175 | 57 | 87.2386 | 0.0035 | 143.1814 | 0.0027 | 133 | 47 | 0.0933 | 0.0073 | 0.29 | 0.68 | 9.0 | 0.53 | 296 | 3.0E-05 | 2 | NoObs |
| 806.02 | 6.6246 | 19862 | 195 | 176.89513 | 0.00088 | 60.32875 | 0.00037 | 78.27 | 0.44 | 0.12509 | 0.00053 | 0.03 | 0.017 | 12.1 | 0.298 | 395 | 2.4E-05 | 2 | NoObs |
| 806.03 | 4.6342 | 1115 | 14 | 83.6924 | 0.0086 | 29.1654 | 0.0012 | 47.3 | 7 | 0.032 | 0.08 | 0.391 | 0.03 | 3.1 | 0.183 | 504 | 4.4E-05 | 2 | NoObs |
| 809.01 | 1.9942 | 16511 | 602 | 103.64747 | 0.00018 | 1.5947453 | 0.0000027 | 7.145002 | 0.000012 | 0.114 | 0.065 | 0.0001 | - | 12.2 | 0.027 | 1512 | 2.2E-05 | 2 | |
| 810.01 | 2.3449 | 1024 | 41 | 103.51 | 0.0018 | 4.782942 | 0.000062 | 14 | 45 | 0.03 | 0.022 | 0.5 | 1.8 | 2.7 | 0.054 | 859 | 4.2E-05 | 2 | |
| 811.01 | 4.2008 | 2267 | 48 | 114.4316 | 0.0021 | 20.50617 | 0.00032 | 39.8 | 0.74 | 0.04186 | 0.0006 | 0.031 | 0.041 | 4.3 | 0.141 | 544 | 4.3E-05 | 2 | |
| 812.01 | 1.8467 | 1757 | 43 | 104.9773 | 0.0015 | 3.34024 | 0.000035 | 13 | 50 | 0.039 | 0.028 | 0.5 | 2.1 | 2.5 | 0.036 | 720 | 4.3E-05 | 2 | |
| 812.02 | 3.2972 | 1491 | 21 | 80.4576 | 0.0054 | 20.06086 | 0.00047 | 50.3 | 2.2 | 0.0352 | 0.0011 | 0.031 | 0.064 | 2.2 | 0.118 | 398 | - | 4 | |
| 812.03 | 4.7450 | 1471 | 14 | 98.2376 | 0.01 | 46.1851 | 0.0019 | 75.7 | 4.2 | 0.0341 | 0.0017 | 0.003 | 0.017 | 2.1 | 0.206 | 301 | - | 4 | |
| 813.01 | 2.2740 | 8958 | 258 | 103.52731 | 0.0003 | 3.8959257 | 0.0000082 | 14.23 | 0.067 | 0.08461 | 0.00029 | 0.0009 | - | 6.5 | 0.048 | 902 | 8.1E-05 | 2 | |
| 814.01 | 5.1665 | 941 | 19 | 108.4529 | 0.005 | 22.367 | 0.00045 | 33 | 173 | 0.028 | 0.031 | 0.4 | 2.5 | 1.8 | 0.15 | 456 | 1.4E-04 | 2 | |
| 815.01 | 2.9262 | 4926 | 62 | 105.6313 | 0.0017 | 34.8442 | 0.00043 | 47 | 14 | 0.101 | 0.013 | 0.94 | 0.28 | 10.4 | 0.209 | 502 | 2.6E-05 | 2 | |
| 816.01 | 3.4728 | 2412 | 70 | 107.9981 | 0.0018 | 7.748017 | 0.000094 | 17.74 | 0.23 | 0.04406 | 0.00045 | 0.02 | 0.022 | 4.6 | 0.078 | 882 | 1.1E-04 | 2 | |
| 817.01 | 3.1504 | 1242 | 12 | 119.2141 | 0.0058 | 23.9716 | 0.0011 | 64.4 | 4.6 | 0.033 | 0.002 | 0.03 | 0.028 | 2.1 | 0.129 | 370 | 2.7E-05 | 3 | NoObs |
| 818.01 | 2.4193 | 1618 | 33 | 109.3409 | 0.0023 | 8.11429 | 0.00013 | 21 | 59 | 0.04 | 0.021 | 0.7 | 1.4 | 3.6 | 0.066 | 591 | 5.0E-05 | 2 | NoObs |
| 821.01 | 4.3866 | 1235 | 21 | 107.0084 | 0.0072 | 21.8131 | 0.0015 | 34 | 171 | 0.034 | 0.033 | 0.6 | 2.2 | 3.9 | 0.154 | 623 | 1.2E-04 | 2 | |
| 822.01 | 3.0811 | 16197 | 257 | 105.18019 | 0.00031 | 7.919371 | 0.000017 | 15.49 | 0.73 | 0.1279 | 0.0015 | 0.73 | 0.18 | 11.5 | 0.078 | 783 | - | 3 | |
| 823.01 | 1.4379 | 5445 | 58 | 103.22827 | 0.00086 | 1.0284369 | 0.0000061 | 4.2 | 1.3 | 0.0753 | 0.0015 | 0.6 | 0.18 | 8.7 | 0.021 | 1874 | - | 3 | |
| 824.01 | 3.7671 | 19284 | 80 | 106.6086 | 0.0014 | 15.3755 | 0.00048 | 39.69 | 0.58 | 0.1221 | 0.0013 | 0.028 | 0.028 | 12.5 | 0.117 | 604 | 1.7E-05 | 2 | |
| 825.01 | 3.2182 | 846 | 21 | 109.9535 | 0.004 | 8.10341 | 0.00023 | 16 | 66 | 0.028 | 0.026 | 0.6 | 1.9 | 2.4 | 0.074 | 671 | 3.2E-05 | 2 | |
| 826.01 | 2.9092 | 765 | 40 | 104.134 | 0.0022 | 6.365997 | 0.000094 | 17.35 | 0.39 | 0.02535 | 0.00045 | 0.017 | - | 1.7 | 0.066 | 749 | 5.3E-05 | 2 | |
| 827.01 | 2.5968 | 860 | 30 | 107.7764 | 0.0028 | 5.97569 | 0.00012 | 12 | 40 | 0.03 | 0.017 | 0.8 | 1.4 | 3.1 | 0.066 | 960 | 1.3E-04 | 2 | |
| 829.01 | 4.2648 | 893 | 27 | 107.7772 | 0.0041 | 18.64902 | 0.00053 | 34.8801 | 0.0027 | 0.03 | 0.34 | 0.015 | 0.01 | 2.6 | 0.14 | 651 | 5.4E-05 | 2 | |
| 829.02 | 4.1408 | 403 | 17 | 71.7806 | 0.0073 | 9.75222 | 0.00032 | 14 | 76 | 0.02 | 0.022 | 0.7 | 2.1 | 1.9 | 0.091 | 807 | - | 4 | |
| 829.03 | 4.5589 | 976 | 22 | 96.8416 | 0.0056 | 38.5596 | 0.001 | 67.1 | 2.6 | 0.02809 | 0.00086 | 0.008 | 0.033 | 2.7 | 0.228 | 510 | - | 4 | |
| 830.01 | 2.6064 | 23232 | 1241 | 103.04717 | 0.00007 | 3.5256346 | 0.0000018 | 11.745 | 0.012 | 0.13434 | 0.0001 | 0.0003 | - | 7.7 | 0.042 | 767 | 2.1E-05 | 2 | 1 |
| 833.01 | 1.5895 | 2413 | 80 | 106.2753 | 0.00075 | 3.951399 | 0.000021 | 13.5 | 4 | 0.04987 | 0.00071 | 0.6 | 0.18 | 4.3 | 0.05 | 1012 | 1.0E-04 | 2 | |
| 834.01 | 8.1433 | 3516 | 156 | 104.3739 | 0.0012 | 23.65422 | 0.00021 | 23.3 | 0.1 | 0.05283 | 0.00022 | 0.02 | 0.014 | 4.9 | 0.163 | 564 | 3.7E-05 | 2 | |
| 834.02 | 6.4410 | 448 | 23 | 73.3322 | 0.0074 | 13.23311 | 0.0004 | 14 | 66 | 0.02 | 0.02 | 0.5 | 2.2 | 1.9 | 0.111 | 683 | - | 4 | |
| 834.03 | 4.6117 | 276 | 18 | 67.8245 | 0.0074 | 6.15542 | 0.0002 | 10.44 | 0.18 | 0.01505 | 0.00052 | 0.03 | 0.01 | 1.4 | 0.066 | 886 | - | 4 | |
| 834.04 | 3.3378 | 141 | 14 | 67.1598 | 0.0078 | 2.090925 | 0.000069 | 5.63 | 0.2 | 0.01504 | 0.00044 | 0.083 | - | 1.4 | 0.032 | 1273 | - | 4 | |
| 835.01 | 2.9410 | 997 | 32 | 113.9362 | 0.0028 | 11.76296 | 0.00024 | 24 | 77 | 0.031 | 0.021 | 0.7 | 1.6 | 1.6 | 0.092 | 485 | 1.3E-04 | 2 | |
| 837.01 | 2.2850 | 884 | 21 | 107.6615 | 0.0035 | 7.95367 | 0.00019 | 26.7 | 1.3 | 0.02694 | 0.00099 | 0.0149 | - | 1.8 | 0.072 | 625 | 5.8E-05 | 2 | |
| 837.02 | 2.3116 | 367 | 13 | 68.9218 | 0.006 | 4.14459 | 0.00011 | 15.96 | 0.96 | 0.0213 | 0.00091 | 0.0707 | - | 1.4 | 0.047 | 774 | - | 4 | |
| 838.01 | 2.0055 | 4715 | 92 | 106.0116 | 0.00083 | 4.859373 | 0.000027 | 12.5 | 3.7 | 0.07226 | 0.00097 | 0.7 | 0.21 | 7.8 | 0.057 | 1066 | - | 3 | |
| 840.01 | 1.9910 | 10350 | 454 | 102.9486 | 0.00017 | 3.0403244 | 0.0000034 | 12 | 1.7 | 0.0957 | 0.0031 | 0.62 | 0.35 | 10.7 | 0.04 | 1098 | 6.8E-05 | 2 | NoObs |
| 841.01 | 3.4700 | 3128 | 35 | 107.0016 | 0.003 | 15.33611 | 0.00033 | 34.98 | 0.96 | 0.0485 | 0.0011 | 0.015 | 0.01 | 4.0 | 0.118 | 583 | 5.5E-05 | 2 | NoObs |
| 841.02 | 4.7489 | 4395 | 41 | 86.4334 | 0.0028 | 31.32865 | 0.00044 | 54.1 | 1.1 | 0.05959 | 0.00099 | 0 | 0.027 | 4.9 | 0.191 | 458 | 4.9E-05 | 2 | NoObs |
| 842.01 | 3.0857 | 1227 | 32 | 108.3511 | 0.0028 | 12.71862 | 0.00026 | 28 | 87 | 0.032 | 0.022 | 0.5 | 1.8 | 2.8 | 0.097 | 565 | 6.2E-05 | 2 | |
| 842.02 | 4.2580 | 1631 | 31 | 131.5836 | 0.0034 | 36.06539 | 0.00066 | 69.7 | 1.9 | 0.03655 | 0.00081 | 0.0255 | - | 3.1 | 0.195 | 399 | - | 4 | |
| 843.01 | 2.8728 | 3081 | 149 | 104.4417 | 0.00059 | 4.1904 | 0.000017 | 9.1 | 4.6 | 0.0531 | 0.0049 | 0.65 | 0.65 | 6.3 | 0.052 | 1169 | - | 2 | |
| 844.01 | 3.0333 | 2096 | 68 | 104.987 | 0.0012 | 3.709881 | 0.000031 | 9.73 | 0.14 | 0.04117 | 0.00047 | 0.0006 | - | 2.8 | 0.046 | 878 | 2.7E-05 | 2 | 3 |
| 845.01 | 5.7133 | 1010 | 35 | 110.2954 | 0.0042 | 16.3294 | 0.00046 | 16 | 26 | 0.0323 | 0.0095 | 0.7 | 1.1 | 3.6 | 0.129 | 701 | 5.3E-05 | 2 | |
| 846.01 | 4.1032 | 26361 | 550 | 119.71309 | 0.00025 | 27.807488 | 0.000048 | 44.82 | 0.76 | 0.16567 | 0.0008 | 0.769 | 0.098 | 15.3 | 0.182 | 534 | 5.5E-05 | 3 | |
| 847.01 | 11.0533 | 3594 | 91 | 136.8967 | 0.003 | 80.8711 | 0.0024 | 59.0063 | 0.0017 | 0.054 | 0.033 | 0.0064 | - | 5.1 | 0.368 | 372 | 2.9E-05 | 2 | |
| 849.01 | 3.3150 | 727 | 28 | 103.9345 | 0.0034 | 10.35545 | 0.00024 | 18 | 62 | 0.027 | 0.018 | 0.7 | 1.5 | 2.8 | 0.093 | 750 | 5.2E-05 | 2 | |
| 850.01 | 2.7054 | 10320 | 317 | 109.52167 | 0.00029 | 10.526305 | 0.000022 | 32.5 | 6.8 | 0.0924 | 0.0042 | 0.45 | 0.49 | 8.7 | 0.093 | 704 | 3.3E-05 | 2 | |
| 851.01 | 2.7220 | 3982 | 58 | 102.9729 | 0.0013 | 4.583526 | 0.000042 | 13.63 | 0.25 | 0.05609 | 0.00073 | 0.028 | 0.035 | 5.5 | 0.055 | 989 | 1.4E-04 | 2 | |
| 852.01 | 2.9127 | 548 | 20 | 104.9044 | 0.0039 | 3.76179 | 0.0001 | 10.2 | 0.44 | 0.0221 | 0.00075 | 0.0256 | - | 2.4 | 0.048 | 1086 | 3.4E-05 | 2 | |
| 853.01 | 2.7276 | 1044 | 29 | 102.6947 | 0.0025 | 8.20371 | 0.00014 | 21 | 79 | 0.029 | 0.023 | 0.5 | 2 | 2.9 | 0.077 | 733 | 3.3E-05 | 2 | |
| 853.02 | 3.2261 | 489 | 11 | 76.4118 | 0.0093 | 14.49679 | 0.00059 | 30 | 377 | 0.021 | 0.054 | 0.5 | 3.6 | 2.1 | 0.112 | 607 | - | 4 | |
| 854.01 | 4.1155 | 1432 | 14 | 134.1698 | 0.0062 | 56.0517 | 0.0027 | 105.3 | 5.3 | 0.0357 | 0.0016 | 0.002 | 0.036 | 1.9 | 0.217 | 248 | 5.1E-05 | 3 | NoObs |
| 855.01 | 5.2392 | 24330 | 571 | 128.78669 | 0.00024 | 41.40846 | 0.00011 | 72.69 | 0.13 | 0.13793 | 0.00019 | 0.018 | 0.01 | 12.5 | 0.232 | 444 | 2.4E-05 | 2 | |
| 856.01 | 5.7434 | 14173 | 335 | 105.85507 | 0.00044 | 39.74897 | 0.00013 | 30 | 1.5 | 0.1396 | 0.0025 | 0.89 | 0.11 | 13.1 | 0.232 | 498 | 3.2E-05 | 3 | |
| 857.01 | 2.5010 | 859 | 34 | 107.8799 | 0.0021 | 5.715374 | 0.000085 | 17 | 81 | 0.027 | 0.027 | 0.3 | 2.4 | 2.2 | 0.061 | 786 | 3.0E-05 | 2 | |
| 858.01 | 2.4580 | 5738 | 97 | 106.98673 | 0.00068 | 13.610127 | 0.000068 | 24.2 | 7.3 | 0.0912 | 0.0027 | 0.86 | 0.26 | 9.9 | 0.113 | 713 | - | 3 | |
| 861.01 | 1.8610 | 341 | 22 | 103.8132 | 0.0033 | 2.237565 | 0.000049 | 9.37 | 0.41 | 0.01774 | 0.00061 | 0.0079 | - | 1.5 | 0.032 | 1090 | 6.5E-05 | 3 | NoObs |
| 863.01 | 2.1284 | 816 | 32 | 105.1473 | 0.0024 | 3.16792 | 0.000051 | 9 | 29 | 0.029 | 0.018 | 0.7 | 1.6 | 2.7 | 0.043 | 1109 | 7.0E-05 | 2 | |
| 864.01 | 2.7477 | 1118 | 45 | 106.5739 | 0.0024 | 4.311802 | 0.000067 | 12.46 | 0.27 | 0.03025 | 0.00051 | 0.026 | 0.036 | 2.2 | 0.051 | 846 | 3.6E-05 | 2 | |
| 864.02 | 4.2836 | 765 | 18 | 121.0857 | 0.0067 | 20.05052 | 0.00067 | 36.1 | 1.6 | 0.0249 | 0.0009 | 0.023 | 0.02 | 1.8 | 0.141 | 509 | - | 4 | |
| 864.03 | 1.4038 | 696 | 14 | 119.5789 | 0.0038 | 9.76742 | 0.00019 | 56 | 5.3 | 0.0253 | 0.0016 | 0.072 | - | 1.8 | 0.087 | 648 | - | 4 | |
| 865.01 | 8.1358 | 7359 | 142 | 155.237 | 0.0014 | 119.0213 | 0.002 | 120.686 | 0.085 | 0.074 | 0.036 | 0.004 | 0.032 | 6.0 | 0.473 | 306 | 4.6E-05 | 2 | |
| 867.01 | 3.7401 | 1584 | 45 | 113.2774 | 0.0021 | 16.08561 | 0.00025 | 34.88 | 0.71 | 0.03503 | 0.00057 | 0.023 | 0.03 | 3.4 | 0.122 | 600 | 5.6E-05 | 2 | |
| †868.01 | 7.6456 | 23042 | 160 | 141.4312 | 0.0034 | 234 | 14 | 171 | 13 | 0.1606 | 0.0025 | 0.84 | 0.14 | 12.5 | 0.629 | 193 | 2.8E-05 | 2 | |
| 869.01 | 2.7105 | 987 | 19 | 107.9535 | 0.0045 | 7.49 | 0.00024 | 18 | 100 | 0.03 | 0.035 | 0.6 | 2.2 | 3.2 | 0.074 | 809 | 6.7E-05 | 2 | NoObs |
| 869.02 | 4.2450 | 1570 | 18 | 134.2345 | 0.0067 | 36.2911 | 0.0019 | 37 | 45 | 0.0413 | 0.0097 | 0.83 | 0.74 | 4.3 | 0.212 | 478 | 5.6E-05 | 2 | NoObs |
| 870.01 | 2.7731 | 900 | 23 | 105.1871 | 0.004 | 5.91213 | 0.00017 | 11.9 | 3.6 | 0.0298 | 0.0013 | 0.53 | 0.16 | 3.6 | 0.062 | 859 | 5.8E-05 | 2 | NoObs |
| 870.02 | 4.3907 | 1071 | 29 | 108.6825 | 0.0043 | 8.98597 | 0.00027 | 21.71 | 0.84 | 0.02837 | 0.00089 | 0.003 | 0.01 | 3.5 | 0.081 | 751 | 5.8E-05 | 2 | NoObs |
| 871.01 | 2.1775 | 40083 | 206 | 112.4222 | 0.00037 | 12.940664 | 0.000033 | 38.6 | 2.5 | 0.2084 | 0.0044 | 0.77 | 0.19 | 10.9 | 0.105 | 531 | 1.8E-05 | 2 | |
| 872.01 | 4.3863 | 6657 | 128 | 119.68414 | 0.00099 | 33.60167 | 0.00025 | 50.2 | 7.4 | 0.0842 | 0.0027 | 0.69 | 0.33 | 7.4 | 0.199 | 456 | 2.0E-05 | 2 | |
| 873.01 | 2.2081 | 363 | 19 | 105.2243 | 0.0038 | 4.34761 | 0.00011 | 15.47 | 0.73 | 0.01882 | 0.00068 | 0.0167 | - | 1.3 | 0.051 | 866 | 7.6E-05 | 2 | |
| 874.01 | 2.1926 | 718 | 28 | 102.9798 | 0.0027 | 4.601803 | 0.000083 | 16.56 | 0.63 | 0.02429 | 0.00071 | 0.0081 | - | 2.2 | 0.053 | 882 | 9.0E-05 | 2 | |
| 875.01 | 2.9703 | 2374 | 99 | 103.6225 | 0.00094 | 4.220936 | 0.000027 | 11.42 | 0.11 | 0.04377 | 0.00034 | 0.014 | 0.02 | 2.2 | 0.042 | 607 | 8.0E-05 | 2 | |
| 876.01 | 1.8643 | 17802 | 89 | 104.89891 | 0.00077 | 6.998077 | 0.000038 | 21.8 | 3.6 | 0.1436 | 0.0066 | 0.81 | 0.28 | 9.2 | 0.07 | 693 | 1.9E-05 | 2 | |
| 877.01 | 2.2818 | 1323 | 32 | 103.9571 | 0.0025 | 5.95487 | 0.0001 | 19 | 78 | 0.034 | 0.026 | 0.4 | 2.2 | 2.5 | 0.055 | 652 | 5.4E-05 | 2 | |
| 877.02 | 2.7065 | 1184 | 21 | 114.2249 | 0.0037 | 12.03957 | 0.00034 | 35.1 | 1.6 | 0.0314 | 0.0011 | 0.02 | 0.042 | 2.3 | 0.088 | 516 | 5.7E-05 | 2 | |
| 878.01 | 4.4967 | 1362 | 33 | 106.8197 | 0.0039 | 23.58879 | 0.00063 | 19.5 | 6.9 | 0.0414 | 0.0031 | 0.9 | 0.31 | 5.2 | 0.158 | 568 | 8.1E-05 | 2 | |
| 880.01 | 4.0223 | 1774 | 40 | 127.1322 | 0.0027 | 26.44302 | 0.00052 | 25.1 | 6.9 | 0.0466 | 0.0024 | 0.89 | 0.28 | 4.9 | 0.176 | 569 | 5.4E-05 | 2 | 1 |
| 880.02 | 6.5130 | 3757 | 93 | 107.1236 | 0.0017 | 51.52991 | 0.00041 | 64.08 | 0.53 | 0.05485 | 0.00041 | 0.018 | 0.017 | 5.8 | 0.274 | 456 | 4.6E-05 | 2 | 1 |
| 880.03 | 2.8108 | 692 | 32 | 69.7841 | 0.0029 | 5.902243 | 0.000072 | 12 | 41 | 0.026 | 0.016 | 0.7 | 1.6 | 2.8 | 0.065 | 936 | - | 4 | |
| 880.04 | 2.2269 | 248 | 16 | 68.0797 | 0.0048 | 2.382948 | 0.000049 | 5 | 25 | 0.019 | 0.016 | 0.8 | 1.5 | 2.0 | 0.035 | 1275 | - | 4 | |
| 881.01 | 3.9739 | 1837 | 35 | 140.6795 | 0.0029 | 21.02147 | 0.00051 | 41.2 | 1 | 0.0387 | 0.00076 | 0.0431 | - | 2.5 | 0.142 | 459 | 8.5E-05 | 2 | NoObs |
| 881.02 | 7.2304 | 3196 | 29 | 140.3562 | 0.0051 | 226.8916 | 0.0062 | 139 | 42 | 0.0598 | 0.0039 | 0.84 | 0.35 | 3.9 | 0.693 | 208 | 7.2E-05 | 2 | NoObs |
| 882.01 | 1.9042 | 24911 | 121 | 103.69391 | 0.00049 | 1.9568099 | 0.0000067 | 7.8 | 2.3 | 0.1511 | 0.0015 | 0.36 | 0.11 | 13.6 | 0.03 | 1176 | 3.6E-05 | 3 | |
| 883.01 | 2.1328 | 36019 | 333 | 103.10127 | 0.00024 | 2.6888995 | 0.0000044 | 11.296 | 0.052 | 0.16675 | 0.00054 | 0 | 0.014 | 10.0 | 0.034 | 829 | 3.5E-05 | 2 | |
| 884.01 | 2.8824 | 3046 | 86 | 110.1842 | 0.0011 | 9.439536 | 0.000073 | 24 | 37 | 0.05 | 0.017 | 0.4 | 1.3 | 3.0 | 0.082 | 564 | 2.5E-05 | 2 | 1 |
| 884.02 | 3.5113 | 2038 | 42 | 111.6901 | 0.0021 | 20.47687 | 0.00024 | 53.6 | 1.2 | 0.04552 | 0.0008 | 0.019 | 0.03 | 2.7 | 0.137 | 436 | 3.0E-05 | 2 | 1 |
| 884.03 | 2.2744 | 442 | 19 | 68.3065 | 0.0042 | 3.336241 | 0.00006 | 11.74 | 0.55 | 0.0207 | 0.00072 | 0.0542 | - | 1.2 | 0.041 | 797 | - | 4 | |
| 886.01 | 2.1724 | 1325 | 25 | 103.1784 | 0.0029 | 8.01281 | 0.00016 | 28 | 256 | 0.035 | 0.04 | 0.4 | 3.4 | 2.0 | 0.059 | 488 | 9.2E-05 | 2 | NoObs |
| 887.01 | 2.5948 | 622 | 30 | 108.3458 | 0.0029 | 7.41121 | 0.00015 | 22.71 | 0.76 | 0.02269 | 0.00058 | 0.0059 | - | 2.3 | 0.075 | 867 | 9.1E-05 | 2 | |
| 889.01 | 2.6190 | 17546 | 311 | 102.99117 | 0.00034 | 8.884903 | 0.00002 | 36.26 | 0.18 | 0.11449 | 0.00035 | 0.03 | 0.017 | 11.7 | 0.083 | 754 | 2.9E-05 | 2 | |
| 890.01 | 4.2560 | 7353 | 219 | 109.62331 | 0.00052 | 8.09888 | 0.00003 | 15.649 | 0.064 | 0.07701 | 0.00026 | 0 | 0.01 | 7.6 | 0.081 | 878 | 1.3E-05 | 2 | |
| 891.01 | 4.9086 | 911 | 38 | 109.9676 | 0.0033 | 10.00634 | 0.00024 | 14 | 40 | 0.028 | 0.015 | 0.6 | 1.7 | 2.7 | 0.093 | 785 | 2.1E-05 | 2 | |
| 892.01 | 2.8869 | 1221 | 31 | 105.6161 | 0.003 | 10.37176 | 0.00021 | 24 | 95 | 0.033 | 0.027 | 0.5 | 2 | 2.8 | 0.09 | 654 | 3.8E-05 | 2 | |
| 893.01 | 4.3828 | 627 | 25 | 105.1843 | 0.0064 | 4.40845 | 0.00028 | 7.85 | 0.23 | 0.0233 | 0.00059 | 0.0248 | - | 2.6 | 0.054 | 1091 | 7.9E-05 | 3 | |
| 895.01 | 3.9799 | 14694 | 73 | 104.8936 | 0.0017 | 4.409394 | 0.000052 | 9.459 | 0.054 | 0.1066 | 0.0011 | 0.01 | - | 12.8 | 0.053 | 1093 | 4.5E-05 | 2 | |
| 896.01 | 4.0317 | 2608 | 64 | 108.5683 | 0.0018 | 16.2398 | 0.00021 | 32.61 | 0.45 | 0.04535 | 0.00052 | 0.015 | 0.022 | 3.9 | 0.123 | 578 | 5.1E-05 | 2 | |
| 896.02 | 3.0443 | 1319 | 47 | 107.0462 | 0.0022 | 6.308146 | 0.000095 | 16.7 | 0.36 | 0.0331 | 0.00056 | 0.0237 | - | 2.8 | 0.066 | 789 | 5.8E-05 | 2 | |
| 897.01 | 2.1494 | 14879 | 663 | 102.88985 | 0.00014 | 2.0523497 | 0.0000019 | 8.527 | 0.019 | 0.10906 | 0.00015 | 0.03 | 0.01 | 12.0 | 0.032 | 1417 | 5.4E-05 | 2 | NoObs |
| 898.01 | 2.5752 | 1950 | 30 | 108.7101 | 0.0026 | 9.77059 | 0.00018 | 31 | 144 | 0.041 | 0.037 | 0.3 | 2.4 | 3.0 | 0.074 | 540 | 1.9E-05 | 2 | NoObs |
| 898.02 | 2.0540 | 1210 | 23 | 105.6293 | 0.0031 | 5.16991 | 0.00011 | 18 | 96 | 0.033 | 0.032 | 0.4 | 2.5 | 2.4 | 0.049 | 664 | 2.3E-05 | 2 | NoObs |
| 898.03 | 3.3063 | 1362 | 17 | 80.9862 | 0.0062 | 20.08923 | 0.00054 | 44 | 278 | 0.034 | 0.041 | 0.4 | 2.7 | 2.5 | 0.12 | 424 | - | 4 | |
| 899.01 | 2.0930 | 858 | 29 | 107.3219 | 0.0023 | 7.11388 | 0.00011 | 24 | 128 | 0.028 | 0.019 | 0.5 | 2.4 | 1.7 | 0.055 | 510 | 8.7E-05 | 2 | NoObs |
| 899.02 | 1.8687 | 518 | 25 | 67.3709 | 0.0028 | 3.306569 | 0.000039 | 13.34 | 0.53 | 0.02123 | 0.00066 | 0.0214 | - | 1.3 | 0.033 | 658 | - | 4 | |
| 899.03 | 2.4349 | 788 | 21 | 80.3944 | 0.0037 | 15.36813 | 0.00026 | 35 | 221 | 0.029 | 0.036 | 0.7 | 2 | 1.7 | 0.091 | 397 | - | 4 | |
| 900.01 | 3.0479 | 1391 | 24 | 105.3395 | 0.0038 | 13.81001 | 0.00056 | 39.4 | 1.7 | 0.0334 | 0.0011 | 0.056 | 0.01 | 4.3 | 0.116 | 798 | 1.9E-05 | 2 | |
| 901.01 | 1.9512 | 6817 | 123 | 109.93915 | 0.00052 | 12.732557 | 0.000046 | 53.09 | 0.54 | 0.07377 | 0.00054 | 0.009 | 0.02 | 4.4 | 0.089 | 463 | 1.6E-05 | 2 | |
| 902.01 | 6.9808 | 8447 | 90 | 169.8066 | 0.0014 | 83.9042 | 0.0017 | 107.2 | 1.1 | 0.07978 | 0.00066 | 0.005 | 0.01 | 5.7 | 0.324 | 270 | 2.1E-05 | 2 | |
| 903.01 | 4.2191 | 7315 | 262 | 106.43312 | 0.00049 | 5.007341 | 0.000017 | 9.815 | 0.034 | 0.07665 | 0.00021 | - | 0.01 | 5.6 | 0.057 | 853 | 6.3E-05 | 2 | |
| 904.01 | 1.8541 | 611 | 20 | 103.1507 | 0.0031 | 2.211073 | 0.000047 | 7 | 33 | 0.027 | 0.028 | 0.7 | 1.8 | 2.1 | 0.029 | 960 | 2.9E-05 | 2 | NoObs |
| 904.02 | 3.2550 | 1667 | 20 | 111.8263 | 0.0045 | 27.93886 | 0.00095 | 58 | 299 | 0.038 | 0.042 | 0.6 | 2.1 | 3.0 | 0.159 | 410 | 2.2E-05 | 2 | NoObs |
| 905.01 | 2.4135 | 1773 | 51 | 105.6967 | 0.0016 | 5.795111 | 0.000065 | 18.83 | 0.39 | 0.03715 | 0.00053 | 0.029 | 0.037 | 2.0 | 0.062 | 698 | 7.8E-05 | 3 | NoObs |
| 906.01 | 2.5473 | 871 | 24 | 107.1339 | 0.0033 | 7.15684 | 0.00016 | 18 | 83 | 0.031 | 0.029 | 0.7 | 1.7 | 2.8 | 0.071 | 759 | 4.1E-05 | 2 | |
| 907.01 | 4.1177 | 961 | 26 | 109.117 | 0.0039 | 16.51385 | 0.0007 | 32.2 | 1 | 0.02811 | 0.00071 | 0.03 | 0.052 | 3.5 | 0.13 | 738 | 8.5E-05 | 2 | NoObs |
| 907.02 | 5.0796 | 875 | 21 | 123.386 | 0.0058 | 30.1324 | 0.002 | 48.4 | 1.7 | 0.02725 | 0.00087 | 0.025 | 0.01 | 3.4 | 0.194 | 604 | 8.9E-05 | 2 | NoObs |
| 907.03 | 2.9275 | 178 | 8.2 | 69.3138 | 0.0097 | 4.79085 | 0.00021 | 8 | 60 | 0.017 | 0.021 | 0.8 | 2 | 2.1 | 0.057 | 1115 | - | 4 | |
| 908.01 | 3.3536 | 8282 | 398 | 104.44572 | 0.00028 | 4.7083263 | 0.0000091 | 12.227 | 0.034 | 0.08075 | 0.00017 | 0.0001 | - | 11.4 | 0.056 | 1140 | 6.7E-05 | 2 | |
| 910.01 | 3.0648 | 1158 | 47 | 104.724 | 0.002 | 5.392096 | 0.000073 | 14.21 | 0.28 | 0.03107 | 0.00049 | 0.0403 | - | 1.9 | 0.057 | 696 | 2.5E-05 | 2 | |
| 911.01 | 2.2475 | 650 | 22 | 104.007 | 0.0033 | 4.093609 | 0.000094 | 14.8 | 0.73 | 0.02338 | 0.00085 | 0.0196 | - | 1.7 | 0.05 | 945 | 4.5E-05 | 2 | |
| 912.01 | 2.8941 | 1632 | 24 | 104.804 | 0.0036 | 10.84847 | 0.00027 | 28 | 121 | 0.037 | 0.031 | 0.4 | 2.3 | 2.6 | 0.081 | 521 | 7.7E-05 | 2 | |
| 913.01 | 3.2508 | 18588 | 1072 | 102.63655 | 0.0001 | 4.0822762 | 0.0000027 | 10.784082 | 0.000007 | 0.1216 | 0.0049 | 0.0002 | - | 9.1 | 0.049 | 902 | - | 3 | |
| 914.01 | 3.2170 | 500 | 23 | 102.7363 | 0.0042 | 3.88667 | 0.00011 | 9.85 | 0.35 | 0.02162 | 0.00061 | 0.0097 | - | 1.2 | 0.047 | 805 | 4.8E-05 | 2 | |
| 916.01 | 1.7644 | 1387 | 67 | 104.31311 | 0.00091 | 3.314908 | 0.000021 | 12 | 23 | 0.037 | 0.013 | 0.7 | 1.2 | 3.9 | 0.044 | 1112 | 4.1E-05 | 2 | |
| 917.01 | 2.1878 | 828 | 23 | 106.3562 | 0.0032 | 6.71972 | 0.00015 | 19 | 69 | 0.03 | 0.02 | 0.7 | 1.6 | 3.2 | 0.071 | 932 | 1.3E-04 | 3 | |
| 918.01 | 6.5105 | 15499 | 228 | 139.58885 | 0.00095 | 39.64552 | 0.00038 | 51.72 | 0.2 | 0.11111 | 0.00035 | 0.0034 | - | 10.6 | 0.226 | 463 | 1.2E-05 | 2 | |
| 920.01 | 2.8954 | 1096 | 23 | 123.4895 | 0.0038 | 21.80587 | 0.0006 | 53.5 | 2.1 | 0.02909 | 0.0008 | 0.031 | 0.084 | 1.9 | 0.148 | 469 | 3.9E-05 | 2 | |
| 921.01 | 3.5805 | 1137 | 29 | 108.3417 | 0.0034 | 10.28175 | 0.00025 | 23.15 | 0.2 | 0.03071 | 0.00069 | 0.023 | 0.01 | 2.3 | 0.089 | 618 | 4.1E-05 | 2 | NoObs |
| 921.02 | 4.0886 | 1625 | 32 | 115.6245 | 0.0036 | 18.11903 | 0.00046 | 36.3 | 1 | 0.03581 | 0.00081 | 0.026 | 0.037 | 2.7 | 0.13 | 512 | 3.6E-05 | 2 | NoObs |
| 921.03 | 1.6334 | 344 | 10 | 66.2771 | 0.0064 | 3.78406 | 0.0001 | 19 | 1.7 | 0.0198 | 0.0015 | 0.0153 | - | 1.5 | 0.046 | 860 | - | 4 | |
| 922.01 | 2.7126 | 652 | 19 | 104.6377 | 0.0049 | 5.15456 | 0.00017 | 12 | 57 | 0.026 | 0.024 | 0.6 | 2 | 2.7 | 0.058 | 951 | 5.3E-05 | 2 | |
| 923.01 | 3.4256 | 1232 | 40 | 107.8993 | 0.0022 | 5.743325 | 0.000091 | 13.24 | 0.3 | 0.03148 | 0.00057 | 0.015 | 0.03 | 2.9 | 0.064 | 911 | 4.0E-05 | 2 | |
| 924.01 | 2.6522 | 1115 | 17 | 106.3084 | 0.0049 | 39.4766 | 0.0014 | 127 | 1723 | 0.03 | 0.073 | 0.3 | 4.1 | 3.1 | 0.233 | 526 | 3.3E-05 | 3 | |
| 926.01 | 3.3262 | 1561 | 75 | 103.964 | 0.0017 | 3.166392 | 0.000035 | 7.537 | 0.037 | 0.03513 | 0.00032 | 0.0107 | - | 3.4 | 0.043 | 1154 | 2.3E-05 | 2 | |
| 928.01 | 1.8191 | 474 | 27 | 103.8588 | 0.0022 | 2.494093 | 0.000027 | 12 | 35 | 0.023 | 0.014 | 0.1 | 2.1 | 2.3 | 0.036 | 1208 | 2.6E-05 | 3 | 3 |
| 929.01 | 4.1458 | 7699 | 252 | 107.63388 | 0.00064 | 6.49162 | 0.000042 | 12.851 | 0.047 | 0.07807 | 0.00023 | 0.0007 | - | 9.0 | 0.07 | 998 | 1.5E-05 | 2 | |
| 931.01 | 3.2164 | 16852 | 806 | 103.67816 | 0.00014 | 3.8556046 | 0.0000038 | 10.241915 | 0.000011 | 0.1162 | 0.0075 | 0.0059 | - | 8.6 | 0.048 | 948 | 4.0E-05 | 2 | |
| 934.01 | 2.9175 | 1662 | 48 | 106.0083 | 0.0018 | 5.826727 | 0.000073 | 15 | 52 | 0.037 | 0.024 | 0.3 | 2.1 | 3.2 | 0.064 | 889 | 2.6E-05 | 2 | |
| 934.02 | 3.3818 | 581 | 12 | 75.543 | 0.0087 | 12.41208 | 0.00047 | 25 | 247 | 0.023 | 0.047 | 0.5 | 3.2 | 2.0 | 0.106 | 691 | - | 4 | |
| 934.03 | 3.9208 | 821 | 15 | 80.1226 | 0.0085 | 18.74711 | 0.00068 | 29 | 220 | 0.028 | 0.042 | 0.6 | 2.6 | 2.4 | 0.14 | 601 | - | 4 | |
| 935.01 | 5.1637 | 1942 | 62 | 113.0119 | 0.002 | 20.85987 | 0.00029 | 25 | 19 | 0.0426 | 0.0054 | 0.64 | 0.8 | 3.6 | 0.152 | 632 | 4.1E-05 | 2 | |
| 935.02 | 6.4005 | 1764 | 43 | 74.1845 | 0.0038 | 42.6329 | 0.00069 | 52.68 | 0.37 | 0.03741 | 0.00055 | 0.016 | 0.02 | 3.2 | 0.246 | 496 | - | 4 | |
| 935.03 | 8.5789 | 1035 | 20 | 67.9393 | 0.0084 | 87.6464 | 0.0036 | 74 | 499 | 0.029 | 0.04 | 0.4 | 2.9 | 2.5 | 0.397 | 391 | - | 4 | |
| 936.01 | 2.5574 | 2215 | 62 | 111.4147 | 0.0012 | 9.467895 | 0.000083 | 31.19 | 0.49 | 0.04437 | 0.00046 | 0.031 | 0.039 | 3.5 | 0.07 | 519 | 1.7E-05 | 2 | NoObs |
| 936.02 | 1.1718 | 729 | 45 | 67.5395 | 0.0011 | 0.8930442 | 0.0000044 | 6 | 12 | 0.026 | 0.011 | 0.2 | 1.7 | 2.0 | 0.014 | 1161 | - | 4 | |
| 937.01 | 4.1031 | 982 | 22 | 109.5797 | 0.0041 | 20.83479 | 0.00058 | 39.5 | 1.4 | 0.02849 | 0.00085 | 0.018 | 0.037 | 2.3 | 0.146 | 530 | 5.1E-05 | 2 | |
| 938.01 | 3.2072 | 1014 | 29 | 104.7015 | 0.0033 | 9.94611 | 0.00022 | 18 | 61 | 0.032 | 0.021 | 0.7 | 1.6 | 2.9 | 0.09 | 719 | 5.2E-05 | 2 | |
| 938.02 | 1.7854 | 234 | 16 | 66.9239 | 0.0043 | 1.0456 | 0.00002 | 4.65 | 0.29 | 0.01405 | 0.00074 | 0.0461 | - | 1.3 | 0.02 | 1525 | - | 4 | |
| 939.01 | 2.6119 | 299 | 17 | 103.5282 | 0.004 | 3.388069 | 0.000089 | 9.74 | 0.14 | 0.01682 | 0.00056 | 0.018 | 0.01 | 1.6 | 0.045 | 1099 | 6.2E-05 | 2 | NoObs |
| 940.01 | 4.7016 | 2114 | 187 | 102.573 | 0.00073 | 6.104843 | 0.00003 | 10.38695 | 0.000051 | 0.041 | 0.024 | 0.0004 | - | 3.5 | 0.064 | 816 | 5.2E-05 | 2 | |

| | | | | | | | | | | | | | | | | | | | | |
|---|---|---|---|---|---|---|---|---|---|---|---|---|---|---|---|---|---|---|---|---|
| 941.01 | 3.2620 | 2355 | 61 | 107.7862 | 0.0016 | 6.581521 | 0.000076 | 16.67 | 0.27 | 0.04256 | 0.00051 | 0.03 | 0.035 | 5.4 | 0.069 | 904 | 5.4E-05 | 2 | NoObs | |
| 941.02 | 1.8622 | 636 | 21 | 103.6646 | 0.0034 | 2.382649 | 0.000056 | 8 | 45 | 0.027 | 0.034 | 0.7 | 2.1 | 3.4 | 0.035 | 1269 | 6.9E-05 | 2 | NoObs | |
| 941.03 | 3.3174 | 2621 | 38 | 122.0188 | 0.0028 | 24.66469 | 0.00051 | 45 | 50 | 0.052 | 0.012 | 0.8 | 0.75 | 6.6 | 0.165 | 585 | 4.8E-05 | 2 | NoObs | |
| 942.01 | 2.1136 | 1332 | 33 | 107.8585 | 0.0022 | 11.51507 | 0.00018 | 37 | 174 | 0.034 | 0.035 | 0.6 | 2.1 | 2.5 | 0.095 | 582 | 2.2E-05 | 2 | | |
| 943.01 | 2.2705 | 840 | 33 | 104.9918 | 0.0023 | 3.601425 | 0.000058 | 9 | 31 | 0.03 | 0.02 | 0.7 | 1.5 | 2.2 | 0.045 | 887 | 2.7E-05 | 2 | NoObs | |
| 944.01 | 2.2498 | 1941 | 89 | 103.24424 | 0.00084 | 3.108254 | 0.000018 | 11.23 | 0.15 | 0.03893 | 0.00038 | 0.023 | 0.027 | 3.9 | 0.041 | 1080 | 4.2E-05 | 2 | | |
| 945.01 | 5.8083 | 530 | 16 | 121.8574 | 0.0074 | 25.8529 | 0.0014 | 37.25 | 0.73 | 0.02154 | 0.00081 | 0.04 | 0.01 | 2.0 | 0.175 | 595 | 7.0E-05 | 3 | | |
| 945.02 | 7.1766 | 754 | 19 | 79.35 | 0.0085 | 40.7193 | 0.0016 | 35 | 96 | 0.028 | 0.015 | 0.7 | 1.4 | 2.6 | 0.238 | 510 | - | 4 | | |
| 947.01 | 3.6722 | 1682 | 31 | 122.9264 | 0.0025 | 28.59891 | 0.00057 | 49 | 15 | 0.03953 | 0.00093 | 0.4 | 0.12 | 2.7 | 0.146 | 353 | 2.6E-05 | 2 | NoObs | |
| 949.01 | 4.0447 | 1144 | 35 | 103.7626 | 0.003 | 12.53274 | 0.00026 | 24.61 | 0.57 | 0.03051 | 0.0006 | 0.0129 | - | 2.5 | 0.106 | 670 | 2.7E-05 | 2 | | |
| 951.01 | 3.7463 | 2332 | 67 | 104.545 | 0.0016 | 13.19712 | 0.00014 | 23 | 29 | 0.046 | 0.012 | 0.6 | 1 | 6.0 | 0.108 | 702 | 3.7E-05 | 2 | | |
| 952.01 | 2.1831 | 1642 | 40 | 104.4075 | 0.0019 | 5.901255 | 0.000076 | 21.08 | 0.53 | 0.03745 | 0.00067 | 0.028 | 0.04 | 2.3 | 0.05 | 575 | 6.4E-05 | 2 | NoObs | |
| 952.02 | 2.2980 | 1379 | 29 | 103.6308 | 0.0026 | 8.75246 | 0.00015 | 20 | 40 | 0.038 | 0.011 | 0.8 | 1 | 2.3 | 0.065 | 504 | 6.6E-05 | 2 | NoObs | |
| 952.03 | 3.1143 | 1897 | 30 | 88.2077 | 0.0029 | 22.78033 | 0.00031 | 57.6 | 1.7 | 0.04023 | 0.00093 | 0.015 | 0.033 | 2.4 | 0.124 | 365 | - | 4 | | |
| 952.04 | 1.6236 | 385 | 12 | 66.6118 | 0.0051 | 2.896029 | 0.000063 | 14.6 | 1.4 | 0.0188 | 0.0015 | 0.0155 | - | 1.1 | 0.031 | 730 | - | 4 | | |
| 953.01 | 2.7191 | 2760 | 98 | 103.4274 | 0.0011 | 3.584109 | 0.000025 | 10.46 | 0.11 | 0.04688 | 0.00038 | 0.0058 | - | 4.4 | 0.046 | 1045 | 2.5E-05 | 2 | | |
| 954.01 | 2.9184 | 798 | 29 | 107.5203 | 0.0029 | 8.11522 | 0.00016 | 20 | 110 | 0.027 | 0.028 | 0.4 | 2.6 | 2.3 | 0.08 | 792 | 2.8E-05 | 2 | NoObs | |
| 954.02 | 4.7151 | 924 | 19 | 107.2189 | 0.0057 | 36.9254 | 0.0014 | 52 | 300 | 0.028 | 0.03 | 0.5 | 2.4 | 2.5 | 0.219 | 479 | 2.6E-05 | 2 | NoObs | |
| 955.01 | 3.9934 | 554 | 28 | 108.7269 | 0.0035 | 7.03918 | 0.00018 | 13 | 84 | 0.021 | 0.024 | 0.3 | 2.9 | 2.3 | 0.074 | 975 | 2.9E-05 | 3 | | |
| 956.01 | 2.8217 | 2199 | 72 | 108.6457 | 0.0012 | 8.36077 | 0.000072 | 21 | 35 | 0.044 | 0.017 | 0.6 | 1.3 | 5.1 | 0.077 | 746 | 1.3E-05 | 2 | | |
| 959.01 | 2.5462 | 36851 | 957 | 108.07184 | 0.00012 | 12.713795 | 0.000017 | 47 | 1.2 | 0.17862 | 0.00065 | 0.29 | 0.18 | 5.3 | 0.063 | 296 | 3.2E-06 | 2 | | |
| 960.01 | 6.1820 | 39678 | 1062 | 110.15256 | 0.0002 | 15.801109 | 0.000035 | 20.85 | 0.22 | 0.18276 | 0.0005 | 0.46 | 0.11 | 13.9 | 0.12 | 555 | 1.7E-05 | 2 | | |
| 961.01 | 0.5391 | 1609 | 50 | 103.48288 | 0.00053 | 1.2137724 | 0.0000044 | 9.3 | 2.8 | 0.053 | 0.0033 | 0.82 | 0.25 | 3.9 | 0.019 | 1106 | - | 3 | | 1 |
| 961.02 | 0.5518 | 1282 | 71 | 66.86865 | 0.00041 | 0.4532882 | 0.0000009 | 2.6 | 0.78 | 0.19431 | 0.00081 | 1.29 | 0.39 | 14.4 | 0.01 | 1524 | - | 4 | | |
| 961.03 | 0.7135 | 982 | 31 | 66.7918 | 0.0011 | 1.8651126 | 0.0000091 | 6.7 | 2 | 0.14 | 0.23 | 1.2 | 0.36 | 10.7 | 0.025 | 964 | - | 4 | | |
| 972.01 | 4.5448 | 356 | 134 | 194.53889 | 0.0006 | 13.118925 | 0.000085 | 16 | 14 | 0.0192 | 0.0025 | 0.76 | 0.74 | 5.3 | 0.126 | 1540 | 3.7E-06 | 2 | | 1 |
| 974.01 | 10.3158 | 164 | 37 | 105.9806 | 0.006 | 53.5067 | 0.002 | 39.95 | 0.57 | 0.01176 | 0.00019 | 0.012 | 0.028 | 1.5 | 0.289 | 542 | 1.8E-05 | 2 | | |
| 975.01 | 3.4387 | 70 | 37 | 193.8369 | 0.0017 | 2.785755 | 0.000034 | 3.4 | 8.7 | 0.0089 | 0.0037 | 0.8 | 1 | 1.1 | 0.039 | 1404 | 1.3E-05 | 2 | | |
| 976.01 | 5.7336 | 23783 | 137 | 117.971 | 0.0011 | 52.56862 | 0.00047 | 62 | 19 | 0.1553 | 0.0015 | 0.55 | 0.17 | 27.5 | 0.337 | 757 | - | 3 | | 1 |
| 977.01 | 3.5914 | 1143 | 42 | 194.4372 | 0.0019 | 1.353659 | 0.000027 | 2.95 | 0.89 | 0.02656 | 0.00079 | 0.051 | 0.015 | 0.8 | 0.014 | 628 | - | 3 | | 1 |
| 978.01 | 13.5255 | 471 | 100 | 195.4346 | 0.0016 | 18.95486 | 0.00049 | 10 | 11 | 0.0199 | 0.0034 | 0 | 1.2 | 3.0 | 0.147 | 857 | - | 4 | | 1 |
| 981.01 | 3.0023 | 104 | 12 | 194.726 | 0.0052 | 3.99942 | 0.00022 | 2.9 | 0.87 | 0.0132 | 0.0012 | 0.96 | 0.29 | 5.9 | 0.055 | 1928 | - | 3 | | 1 |
| 984.01 | 2.8947 | 717 | 12 | 195.0462 | 0.0047 | 4.28899 | 0.00022 | 7.9 | 2.4 | 0.027 | 0.0021 | 0.58 | 0.17 | 4.4 | 0.054 | 1358 | - | 3 | | 1 |
| 986.01 | 3.1093 | 553 | 19 | 193.8655 | 0.0034 | 8.18758 | 0.0003 | 19 | 157 | 0.022 | 0.039 | 0.4 | 3.1 | 1.3 | 0.077 | 626 | 6.9E-05 | 2 | | |
| 987.01 | 1.0682 | 176 | 13 | 194.8413 | 0.0023 | 3.179301 | 0.000077 | 23 | 165 | 0.014 | 0.021 | 0.6 | 2.5 | 1.3 | 0.042 | 1041 | 6.7E-05 | 2 | | |
| 988.01 | 3.1800 | 807 | 9.7 | 201.0204 | 0.0038 | 10.38143 | 0.00044 | 11.8 | 3.5 | 0.0328 | 0.0021 | 0.84 | 0.25 | 6.3 | 0.097 | 947 | 5.6E-05 | 2 | | |
| 991.01 | 2.3459 | 298 | 24 | 71.2251 | 0.0035 | 12.06208 | 0.00018 | 29 | 179 | 0.018 | 0.021 | 0.8 | 1.8 | 3.2 | 0.108 | 981 | - | 2 | | |
| 992.01 | 3.9772 | 416 | 12 | 69.4505 | 0.0098 | 9.93167 | 0.00042 | 18.8 | 1.2 | 0.01897 | 0.00098 | 0.024 | 0.014 | 1.6 | 0.091 | 735 | - | 3 | | 1 |
| 993.01 | 3.3032 | 406 | 14 | 77.3278 | 0.0074 | 21.85242 | 0.0007 | 44 | 418 | 0.019 | 0.036 | 0.5 | 3.1 | 1.4 | 0.152 | 519 | 8.8E-05 | 3 | | 1 |
| 994.01 | 2.1635 | 212 | 11 | 65.8649 | 0.0069 | 4.29889 | 0.00013 | 11.4 | 3.4 | 0.0142 | 0.0011 | 0.46 | 0.14 | 1.5 | 0.052 | 1032 | 1.1E-04 | 3 | | 1 |
| 998.01 | 5.2630 | 88851 | 564 | 147.03869 | 0.00025 | 161.78801 | 0.00019 | 297.87 | 0.69 | 0.26748 | 0.00042 | 0.001 | 0.014 | 25.0 | 0.592 | 308 | - | 3 | | 1 |

| | | | | | | | | | | | | | | | | | | | |
|---|---|---|---|---|---|---|---|---|---|---|---|---|---|---|---|---|---|---|---|
| 999.01 | 4.0767 | 1222 | 23 | 79.1504 | 0.0052 | 16.56815 | 0.00039 | 30 | 215 | 0.032 | 0.048 | 0.4 | 3 | 2.9 | 0.125 | 583 | 8.7E-05 | 2 | |
| 1001.01 | 12.8302 | 136 | 19 | 68.321 | 0.014 | 20.4024 | 0.0013 | 12.28752 | 0.00076 | 0.01033 | - | - | - | 2.6 | 0.159 | 1006 | 8.1E-05 | 3 | 1 |
| 1002.01 | 1.8181 | 155 | 14 | 65.9865 | 0.0046 | 3.481678 | 0.000066 | 18.11 | 0.9 | 0.01455 | 0.00054 | 0.0848 | - | 1.0 | 0.043 | 842 | 9.5E-05 | 3 | 1 |
| 1003.01 | 7.3074 | 24649 | 158 | 105.2651 | 0.0014 | 8.360619 | 0.000057 | 9.94 | 0.06 | 0.1409 | 0.00067 | 0.0047 | - | 14.1 | 0.08 | 766 | - | 3 | 1 |
| 1005.01 | 8.5483 | 4926 | 105 | 130.0317 | 0.0021 | 35.61842 | 0.00039 | 33.93486 | 0.00038 | 0.062 | 0.043 | 0.0023 | - | 6.5 | 0.208 | 469 | - | 4 | |
| 1008.01 | 5.4787 | 31943 | 189 | 181.92435 | 0.0007 | 300 | - | 361 | 108 | 1.46567 | - | 5.3 | 1.6 | 204.8 | 0.895 | 283 | - | 4 | |
| 1010.01 | 18.1142 | 228 | 12 | 78.953 | 0.036 | 110.645 | 0.021 | 26 | 48 | 0.0162 | 0.0055 | 0.85 | 0.86 | 2.7 | 0.48 | 506 | - | 4 | |
| 1013.01 | 0.7714 | 862 | 62 | 66.661 | 0.00064 | 0.5187505 | 0.0000015 | 6.51 | 0.21 | 0.02606 | 0.00049 | 0.014 | 0.03 | 1.6 | 0.012 | 1641 | - | 4 | |
| 1014.01 | 3.1358 | 1437 | 19 | 75.3548 | 0.0048 | 17.31731 | 0.0004 | 45 | 384 | 0.035 | 0.061 | 0.2 | 3.4 | 2.7 | 0.12 | 502 | - | 4 | |
| 1015.01 | 4.4039 | 552 | 26 | 73.1092 | 0.0052 | 9.42869 | 0.00018 | 16 | 108 | 0.021 | 0.029 | 0.3 | 3 | 2.4 | 0.091 | 921 | - | 4 | |
| 1015.02 | 3.3973 | 151 | 9.6 | 68.752 | 0.011 | 4.08909 | 0.00017 | 6 | 42 | 0.015 | 0.02 | 0.8 | 1.9 | 1.6 | 0.052 | 1218 | - | 4 | |
| 1017.01 | 4.1116 | 741 | 26 | 74.676 | 0.0046 | 17.44529 | 0.00036 | 23 | 84 | 0.027 | 0.02 | 0.7 | 1.6 | 3.0 | 0.133 | 659 | - | 4 | |
| 1019.01 | 2.5586 | 39 | 8.2 | 68.32 | 0.01 | 2.49677 | 0.00011 | 7.94 | 0.47 | 0.00728 | 0.00041 | 0.018 | - | 1.2 | 0.037 | 1364 | - | 4 | |
| 1020.01 | 6.3051 | 10344 | 300 | 97.06991 | 0.00072 | 54.35611 | 0.0002 | 42 | 12 | 0.1318 | 0.0024 | 0.88 | 0.26 | 21.9 | 0.294 | 580 | - | 4 | 1 |
| 1022.01 | 4.6376 | 1003 | 14 | 129.1832 | 0.008 | 18.82778 | 0.00079 | 32 | 1.7 | 0.0289 | 0.0013 | 0.025 | 0.05 | 2.8 | 0.139 | 612 | - | 4 | |
| 1024.01 | 1.8696 | 783 | 26 | 66.2982 | 0.0027 | 5.747732 | 0.000065 | 25.2 | 1.1 | 0.02563 | 0.00082 | 0.073 | - | 1.7 | 0.052 | 635 | - | 4 | |
| 1026.01 | 18.9471 | 724 | 19 | 118.039 | 0.018 | 94.1023 | 0.0097 | 38.8 | 1.1 | 0.02401 | 0.0008 | 0.019 | 0.076 | 1.8 | 0.325 | 242 | - | 4 | |
| 1029.01 | 5.5950 | 668 | 15 | 66.871 | 0.011 | 32.3113 | 0.0014 | 47.1 | 2.3 | 0.0235 | 0.001 | 0.007 | 0.014 | 2.5 | 0.203 | 558 | - | 4 | |
| 1030.01 | 2.8480 | 418 | 10 | 71.431 | 0.009 | 9.22978 | 0.00036 | 24.8 | 1.5 | 0.0223 | 0.0011 | 0.089 | 0.01 | 1.8 | 0.088 | 799 | - | 4 | |
| 1031.01 | 6.1479 | 304 | 8.6 | 68.978 | 0.019 | 14.5563 | 0.0012 | 17 | 146 | 0.018 | 0.032 | 0.4 | 3.2 | 2.0 | 0.119 | 728 | - | 4 | |
| †1032.01 | 24.0138 | 4165 | 87 | 109.1034 | 0.0044 | 615.3 | 4.3 | 106 | 32 | 0.0764 | 0.0015 | 0.86 | 0.26 | 26.2 | 1.558 | 300 | - | 4 | 1 |
| 1050.01 | 1.5002 | 349 | 37 | 66.3422 | 0.0016 | 1.2690943 | 0.0000089 | 4 | 16 | 0.021 | 0.015 | 0.8 | 1.3 | 2.3 | 0.022 | 1462 | - | 4 | |
| 1051.01 | 2.9340 | 429 | 18 | 71.1168 | 0.0054 | 6.79673 | 0.00016 | 19.8 | 0.93 | 0.02049 | 0.00075 | 0.0435 | - | 1.7 | 0.071 | 841 | - | 4 | |
| 1052.01 | 4.3752 | 546 | 15 | 76.3472 | 0.008 | 17.0282 | 0.00058 | 18 | 65 | 0.025 | 0.017 | 0.8 | 1.3 | 2.6 | 0.133 | 719 | - | 4 | |
| 1053.01 | 1.6860 | 230 | 14 | 66.2102 | 0.0048 | 1.224848 | 0.000025 | 3 | 15 | 0.021 | 0.018 | 0.8 | 1.4 | 1.9 | 0.022 | 1466 | - | 4 | |
| 1054.01 | 3.7930 | 219 | 13 | 67.384 | 0.0091 | 3.32361 | 0.00013 | 7.05 | 0.28 | 0.01749 | 0.00055 | 0.027 | 0.048 | 2.0 | 0.043 | 1116 | - | 4 | |
| 1059.01 | 1.4113 | 153 | 15 | 66.4557 | 0.0039 | 1.022666 | 0.000017 | 8.259 | 0.023 | 0.01997 | 0.0004 | 0.134 | - | 1.4 | 0.019 | 1400 | - | 4 | |
| 1060.01 | 5.5786 | 249 | 17 | 73.2117 | 0.0084 | 12.10963 | 0.00045 | 17.02 | 0.69 | 0.01436 | 0.00053 | 0.008 | 0.01 | 1.5 | 0.107 | 856 | - | 4 | |
| 1060.02 | 4.8600 | 127 | 13 | 70.697 | 0.01 | 4.75793 | 0.00021 | 5 | 25 | 0.011 | 0.012 | 0.8 | 1.6 | 1.2 | 0.058 | 1163 | - | 4 | |
| 1061.01 | 5.8923 | 406 | 14 | 75.732 | 0.01 | 41.818 | 0.0021 | 34 | 162 | 0.021 | 0.019 | 0.8 | 1.6 | 2.3 | 0.241 | 520 | - | 4 | |
| 1063.01 | 5.0997 | 266763 | 7754 | 109.30531 | 0.00006 | 89.69815 | 0.000024 | 198 | 59 | 0.4741 | 0.00024 | - | - | 93.4 | 0.418 | 547 | - | 3 | 1 |
| 1064.01 | 3.2021 | 19234 | 310 | 66.46468 | 0.00033 | 1.1873532 | 0.0000017 | 3.25 | 0.97 | 0.12338 | 0.00039 | - | - | 15.7 | 0.023 | 1939 | - | 4 | |
| 1065.01 | 3.8421 | 21849 | 257 | 66.63778 | 0.00048 | 4.0206268 | 0.0000084 | 5.07 | 0.41 | 0.1743 | 0.0053 | 0.87 | 0.16 | 18.9 | 0.051 | 1160 | - | 4 | |
| 1072.01 | 3.6243 | 465 | 19 | 72.112 | 0.006 | 10.12804 | 0.00026 | 19.22 | 0.34 | 0.01809 | 0.00099 | 0.467 | 0.01 | 2.1 | 0.094 | 881 | - | 4 | |
| 1075.01 | 1.7654 | 4752 | 168 | 66.27522 | 0.0004 | 1.343764 | 0.0000023 | 5.4 | 1.6 | 0.06472 | 0.00038 | 0.297 | 0.089 | 7.4 | 0.024 | 1736 | - | 4 | |
| 1078.01 | 1.4965 | 1305 | 30 | 67.8711 | 0.002 | 3.353682 | 0.000029 | 18.96 | 0.81 | 0.0349 | 0.001 | 0.0461 | - | 1.9 | 0.034 | 660 | - | 4 | |
| 1081.01 | 3.2893 | 619 | 24 | 67.3026 | 0.0042 | 9.95693 | 0.00018 | 21.7 | 1.2 | 0.01779 | 0.00081 | 0.052 | 0.01 | 1.4 | 0.092 | 732 | - | 4 | |
| 1082.01 | 2.5360 | 521 | 12 | 68.2737 | 0.0071 | 6.50318 | 0.0002 | 20.98 | 0.25 | 0.0227 | 0.0011 | 0.0542 | - | 1.9 | 0.066 | 763 | - | 4 | |
| 1083.01 | 3.5414 | 338 | 14 | 71.5846 | 0.0079 | 7.33679 | 0.00025 | 16.4 | 0.96 | 0.01694 | 0.00082 | 0.032 | 0.01 | 1.6 | 0.075 | 834 | - | 4 | |
| 1085.01 | 1.9381 | 428 | 10 | 72.1556 | 0.0069 | 7.71794 | 0.00023 | 30.9 | 4 | 0.0167 | 0.0019 | 0.015 | 0.01 | 1.2 | 0.063 | 573 | - | 4 | |
| 1086.01 | 5.9559 | 438 | 17 | 77.8771 | 0.0092 | 27.6625 | 0.0012 | 24 | 93 | 0.021 | 0.016 | 0.8 | 1.5 | 2.5 | 0.184 | 623 | - | 4 | |

| | | | | | | | | | | | | | | | | | | |
|---|---|---|---|---|---|---|---|---|---|---|---|---|---|---|---|---|---|---|
| 1089.01 | 10.1641 | 8682 | 199 | 108.5986 | 0.0011 | 86.67747 | 0.00047 | 70.54 | 0.26 | 0.0827 | 0.00028 | 0.002 | 0.022 | 9.6 | 0.395 | 429 | - | 4 |
| 1089.02 | 2.8501 | 1893 | 61 | 73.321 | 0.0015 | 12.21822 | 0.000083 | 19 | 12 | 0.0488 | 0.006 | 0.86 | 0.48 | 5.7 | 0.107 | 824 | - | 4 |
| 1094.01 | 4.7951 | 847 | 24 | 68.5851 | 0.0056 | 6.10027 | 0.00015 | 9.95 | 0.32 | 0.02633 | 0.00072 | 0.0201 | - | 2.4 | 0.066 | 893 | - | 4 |
| 1095.01 | 2.7384 | 5841 | 50 | 136.9783 | 0.0019 | 51.59825 | 0.00052 | 70 | 21 | 0.2841 | 0.0019 | 1.41 | 0.42 | 25.8 | 0.272 | 422 | - | 4 |
| 1099.01 | 3.7160 | 4771 | 34 | 131.003 | 0.0027 | 161.5252 | 0.0021 | 352.6 | 9.6 | 0.0613 | 0.0012 | 0.031 | 0.056 | 3.7 | 0.573 | 244 | - | 4 |
| 1101.01 | 2.2767 | 362 | 11 | 67.8593 | 0.0075 | 2.847635 | 0.000091 | 10.01 | 0.88 | 0.0177 | 0.0012 | 0.0617 | - | 1.1 | 0.036 | 865 | - | 4 |
| 1102.01 | 3.8679 | 543 | 18 | 70.6093 | 0.0063 | 12.3318 | 0.00034 | 14 | 36 | 0.025 | 0.013 | 0.8 | 1.1 | 3.2 | 0.108 | 841 | - | 4 |
| 1102.02 | 3.8409 | 408 | 17 | 73.563 | 0.0067 | 8.14561 | 0.00024 | 11.4 | 1.2 | 0.0074 | 0.0031 | 0.744 | 0.022 | 0.9 | 0.082 | 966 | - | 4 |
| 1106.01 | 3.6049 | 366 | 15 | 72.9022 | 0.0069 | 7.42603 | 0.00022 | 14.1 | 1.6 | 0.0093 | 0.0013 | 0.015 | 0.014 | 0.9 | 0.076 | 905 | - | 4 |
| 1108.01 | 3.6924 | 369 | 18 | 73.0373 | 0.0063 | 9.46255 | 0.00026 | 20.15 | 0.89 | 0.01828 | 0.00064 | 0.012 | 0.01 | 1.4 | 0.086 | 672 | - | 4 |
| 1109.01 | 6.4230 | 231 | 11 | 67.738 | 0.015 | 6.72233 | 0.00043 | 8 | 83 | 0.013 | 0.03 | 0.4 | 3.6 | 1.4 | 0.068 | 822 | - | 4 |
| 1110.01 | 3.5692 | 339 | 11 | 68.093 | 0.01 | 8.73492 | 0.00039 | 20.4 | 1.4 | 0.0182 | 0.0011 | 0.031 | 0.01 | 1.9 | 0.085 | 837 | - | 4 |
| 1111.01 | 3.7924 | 358 | 10 | 69.5 | 0.011 | 10.26494 | 0.00051 | 21.3553 | 0.0018 | 0.02 | 0.83 | 0.041 | 0.01 | 1.6 | 0.092 | 689 | - | 4 |
| 1112.01 | 7.9162 | 436 | 14 | 91.323 | 0.013 | 37.8122 | 0.0023 | 36.4 | 1.6 | 0.01946 | 0.00084 | 0.053 | 0.014 | 2.2 | 0.226 | 547 | - | 4 |
| 1113.01 | 4.6584 | 412 | 18 | 82.7491 | 0.0072 | 25.93496 | 0.00086 | 23 | 84 | 0.022 | 0.014 | 0.8 | 1.2 | 2.6 | 0.178 | 676 | - | 4 |
| 1113.02 | 7.0361 | 491 | 15 | 91.699 | 0.01 | 83.4411 | 0.0042 | 52 | 141 | 0.023 | 0.012 | 0.8 | 1.1 | 2.8 | 0.388 | 458 | - | 4 |
| 1114.01 | 2.3308 | 304 | 11 | 72.7953 | 0.0073 | 7.0472 | 0.00022 | 14.1 | 4.2 | 0.0181 | 0.0014 | 0.66 | 0.2 | 2.2 | 0.073 | 960 | - | 4 |
| 1115.01 | 6.0624 | 232 | 17 | 69.3262 | 0.0097 | 12.99172 | 0.00055 | 16.66 | 0.45 | 0.01371 | 0.00052 | 0.012 | 0.014 | 2.0 | 0.111 | 837 | - | 4 |
| 1116.01 | 1.9015 | 187 | 21 | 68.1021 | 0.0033 | 3.749224 | 0.000053 | 14 | 68 | 0.013 | 0.013 | 0.4 | 2.3 | 1.4 | 0.048 | 1144 | - | 4 |
| 1117.01 | 6.1980 | 148 | 21 | 74.7572 | 0.0081 | 11.08977 | 0.00039 | 11 | 41 | 0.0119 | 0.009 | 0.6 | 1.8 | 1.2 | 0.101 | 842 | - | 4 |
| 1118.01 | 1.7024 | 198 | 12 | 70.6009 | 0.0056 | 7.37325 | 0.00018 | 36.4 | 3.1 | 0.01365 | 0.00097 | 0.023 | 0.01 | 1.8 | 0.077 | 1049 | - | 4 |
| 1121.01 | 4.0277 | 58683 | 441 | 73.69733 | 0.00035 | 14.154037 | 0.000021 | 27.2 | 8.2 | 0.2488 | 0.0012 | 0.54 | 0.16 | 24.7 | 0.115 | 676 | - | 4 |
| 1123.01 | 1.1651 | 1856 | 116 | 66.58381 | 0.00044 | 0.8484851 | 0.0000016 | 4.1 | 1.2 | 0.04351 | 0.00044 | 0.56 | 0.17 | 4.4 | 0.018 | 1784 | - | 4 |
| 1128.01 | 1.7084 | 192 | 44 | 66.0736 | 0.0015 | 0.9748817 | 0.0000064 | 4.493 | 0.078 | 0.01342 | 0.0003 | 0.0103 | - | 1.0 | 0.019 | 1378 | - | 4 |
| 1129.01 | 1.4556 | 336 | 9.1 | 66.3832 | 0.0063 | 4.89768 | 0.00013 | 24 | 145 | 0.022 | 0.027 | 0.7 | 2.1 | 2.7 | 0.056 | 959 | - | 4 |
| 1134.02 | 7.7281 | 12103 | 109 | 389.8516 | 0.0025 | 200.623 | 0.016 | 169 | 51 | 0.1085 | 0.0021 | 0.5 | 0.15 | 9.7 | 0.646 | 245 | - | 4 |
| 1140.01 | 0.6290 | 1146 | 120 | 66.44743 | 0.00023 | 0.5532587 | 0.0000006 | 6.888 | 0.071 | 0.03185 | 0.00021 | 0.028 | 0.017 | 11.2 | 0.015 | 3538 | - | 4 |
| 1141.01 | 2.7088 | 765 | 15 | 68.1853 | 0.0063 | 5.72796 | 0.00015 | 16.43 | 0.91 | 0.0273 | 0.0012 | 0.0488 | - | 2.0 | 0.052 | 634 | - | 4 |
| 1142.01 | 2.0797 | 465 | 16 | 66.4288 | 0.0046 | 3.755719 | 0.000073 | 13.9 | 4.2 | 0.01927 | 0.00098 | 0.031 | 0.009 | 1.5 | 0.046 | 892 | - | 4 |
| 1144.01 | 3.4602 | 225 | 14 | 67.5826 | 0.0077 | 2.441836 | 0.000078 | 5.44 | 0.21 | 0.01666 | 0.00054 | 0.0224 | - | 1.6 | 0.036 | 1250 | - | 4 |
| 1145.01 | 5.3510 | 486 | 23 | 95.8746 | 0.0052 | 30.5908 | 0.00075 | 45.6 | 1.4 | 0.02011 | 0.00055 | 0.023 | 0.014 | 2.0 | 0.195 | 549 | - | 4 |
| 1146.01 | 2.0887 | 403 | 12 | 69.3494 | 0.0063 | 7.09695 | 0.0002 | 22 | 201 | 0.024 | 0.044 | 0.7 | 2.4 | 1.3 | 0.056 | 512 | - | 4 |
| 1148.01 | 4.7868 | 202 | 15 | 68.4025 | 0.009 | 11.47566 | 0.00044 | 10 | 44 | 0.015 | 0.012 | 0.8 | 1.3 | 2.2 | 0.104 | 952 | - | 4 |
| 1149.01 | 1.7338 | 269 | 7.3 | 69.2328 | 0.0088 | 7.16918 | 0.00027 | 18.3 | 5.5 | 0.0173 | 0.0021 | 0.7 | 0.21 | 1.6 | 0.072 | 783 | - | 4 |
| 1150.01 | 1.9085 | 82 | 20 | 66.9456 | 0.0036 | 0.677375 | 0.000011 | 2.78 | 0.22 | 0.00648 | 0.00053 | 0.0522 | - | 0.7 | 0.015 | 1940 | - | 4 |
| 1151.01 | 3.3203 | 113 | 14 | 67.8179 | 0.007 | 5.21785 | 0.00015 | 11.62 | 0.54 | 0.0107 | 0.00041 | 0.0566 | - | 1.2 | 0.059 | 989 | - | 4 |
| 1151.02 | 3.4924 | 116 | 12 | 68.7486 | 0.008 | 7.41084 | 0.00025 | 18.1 | 1 | 0.0109 | 0.00051 | 0.057 | 0.01 | 1.2 | 0.075 | 877 | - | 4 |
| 1152.01 | 3.4095 | 84601 | 309 | 111.24288 | 0.00037 | 4.7222503 | 0.0000086 | 12.99 | 0.5 | 0.2689 | 0.003 | 0.41 | 0.21 | 19.2 | 0.046 | 676 | - | 4 |
| 1153.01 | 1.6005 | 36772 | 273 | 67.01944 | 0.00053 | 0.6350732 | 0.0000029 | 2.39 | 0.061 | 0.2025 | 0.0018 | 0.79 | 0.11 | 20.8 | 0.015 | 2137 | - | 4 |
| 1154.01 | 7.8146 | 14832 | 713 | 69.04584 | 0.00031 | 6.8108256 | 0.000009 | 7.3972 | 0.009 | 0.10842 | 0.00011 | - | - | 22.2 | 0.077 | 1522 | - | 4 |
| 1156.01 | 1.3680 | 2964 | 112 | 67.36536 | 0.00048 | 1.8724219 | 0.0000038 | 6.7 | 2 | 0.05836 | 0.00059 | 0.71 | 0.21 | 6.0 | 0.029 | 1289 | - | 4 |
| 1157.01 | 5.0609 | 1474 | 377 | 66.38704 | 0.00048 | 0.9337473 | 0.0000019 | 1.2 | 0.36 | 0.04357 | 0.00013 | 0.83 | 0.25 | 17.8 | 0.021 | 3067 | - | 4 |
| 1159.01 | 5.0986 | 2444 | 34 | 99.1389 | 0.0072 | 64.6198 | 0.002 | 54 | 18 | 0.0538 | 0.0039 | 0.87 | 0.34 | 5.3 | 0.304 | 372 | - | 4 |
| 1160.01 | 3.1995 | 1083 | 19 | 73.5808 | 0.0053 | 13.21436 | 0.00029 | 32.4 | 1.6 | 0.0296 | 0.0011 | 0.013 | 0.044 | 1.8 | 0.104 | 529 | - | 4 |
| 1161.01 | 3.7439 | 371 | 21 | 68.9164 | 0.0054 | 6.05767 | 0.00014 | 9 | 46 | 0.02 | 0.019 | 0.7 | 1.9 | 2.3 | 0.065 | 925 | - | 4 |
| 1162.01 | 11.7092 | 934 | 65 | 106.9379 | 0.0033 | 158.6951 | 0.0026 | 90 | 86 | 0.0285 | 0.0056 | 0.54 | 0.98 | 3.9 | 0.594 | 375 | - | 4 |
| 1163.01 | 1.8096 | 347 | 20 | 68.7149 | 0.0034 | 2.936566 | 0.000043 | 12.94 | 0.56 | 0.01869 | 0.00061 | 0.0677 | - | 1.9 | 0.04 | 1153 | - | 4 |
| 1163.02 | 3.4700 | 348 | 17 | 67.8239 | 0.0066 | 8.01533 | 0.00022 | 18.36 | 0.88 | 0.01746 | 0.0007 | 0.027 | 0.01 | 1.8 | 0.079 | 821 | - | 4 |
| 1164.01 | 3.5820 | 224 | 15 | 66.3891 | 0.0073 | 2.801128 | 0.000088 | 5.95 | 0.26 | 0.01489 | 0.00055 | 0.0147 | - | 1.2 | 0.032 | 829 | - | 4 |
| 1165.01 | 1.8463 | 513 | 35 | 69.3531 | 0.002 | 7.05404 | 0.00006 | 34.8 | 1.3 | 0.02087 | 0.0006 | 0.0053 | - | 2.9 | 0.074 | 1009 | - | 4 |
| 1166.01 | 1.5231 | 557 | 15 | 66.2153 | 0.004 | 7.67503 | 0.00013 | 30 | 221 | 0.028 | 0.04 | 0.8 | 2 | 2.8 | 0.078 | 880 | - | 4 |
| †1168.01 | 23.2431 | 845 | 28 | 161.4507 | 0.008 | 458 | 2054 | 154 | 693 | 0.02577 | - | - | - | 3.9 | 1.221 | 291 | - | 4 |
| 1169.01 | 1.5932 | 196 | 48 | 66.2862 | 0.0013 | 0.6892091 | 0.000004 | 3.4 | 1 | 0.01247 | 0.00026 | - | - | 1.2 | 0.015 | 1895 | - | 4 |
| 1170.01 | 1.2491 | 510 | 19 | 70.3023 | 0.0028 | 7.343619 | 0.00009 | 29 | 185 | 0.024 | 0.029 | 0.8 | 1.9 | 1.4 | 0.073 | 670 | - | 4 |
| 1171.01 | 1.7492 | 182 | 21 | 66.5297 | 0.0024 | 0.4452672 | 0.000005 | 2.112 | 0.058 | 0.01437 | 0.00043 | 0.0092 | - | 1.1 | 0.012 | 2046 | - | 4 |
| 1175.01 | 12.4129 | 115 | 13 | 75.956 | 0.02 | 31.5953 | 0.0028 | 19.76 | 0.73 | 0.00972 | 0.00043 | 0.025 | 0.02 | 0.9 | 0.195 | 493 | - | 4 |
| 1176.01 | 1.8347 | 30840 | 735 | 111.6887 | 0.0001 | 1.9737605 | 0.000001 | 9.635 | 0.024 | 0.15708 | 0.00029 | 0.001 | 0.01 | 11.1 | 0.028 | 974 | - | 4 |
| 1177.01 | 2.5220 | 21340 | 46 | 113.6797 | 0.0018 | 3.3056 | 0.00003 | 11.3 | 3.4 | 0.1301 | 0.003 | - | - | 9.8 | 0.043 | 945 | - | 4 |
| 1178.01 | 7.6265 | 13791 | 132 | 67.3695 | 0.0016 | 4.800633 | 0.000032 | 5.3 | 1.6 | 0.10425 | 0.00069 | - | - | 13.7 | 0.059 | 1293 | - | 4 |
| 1180.01 | 10.2896 | 18755 | 488 | 96.01592 | 0.00055 | 34.820008 | 0.000099 | 22.65 | 0.45 | 0.13235 | 0.00065 | 0.64 | 0.13 | 13.8 | 0.205 | 478 | - | 4 |
| 1185.01 | 1.4882 | 1707 | 108 | 66.86259 | 0.00047 | 1.6657816 | 0.0000034 | 5.2 | 1.6 | 0.04466 | 0.00039 | 0.74 | 0.22 | 5.8 | 0.029 | 1699 | - | 4 |
| 1187.01 | 0.7778 | 1835 | 187 | 66.79903 | 0.00021 | 0.3705285 | 0.0000004 | 3.809 | 0.035 | 0.03961 | 0.00024 | 0.0263 | - | 2.5 | 0.01 | 1789 | - | 4 |
| 1190.01 | 1.3773 | 727 | 34 | 111.5863 | 0.0011 | 0.3937291 | 0.0000022 | 2.32 | 0.69 | 0.02411 | 0.00056 | - | - | 2.0 | 0.01 | 2118 | - | 4 |
| †1192.01 | 17.0021 | 1399 | 31 | 105.5114 | 0.0099 | 123.2759 | 0.0077 | 57.3 | 0.19 | 0.033 | 0.055 | 0.017 | 0.069 | 4.1 | 0.491 | 362 | - | 4 |
| 1193.01 | 3.2991 | 2692 | 33 | 82.2193 | 0.0035 | 59.5309 | 0.0011 | 59 | 18 | 0.106 | 0.038 | 1.04 | 0.31 | 12.9 | 0.303 | 463 | - | 4 |
| 1198.01 | 5.4692 | 660 | 17 | 72.4887 | 0.0086 | 16.08903 | 0.00063 | 20 | 152 | 0.024 | 0.037 | 0.5 | 2.8 | 2.0 | 0.128 | 679 | - | 4 |
| 1198.02 | 5.2647 | 303 | 9.9 | 75.726 | 0.014 | 10.30133 | 0.00063 | 16.75 | 0.97 | 0.0184 | 0.00095 | 0.038 | 0.01 | 1.5 | 0.095 | 788 | - | 4 |
| 1199.01 | 5.5648 | 1056 | 23 | 80.864 | 0.0062 | 53.5296 | 0.0015 | 76.9 | 2.5 | 0.02917 | 0.00083 | 0.007 | 0.028 | 2.7 | 0.263 | 379 | - | 4 |
| 1200.01 | 1.4296 | 424 | 41 | 66.3066 | 0.0011 | 0.3937319 | 0.000002 | 1.4 | 1.1 | 0.0242 | 0.0037 | 0.71 | 0.75 | 2.4 | 0.011 | 2399 | - | 4 |
| 1201.01 | 0.9997 | 616 | 12 | 66.0546 | 0.0039 | 2.757529 | 0.000046 | 13 | 3.9 | 0.0262 | 0.0028 | 0.67 | 0.2 | 1.4 | 0.03 | 686 | - | 4 |
| 1202.01 | 1.3660 | 414 | 15 | 66.8036 | 0.0037 | 0.928308 | 0.000015 | 6.04 | 0.28 | 0.02657 | 0.00085 | 0.059 | - | 1.5 | 0.015 | 1104 | - | 4 |
| 1203.01 | 5.2656 | 813 | 15 | 93.2296 | 0.0092 | 31.8811 | 0.0014 | 46.8 | 2.4 | 0.0254 | 0.0011 | 0.03 | 0.07 | 2.5 | 0.201 | 544 | - | 4 |
| 1203.02 | 3.3983 | 516 | 12 | 75.9003 | 0.0086 | 14.12823 | 0.00054 | 21 | 146 | 0.025 | 0.033 | 0.8 | 2 | 2.4 | 0.117 | 713 | - | 4 |
| 1204.01 | 6.4828 | 326 | 13 | 69.33 | 0.013 | 8.39776 | 0.00048 | 9.82 | 0.3 | 0.01683 | 0.00073 | 0.047 | 0.01 | 1.7 | 0.083 | 906 | - | 4 |
| 1205.01 | 5.2040 | 378 | 19 | 72.9508 | 0.0078 | 8.63851 | 0.00029 | 8 | 34 | 0.02 | 0.016 | 0.8 | 1.6 | 2.1 | 0.085 | 888 | - | 4 |
| 1207.01 | 2.1368 | 718 | 16 | 75.3282 | 0.0046 | 13.73459 | 0.00028 | 29 | 158 | 0.029 | 0.029 | 0.9 | 1.4 | 2.9 | 0.111 | 650 | - | 4 |
| †1208.01 | 8.5880 | 3312 | 74 | 182.4388 | 0.0017 | 300 | - | 182.55974 | - | 0.05951 | - | 0.8091 | - | 7.0 | 0.913 | 301 | - | 4 |
| 1210.01 | 5.8162 | 286 | 17 | 67.0989 | 0.0091 | 14.55495 | 0.00057 | 19.4 | 0.8 | 0.01534 | 0.00058 | 0.009 | 0.01 | 1.2 | 0.119 | 673 | - | 4 |
| 1212.01 | 3.8647 | 274 | 9.4 | 67.918 | 0.013 | 11.30143 | 0.00059 | 24.8 | 2.1 | 0.0159 | 0.0012 | 0.214 | 0.01 | 1.8 | 0.101 | 798 | - | 4 |
| 1214.01 | 2.8863 | 177 | 9.8 | 66.4458 | 0.0095 | 4.24176 | 0.00017 | 9.81878 | 0.0004 | 0.01236 | - | 0.5139 | - | 1.5 | 0.052 | 1146 | - | 4 |
| 1215.01 | 7.4620 | 228 | 25 | 75.118 | 0.0069 | 17.32298 | 0.00055 | 13 | 38 | 0.015 | 0.0086 | 0.7 | 1.5 | 2.2 | 0.138 | 869 | - | 4 |
| 1215.02 | 7.2268 | 255 | 20 | 78.3889 | 0.0084 | 33.0058 | 0.0015 | 35 | 289 | 0.014 | 0.024 | 0.1 | 3.4 | 2.1 | 0.212 | 701 | - | 4 |
| 1216.01 | 4.0428 | 181 | 16 | 67.3671 | 0.0074 | 11.13127 | 0.00035 | 12 | 51 | 0.015 | 0.012 | 0.8 | 1.3 | 2.0 | 0.101 | 883 | - | 4 |
| 1218.01 | 4.7972 | 291 | 21 | 66.6822 | 0.006 | 29.61863 | 0.00078 | 28 | 91 | 0.018 | 0.011 | 0.8 | 1.2 | 2.2 | 0.192 | 593 | - | 4 |
| 1219.01 | 3.0841 | 97 | 7.8 | 67.686 | 0.012 | 3.50394 | 0.00018 | 8.93 | 0.24 | 0.01088 | 0.00063 | 0.034 | 0.01 | 1.2 | 0.044 | 1031 | - | 4 |
| 1220.01 | 2.3617 | 129 | 14 | 69.8892 | 0.0058 | 6.40068 | 0.00016 | 15.6 | 4.7 | 0.011 | 0.00063 | 0.45 | 0.14 | 1.1 | 0.065 | 797 | - | 4 |
| 1221.01 | 7.8099 | 149 | 16 | 71.535 | 0.013 | 30.1562 | 0.0016 | 27 | 149 | 0.012 | 0.013 | 0.6 | 2.3 | 5.0 | 0.212 | 944 | - | 4 |
| 1221.02 | 12.5482 | 131 | 15 | 85.316 | 0.016 | 51.0917 | 0.0039 | 18 | 46 | 0.0124 | 0.0059 | 0.8 | 1 | 5.3 | 0.301 | 792 | - | 4 |
| 1222.01 | 3.2088 | 59 | 8.7 | 67.499 | 0.011 | 4.28545 | 0.00019 | 10.81 | 0.53 | 0.00955 | 0.00042 | 0.018 | 0.01 | 0.8 | 0.048 | 841 | - | 4 |
| 1225.01 | 1.7802 | 28197 | 668 | 66.31535 | 0.00011 | 1.714272 | 0.0000008 | 6.6 | 2 | 0.17318 | 0.00045 | 0.58 | 0.17 | 14.3 | 0.023 | 1002 | - | 4 |
| 1226.01 | 7.8230 | 83463 | 1238 | 106.16362 | 0.00021 | 137.75992 | 0.00016 | 169 | 51 | 0.25847 | 0.00028 | - | - | 26.6 | 0.514 | 301 | - | 4 |
| 1227.01 | 1.7476 | 18166 | 105 | 66.57494 | 0.00064 | 2.1552826 | 0.0000059 | 10.6 | 3.2 | 0.1204 | 0.0013 | - | - | 8.3 | 0.032 | 1066 | - | 4 |
| 1228.01 | 2.4024 | 17248 | 376 | 68.26916 | 0.00023 | 3.6613278 | 0.0000036 | 8.82 | 0.27 | 0.139 | 0.0012 | 0.81 | 0.12 | 14.8 | 0.046 | 1058 | - | 4 |
| 1229.01 | 1.5996 | 10490 | 210 | 66.29515 | 0.00028 | 0.7697382 | 0.0000009 | 4.1 | 1.2 | 0.09151 | 0.00047 | - | - | 7.8 | 0.016 | 1669 | - | 4 |
| 1230.01 | 31.8501 | 6463 | 313 | 147.034 | 0.0028 | 165.7537 | 0.004 | 43.5 | 2.4 | 0.07472 | 0.00091 | 0.62 | 0.22 | 50.2 | 0.67 | 650 | - | 4 |
| 1232.01 | 9.3600 | 18640 | 289 | 165.64948 | 0.00088 | 238.8147 | 0.0012 | 131 | 39 | 0.3618 | 0.00076 | 1.4 | 0.42 | 48.6 | 0.75 | 277 | - | 4 |
| 1236.01 | 7.8927 | 748 | 39 | 84.0494 | 0.005 | 35.74468 | 0.00084 | 35.69 | 0.62 | 0.02448 | 0.0004 | 0.028 | 0.033 | 2.8 | 0.222 | 627 | - | 4 |
| 1236.02 | 5.6648 | 194 | 20 | 70.6889 | 0.0076 | 6.15488 | 0.00021 | 8.25 | 0.21 | 0.01465 | 0.00034 | 0.0146 | - | 1.7 | 0.069 | 1125 | - | 4 |
| 1238.01 | 7.1034 | 532 | 24 | 72.1086 | 0.0079 | 27.07239 | 0.00092 | 29.55 | 0.86 | 0.02088 | 0.00055 | 0.021 | 0.022 | 2.1 | 0.177 | 543 | - | 4 |
| 1240.01 | 2.2251 | 256 | 26 | 66.6516 | 0.003 | 2.13957 | 0.000027 | 7.74 | 0.27 | 0.01517 | 0.00042 | 0.0088 | - | 1.5 | 0.033 | 1272 | - | 4 |
| 1241.01 | 10.5451 | 342 | 30 | 80.0705 | 0.009 | 21.40827 | 0.00083 | 9 | 10 | 0.0198 | 0.0039 | 0.84 | 0.67 | 10.4 | 0.17 | 1142 | - | 4 |
| 1241.02 | 12.6584 | 153 | 22 | 68.746 | 0.013 | 10.49447 | 0.00056 | 4.2 | 8.5 | 0.0133 | 0.0051 | 0.8 | 1.1 | 7.0 | 0.106 | 1446 | - | 4 |
| 1242.01 | 4.6572 | 2822 | 85 | 216.2398 | 0.0015 | 99.642 | 0.0011 | 95 | 29 | 0.05844 | 0.00065 | 0.78 | 0.23 | 6.3 | 0.436 | 412 | - | 4 |
| 1244.01 | 3.0620 | 327 | 18 | 67.0974 | 0.0056 | 10.80458 | 0.00025 | 23 | 178 | 0.017 | 0.027 | 0.5 | 2.8 | 1.5 | 0.097 | 746 | - | 4 |
| 1245.01 | 5.3201 | 260 | 18 | 73.7271 | 0.0084 | 13.72029 | 0.00051 | 20.1 | 0.8 | 0.01464 | 0.00051 | 0.005 | 0.01 | 1.5 | 0.116 | 774 | - | 4 |
| 1246.01 | 3.5116 | 354 | 13 | 81.5517 | 0.0086 | 19.03726 | 0.00071 | 23 | 97 | 0.021 | 0.016 | 0.9 | 1.2 | 2.4 | 0.144 | 723 | - | 4 |
| 1247.01 | 3.9417 | 22189 | 455 | 67.54977 | 0.00027 | 2.7398736 | 0.0000032 | 5.988 | 0.012 | 0.13277 | 0.00021 | 0.001 | 0.01 | 19.7 | 0.04 | 1617 | - | 4 |
| 1251.01 | 1.2398 | 9222 | 383 | 67.10443 | 0.00012 | 0.576082 | 0.0000003 | 3.9 | 1.2 | 0.0862 | 0.00025 | - | - | 6.8 | 0.013 | 1808 | - | 4 |
| 1257.01 | 4.4410 | 8084 | 110 | 106.7926 | 0.0011 | 86.64814 | 0.00043 | 164.8 | 1.4 | 0.07981 | 0.00052 | 0.026 | 0.02 | 10.4 | 0.385 | 398 | - | 4 |
| 1258.01 | 6.3287 | 3080 | 58 | 93.3913 | 0.0028 | 36.33921 | 0.00048 | 37 | 22 | 0.0533 | 0.0066 | 0.63 | 0.72 | 4.2 | 0.214 | 446 | - | 4 |
| 1261.01 | 11.5198 | 5135 | 90 | 149.0435 | 0.0024 | 133.4589 | 0.0018 | 94.43 | 0.75 | 0.06354 | 0.00046 | 0.026 | 0.017 | 6.3 | 0.521 | 335 | - | 4 |
| 1263.01 | 9.5353 | 1074 | 27 | 66.8815 | 0.0088 | 31.9716 | 0.0012 | 25.99 | 0.61 | 0.02907 | 0.00067 | 0.029 | 0.039 | 2.0 | 0.182 | 393 | - | 4 |
| 1264.01 | 3.5197 | 1122 | 25 | 75.2628 | 0.0044 | 14.13146 | 0.00027 | 31.8 | 1.1 | 0.03033 | 0.00087 | 0.003 | 0.01 | 2.7 | 0.114 | 637 | - | 4 |
| 1266.01 | 3.9623 | 886 | 30 | 70.6651 | 0.0041 | 11.41944 | 0.0002 | 23.21 | 0.66 | 0.02688 | 0.00062 | 0.029 | 0.047 | 2.0 | 0.087 | 543 | - | 4 |
| †1268.01 | 10.3358 | 6430 | 76 | 293.0702 | 0.0019 | 191 | 14 | 135.9 | 9.9 | 0.07383 | - | 0.4742 | - | 8.6 | 0.671 | 336 | - | 4 |
| 1270.01 | 1.2127 | 885 | 29 | 71.5555 | 0.0017 | 5.729434 | 0.000043 | 20 | 65 | 0.033 | 0.02 | 0.9 | 1.1 | 2.0 | 0.06 | 698 | - | 4 |
| 1272.01 | 2.3435 | 201 | 60 | 66.8481 | 0.0014 | 0.5309685 | 0.0000033 | 1.924 | 0.025 | 0.01393 | 0.00016 | 0.0147 | - | 1.9 | 0.013 | 2667 | - | 4 |
| 1273.01 | 5.3917 | 1155 | 35 | 102.9068 | 0.004 | 40.05949 | 0.00075 | 58.8 | 1.3 | 0.03022 | 0.00057 | 0.028 | 0.039 | 2.8 | 0.23 | 460 | - | 4 |
| 1275.01 | 6.3429 | 1053 | 50 | 100.6306 | 0.0041 | 50.28661 | 0.0009 | 56 | 150 | 0.03 | 0.016 | 0.4 | 1.8 | 3.3 | 0.269 | 466 | - | 4 |
| 1276.01 | 5.1517 | 561 | 26 | 71.6997 | 0.0053 | 22.78864 | 0.00052 | 26 | 94 | 0.023 | 0.016 | 0.7 | 1.7 | 1.5 | 0.154 | 474 | - | 4 |
| 1278.01 | 5.8956 | 285 | 14 | 78.079 | 0.011 | 12.40287 | 0.00061 | 16.24 | 0.72 | 0.01629 | 0.00068 | 0.061 | 0.01 | 1.6 | 0.107 | 744 | - | 4 |
| 1278.02 | 6.2243 | 740 | 20 | 94.3182 | 0.0078 | 44.3474 | 0.0016 | 56.8 | 2 | 0.02488 | 0.0008 | 0.042 | 0.01 | 2.4 | 0.25 | 487 | - | 4 |
| 1279.01 | 4.9091 | 317 | 25 | 71.2073 | 0.0056 | 14.37428 | 0.00035 | 22.86 | 0.67 | 0.016 | 0.00042 | 0.015 | 0.033 | 1.1 | 0.115 | 584 | - | 4 |
| 1281.01 | 2.8618 | 568 | 15 | 74.8243 | 0.0056 | 49.4787 | 0.0012 | 134 | 1323 | 0.022 | 0.045 | 0.4 | 3.4 | 2.0 | 0.265 | 431 | - | 4 |
| 1282.01 | 8.5310 | 230 | 30 | 95.883 | 0.006 | 30.86317 | 0.00087 | 28.97 | 0.62 | 0.0138 | 0.00029 | 0.027 | 0.01 | 2.0 | 0.201 | 688 | - | 4 |
| 1283.01 | 5.8824 | 58 | 20 | 71.0524 | 0.0075 | 8.09166 | 0.00026 | 10.16 | 0.31 | 0.00717 | 0.00021 | 0.028 | 0.01 | 0.7 | 0.077 | 883 | - | 4 |
| 1284.01 | 2.3467 | 8533 | 426 | 66.12539 | 0.00024 | 1.5585457 | 0.0000016 | 5.5 | 1.7 | 0.08219 | 0.00025 | - | - | 10.8 | 0.028 | 1800 | - | 4 |
| 1285.01 | 1.6718 | 4957 | 94 | 67.72319 | 0.00069 | 0.937416 | 0.0000028 | 2.12 | 0.28 | 0.0814 | 0.0029 | 0.9 | 0.19 | 8.0 | 0.019 | 1606 | - | 4 |
| 1288.01 | 6.6527 | 9044 | 135 | 152.7044 | 0.0013 | 117.93 | 0.001 | 170.9 | 1.2 | 0.08401 | 0.00047 | 0.008 | 0.014 | 9.8 | 0.488 | 400 | - | 4 |
| 1290.01 | 3.2832 | 3485 | 78 | 77.7765 | 0.0013 | 11.973777 | 0.000072 | 23.8 | 7.2 | 0.05627 | 0.00059 | 0.38 | 0.12 | 5.8 | 0.105 | 786 | - | 4 |
| 1293.01 | 1.8991 | 3999 | 88 | 66.72149 | 0.00086 | 11.703074 | 0.000043 | 25.5 | 7.7 | 0.0757 | 0.0019 | 0.86 | 0.26 | 7.1 | 0.103 | 732 | - | 4 |
| 1295.01 | 1.6040 | 684 | 51 | 66.3081 | 0.0012 | 1.577794 | 0.0000082 | 7.86 | 0.18 | 0.0243 | 0.00044 | 0.0138 | - | 2.4 | 0.027 | 1488 | - | 4 |
| 1296.01 | 4.0478 | 34583 | 1371 | 68.32449 | 0.00005 | 2.3808626 | 0.0000005 | 4.344 | 0.027 | 0.17956 | 0.0003 | 0.635 | 0.071 | 23.6 | 0.036 | 1409 | - | 4 |
| 1298.01 | 2.0927 | 1509 | 28 | 76.2253 | 0.0027 | 11.00823 | 0.00013 | 22 | 20 | 0.0445 | 0.0079 | 0.9 | 0.5 | 3.7 | 0.084 | 554 | - | 4 |
| 1299.01 | 15.1959 | 856 | 68 | 104.5278 | 0.0048 | 52.4989 | 0.0013 | 28 | 0.24 | 0.02601 | 0.00024 | 0.021 | 0.017 | 10.2 | 0.305 | 748 | - | 4 |
| 1300.01 | 1.1440 | 452 | 57 | 66.94395 | 0.00089 | 0.6313314 | 0.0000025 | 4.779 | 0.012 | 0.02036 | 0.00025 | 0.07 | - | 1.7 | 0.013 | 1485 | - | 4 |
| 1301.01 | 2.6177 | 763 | 13 | 117.9894 | 0.0064 | 12.69902 | 0.00037 | 33 | 379 | 0.027 | 0.064 | 0.6 | 3.4 | 1.9 | 0.104 | 588 | - | 4 |
| 1301.02 | 3.5078 | 1134 | 13 | 122.2828 | 0.0083 | 37.5214 | 0.0014 | 60 | 432 | 0.034 | 0.048 | 0.7 | 2.2 | 2.3 | 0.215 | 409 | - | 4 |
| 1302.01 | 6.8896 | 873 | 28 | 120.558 | 0.0049 | 55.6394 | 0.0014 | 63.6 | 1.6 | 0.02629 | 0.0006 | 0.004 | 0.01 | 2.4 | 0.289 | 427 | - | 4 |
| 1303.01 | 18.2696 | 416 | 24 | 76.689 | 0.014 | 34.2936 | 0.0021 | 14.62 | 0.32 | 0.01818 | 0.00046 | 0.026 | 0.03 | 2.6 | 0.21 | 579 | - | 4 |
| 1304.01 | 3.1186 | 322 | 11 | 69.4812 | 0.0089 | 4.59723 | 0.00018 | 9 | 60 | 0.021 | 0.029 | 0.7 | 2.3 | 1.8 | 0.055 | 964 | - | 4 |
| 1305.01 | 2.2079 | 284 | 15 | 67.5591 | 0.0051 | 2.633895 | 0.000057 | 9.19 | 0.11 | 0.0174 | 0.00058 | 0.0493 | - | 1.3 | 0.036 | 1020 | - | 4 |
| 1306.01 | 1.7321 | 259 | 14 | 66.327 | 0.0047 | 1.796375 | 0.000036 | 10.201 | 0.054 | 0.02101 | 0.00057 | 0.0975 | - | 2.1 | 0.029 | 1431 | - | 4 |
| 1306.02 | 2.2614 | 287 | 13 | 69.007 | 0.0059 | 3.468005 | 0.000088 | 6 | 28 | 0.022 | 0.018 | 0.9 | 1.3 | 2.2 | 0.046 | 1136 | - | 4 |
| 1306.03 | 3.7452 | 234 | 11 | 66.458 | 0.011 | 5.91495 | 0.00027 | 9 | 63 | 0.018 | 0.025 | 0.7 | 2.3 | 1.8 | 0.065 | 956 | - | 4 |
| 1307.01 | 3.8988 | 799 | 23 | 105.4744 | 0.0045 | 44.84967 | 0.00096 | 96 | 1068 | 0.025 | 0.058 | 0.1 | 4 | 3.0 | 0.252 | 505 | - | 4 |
| 1307.02 | 2.3012 | 556 | 18 | 70.6975 | 0.0042 | 20.342 | 0.00039 | 58 | 404 | 0.024 | 0.033 | 0.7 | 2.2 | 2.8 | 0.149 | 657 | - | 4 |
| 1308.01 | 5.5317 | 415 | 24 | 78.9578 | 0.0056 | 23.58427 | 0.00063 | 28 | 134 | 0.019 | 0.019 | 0.6 | 2.2 | 2.0 | 0.164 | 607 | - | 4 |
| 1309.01 | 4.6916 | 289 | 25 | 70.4515 | 0.0053 | 10.11672 | 0.00022 | 11 | 32 | 0.0175 | 0.01 | 0.8 | 1.3 | 2.5 | 0.098 | 1117 | - | 4 |
| 1310.01 | 3.6617 | 438 | 16 | 72.4384 | 0.0063 | 19.12903 | 0.00057 | 36 | 330 | 0.02 | 0.037 | 0.5 | 3.1 | 2.0 | 0.143 | 657 | - | 4 |
| 1311.01 | 7.2596 | 625 | 32 | 135.4621 | 0.005 | 83.5755 | 0.0021 | 93.4 | 2 | 0.0223 | 0.00045 | 0.026 | 0.032 | 3.4 | 0.391 | 494 | - | 4 |
| 1312.01 | 2.6895 | 283 | 15 | 68.0665 | 0.0055 | 6.14681 | 0.00015 | 19.34 | 0.19 | 0.01764 | 0.00062 | 0.084 | - | 1.6 | 0.067 | 950 | - | 4 |
| 1314.01 | 7.8679 | 139 | 18 | 73.143 | 0.01 | 8.57549 | 0.00038 | 6 | 22 | 0.012 | 0.0086 | 0.7 | 1.6 | 3.5 | 0.089 | 1213 | - | 4 |
| 1315.01 | 3.4409 | 148 | 21 | 66.7627 | 0.0054 | 6.84639 | 0.00015 | 7 | 23 | 0.0142 | 0.0077 | 0.89 | 0.96 | 1.9 | 0.074 | 1106 | - | 4 |
| 1316.01 | 6.4621 | 54 | 16 | 71.33 | 0.01 | 7.64981 | 0.00033 | 9.1 | 2.7 | 0.00653 | 0.00032 | - | - | 0.7 | 0.075 | 936 | - | 4 |
| 1318.01 | 2.5645 | 19619 | 195 | 66.96805 | 0.00043 | 1.6346139 | 0.0000031 | 5.5 | 1.7 | 0.12474 | 0.00065 | - | - | 13.7 | 0.028 | 1552 | - | 4 |
| 1321.01 | 1.5476 | 6016 | 245 | 67.22337 | 0.00026 | 0.711965 | 0.0000008 | 3.14 | 0.94 | 0.07497 | 0.00034 | 0.41 | 0.12 | 8.8 | 0.016 | 1915 | - | 4 |
| 1324.01 | 1.4501 | 1448 | 113 | 66.1377 | 0.00053 | 0.5220586 | 0.0000012 | 2.94 | 0.035 | 0.03508 | 0.00029 | 0.013 | 0.022 | 3.6 | 0.013 | 1951 | - | 4 |
| 1325.01 | 3.4427 | 2926 | 93 | 70.4706 | 0.0012 | 10.035392 | 0.000051 | 23.55971 | 0.00012 | 0.049 | 0.037 | 0.0227 | - | 3.3 | 0.09 | 652 | - | 4 |
| 1326.01 | 3.5523 | 13485 | 847 | 68.97114 | 0.00027 | 53.100927 | 0.000065 | 70 | 21 | 0.44508 | 0.00036 | 1.71 | 0.51 | 40.5 | 0.272 | 399 | - | 4 |
| 1328.01 | 3.1414 | 4981 | 25 | 93.9747 | 0.0035 | 80.966 | 0.0014 | 266 | 15 | 0.061 | 0.0025 | 0.024 | 0.055 | 4.8 | 0.362 | 338 | - | 4 |
| 1329.01 | 5.1611 | 1074 | 24 | 86.6605 | 0.0057 | 33.19959 | 0.00088 | 50.6 | 1.7 | 0.02911 | 0.0008 | 0.027 | 0.042 | 2.5 | 0.198 | 457 | - | 4 |
| 1335.01 | 11.2408 | 1862 | 40 | 226.4189 | 0.0051 | 127.8295 | 0.0072 | 84 | 221 | 0.04 | 0.021 | 0.5 | 1.7 | 7.6 | 0.528 | 480 | - | 4 |
| 1336.01 | 3.6241 | 458 | 15 | 69.8593 | 0.0072 | 10.21869 | 0.00031 | 22 | 1.1 | 0.02046 | 0.00081 | 0.026 | 0.049 | 2.5 | 0.095 | 884 | - | 4 |
| 1337.01 | 1.8187 | 237 | 13 | 67.5043 | 0.0052 | 1.922787 | 0.000043 | 7 | 39 | 0.016 | 0.017 | 0.5 | 2.4 | 1.4 | 0.03 | 1177 | - | 4 |
| 1338.01 | 3.0650 | 229 | 16 | 111.0667 | 0.0059 | 3.223004 | 0.000091 | 8.24 | 0.36 | 0.01526 | 0.00058 | 0.0451 | - | 1.7 | 0.044 | 1195 | - | 4 |
| 1339.01 | 2.8821 | 267 | 15 | 67.9005 | 0.0063 | 4.16804 | 0.00011 | 8 | 66 | 0.016 | 0.025 | 0.7 | 2.2 | 2.0 | 0.052 | 1150 | - | 4 |
| 1341.01 | 2.9285 | 232 | 14 | 67.0276 | 0.0067 | 4.51434 | 0.00013 | 11.49 | 0.18 | 0.0133 | 0.00076 | 0.2746 | - | 1.6 | 0.055 | 1165 | - | 4 |
| 1342.01 | 3.1222 | 174 | 17 | 67.0578 | 0.0058 | 3.773707 | 0.000093 | 9.94 | 0.4 | 0.01364 | 0.00046 | 0.0572 | - | 1.6 | 0.049 | 1242 | - | 4 |
| 1344.01 | 2.6231 | 127 | 15 | 66.3603 | 0.0055 | 4.48761 | 0.0001 | 15.27 | 0.15 | 0.01191 | 0.0004 | 0.047 | - | 1.1 | 0.054 | 1021 | - | 4 |
| 1345.01 | 2.0859 | 48165 | 262 | 66.6039 | 0.00028 | 0.3230201 | 0.0000004 | 1.6 | 0.48 | 0.19676 | 0.00088 | - | - | 18.2 | 0.009 | 2237 | - | 4 |
| 1346.01 | 2.7042 | 52057 | 318 | 70.07146 | 0.0003 | 4.7081251 | 0.0000063 | 13.5 | 4 | 0.2291 | 0.0012 | 0.49 | 0.15 | 22.2 | 0.056 | 989 | - | 4 |
| 1348.01 | 2.9866 | 16949 | 93 | 70.5634 | 0.0011 | 7.702363 | 0.000036 | 22 | 6.6 | 0.1158 | 0.0013 | - | - | 7.7 | 0.075 | 676 | - | 4 |
| 1349.01 | 3.8745 | 22623 | 107 | 75.187 | 0.0013 | 16.6621 | 0.0001 | 26.2 | 7.9 | 0.1653 | 0.0036 | 0.72 | 0.22 | 19.6 | 0.133 | 783 | - | 4 |
| 1353.01 | 10.8183 | 13387 | 72 | 169.6578 | 0.0039 | 125.8649 | 0.003 | 115.5 | 1.4 | 0.1025 | 0.0011 | 0.029 | 0.025 | 18.1 | 0.52 | 469 | - | 4 |
| 1355.01 | 5.1862 | 3085 | 35 | 84.9328 | 0.0039 | 51.92904 | 0.0009 | 80.9 | 1.9 | 0.04945 | 0.00097 | 0.023 | 0.03 | 2.8 | 0.266 | 342 | - | 4 |
| 1360.01 | 4.5398 | 1494 | 25 | 97.298 | 0.0049 | 36.76944 | 0.00082 | 62 | 441 | 0.035 | 0.051 | 0.2 | 3.1 | 2.7 | 0.207 | 408 | - | 4 |
| 1360.02 | 2.3547 | 907 | 17 | 74.93 | 0.0045 | 14.58951 | 0.00029 | 39 | 361 | 0.029 | 0.054 | 0.6 | 2.9 | 2.3 | 0.112 | 555 | - | 4 |
| 1361.01 | 4.8039 | 1478 | 32 | 84.1831 | 0.0039 | 59.8791 | 0.0011 | 100.4 | 2.5 | 0.03422 | 0.00072 | 0.017 | 0.032 | 2.2 | 0.243 | 279 | - | 4 |
| 1363.01 | 2.3642 | 494 | 16 | 67.8975 | 0.0051 | 3.546626 | 0.000077 | 11.87 | 0.74 | 0.0202 | 0.00095 | 0.0643 | - | 1.8 | 0.046 | 1052 | - | 4 |
| 1364.01 | 4.0375 | 924 | 13 | 119.0465 | 0.0076 | 20.83299 | 0.00082 | 41.3 | 2.6 | 0.0274 | 0.0013 | 0.03 | 0.11 | 2.9 | 0.148 | 596 | - | 4 |
| 1364.02 | 2.8321 | 785 | 16 | 114.9246 | 0.0051 | 7.05519 | 0.00018 | 20.6 | 1.2 | 0.0263 | 0.0012 | 0.035 | 0.01 | 2.7 | 0.072 | 854 | - | 4 |
| 1366.01 | 4.5442 | 962 | 28 | 97.7735 | 0.0042 | 19.25493 | 0.0004 | 33.04 | 0.94 | 0.02751 | 0.00066 | 0.025 | 0.039 | 2.4 | 0.141 | 582 | - | 4 |
| 1367.01 | 0.9983 | 342 | 33 | 67.06 | 0.0014 | 0.5678602 | 0.0000036 | 5 | 31 | 0.015 | 0.018 | 0.2 | 2.9 | 1.2 | 0.013 | 1639 | - | 4 |
| 1369.01 | 2.7220 | 276 | 19 | 66.8903 | 0.0047 | 3.016111 | 0.00006 | 9.07 | 0.11 | 0.01642 | 0.00047 | 0.0586 | - | 1.8 | 0.042 | 1239 | - | 4 |
| 1370.01 | 2.3585 | 436 | 12 | 68.2586 | 0.0065 | 6.88354 | 0.00019 | 22.7 | 1.7 | 0.0198 | 0.0011 | 0 | 0.01 | 1.7 | 0.071 | 795 | - | 4 |
| 1372.01 | 10.4241 | 530 | 12 | 109.049 | 0.017 | 69.7118 | 0.0057 | 43 | 265 | 0.022 | 0.027 | 0.6 | 2.4 | 2.2 | 0.338 | 416 | - | 4 |
| 1375.01 | 4.7760 | 2569 | 61 | 139.9267 | 0.0022 | 321.2161 | 0.0031 | 207 | 62 | 0.141 | 0.017 | 1.14 | 0.34 | 17.9 | 0.958 | 300 | - | 4 |
| 1376.01 | 2.6400 | 297 | 21 | 117.2434 | 0.0038 | 7.13906 | 0.00014 | 21.28 | 0.8 | 0.01706 | 0.00051 | 0.0486 | - | 2.8 | 0.085 | 1337 | - | 4 |
| 1377.01 | 5.0697 | 193 | 12 | 67.557 | 0.012 | 11.29698 | 0.00059 | 17.91 | 0.96 | 0.01362 | 0.00066 | 0.03 | 0.014 | 1.6 | 0.102 | 868 | - | 4 |
| 1378.01 | 4.8532 | 237 | 23 | 66.3534 | 0.0055 | 19.30204 | 0.00049 | 30 | 181 | 0.014 | 0.018 | 0.3 | 2.8 | 1.3 | 0.13 | 590 | - | 4 |
| 1379.01 | 2.4909 | 162 | 20 | 69.8878 | 0.004 | 5.621614 | 0.000095 | 18.43 | 0.85 | 0.01196 | 0.00044 | 0.0058 | - | 0.8 | 0.062 | 790 | - | 4 |
| 1381.01 | 3.8584 | 64104 | 411 | 67.58585 | 0.0003 | 5.1174029 | 0.0000068 | 9.54 | 0.24 | 0.2588 | 0.0023 | 0.69 | 0.13 | 25.0 | 0.059 | 1004 | - | 4 |
| 1382.01 | 3.7694 | 49487 | 148 | 112.53263 | 0.00071 | 4.202347 | 0.000015 | 10.271 | 0.082 | 0.2003 | 0.0012 | 0.0001 | - | 24.1 | 0.053 | 1191 | - | 4 |
| 1383.01 | 3.0149 | 45922 | 317 | 69.14077 | 0.00033 | 3.2217873 | 0.0000046 | 6.74 | 0.49 | 0.2402 | 0.0067 | 0.82 | 0.18 | 28.4 | 0.044 | 1313 | - | 4 |
| 1384.01 | 1.6620 | 38858 | 415 | 66.04537 | 0.00016 | 0.62438 | 0.0000004 | 3.25 | 0.97 | 0.183 | 0.00054 | 0.191 | 0.057 | 18.6 | 0.014 | 1818 | - | 4 |
| 1385.01 | 4.0744 | 49147 | 772 | 72.08453 | 0.00018 | 18.610068 | 0.000015 | 42 | 13 | 0.19775 | 0.0003 | - | - | 15.1 | 0.138 | 581 | - | 4 |
| 1386.01 | 2.0583 | 16933 | 537 | 66.80499 | 0.00016 | 1.1375241 | 0.0000008 | 4.7 | 1.4 | 0.11714 | 0.00024 | 0.06 | 0.018 | 12.9 | 0.022 | 1685 | - | 4 |
| 1387.01 | 4.8792 | 59843 | 1404 | 85.56423 | 0.00014 | 23.799979 | 0.000015 | 36 | 11 | 0.2885 | 0.0017 | 0.73 | 0.22 | 31.1 | 0.165 | 615 | - | 4 |
| 1389.01 | 2.2375 | 21642 | 162 | 67.79932 | 0.00052 | 4.3500866 | 0.0000096 | 10.3 | 2.5 | 0.176 | 0.016 | 0.88 | 0.26 | 16.3 | 0.053 | 1021 | - | 4 |
| 1390.01 | 1.5474 | 9692 | 124 | 67.50781 | 0.0005 | 1.7441009 | 0.0000038 | 8.1 | 4 | 0.094 | 0.01 | 0.59 | 0.68 | 6.9 | 0.024 | 976 | - | 4 |
| 1391.01 | 1.9404 | 4574 | 90 | 66.8021 | 0.00089 | 7.981177 | 0.000031 | 17.9 | 5.4 | 0.079 | 0.0016 | 0.84 | 0.25 | 8.3 | 0.08 | 921 | - | 4 |
| 1395.01 | 1.5514 | 1579 | 20 | 114.2969 | 0.0029 | 6.230288 | 0.000089 | 35.1 | 2.7 | 0.0378 | 0.0022 | 0.0253 | - | 2.5 | 0.064 | 718 | - | 4 |
| 1396.01 | 3.1586 | 844 | 21 | 116.9867 | 0.0045 | 6.62641 | 0.00015 | 16.7 | 0.72 | 0.02646 | 0.00092 | 0.034 | 0.01 | 2.5 | 0.07 | 884 | - | 4 |
| 1396.02 | 2.5938 | 368 | 11 | 113.2709 | 0.0075 | 3.70128 | 0.00014 | 11.85 | 0.85 | 0.0194 | 0.0011 | 0.0099 | - | 1.9 | 0.048 | 1067 | - | 4 |
| 1400.01 | 3.3141 | 783 | 70 | 75.3301 | 0.0015 | 9.414683 | 0.000063 | 22.76 | 0.32 | 0.02523 | 0.00027 | 0.029 | 0.032 | 2.9 | 0.091 | 932 | - | 4 |
| 1401.01 | 2.6369 | 261 | 64 | 66.8031 | 0.0014 | 0.5667194 | 0.0000035 | 1.773 | 0.018 | 0.01408 | 0.00018 | 0 | 0.025 | 2.0 | 0.014 | 2820 | - | 4 |
| 1402.01 | 3.8080 | 461 | 14 | 70.1876 | 0.0087 | 7.13962 | 0.00025 | 14.72 | 0.7 | 0.02196 | 0.00087 | 0.0547 | - | 1.7 | 0.073 | 792 | - | 4 |
| 1403.01 | 3.0231 | 824 | 15 | 74.7225 | 0.0063 | 18.75409 | 0.00051 | 40 | 408 | 0.028 | 0.057 | 0.6 | 3.1 | 2.0 | 0.117 | 444 | - | 4 |
| 1404.01 | 3.0415 | 467 | 11 | 68.2404 | 0.0092 | 6.66205 | 0.00026 | 13 | 98 | 0.024 | 0.036 | 0.7 | 2.4 | 1.5 | 0.06 | 603 | - | 4 |
| 1405.01 | 4.0389 | 536 | 11 | 119.3 | 0.0097 | 11.41969 | 0.00057 | 23.5 | 1.5 | 0.023 | 0.0012 | 0.085 | 0.01 | 2.1 | 0.102 | 767 | - | 4 |
| 1406.01 | 3.9013 | 522 | 27 | 69.9238 | 0.0047 | 11.36121 | 0.00022 | 16 | 59 | 0.023 | 0.017 | 0.7 | 1.6 | 2.5 | 0.102 | 828 | - | 4 |
| 1407.01 | 2.6597 | 217 | 14 | 67.2441 | 0.0064 | 1.387195 | 0.000039 | 4 | 52 | 0.01 | 0.027 | 0.5 | 3.7 | 0.7 | 0.024 | 1245 | - | 4 |
| 1408.01 | 3.2050 | 565 | 16 | 78.4231 | 0.0062 | 14.53451 | 0.00039 | 36.3 | 2 | 0.02179 | 0.00095 | 0.021 | 0.01 | 1.8 | 0.098 | 492 | - | 4 |
| 1409.01 | 2.4035 | 731 | 19 | 66.4929 | 0.0045 | 16.56074 | 0.00031 | 25.3 | 7.6 | 0.0307 | 0.0014 | 0.83 | 0.25 | 3.7 | 0.13 | 719 | - | 4 |
| 1410.01 | 4.5646 | 597 | 21 | 83.2427 | 0.0062 | 15.74997 | 0.00043 | 26.6 | 2 | 0.01582 | 0.00097 | 0.057 | 0.017 | 1.6 | 0.126 | 709 | - | 4 |
| 1412.01 | 8.7616 | 379 | 22 | 67.6756 | 0.0089 | 37.8118 | 0.0016 | 33.8 | 1 | 0.01737 | 0.0005 | 0.007 | 0.01 | 2.1 | 0.228 | 580 | - | 4 |
| 1413.01 | 7.9911 | 185 | 16 | 71.634 | 0.012 | 12.6449 | 0.00064 | 9 | 39 | 0.014 | 0.012 | 0.7 | 1.8 | 1.5 | 0.107 | 730 | - | 4 |
| 1415.01 | 1.7411 | 34162 | 250 | 66.80918 | 0.00027 | 0.3129426 | 0.0000004 | 1.75 | 0.53 | 0.16573 | 0.00082 | - | - | 18.8 | 0.009 | 2812 | - | 4 |
| 1416.01 | 4.3558 | 26075 | 154 | 68.335 | 0.00087 | 2.4958009 | 0.0000094 | 5.026097 | 0.000027 | 0.145 | 0.043 | 0.0081 | - | 16.9 | 0.037 | 1366 | - | 4 |
| 1419.01 | 1.3238 | 2118 | 95 | 111.25851 | 0.00052 | 1.3361033 | 0.0000035 | 4.1 | 1.3 | 0.0529 | 0.0035 | 0.9 | 0.3 | 5.8 | 0.024 | 1674 | - | 4 |
| 1422.01 | 2.0371 | 1361 | 26 | 68.9225 | 0.0027 | 5.841618 | 0.000071 | 17 | 94 | 0.038 | 0.041 | 0.7 | 2 | 3.1 | 0.051 | 627 | - | 4 |
| 1422.02 | 2.9373 | 1574 | 20 | 66.6483 | 0.0042 | 19.85037 | 0.00038 | 57.3 | 2.8 | 0.0358 | 0.0013 | 0.02 | 0.044 | 2.9 | 0.116 | 416 | - | 4 |
| 1422.03 | 1.5785 | 461 | 10 | 67.7522 | 0.0057 | 3.621387 | 0.000093 | 18.6 | 1.6 | 0.0247 | 0.0016 | 0.0002 | - | 2.0 | 0.037 | 736 | - | 4 |
| 1423.01 | 4.0874 | 4570 | 30 | 83.5949 | 0.0035 | 124.4198 | 0.0019 | 277.7 | 9.5 | 0.0599 | 0.0015 | 0.031 | 0.062 | 4.3 | 0.475 | 274 | - | 4 |
| 1424.01 | 1.5471 | 486 | 35 | 66.4598 | 0.0018 | 1.2195667 | 0.0000092 | 7.26 | 0.22 | 0.02292 | 0.00049 | 0.0673 | - | 1.6 | 0.02 | 1151 | - | 4 |
| 1425.01 | 2.1991 | 514 | 29 | 66.7336 | 0.0027 | 2.053893 | 0.000024 | 5 | 59 | 0.012 | 0.028 | 0.8 | 2.5 | 0.8 | 0.031 | 1106 | - | 4 |
| 1426.01 | 6.7799 | 924 | 29 | 104.2748 | 0.0046 | 38.87641 | 0.00096 | 46.5 | 1 | 0.02863 | 0.00056 | 0.017 | 0.01 | 3.3 | 0.231 | 551 | - | 4 |
| 1426.02 | 5.2850 | 4370 | 88 | 58.0562 | 0.0022 | 74.91443 | 0.00065 | 86 | 17 | 0.0673 | 0.0028 | 0.79 | 0.32 | 7.7 | 0.357 | 443 | - | 4 |
| 1426.03 | 4.8985 | 4322 | 76 | 157.6733 | 0.0019 | 150.0341 | 0.0015 | 103 | 31 | 0.314 | 0.001 | 1.53 | 0.46 | 36.1 | 0.568 | 352 | - | 4 |
| 1427.01 | 1.7776 | 564 | 18 | 66.1119 | 0.0036 | 2.613011 | 0.000041 | 11.48 | 0.78 | 0.02063 | 0.00098 | 0.0377 | - | 1.5 | 0.031 | 830 | - | 4 |
| 1428.01 | 1.4407 | 476 | 60 | 66.90506 | 0.00098 | 0.9278604 | 0.000004 | 5.068 | 0.098 | 0.02028 | 0.00034 | 0.017 | - | 1.9 | 0.018 | 1439 | - | 4 |
| 1429.01 | 10.1791 | 2462 | 33 | 185.7365 | 0.0056 | 205.9317 | 0.0078 | 161.6 | 3.4 | 0.04413 | 0.00087 | 0.008 | 0.074 | 4.2 | 0.69 | 276 | - | 4 |
| 1430.01 | 2.3132 | 1055 | 22 | 85.6439 | 0.0034 | 10.4753 | 0.00017 | 35 | 300 | 0.03 | 0.054 | 0.3 | 3.3 | 2.3 | 0.084 | 577 | - | 4 |
| 1432.01 | 3.4501 | 455 | 19 | 71.4089 | 0.0056 | 6.88589 | 0.00016 | 14 | 91 | 0.021 | 0.029 | 0.5 | 2.6 | 1.9 | 0.072 | 833 | - | 4 |
| 1433.01 | 3.6136 | 705 | 13 | 72.3786 | 0.0074 | 19.80752 | 0.0007 | 37 | 391 | 0.025 | 0.055 | 0.5 | 3.3 | 1.7 | 0.142 | 516 | - | 4 |
| 1434.01 | 2.1140 | 282 | 14 | 66.2056 | 0.005 | 2.34324 | 0.000051 | 8.4 | 0.14 | 0.01481 | 0.00076 | 0.1399 | - | 1.4 | 0.033 | 1084 | - | 4 |
| 1435.01 | 8.4693 | 519 | 26 | 79.5568 | 0.0073 | 40.7174 | 0.0014 | 36 | 174 | 0.02 | 0.02 | 0.3 | 2.5 | 1.7 | 0.234 | 454 | - | 4 |
| 1436.01 | 2.4834 | 230 | 16 | 68.0097 | 0.0049 | 2.508513 | 0.000055 | 8.09 | 0.4 | 0.01464 | 0.00063 | 0.0331 | - | 1.5 | 0.037 | 1240 | - | 4 |
| 1437.01 | 3.2369 | 287 | 8.8 | 66.896 | 0.011 | 7.01738 | 0.00034 | 11 | 90 | 0.02 | 0.031 | 0.8 | 2.2 | 2.2 | 0.073 | 951 | - | 4 |
| 1438.01 | 5.6224 | 199 | 25 | 69.6485 | 0.0061 | 6.91126 | 0.00018 | 9.59 | 0.27 | 0.01294 | 0.00032 | 0.021 | 0.033 | 1.5 | 0.072 | 939 | - | 4 |
| †1439.01 | 23.0602 | 1176 | 72 | 110.8684 | 0.0034 | 1155 | 26 | 393.3 | 9 | 0.03041 | - | 0.0686 | - | 4.1 | 2.235 | 195 | - | 4 |
| 1440.01 | 3.0898 | 266 | 11 | 69.0574 | 0.009 | 7.19298 | 0.00027 | 19.1 | 1.3 | 0.01675 | 0.00098 | 0.032 | 0.01 | 1.4 | 0.074 | 851 | - | 4 |
| 1441.01 | 3.1188 | 283 | 12 | 67.646 | 0.0087 | 8.50695 | 0.00031 | 12 | 86 | 0.02 | 0.025 | 0.8 | 1.7 | 2.3 | 0.083 | 871 | - | 4 |
| 1442.01 | 1.5890 | 127 | 36 | 67.1309 | 0.0016 | 0.6693109 | 0.0000049 | 2.54 | 0.76 | 0.01096 | 0.00026 | 0.46 | 0.14 | 1.6 | 0.015 | 2242 | - | 4 |
| 1444.01 | 6.4860 | 413 | 19 | 73.2196 | 0.0089 | 44.9279 | 0.0018 | 34 | 117 | 0.021 | 0.014 | 0.8 | 1.4 | 3.0 | 0.259 | 611 | - | 4 |
| 1445.01 | 4.6119 | 100 | 18 | 68.682 | 0.0065 | 7.16875 | 0.0002 | 12.62 | 0.49 | 0.00943 | 0.00033 | 0.027 | 0.01 | 1.2 | 0.076 | 1099 | - | 4 |
| 1446.01 | 2.2971 | 44429 | 235 | 66.78619 | 0.00037 | 1.2277586 | 0.0000019 | 4.6 | 1.4 | 0.19414 | 0.00096 | 0.168 | 0.05 | 22.7 | 0.023 | 1790 | - | 4 |
| 1447.01 | 7.1143 | 148151 | 303 | 94.30528 | 0.00069 | 40.24666 | 0.00013 | 52 | 16 | 0.55 | 0.27 | 0.97 | 0.29 | 69.3 | 0.239 | 594 | - | 4 |
| 1447.02 | 5.7569 | 15260 | 205 | 66.63923 | 0.0008 | 2.279999 | 0.0000078 | 3.4 | 1 | 0.10971 | 0.00048 | - | - | 13.8 | 0.035 | 1553 | - | 4 |
| 1448.01 | 2.9237 | 44947 | 86 | 67.1094 | 0.0012 | 2.486588 | 0.000013 | 8.24 | 0.13 | 0.1899 | 0.0021 | 0.025 | 0.028 | 23.5 | 0.037 | 1383 | - | 4 |
| 1449.01 | 3.8021 | 49724 | 929 | 70.1369 | 0.00015 | 10.9802481 | 0.0000072 | 22.25 | 0.31 | 0.228 | 0.0011 | 0.739 | 0.092 | 24.9 | 0.099 | 801 | - | 4 |
| 1450.01 | 4.0904 | 18255 | 225 | 66.98404 | 0.00057 | 2.1446308 | 0.0000052 | 4.5 | 1.4 | 0.1201 | 0.00052 | - | - | 17.7 | 0.034 | 1663 | - | 4 |
| 1451.01 | 5.5358 | 78850 | 1241 | 92.74682 | 0.00017 | 27.322068 | 0.000024 | 47.197 | 0.066 | 0.25125 | 0.00024 | 0.0072 | - | 17.5 | 0.174 | 459 | - | 4 |
| 1452.01 | 2.3632 | 13235 | 76 | 66.2764 | 0.0012 | 1.1522207 | 0.0000057 | 3.01 | 0.9 | 0.1211 | 0.0021 | 0.67 | 0.2 | 22.0 | 0.023 | 2562 | - | 4 |
| 1454.01 | 9.6294 | 11369 | 35 | 70.3035 | 0.00031 | 121.59089 | 0.00017 | 60.88 | 0.21 | 0.10541 | 0.00026 | 0.824 | 0.039 | 9.2 | 0.487 | 327 | - | 4 |
| 1459.01 | 1.0700 | 4809 | 116 | 66.11011 | 0.00041 | 0.692023 | 0.0000013 | 3.4 | 1 | 0.0754 | 0.001 | 0.72 | 0.22 | 6.9 | 0.013 | 1435 | - | 4 |
| 1461.01 | 2.1651 | 5706 | 68 | 73.7078 | 0.0011 | 7.946693 | 0.000038 | 23.9 | 7.2 | 0.073 | 0.001 | 0.43 | 0.13 | 5.1 | 0.071 | 625 | - | 4 |
| †1463.01 | 11.9788 | 22843 | 1036 | 77.08517 | 0.00018 | 253.0083 | 0.0055 | 188.3 | 2.5 | 0.13679 | 0.00042 | 0.22 | 0.13 | 16.3 | 0.795 | 311 | - | 4 |
| 1465.01 | 1.6975 | 4866 | 61 | 68.5853 | 0.001 | 9.771425 | 0.000044 | 30.9 | 9.3 | 0.0728 | 0.0014 | 0.66 | 0.2 | 4.9 | 0.089 | 653 | - | 4 |
| 1468.01 | 6.2072 | 1392 | 65 | 68.8407 | 0.0025 | 8.480842 | 0.000093 | 10.89 | 0.12 | 0.03339 | 0.00034 | 0.019 | 0.02 | 3.7 | 0.083 | 873 | - | 4 |
| 1472.01 | 6.5891 | 4430 | 93 | 94.0088 | 0.0018 | 85.35029 | 0.00072 | 104.42 | 0.89 | 0.05905 | 0.00044 | 0.03 | 0.022 | 3.6 | 0.37 | 295 | - | 4 |
| 1474.01 | 5.8511 | 4548 | 125 | 129.0525 | 0.0014 | 69.74538 | 0.00075 | 149.2 | 1.4 | 0.06164 | 0.00046 | 0.001 | 0.028 | 11.3 | 0.356 | 622 | - | 4 |
| 1475.01 | 1.5907 | 816 | 18 | 111.9412 | 0.0033 | 1.609323 | 0.000024 | 7.5 | 2.3 | 0.0259 | 0.0013 | 0.106 | 0.032 | 1.8 | 0.022 | 969 | - | 4 |
| 1475.02 | 3.0303 | 1197 | 15 | 118.4738 | 0.0059 | 9.51248 | 0.00025 | 24.8 | 1.5 | 0.0317 | 0.0015 | 0.006 | 0.01 | 2.2 | 0.073 | 532 | - | 4 |
| 1476.01 | 5.7031 | 2770 | 24 | 155.2876 | 0.0058 | 56.3647 | 0.0024 | 80.6 | 2.7 | 0.0467 | 0.0013 | 0.001 | 0.01 | 4.6 | 0.286 | 413 | - | 4 |
| †1477.01 | 8.3090 | 15017 | 84 | 161.9327 | 0.0016 | 400 | - | 315.54302 | - | 0.12163 | - | 0.7224 | - | 9.4 | 1.044 | 195 | - | 4 |
| 1478.01 | 8.0856 | 3028 | 150 | 132.4853 | 0.001 | 76.13333 | 0.00047 | 75.54 | 0.34 | 0.04886 | 0.00022 | 0.002 | 0.017 | 3.7 | 0.348 | 341 | - | 4 |
| 1480.01 | 3.9475 | 1498 | 26 | 75.3196 | 0.0047 | 20.38139 | 0.0004 | 41.2 | 1.4 | 0.03472 | 0.00094 | 0 | 0.014 | 2.5 | 0.138 | 475 | - | 4 |
| 1486.01 | 7.0829 | 8289 | 75 | 96.8928 | 0.0021 | 254.5598 | 0.003 | 206.4403 | 0.0024 | 0.09359 | - | 0.7941 | - | 8.5 | 0.796 | 256 | - | 4 |
| 1486.02 | 5.4993 | 849 | 18 | 79.6407 | 0.0083 | 30.1839 | 0.0012 | 43.3 | 1.8 | 0.0263 | 0.00098 | 0.038 | 0.025 | 2.4 | 0.192 | 521 | - | 4 |
| 1488.01 | 1.6191 | 827 | 24 | 67.6076 | 0.0026 | 3.949663 | 0.000044 | 20.809 | 0.089 | 0.0273 | 0.00075 | 0.0355 | - | 2.6 | 0.048 | 941 | - | 4 |
| 1489.01 | 3.8636 | 982 | 20 | 73.4823 | 0.0057 | 16.00474 | 0.00042 | 32.5 | 1.4 | 0.02796 | 0.00095 | 0.028 | 0.048 | 2.4 | 0.12 | 564 | - | 4 |
| 1494.01 | 2.9005 | 767 | 13 | 69.6943 | 0.0092 | 8.19571 | 0.00026 | 17 | 162 | 0.028 | 0.054 | 0.7 | 2.7 | 2.4 | 0.073 | 659 | - | 4 |
| 1495.01 | 5.0903 | 803 | 26 | 70.5618 | 0.0057 | 15.5947 | 0.00037 | 24.32 | 0.74 | 0.02528 | 0.00066 | 0.024 | 0.037 | 2.6 | 0.124 | 693 | - | 4 |
| 1498.01 | 3.9280 | 502 | 19 | 71.7503 | 0.0061 | 5.83375 | 0.00016 | 11.59 | 0.46 | 0.02101 | 0.00071 | 0.0229 | - | 1.8 | 0.064 | 917 | - | 4 |
| 1499.01 | 4.2960 | 737 | 41 | 73.59 | 0.0033 | 14.16394 | 0.00019 | 21 | 66 | 0.026 | 0.016 | 0.6 | 1.7 | 3.0 | 0.115 | 708 | - | 4 |
| 1501.01 | 2.1661 | 442 | 16 | 67.0247 | 0.0048 | 2.61709 | 0.000053 | 8.9 | 2.7 | 0.01899 | 0.00098 | 0.103 | 0.031 | 1.7 | 0.035 | 1003 | - | 4 |
| 1502.01 | 1.5056 | 448 | 19 | 66.1275 | 0.0032 | 1.876417 | 0.000025 | 10.2 | 0.48 | 0.02283 | 0.0008 | 0.0573 | - | 1.9 | 0.029 | 1131 | - | 4 |
| 1503.01 | 10.8923 | 2445 | 36 | 71.289 | 0.0057 | 150.2421 | 0.005 | 110 | 2.2 | 0.04384 | 0.00077 | 0.001 | 0.059 | 2.7 | 0.535 | 242 | - | 4 |
| 1505.01 | 3.1205 | 450 | 14 | 70.3243 | 0.007 | 5.03266 | 0.00015 | 13.09 | 0.63 | 0.02214 | 0.00087 | 0.074 | - | 2.2 | 0.058 | 999 | - | 4 |
| 1506.01 | 6.3551 | 850 | 19 | 72.5005 | 0.0081 | 40.4291 | 0.0015 | 50.8 | 1.9 | 0.02622 | 0.00088 | 0.019 | 0.014 | 2.3 | 0.232 | 462 | - | 4 |
| 1507.01 | 6.0774 | 639 | 17 | 68.0439 | 0.0094 | 21.36041 | 0.00093 | 27.6 | 1.2 | 0.02269 | 0.00087 | 0.023 | 0.01 | 2.3 | 0.154 | 634 | - | 4 |
| 1508.01 | 4.5912 | 735 | 15 | 69.4735 | 0.0085 | 22.04698 | 0.00079 | 38.888 | 0.0057 | 0.03 | 0.53 | 0.022 | 0.014 | 1.6 | 0.152 | 483 | - | 4 |
| 1509.01 | 14.3572 | 650 | 64 | 86.043 | 0.0043 | 49.6443 | 0.001 | 26.97 | 0.24 | 0.0226 | 0.00022 | 0.02 | 0.017 | 3.2 | 0.276 | 580 | - | 4 |
| 1510.01 | 1.1235 | 509 | 19 | 111.3478 | 0.0025 | 0.839992 | 0.00001 | 7 | 21 | 0.028 | 0.018 | 0.2 | 2 | 1.8 | 0.016 | 1275 | - | 4 |
| 1511.01 | 2.2855 | 360 | 23 | 66.2205 | 0.0036 | 2.578886 | 0.000039 | 9.04 | 0.36 | 0.01828 | 0.00056 | 0.0019 | - | 1.6 | 0.037 | 1150 | - | 4 |
| 1512.01 | 2.0692 | 701 | 19 | 70.867 | 0.0039 | 9.04184 | 0.00015 | 25 | 167 | 0.028 | 0.037 | 0.7 | 2.1 | 2.1 | 0.082 | 652 | - | 4 |
| 1515.01 | 1.5355 | 331 | 25 | 67.7191 | 0.0025 | 1.937029 | 0.000021 | 9.96 | 0.4 | 0.01774 | 0.00056 | 0.014 | 0.044 | 1.3 | 0.025 | 927 | - | 4 |
| 1516.01 | 5.1817 | 676 | 22 | 72.5445 | 0.006 | 20.55453 | 0.00052 | 31 | 237 | 0.024 | 0.037 | 0.1 | 3.3 | 2.3 | 0.151 | 656 | - | 4 |
| 1517.01 | 6.0046 | 1010 | 37 | 84.8066 | 0.0041 | 40.06897 | 0.00076 | 53.1 | 1.1 | 0.02823 | 0.00051 | 0.029 | 0.036 | 3.3 | 0.235 | 544 | - | 4 |
| 1518.01 | 5.2245 | 554 | 17 | 84.8435 | 0.008 | 27.50658 | 0.00096 | 29 | 158 | 0.023 | 0.025 | 0.7 | 2 | 2.2 | 0.18 | 541 | - | 4 |
| 1519.01 | 2.2369 | 349 | 11 | 66.8256 | 0.0067 | 5.1445 | 0.00015 | 17 | 127 | 0.019 | 0.03 | 0.4 | 3 | 1.5 | 0.056 | 785 | - | 4 |
| 1520.01 | 3.2262 | 542 | 21 | 71.162 | 0.0048 | 18.45865 | 0.00037 | 40 | 328 | 0.022 | 0.036 | 0.4 | 3.1 | 2.1 | 0.136 | 597 | - | 4 |
| 1521.01 | 4.0344 | 594 | 17 | 89.1984 | 0.0074 | 25.94116 | 0.00083 | 49 | 559 | 0.022 | 0.052 | 0.3 | 3.8 | 2.4 | 0.168 | 532 | - | 4 |
| 1522.01 | 5.0713 | 569 | 20 | 81.9293 | 0.0074 | 33.3857 | 0.0011 | 52.6 | 2.1 | 0.02135 | 0.00072 | 0.023 | 0.042 | 2.5 | 0.207 | 562 | - | 4 |
| 1523.01 | 5.3847 | 260 | 15 | 68.8682 | 0.01 | 8.47997 | 0.00038 | 7 | 28 | 0.019 | 0.013 | 0.8 | 1.4 | 2.2 | 0.082 | 835 | - | 4 |
| 1525.01 | 4.4884 | 219 | 22 | 66.5555 | 0.0058 | 7.71467 | 0.00019 | 12.78 | 0.4 | 0.01394 | 0.00037 | 0.006 | 0.032 | 2.1 | 0.081 | 1214 | - | 4 |
| 1526.01 | 3.0765 | 200 | 10 | 66.4632 | 0.0094 | 4.44448 | 0.00017 | 12.62 | 0.87 | 0.01487 | 0.00086 | 0.0388 | - | 1.3 | 0.054 | 1003 | - | 4 |
| 1527.01 | 5.6836 | 1117 | 18 | 95.8719 | 0.008 | 192.674 | 0.011 | 222 | 705 | 0.034 | 0.021 | 0.8 | 1.3 | 4.9 | 0.67 | 337 | - | 4 |
| 1528.01 | 1.5604 | 211 | 17 | 67.0984 | 0.0035 | 3.989558 | 0.000057 | 21.7 | 1.3 | 0.01431 | 0.00079 | 0.0163 | - | 0.8 | 0.046 | 746 | - | 4 |
| 1529.01 | 3.5244 | 301 | 15 | 80.7485 | 0.0074 | 17.97619 | 0.00056 | 36 | 320 | 0.017 | 0.03 | 0.5 | 3.1 | 1.7 | 0.138 | 699 | - | 4 |
| 1530.01 | 3.5406 | 260 | 20 | 72.7821 | 0.0048 | 12.98486 | 0.00028 | 16 | 70 | 0.017 | 0.014 | 0.8 | 1.4 | 2.0 | 0.112 | 819 | - | 4 |
| 1531.01 | 2.3874 | 171 | 20 | 69.5839 | 0.0042 | 5.69927 | 0.0001 | 18 | 103 | 0.012 | 0.014 | 0.2 | 2.8 | 1.4 | 0.064 | 1049 | - | 4 |
| 1532.01 | 4.9255 | 208 | 18 | 80.4987 | 0.0085 | 18.11417 | 0.00067 | 17 | 70 | 0.015 | 0.011 | 0.8 | 1.5 | 2.0 | 0.141 | 816 | - | 4 |
| 1533.01 | 3.0841 | 155 | 16 | 67.8854 | 0.006 | 6.24151 | 0.00016 | 15 | 99 | 0.012 | 0.016 | 0.3 | 2.9 | 1.4 | 0.069 | 1062 | - | 4 |
| 1534.01 | 5.9455 | 170 | 16 | 79.9832 | 0.0099 | 20.42238 | 0.00086 | 16 | 65 | 0.014 | 0.01 | 0.8 | 1.4 | 1.6 | 0.152 | 730 | - | 4 |
| 1535.01 | 6.6899 | 379 | 18 | 121.5617 | 0.0078 | 70.699 | 0.0029 | 66 | 314 | 0.019 | 0.018 | 0.6 | 2 | 2.4 | 0.347 | 485 | - | 4 |
| 1536.01 | 2.9231 | 65 | 13 | 66.5632 | 0.0071 | 3.74438 | 0.00011 | 10.32 | 0.47 | 0.00854 | 0.00036 | 0.0514 | - | 1.0 | 0.049 | 1209 | - | 4 |
| 1537.01 | 6.1963 | 64 | 14 | 69.678 | 0.012 | 10.19201 | 0.00049 | 12.83 | 0.34 | 0.00785 | 0.0003 | 0.029 | 0.01 | 1.0 | 0.094 | 922 | - | 4 |
| 1539.01 | 4.1246 | 72512 | 205 | 66.8696 | 0.00064 | 2.8194478 | 0.0000078 | 5.63 | 0.25 | 0.2574 | 0.0036 | 0.53 | 0.21 | 32.7 | 0.04 | 1313 | - | 4 |
| 1540.01 | 2.9953 | 48706 | 207 | 111.21359 | 0.00049 | 1.2078522 | 0.0000029 | 3.8 | 1.1 | 0.1971 | 0.0011 | - | - | 19.3 | 0.022 | 1516 | - | 4 |
| 1541.01 | 3.2386 | 49845 | 549 | 66.6509 | 0.00021 | 2.37928 | 0.0000021 | 6.760957 | 0.000011 | 0.2 | 0.013 | 0.0001 | - | 20.6 | 0.036 | 1394 | - | 4 |
| 1543.01 | 4.7888 | 27498 | 349 | 69.02882 | 0.00042 | 3.9643332 | 0.000007 | 7.3 | 0.021 | 0.14827 | 0.00032 | 0.0016 | - | 14.9 | 0.05 | 1102 | - | 4 |
| 1546.01 | 1.7236 | 12880 | 105 | 66.9341 | 0.00064 | 0.9175586 | 0.0000026 | 4.14 | 0.042 | 0.10122 | 0.00077 | 0.026 | 0.022 | 5.8 | 0.018 | 1309 | - | 4 |
| 1548.01 | 2.1552 | 5437 | 114 | 68.06092 | 0.00069 | 2.1393302 | 0.0000063 | 8.1 | 2.4 | 0.06561 | 0.0006 | - | - | 4.9 | 0.029 | 986 | - | 4 |
| 1549.01 | 3.7446 | 12205 | 254 | 67.22398 | 0.00053 | 29.481036 | 0.000071 | 36 | 11 | 0.26867 | 0.00068 | 1.24 | 0.37 | 32.8 | 0.189 | 579 | - | 4 |
| 1553.01 | 6.6202 | 5978 | 78 | 89.3039 | 0.0022 | 52.75935 | 0.00051 | 65.33 | 0.67 | 0.06878 | 0.0006 | 0 | 0.022 | 7.1 | 0.282 | 481 | - | 4 |
| 1557.01 | 1.9833 | 1794 | 54 | 66.9645 | 0.0014 | 3.295711 | 0.000019 | 13 | 58 | 0.039 | 0.035 | 0.4 | 2.3 | 5.2 | 0.043 | 1123 | - | 4 |
| 1560.01 | 2.7667 | 4350 | 95 | 91.2984 | 0.0011 | 31.56915 | 0.00016 | 39 | 12 | 0.24781 | 0.00089 | 1.35 | 0.41 | 29.6 | 0.2 | 580 | - | 4 |
| 1561.01 | 2.0755 | 2226 | 34 | 115.081 | 0.002 | 9.085928 | 0.000088 | 19.1 | 5.7 | 0.0521 | 0.0016 | 0.78 | 0.23 | 5.6 | 0.087 | 836 | - | 4 |
| 1564.01 | 5.3239 | 3166 | 70 | 82.2007 | 0.0022 | 53.44931 | 0.00052 | 80.68 | 0.97 | 0.05007 | 0.0005 | 0.0032 | - | 3.1 | 0.275 | 360 | - | 4 |
| 1569.01 | 2.7366 | 1192 | 21 | 76.0275 | 0.0042 | 13.75242 | 0.00026 | 41.4 | 2 | 0.0314 | 0.0012 | 0.025 | 0.053 | 2.5 | 0.103 | 543 | - | 4 |
| 1573.01 | 3.4077 | 1891 | 63 | 89.521 | 0.0015 | 24.80762 | 0.00017 | 61.33 | 0.92 | 0.03919 | 0.00046 | 0.021 | 0.025 | 3.8 | 0.17 | 589 | - | 4 |
| 1574.01 | 11.5744 | 4848 | 134 | 98.1547 | 0.0021 | 114.7316 | 0.001 | 80.46 | 0.4 | 0.06184 | 0.0003 | 0 | 0.022 | 5.8 | 0.465 | 331 | - | 4 |
| 1576.01 | 2.7055 | 840 | 37 | 76.0467 | 0.0022 | 10.41565 | 0.00011 | 29 | 163 | 0.026 | 0.03 | 0.3 | 2.7 | 3.2 | 0.095 | 828 | - | 4 |
| 1577.01 | 1.7039 | 558 | 13 | 66.7309 | 0.005 | 2.806213 | 0.00006 | 11.2 | 3.4 | 0.022 | 0.0015 | 0.252 | 0.076 | 1.5 | 0.032 | 810 | - | 4 |
| 1581.01 | 10.6245 | 629 | 15 | 70.537 | 0.017 | 29.5511 | 0.0022 | 16 | 60 | 0.025 | 0.018 | 0.7 | 1.7 | 2.3 | 0.186 | 499 | - | 4 |
| 1582.01 | 4.7382 | 3809 | 47 | 79.9795 | 0.0025 | 186.3827 | 0.0019 | 210 | 63 | 0.0636 | 0.0012 | 0.64 | 0.19 | 4.5 | 0.626 | 240 | - | 4 |
| 1583.01 | 4.5321 | 482 | 18 | 70.2059 | 0.0072 | 8.04725 | 0.00025 | 13.64 | 0.58 | 0.01984 | 0.00072 | 0.03 | 0.01 | 1.8 | 0.079 | 797 | - | 4 |
| 1584.01 | 2.3611 | 553 | 14 | 70.2416 | 0.0061 | 5.87084 | 0.00015 | 23.3 | 1.5 | 0.0248 | 0.0011 | 0.0676 | - | 1.9 | 0.054 | 667 | - | 4 |
| 1585.01 | 4.9613 | 833 | 21 | 72.7384 | 0.0065 | 19.1797 | 0.0006 | 30.7 | 0.47 | 0.02583 | 0.0008 | 0.062 | 0.014 | 2.2 | 0.14 | 573 | - | 4 |
| 1586.01 | 2.0622 | 607 | 25 | 68.4635 | 0.003 | 6.991262 | 0.000091 | 20 | 125 | 0.024 | 0.03 | 0.7 | 2.2 | 1.7 | 0.066 | 647 | - | 4 |
| 1587.01 | 3.1788 | 4196 | 35 | 92.0623 | 0.003 | 52.97102 | 0.00077 | 68 | 20 | 0.0795 | 0.0067 | 0.88 | 0.26 | 7.4 | 0.269 | 393 | - | 4 |
| 1588.01 | 1.5328 | 499 | 20 | 68.5461 | 0.003 | 3.517485 | 0.000045 | 19.21441 | 0.00025 | 0 | 1.8 | 0.0582 | - | 1.5 | 0.037 | 751 | - | 4 |
| 1589.01 | 4.2548 | 440 | 22 | 71.7748 | 0.0064 | 8.72548 | 0.00024 | 16.32 | 0.57 | 0.01933 | 0.00058 | 0.017 | 0.01 | 2.2 | 0.085 | 895 | - | 4 |
| 1589.02 | 4.0724 | 456 | 19 | 68.2065 | 0.0061 | 12.88195 | 0.00033 | 24.9 | 1 | 0.01983 | 0.00068 | 0.011 | 0.01 | 2.3 | 0.11 | 787 | - | 4 |
| 1590.01 | 3.6462 | 779 | 13 | 134.2278 | 0.0062 | 25.78004 | 0.00084 | 34 | 106 | 0.029 | 0.017 | 0.8 | 1.3 | 2.8 | 0.163 | 494 | - | 4 |
| 1590.02 | 1.5004 | 331 | 11 | 110.8828 | 0.0051 | 2.355804 | 0.000058 | 13.15 | 0.92 | 0.0198 | 0.001 | 0.0393 | - | 1.9 | 0.033 | 1098 | - | 4 |
| 1591.01 | 2.4255 | 829 | 17 | 73.2918 | 0.0051 | 19.65703 | 0.00044 | 67.4 | 4 | 0.027 | 0.0013 | 0.029 | 0.01 | 1.5 | 0.134 | 434 | - | 4 |
| 1593.01 | 2.6164 | 600 | 12 | 115.4672 | 0.0075 | 9.69448 | 0.00035 | 18 | 152 | 0.026 | 0.041 | 0.8 | 2.2 | 2.5 | 0.09 | 794 | - | 4 |
| 1595.01 | 5.1212 | 786 | 21 | 75.0828 | 0.007 | 40.1088 | 0.0012 | 41 | 158 | 0.028 | 0.021 | 0.7 | 1.5 | 2.9 | 0.233 | 502 | - | 4 |
| 1596.01 | 2.7169 | 390 | 17 | 67.6846 | 0.0053 | 5.9236 | 0.00013 | 12 | 65 | 0.021 | 0.022 | 0.7 | 1.9 | 2.3 | 0.061 | 825 | - | 4 |
| 1596.02 | 4.3349 | 1185 | 16 | 71.6868 | 0.0073 | 105.3551 | 0.0039 | 205 | 2218 | 0.032 | 0.071 | 0.5 | 3.5 | 3.5 | 0.416 | 316 | - | 4 |
| 1597.01 | 4.6377 | 342 | 31 | 67.42 | 0.0042 | 7.79663 | 0.00014 | 13.1 | 0.31 | 0.01668 | 0.00035 | 0.019 | 0.032 | 2.1 | 0.08 | 1043 | - | 4 |
| 1598.01 | 5.6905 | 1120 | 37 | 76.8112 | 0.0039 | 56.4754 | 0.0011 | 78.7 | 1.6 | 0.02968 | 0.00055 | 0.002 | 0.025 | 3.0 | 0.292 | 437 | - | 4 |
| 1599.01 | 4.8627 | 484 | 17 | 72.9972 | 0.0075 | 20.42116 | 0.00068 | 19 | 56 | 0.023 | 0.013 | 0.8 | 1.2 | 2.5 | 0.149 | 636 | - | 4 |
| 1601.01 | 6.2071 | 286 | 20 | 66.7171 | 0.0079 | 10.35066 | 0.00036 | 11 | 54 | 0.016 | 0.016 | 0.5 | 2.2 | 1.5 | 0.093 | 726 | - | 4 |
| 1602.01 | 5.8532 | 263 | 12 | 70.223 | 0.013 | 9.97788 | 0.00058 | 8 | 42 | 0.018 | 0.017 | 0.8 | 1.7 | 1.7 | 0.092 | 757 | - | 4 |
| 1603.01 | 2.7672 | 194 | 20 | 66.485 | 0.0046 | 3.02153 | 0.000059 | 5 | 21 | 0.016 | 0.013 | 0.8 | 1.4 | 1.4 | 0.042 | 1146 | - | 4 |
| 1605.01 | 1.4484 | 361 | 17 | 68.0478 | 0.0035 | 4.939157 | 0.000074 | 16 | 1 | 0.01413 | 0.00073 | 0.017 | 0.053 | 1.8 | 0.058 | 1126 | - | 4 |
| 1606.01 | 1.9798 | 241 | 18 | 66.2173 | 0.004 | 5.082573 | 0.000087 | 18.52 | 0.14 | 0.0131 | 0.00066 | 0.2899 | - | 1.2 | 0.058 | 908 | - | 4 |
| 1608.01 | 4.4061 | 206 | 17 | 71.7175 | 0.0074 | 9.1759 | 0.00028 | 15.85 | 0.66 | 0.01344 | 0.00049 | 0.047 | 0.01 | 1.6 | 0.089 | 943 | - | 4 |
| 1609.01 | 5.4671 | 376 | 20 | 102.5701 | 0.0077 | 41.6984 | 0.0015 | 31 | 74 | 0.0206 | 0.009 | 0.86 | 0.95 | 2.3 | 0.243 | 551 | - | 4 |

**Table 3.**
Notes to table of Planet Candidate Characteristics

Key:

| | |
|---|---|
| APO | Active pixel offset. The pixel that actually dims during a transit is offset from the position of the target star implying a background variable star. |
| Double star | There is within 4" an object evident in images that has not been ruled out as the source of the transit. |
| V-shaped | The transit light curve is "V" shaped, a possible indication of an eclipsing binary |
| Odd-even | Transit depths are alternately deeper and shallower, an indication of an eclipsing binary |
| Occultation | Evidence of secondary eclipse, implying possible EB or self luminous planet |
| SB1 | Spectroscopic binary. RV varies by over 1 km/s in low SNR reconnaissance spectra. Double lines not seen. |
| SB2 | Spectroscopic binary. Double lines seen in spectrum. |

| KOI | Note |
|---|---|
| 1.01 | TrES-2; O'Donovan et al. 2006 ApJ 650 L61 |
| 2.01 | HAT-P-7b; Kashyap et al. 2008 ApJ 687 1339 |
| 3.01 | HAT-P-11b; Dittman et al. 2009 ApJ 699 L48 |
| 4.01 | Rapid rotator Vrot = 40 km/s |
| 5.01 | Double star; 0.16" NE; delta_m=3.1 at 692 nm |
| 7.01 | Kepler-4b; Borucki et al. 2010 ApJ 713 L126 |
| 10.01 | Kepler-8b; Jenkins et al. 2010 ApJ 724 1108 |
| 12.01 | Marginally saturated |
| 13.01 | Double star; 0.8" E; delta_m=0.4 mag at 692 nm |
| 17.01 | Kepler-6b; Dunham et al. 2010 ApJ 713 L136 |
| 18.01 | Kepler-5b; Koch et al. 2010 ApJ 713 L131 |
| 44.01 | Variable transit depths |
| 51.01 | Light curve has spot/rotation modulation |
| 63.01 | Radial velocity variations have a dispersion 23 m/s |
| 64.01 | May be an F-M binary |
| 69.01 | Saturated. Double star; 0.05" NW; delta_mag=1.4 mag |
| 72.01 | Kepler-10b; Batalha et al. 2011 ApJ accepted |
| 97.01 | Kepler-7b; Latham et al. 2010 ApJ 713 L140 |
| 99.01 | Double star; 4" SE |
| 100.01 | Rapid rotator; Vrot=35 km/s |
| 102.01 | Double star; 2.5" SW |
| 112.01 | Double star; 0.09"; delta_m = 2.7 at 692 nm |
| 117.02 | Possible APO |
| 117.03 | Possible APO |
| 119.01 | Possible SB1 |
| 131.01 | Possible APO |
| 135.01 | Centroid analysis clean |
| 144.01 | KIC radius likely overestimated |
| 151.01 | V-shaped; may be triple system |
| 155.01 | Double star; 2" W |
| 157.01 | Kepler-11b; Lissauer et al. 2011 Nature accepted |
| 157.02 | Kepler-11c; Lissauer et al. 2011 Nature accepted |
| 157.03 | Kepler-11d; Lissauer et al. 2011 Nature accepted |
| 157.04 | Kepler-11e; Lissauer et al. 2011 Nature accepted |
| 157.05 | Kepler-11f; Lissauer et al. 2011 Nature accepted |
| 157.06 | Kepler-11g; Lissauer et al. 2011 Nature accepted |
| 179.01 | Double Star; 4” E |
| 180.01 | Variable star |
| 184.01 | Odd-even |
| 191.01 | Double star; 1” E |
| 191.02 | Possible APO; Double star 1” E |
| 208.01 | Variable star with possible spots |
| 225.01 | Possible ellipsoidal variations |
| 226.01 | Possible APO |
| 254.01 | 5% primary transit |
| 256.01 | KIC stellar radius may be too large |
| 258.01 | V-shaped; Multiple stars 1” and 2” E |
| 263.01 | Double star; 4” E |
| 268.01 | Multiple Stars: 2” S and 3” SE |
| 271.02 | Possible Odd-even |
| 274.01 | Possible APO |
| 284.01 | Double star; 0.9” E |
| 340.01 | Radius large; but log g may be too low in the KIC |
| 377.01 | Kepler-9b; Holman et al. 2010 Science 330 51 |
| 377.02 | Kepler-9c; Holman et al. 2010 Science 330 51 |
| 377.03 | Kepler-9d; Torres et al. 2010 arXiv:1008.4393 |
| 531.01 | Strange light curve; worth follow-up. |
| 607.01 | Odd light curve; worth follow-up. |
| 687.01 | Varying depths; possible encroaching companion. |
| 741.01 | Slight V shape and deep; no APO. |
| 774.01 | Possible occultation |
| 961.01 | Short duration, under sampled transit |
| 962.01 | Weak transit signal; possible low radius planet |
| 968.01 | Not convincing transit |
| 972.01 | Pulsating star |
| 973.01 | Possible APO; poor light curve |
| 976.01 | V-shaped; poor fit |
| 977.01 | Phase-correlated variations; saturated |
| 978.01 | Possibly spurious |
| 981.01 | V-shaped; saturated |
| 984.01 | V-shaped |
| 992.01 | Poor fit to light curve |
| 993.01 | Possible APO |
| 994.01 | Possible APO |
| 998.01 | Eccentric eclipsing binary |
| 1063.01 | V-shaped; large planet radius (2.1 RJ) |

**Table 4**
Very Probable False Positives

Key:

| | |
|---|---|
| $t_0$ | Time of a transit center based a linear fit to all observed transits and its uncertainty |
| Period | Average interval between transits based on a linear fit to all observed transits and uncertainty |
| APO | Active pixel offset. The pixel that actually dims during a transit is offset from the position of the target star implying a background variable star. |
| Double star | There is within 4” an object evident in images that has not been ruled out as the source of the transit. |
| V-shaped | The transit light curve is “V” shaped, a possible indication of an eclipsing binary |
| Odd-even | Transit depths are alternately deeper and shallower, an indication of an eclipsing binary |
| Occultation | Evidence of secondary eclipse, implying possible EB or self luminous planet |
| SB1 | Single-line eclipsing binary star. RV varies by over 1 km/s in low SNR reconnaissance spectra. Double lines not seen. |
| SB2 | Double-line eclipsing binary  Double lines seen in spectrum. |

| KOI | Kepler ID | $t_0$ (BJD-2454900) | Period (days) | Depth (ppm) | SNR | Comment |
|---|---|---|---|---|---|---|
| 6.01 | 3248033 | 66.69954 | 1.334103 | 397 | 97 | APO Binary |
| 8.01 | 5903312 | 54.70223 | 1.160154 | 399 | 41 | APO Binary |
| 9.01 | 11553706 | 68.06724 | 3.719813 | 3423 | 380 | APO Binary |
| 11.01 | 11913073 | 104.65803 | 3.748075 | 547 | 65 | APO Binary |
| 14.01 | 7684873 | 104.53055 | 2.947317 | 302 | 59 | Rapid rotator; Vrot = 90 km/s; Secondary eclipse |
| 15.01 | 3964562 | 68.25804 | 3.012481 | 1599 | 301 | APO Binary |
| 16.01 | 9110357 | 66.40566 | 0.895298 | 1527 | 283 | APO Binary |
| 19.01 | 7255336 | 66.93003 | 1.203197 | 2472 | 92 | Binary, Odd-even |
| 21.01 | 10125352 | 54.97329 | 4.288459 | 3127 | 246 | Binary |
| 23.01 | 9071386 | 69.86191 | 4.693309 | 14756 | 1443 | SB1; 18 km/s radial velocity amplitude; secondary eclipse in light curve |
| 24.01 | 4743513 | 103.98992 | 2.086268 | 10806 | 421 | APO Binary |
| 25.01 | 10593759 | 69.00948 | 3.132604 | 7879 | 122 | Binary |
| 26.01 | 5021737 | 77.1360 | 15.03952 | ~10000 | 168 | Binary |
| 27.01 | 3832716 | 103.33754 | 1.141879 | 291300 | 189 | Multiple stellar eclipses |
| 28.01 | 4247791 | 101.14347 | 4.100902 | 111224 | 97 | Multiple stellar eclipses |
| 31.01 | 6956014 | 102.7 | 0.925516 | 742 | 26 | Binary |
| 33.01 | 5725087 | 66.55824 | 0.366201 | 356 | 52 | Probable binary star (Per=0.2 d) |
| 43.01 | 9025922 | 110.08114 | 11.320908 | 2518 | 39 | APO Binary |
| 45.01 | 3742855 | 107.36379 | 6.397234 | 18199 | 287 | APO Binary |
| 48.01 | 7837302 | 106.7648 | 23.836924 | 27844 | 1008 | Binary |
| 52.01 | 3558981 | 101.0114 | 2.987841 | 46738 | 127 | Binary |
| 53.01 | 2445975 | 105.26025 | 3.388834 | 8807 | 55 | APO Binary |
| 61.01 | 8248939 | 114.8623 | 1.633372 | 623 | 16 | APO Binary |
| 66.01 | 10620329 | 103.68382 | 1.308783 | 691 | 206 | Binary |
| 68.01 | 8669092 | 1.63959 | 1.000977 | 2518 | 53 | Double star: 0.83arcsec SE; delta_m=2.7 mag at 692 nm; Light curve shows modulation in phase with the transit. |
| 74.01 | 6889235 | 58.87577 | 5.188712 | 790 | 50 | Binary |
| 76.01 | 9955262 | 87.75975 | 77.451216 | 914 | 27 | Saturated star; Variable star |
| 80.01 | 9552608 | 68.41332 | 9.250714 | 988 | 193 | SB1: V-shaped; Secondary eclipse |
| 81.01 | 8823868 | 76.07155 | 23.875999 | 1640 | 37 | Binary |
| 88.01 | 7700871 | 66.88714 | 2.589751 | 236 | 70 | APO Binary |
| 90.01 | 9210823 | 68.72067 | 0.828212 | 240 | 50 | APO Binary |
| 106.01 | 10489525 | 64.86603 | 1.612021 | 240 | 45 | APO Binary 12arcsec S |
| 109.01 | 4752451 | 65.86384 | 6.4149 | 414 | 61 | APO Binary |
| 114.01 | 6721123 | 65.26635 | 7.360901 | 297 | 11 | APO Binary |
| 120.01 | 11869052 | 70.89987 | 20.546581 | 1860 | 141 | Binary |
| 121.01 | 3247396 | 69.45725 | 8.810982 | 389 | 52 | APO Binary 1.2arcsec SW |
| 125.01 | 11449844 | 84.85057 | 38.479316 | 23913 | 284 | SB1 |
| 126.01 | 5897826 | 135.33287 | 33.77925 | 17432 | 1050 | Hierarchical triple |
| 129.01 | 11974540 | 65.88433 | 24.666561 | 4378 | 171 | Binary |
| 130.01 | 5297298 | 88.29605 | 34.193562 | 14449 | 1123 | SB1 |
| 132.01 | 8892910 | 65.95236 | 10.810049 | 5811 | 366 | Binary |
| 133.01 | 11673674 | 66.38896 | 4.618688 | 6661 | 542 | APO Binary |
| 134.01 | 9032900 | 86.18499 | 67.179946 | 5056 | 167 | SB1 |
| 136.01 | 7601633 | 80.39962 | 15.66349 | 5055 | 128 | SB1; V-shaped |
| 140.01 | 5130369 | 70.17318 | 19.979085 | 1132 | 59 | APO Binary 6arcsec N |
| 143.01 | 4649305 | 73.33962 | 22.650871 | 2388 | 141 | SB1 |
| 145.01 | 9904059 | 91.0779 | 45.002812 | 1612 | 20 | Binary |
| 146.01 | 9048161 | 70.34138 | 8.667811 | 4611 | 407 | APO Binary |
| 147.01 | 1996679 | 79.05365 | 20 | 2697 | 22 | APO Binary |
| 154.01 | 9970525 | 72.72903 | 30 | 1078 | 57 | APO Binary |
| 158.01 | 10555375 | 70.57828 | 5.801762 | 527 | 55 | APO Binary |
| 160.01 | 6631721 | 67.43403 | 13.738118 | 625 | 39 | APO Binary 4arcsec N; Odd-even |
| 164.01 | 5652237 | 65.31493 | 4.464747 | 202 | 20 | APO Binary |
| 169.01 | 6185711 | 84.13743 | 11.700984 | 537 | 31 | APO Binary |
| 170.01 | 11044770 | 77.29725 | 15.608935 | 233 | 22 | APO Binary |
| 175.01 | 8323753 | 67.31485 | 6.714228 | 457 | 49 | APO Binary 8arcsec NE; 3% transit on nearby KIC 8323764 |
| 178.01 | 11455491 | 68.159 | 6.143084 | 179 | 28 | APO Binary MAST FP |
| 181.01 | 12504988 | 72.71686 | 5.093946 | 26156 | 1050 | Binary |
| 182.01 | 5376836 | 69.79714 | 3.479294 | 19156 | 205 | SB1: secondary eclipse; 30 km/s variation RV variation |
| 184.01 | 7972785 | 66.56668 | 7.300705 | 13378 | 864 | Binary Brown Dwarf Odd-even |
| 185.01 | 4178389 | 67.38638 | 23.210439 | 29037 | 792 | Binary |
| 198.01 | 10666242 | 86.36912 | 87.233068 | 20687 | 403 | SB1: V-shaped; 9 km/s RV variation |
| 210.01 | 10602291 | 72.32516 | 20.927351 | 8264 | 276 | SB1: V-shaped; 10 km/s RV variation |
| 213.01 | 9164836 | 103.83962 | 48.118647 | 78601 | 1240 | SB1 |
| 215.01 | 12508335 | 88.20608 | 42.943545 | 9709 | 158 | V-shaped Double star: 2arcsec N |
| 218.01 | 9838975 | 76.83238 | 18.692915 | 56736 | 648 | Binary |
| 224.01 | 5547480 | 65.07193 | 3.979789 | 1016 | 51 | APO Binary 4arcsec S |
| 230.01 | 3862246 | 69.3061 | 4.70253 | 4069 | 90 | Binary |
| 231.01 | 4043443 | 95 | 119.7 | 6424 | 60 | Binary |
| 233.01 | 5023956 | 80.48307 | 1.824639 | 8311 | 685 | APO Binary |
| 236.01 | 8453211 | 76.0925 | 5.776826 | 3961 | 323 | APO Binary |
| 243.01 | 9592579 | 67.21109 | 2.637587 | 5743 | 359 | APO Binary |
| 259.01 | 5790807 | 140.12541 | 79.996026 | 24236 | 667 | Binary |
| 264.01 | 3097346 | 103.6909 | 4.029783 | 177 | 61 | APO Binary SB1; 10arcsec E |
| 266.01 | 7375348 | 104.51168 | 25.308485 | 133 | 26 | APO Binary 2arcsec NW |
| 267.01 | 8167959 | 140.33404 | 170.564783 | 118 | 9.2 | Spurious Detection Light curve artifact |
| 272.01 | 5716763 | 102.6666 | 1.281318 | 490 | 101 | APO Binary |
| 286.01 | 8258171 | 106.63073 | 23.631011 | 126 | 28 | Binary; V-shaped, secondary eclipse |
| 287.01 | 8703887 | 108.60714 | 14.170948 | 9409 | 268 | APO: 8arcsec NE |
| 290.01 | 10488450 | 104.42464 | 2.683386 | 327 | 53 | APO Binary |
| 293.01 | 11200415 | 105.21434 | 4.639598 | 144 | 31 | APO: 8arcsec W |
| 300.01 | 3438975 | 104.66069 | 2.976105 | 111 | 28 | APO: 2arcsec S |
| 309.01 | 7024222 | 103.19735 | 1.633091 | 64 | 12 | APO: 8arcsec NE, V-shaped, Secondary eclipse possibly on nearby KIC 7024229 |
| 311.01 | 7024511 | 108.64811 | 66.155811 | 3579 | 30 | APO Binary |
| 320.01 | 8700558 | 106.51072 | 4.791957 | 137 | 32 | Binary |
| 322.01 | 8948424 | 106.58265 | 5.888834 | 15417 | 110 | Binary |
| 324.01 | 9641041 | 104.2285 | 1.089083 | 139 | 62 | Binary |
| 325.01 | 9724984 | 104.00304 | 7.863267 | 449 | 45 | Binary |
| 328.01 | 9895004 | 103.32188 | 2.250817 | 545 | 49 | APO: 3arcsec S, Double star: 2" S, delta_mag > 5 mag at I filter |
| 329.01 | 10031885 | 107.94672 | 8.590949 | 113 | 14 | APO: 8arcsec E |
| 334.01 | 10383687 | 109.93541 | 8.487801 | 363 | 27 | APO Binary |
| 336.01 | 10518725 | 108.25221 | 19.506989 | 173 | 12 | APO Binary |
| 347.01 | 11189127 | 102.85301 | 2.671944 | 15250 | 142 | Binary |
| 358.01 | 12017140 | 105.8018 | 22.845232 | 46728 | 2495 | Binary |
| 359.01 | 12106929 | 104.5856 | 5.936699 | 348 | 24 | APO: 10arcsec S |
| 362.01 | 1571511 | 110.5945 | 14.022451 | 20957 | 2270 | Binary |
| 363.01 | 2438070 | 104.0968 | 2.442946 | 989 | 130 | Binary |
| 376.01 | 12643589 | 144.53871 | 220.7246 | 3856 | 50 | Binary |
| 376.02 | 12643589 | 111.26218 | 1.411632 | 513 | 36 | Binary; deep transit |
| 378.01 | 2449074 | 106.75081 | 4.943872 | 928 | 57 | APO Binary |
| 380.01 | 2452450 | 103.8115 | 8.09694 | 1108 | 109 | APO Binary |
| 381.01 | 3230578 | 103.18917 | 6.337653 | 1734 | 99 | SB1: 90 km/s RV variation, secondary eclipse |
| 382.01 | 3231137 | 105.36798 | 3.900201 | 883 | 85 | APO Binary |
| 389.01 | 3847708 | 104.70781 | 3.741174 | 862 | 86 | APO |
| 390.01 | 3849155 | 103.60189 | 1.168317 | 308 | 40 | APO: 3arcsec S |
| 391.01 | 3858804 | 107.83701 | 25.958842 | 405 | 31 | APO: 7arcsec E |
| 394.01 | 4159347 | 107.35772 | 12.28406 | 511 | 27 | APO: 2arcsec S |
| 395.01 | 4165960 | 104.01164 | 6.774472 | 548 | 38 | APO: 4arcsec S |
| 396.01 | 4252322 | 113.84957 | 14.591555 | 1296 | 50 | Binary |
| 397.01 | 4376644 | 106.79527 | 27.67775 | 8384 | 386 | SB2 |
| 399.01 | 7289157 | 106.82975 | 5.266478 | 72027 | 523 | SB2, V-shaped, Secondary eclipse |
| 400.01 | 2695110 | 142.98685 | 44.190443 | 4005 | 63 | APO |
| 402.01 | 3342592 | 103.54579 | 17.17733 | 24757 | 972 | Binary |
| 404.01 | 4949751 | 119.38611 | 31.805916 | 3360 | 107 | APO Binary |
| 405.01 | 5003117 | 123.6832 | 37.617744 | 29139 | 1119 | Binary |
| 406.01 | 5035972 | 124.18974 | 49.266722 | 8503 | 215 | Binary |
| 407.01 | 5218441 | 104.42236 | 3.613743 | 5160 | 362 | Binary |
| 411.01 | 5478055 | 107.23072 | 15.851644 | 721 | 21 | APO Binary |
| 414.01 | 5872150 | 108.3429 | 20.355117 | 27585 | 1154 | SB1: 20 km/s RV variation, Possible Odd-even, Possible secondary eclipse |
| 414.02 | 5872150 | 106.82314 | 5.922128 | 366 | 19 | SB1: 20 km/s RV variation, Possible Odd-even, Possible secondary eclipse |
| 424.01 | 9597411 | 103.39252 | 1.575632 | 24382 | 482 | Binary |
| 434.01 | 11656302 | 106.10122 | 22.265052 | 19044 | 783 | Transit depth too deep, Double star: 3" SE at J |
| 436.01 | 11805075 | 158.3578 | 200 | 32503 | 327 | Binary; deep transit depth |
| 437.01 | 11824222 | 110.1409 | 15.84134 | 22106 | 139 | Binary |
| 441.01 | 3340312 | 106.91384 | 30.547936 | 632 | 18 | APO: 2arcsec S |
| 445.01 | 4384675 | 191.3437 | 200 | 4254 | 33 | Partial Single Transit |
| 447.01 | 5021176 | 105.68099 | 4.045084 | 869 | 51 | Binary |
| 449.01 | 5779852 | 173.23463 | 252.079331 | 4341 | 73 | Binary |
| 450.01 | 6042214 | 104.95475 | 27.046295 | 1062 | 37 | APO: 4arcsec NE |
| 451.01 | 6200715 | 105.18015 | 3.723577 | 585 | 39 | APO: 8arcsec W |
| 453.01 | 6758917 | 102.64822 | 2.23609 | 29399 | 58 | Binary |
| 455.01 | 7101828 | 126.20207 | 47.880541 | 923 | 24 | APO Binary |
| 461.01 | 8621348 | 105.84204 | 11.344441 | 916 | 44 | Binary |
| 462.01 | 8773869 | 103.88046 | 1.576334 | 839 | 132 | APO Binary |
| 482.01 | 11255761 | 102.55198 | 4.992736 | 771 | 38 | APO:4arcsec ESE |
| 485.01 | 12316431 | 108.55026 | 17.908864 | 1061 | 46 | APO |
| 489.01 | 2576197 | 104.69565 | 2.217017 | 683 | 42 | APO |
| 491.01 | 3541800 | 102.66617 | 4.661868 | 354 | 22 | APO: 8 arcsec S |
| 493.01 | 3834360 | 103.12284 | 2.908459 | 678 | 39 | APO: 4 arcsec SE |
| 495.01 | 4049108 | 102.63841 | 4.804379 | 677 | 39 | APO: 12 arcsec E |
| 498.01 | 4833135 | 110.53567 | 8.660657 | 111 | 7 | Spurious Detection |
| 502.01 | 5282051 | 104.15518 | 5.910368 | 232 | 19 | APO: 10 arcsec S |
| 514.01 | 7602070 | 109.0658 | 11.756019 | 788 | 36 | APO: 8 arcsec SE, transit depth changes |
| 515.01 | 7812179 | 102.60907 | 17.792292 | 1310 | 54 | APO Binary |
| 516.01 | 7840044 | 104.09574 | 13.542045 | 607 | 32 | APO |
| 527.01 | 9636569 | 104.68827 | 10.636614 | 330 | 24 | APO Binary |
| 529.01 | 10068030 | 103.51226 | 2.023127 | 1277 | 48 | Binary |
| 539.01 | 11246364 | 104.24141 | 200 | 362 | 8.3 | Spurious Detection |
| 540.01 | 11521048 | 127.8248 | 25.702616 | 5252 | 54 | APO: 4 arcsec NW, V-shaped, Odd-even |
| 544.01 | 11913012 | 104.66417 | 3.747895 | 389 | 27 | APO: 10 arcsec NE |
| 545.01 | 11972666 | 103.37698 | 1.091763 | 195 | 16 | APO: 10 arcsec E |
| 549.01 | 3437776 | 126.50951 | 42.899607 | 758 | 13 | APO |
| 549.02 | 3437776 | 66.41894 | 0.635578 | 407 | 63 | APO |
| 553.01 | 5303551 | 104.45359 | 2.399009 | 326 | 20 | APO: 6 arcsec S |
| 556.01 | 5738496 | 108.65576 | 9.503451 | 437 | 17 | Binary |
| 565.01 | 7025846 | 103.19693 | 2.340506 | 181 | 22 | APO: 8 arcsec N |
| 570.01 | 8106610 | 105.78106 | 12.398394 | 583 | 24 | APO: 8 arcsec S |
| 576.01 | 8474898 | 173.84915 | 199.444158 | 5745 | 82 | APO: 8 arcsec W |
| 591.01 | 9886221 | 103.79579 | 2.992808 | 140 | 10 | APO |
| 595.01 | 10294465 | 183.93512 | 200 | 4037 | 40 | Spurious Detection |
| 603.01 | 2441151 | 104.6703 | 2.19201 | 1490 | 105 | APO Binary |
| 604.01 | 3970233 | 107.4336 | 8.254955 | 20142 | 662 | Binary |
| 606.01 | 5014753 | 105.04683 | 3.170623 | 14181 | 140 | Binary |
| 608.01 | 5562784 | 125.90668 | 25.337283 | 2035 | 74 | APO |
| 613.01 | 6960456 | 104.55631 | 5.074714 | 566 | 23 | APO Binary |
| 615.01 | 8374580 | 125.50459 | 176.239818 | 6046 | 100 | APO: 3 arcsec SW |
| 616.01 | 9714696 | 102.84973 | 1.433356 | 415 | 28 | Binary |
| 619.01 | 10384962 | 103.7099 | 2.879242 | 49699 | 232 | Binary |
| 621.01 | 12251650 | 107.18212 | 17.762041 | 22623 | 466 | Binary |
| 630.01 | 4659405 | 104.92727 | 4.532367 | 425 | 41 | APO Binary |
| 631.01 | 4742414 | 106.78138 | 15.458069 | 4358 | 693 | SB1 |
| 634.01 | 4861736 | 105.57831 | 6.277803 | 1838 | 162 | APO Binary |
| 636.01 | 5090690 | 109.63508 | 12.011655 | 13445 | 1336 | SB1 |
| 637.01 | 5098444 | 110.97822 | 26.948407 | 16053 | 162 | SB2 |
| 642.01 | 5181817 | 104.43335 | 4.350379 | 183 | 20 | APO Binary |
| 643.01 | 5309353 | 102.73079 | 1.376372 | 309 | 26 | SB2 |
| 646.01 | 5384802 | 103.49023 | 3.041456 | 19146 | 1237 | Hierarchical triple |
| 648.01 | 5596440 | 105.11995 | 10.474877 | 2184 | 81 | Secondary eclipse; radius 7 Rjup - too large. |
| 651.01 | 5796186 | 137.12803 | 42.559655 | 2070 | 82 | APO Binary |
| 653.01 | 5893123 | 102.632 | 1.125969 | 144 | 24 | APO: 8 arcsec SW |
| 656.01 | 5966660 | 103.06339 | 1.906628 | 530 | 110 | APO Binary |
| 668.01 | 6805146 | 111.68746 | 13.7797 | 26160 | 2903 | Binary |
| 669.01 | 6960445 | 104.56468 | 5.074351 | 933 | 88 | APO Binary |
| 675.01 | 7385509 | 102.54614 | 1.655473 | 1999 | 139 | APO: 6 arcsec W; V-shaped |
| 677.01 | 7466863 | 104.07983 | 11.971853 | 454 | 17 | APO: 4 arcsec |
| 681.01 | 7598128 | 121.49531 | 44.258089 | 20356 | 1032 | SB1 |
| 690.01 | 8409588 | 102.70818 | 1.360839 | 1361 | 119 | Binary |
| 696.01 | 8869680 | 107.15636 | 7.033951 | 105 | 20 | APO |
| 699.01 | 8908102 | 107.60562 | 5.414584 | 4914 | 172 | SB1; V-shaped |
| 702.01 | 9053112 | 102.867 | 1.27484 | 359 | 82 | APO Binary |
| 705.01 | 9300285 | 103.18743 | 1.012676 | 1190 | 146 | APO Binary |
| 706.01 | 9426071 | 115.60821 | 28.714155 | 43504 | 1755 | Binary |
| 713.01 | 9640985 | 104.22897 | 2.178174 | 410 | 56 | Binary |
| 715.01 | 9834719 | 103.35853 | 1.621665 | 6413 | 56 | SB2 |
| 724.01 | 10005020 | 107.70358 | 6.970892 | 477 | 25 | APO Binary |
| 726.01 | 10157573 | 106.26507 | 5.11578 | 960 | 39 | APO: 4 arcsec E |
| 727.01 | 10191070 | 103.58189 | 1.213737 | 1772 | 130 | APO Binary |
| 729.01 | 10225800 | 102.67394 | 1.423802 | 2091 | 91 | APO: 2 arcsec N |
| 731.01 | 10259031 | 102.86523 | 7.061031 | 3608 | 99 | APO Binary |
| 742.01 | 10419211 | 103.83978 | 11.52139 | 17206 | 446 | Binary |
| 744.01 | 10480982 | 117.55132 | 19.221399 | 72570 | 1554 | Binary |
| 748.01 | 10583180 | 103.73521 | 2.696436 | 383 | 25 | Binary |
| 754.01 | 10848459 | 103.3066 | 1.736957 | 7825 | 331 | Binary |
| 761.01 | 11152159 | 103.69418 | 2.701326 | 1176 | 39 | APO: 6 arcsec NE |
| 768.01 | 11442465 | 119.67371 | 33.93387 | 8042 | 170 | APO |
| 770.01 | 11463211 | 103.9971 | 1.506357 | 2287 | 109 | APO: 3 arcsec NE |
| 789.01 | 12459725 | 104.49591 | 14.180511 | 462 | 7.4 | APO: 10 arcsec E |
| 792.01 | 2440757 | 103.40219 | 1.433833 | 1890 | 75 | Binary |
| 793.01 | 2445129 | 106.31213 | 10.318845 | 704 | 21 | APO |
| 796.01 | 3114661 | 102.86259 | 1.332894 | 12019 | 87 | APO: 10 arcsec S |
| 798.01 | 3120431 | 104.64252 | 3.34192 | 1045 | 33 | Binary |
| 803.01 | 3554600 | 104.78318 | 7.546344 | 382 | 11 | APO |
| 807.01 | 3836375 | 103.50377 | 1.540409 | 1667 | 54 | APO: 12 arcsec SE |
| 808.01 | 3838486 | 104.98512 | 2.990307 | 646 | 29 | APO |
| 819.01 | 4932348 | 129.93326 | 38.03697 | 87384 | 1478 | Binary |
| 820.01 | 4936180 | 106.72046 | 4.640905 | 4430 | 103 | Binary, V-shaped, Possible ellipsoidal variations |
| 828.01 | 5287983 | 104.00036 | 2.507109 | 1886 | 64 | APO |
| 831.01 | 5370302 | 106.39528 | 3.904278 | 16419 | 521 | Binary |
| 832.01 | 5372966 | 104.82031 | 9.286365 | 46813 | 1422 | Binary |
| 836.01 | 5481416 | 104.58344 | 2.384036 | 1269 | 24 | APO: 6 arcsec E, Odd-even |
| 839.01 | 5649215 | 103.67139 | 2.4467 | 3144 | 209 | APO Binary |
| 848.01 | 6267425 | 105.00138 | 3.166479 | 4834 | 124 | Binary |
| 859.01 | 6675056 | 108.40596 | 10.443261 | 859 | 23 | APO: 12 arcsec SE |
| 862.01 | 6756669 | 106.82149 | 5.851534 | 35214 | 1170 | Binary |
| 866.01 | 6862603 | 104.64695 | 2.861178 | 1298 | 46 | Binary |
| 888.01 | 7552344 | 102.96884 | 1.000786 | 1889 | 107 | Binary |
| 894.01 | 7708215 | 107.5174 | 7.943013 | 2911 | 81 | APO Binary |
| 909.01 | 8256049 | 105.76151 | 16.371955 | 9842 | 178 | APO Binary |
| 915.01 | 8605074 | 127.71347 | 37.601454 | 64858 | 742 | Binary |
| 919.01 | 8686150 | 106.53537 | 51.426011 | 71051 | 1213 | Binary |
| 925.01 | 9016295 | 104.62927 | 19.974485 | 39181 | 762 | Binary |
| 927.01 | 9097120 | 121.98166 | 23.899733 | 21886 | 428 | APO Binary, V-shaped |
| 930.01 | 9159275 | 105.00944 | 3.044891 | 1431 | 103 | APO Binary |
| 932.01 | 9166870 | 103.68018 | 3.855545 | 1497 | 89 | APO |
| 933.01 | 9171801 | 104.6591 | 3.185933 | 1094 | 13 | APO |
| 946.01 | 9661877 | 109.71686 | 20.427268 | 2285 | 58 | APO Binary |
| 948.01 | 9761882 | 106.70756 | 24.586099 | 984 | 21 | APO: 6 arcsec W |
| 950.01 | 9772531 | 116.40765 | 31.201504 | 32363 | 396 | Binary |
| 957.01 | 7661409 | 111.46762 | 3.140565 | 1971 | 120 | Binary |
| 958.01 | 1026957 | 99.5413 | 21.761045 | 933 | 8.9 | Grazing EB |
| 959.01 | 10002261 | 108.07184 | 12.713795 | 36851 | 957 | White Dwarf |
| 964.01 | 10657664 | 104.23836 | 3.273699 | 7811 | 58 | Possible white dwarf |
| 965.01 | 3337351 | 108.0321 | 7.047115 | 41819 | 452 | APO |
| 967.01 | 6579806 | 106.60816 | 9.880483 | 17819 | 299 | Binary star |
| 968.01 | 3560301 | 194.28976 | 4.649301 | 122 | 42 | Spurious Detection |
| 970.01 | 11502218 | 104.01865 | 3.988635 | 729 | 28 | Occultation |
| 971.01 | 11180361 | 104.2261 | 0.533058 | 1237 | 102 | Binary |
| 978.01 | 11494130 | 195.43461 | 18.954856 | 471 | 100 | Binary |
| 980.01 | 12167361 | 112.0942 | 47.931219 | 1826 | 137 | Binary |
| 982.01 | 1433962 | 66.91651 | 1.592683 | 931 | 58 | 5.5-sig odd-even |
| 983.01 | 11607193 | 199.05683 | 7.15466 | 985 | 18 | Spurious Detection |
| 985.01 | 10227501 | 194.95809 | 2.002925 | 363 | 20 | Spurious Detection Poor fit to light curve |
| 989.01 | 10743597 | 117.30935 | 81.192668 | 14022 | 72 | Planet radius too large |
| 989.02 | 10743597 | 65.83445 | 0.817026 | 2431 | 111 | APO Binary |
| 990.01 | 10015516 | 71.53571 | 67.684214 | 23489 | 39 | Planet radius too large |
| 995.01 | 3858949 | 87.9233 | 25.946596 | 671 | 27 | Secondary eclipse; eccentric |
| 996.01 | 3858824 | 87.88238 | 25.952109 | 1733 | 45 | Secondary eclipse; eccentric |
| 997.01 | 2157247 | 66.01327 | 5.686521 | 4902 | 113 | APO Binary Contact binary; transit on nearby object |
| 1000.01 | 2441728 | 66.6694 | 0.856917 | 97 | 22 | APO |
| 1004.01 | 2309585 | 285.71837 | 1.838472 | 949 | 33 | APO |
| 1006.01 | 5738346 | 307.63528 | 30.607094 | 4889 | 28 | APO |
| 1008.01 | 1722276 | 181.92435 | 300 | 31943 | 189 | V-shaped Binary |
| 1009.01 | 892772 | 290.53448 | 5.092371 | 272 | 8.8 | APO |
| 1011.01 | 5728283 | 116.37238 | 6.198276 | 51074 | 1568 | Secondary eclipse; eccentric |
| 1012.01 | 8127639 | 287.8211 | 1.023438 | 1603 | 57 | APO |
| 1016.01 | 8176653 | 67.06724 | 2.866656 | 761 | 26 | APO |
| 1018.01 | 8183911 | 70.07805 | 8.307384 | 160 | 12 | APO |
| 1021.01 | 2558363 | 111.56191 | 0.546248 | 671 | 29 | APO |
| 1023.01 | 2445154 | 72.67279 | 8.410946 | 702 | 25 | APO |
| 1025.01 | 2574201 | 90.33265 | 37.475525 | 1555 | 17 | APO |
| 1028.01 | 2166206 | 73.73622 | 8.0974 | 425 | 30 | APO |
| 1034.01 | 5899544 | 287.88718 | 1.739454 | 11695 | 76 | APO |
| 1035.01 | 5963222 | 66.32803 | 1.217267 | 7546 | 486 | APO |
| 1036.01 | 5982353 | 67.35396 | 19.563101 | 12463 | 1216 | 40-sigma secondary eclipse |
| 1037.01 | 6205468 | 112.30246 | 3.722924 | 2947 | 204 | APO |
| 1038.01 | 6153201 | 287.98492 | 0.530301 | 2537 | 108 | APO Binary 23-sigma secondary eclipse |
| 1039.01 | 5802486 | 110.92264 | 1.07392 | 2365 | 133 | APO |
| 1040.01 | 5817553 | 69.44957 | 4.206046 | 1539 | 137 | APO Binary |
| 1041.01 | 5982368 | 302.11007 | 19.564094 | 3526 | 71 | APO |
| 1042.01 | 5816811 | 67.45204 | 2.227715 | 1457 | 118 | APO |
| 1043.01 | 5816165 | 288.35413 | 0.591908 | 1091 | 34 | 13-sigma secondary eclipse; odd-even |
| 1044.01 | 5802246 | 66.74212 | 0.525157 | 962 | 90 | APO Binary 48-sigma secondary eclipse |
| 1045.01 | 6066403 | 67.21317 | 1.303856 | 439 | 75 | APO Binary 11-sigma odd-even |
| 1046.01 | 6209637 | 66.44315 | 0.734491 | 342 | 164 | Contact binary |
| 1047.01 | 5988031 | 67.4955 | 2.5555 | 329 | 41 | APO Binary V-shaped |
| 1048.01 | 5820218 | 66.17894 | 3.411778 | 588 | 21 | APO |
| 1049.01 | 5876368 | 66.18536 | 0.525428 | 329 | 37 | APO Binary 7.2-sigma odd-even |
| 1055.01 | 5866099 | 66.63221 | 36.976706 | 1000 | 78 | APO Binary Secondary Eclipse; eccentric binary |
| 1056.01 | 5964985 | 66.26465 | 1.850845 | 117 | 27 | Contact binary |
| 1057.01 | 6066416 | 67.20831 | 1.303879 | 145 | 78 | APO Binary 6-sigma secondary eclipse |
| 1058.01 | 6124941 | 69.20124 | 5.670144 | 607 | 29 | APO |
| 1062.01 | 6147122 | 75.21795 | 15.450994 | 241 | 38 | APO Binary Secondary eclipse; eccentric |
| 1063.01 | 8257407 | 109.30531 | 89.69815 | 266763 | 7754 | V-shaped Binary V-shaped; large planet radius (2.1 RJ) |
| 1064.01 | 8218274 | 66.46468 | 1.187353 | 19234 | 310 | Binary |
| 1065.01 | 8242681 | 66.63778 | 4.020627 | 21849 | 257 | Binary |
| 1068.01 | 8264070 | 383.76246 | 2.897046 | 1148 | 30 | APO |
| 1071.01 | 8244190 | 66.18984 | 1.092087 | 220 | 28 | APO |
| 1073.01 | 8262210 | 111.53907 | 1.612925 | 152 | 13 | Contact binary |
| 1075.01 | 10232123 | 66.27522 | 1.343764 | 4752 | 168 | Binary |
| 1076.01 | 10223616 | 75.16079 | 29.122922 | 6778 | 102 | Stellar binary - TTV |
| 1077.01 | 10268907 | 287.58283 | 1.103981 | 2351 | 62 | APO Likely a blend |
| 1079.01 | 10153827 | 66.58797 | 0.293626 | 831 | 68 | Contact binary |
| 1080.01 | 10158990 | 66.60619 | 1.09661 | 257 | 27 | APO 4.7-sigma secondary |
| 1084.01 | 10148521 | 67.05846 | 1.204265 | 218 | 22 | APO Binary 3.8-sigma secondary |
| 1087.01 | 3124412 | 67.36639 | 0.948955 | 3595 | 200 | APO Binary Blend; eccentric binary |
| 1088.01 | 3113266 | 66.34181 | 1.493792 | 4881 | 99 | APO |
| 1090.01 | 3232859 | 71.90254 | 8.387211 | 5448 | 220 | APO |
| 1091.01 | 3098184 | 302.53435 | 15.243206 | 2244 | 48 | APO Secondary eclipse; eccentric |
| 1092.01 | 2720309 | 66.20765 | 1.240024 | 1636 | 57 | APO Contact binary |
| 1093.01 | 3239636 | 66.3844 | 0.528753 | 463 | 72 | APO 12-sigma odd/even |
| 1097.01 | 3340070 | 75.61968 | 10.904413 | 647 | 22 | APO Binary |
| 1098.01 | 3240706 | 67.1046 | 5.489896 | 517 | 19 | APO Binary Secondary eclipse; eccentric |
| 1100.01 | 3228824 | 288.9061 | 0.730941 | 240 | 13 | Contact binary |
| 1104.01 | 2851100 | 66.2646 | 0.890105 | 426 | 15 | APO Binary Secondary eclipse |
| 1105.01 | 3130300 | 66.31262 | 5.765791 | 378 | 22 | APO |
| 1107.01 | 3228959 | 66.70972 | 0.730867 | 212 | 14 | Contact binary |
| 1119.01 | 3003992 | 72.09994 | 7.244998 | 72 | 21 | APO |
| 1120.01 | 6307537 | 90.41441 | 29.744338 | 161172 | 451 | Stellar binary |
| 1121.01 | 6359798 | 73.69733 | 14.154037 | 58683 | 441 | Binary |
| 1122.01 | 6311681 | 67.21665 | 0.844784 | 2986 | 189 | APO Binary Stellar binary |
| 1123.01 | 6365321 | 66.58381 | 0.848485 | 1856 | 116 | Binary |
| 1124.01 | 6301035 | 76.27948 | 11.991361 | 1982 | 75 | APO |
| 1125.01 | 6292162 | 290.51705 | 7.815272 | 2065 | 45 | Secondary eclipse |
| 1126.01 | 6307521 | 90.4045 | 29.743666 | 585 | 25 | APO |
| 1130.01 | 8279765 | 68.23281 | 2.757787 | 37543 | 362 | Secondary eclipse |
| 1132.01 | 8330548 | 67.64426 | 1.91416 | 7236 | 122 | APO |
| 1133.01 | 8374494 | 290.32645 | 5.251691 | 6185 | 151 | APO |
| 1134.01 | 8414907 | 210.0594 | 200.611031 | 23227 | 188 | APO |
| 1134.02 | 8414907 | 389.85157 | 200.622704 | 12103 | 109 | APO |
| 1135.01 | 8397446 | 381.592 | 0.986617 | 2226 | 69 | APO |
| 1136.01 | 8386035 | 67.51314 | 1.634815 | 948 | 82 | APO |
| 1138.01 | 8415745 | 299.70512 | 31.827907 | 4644 | 51 | APO |
| 1139.01 | 8378634 | 67.70132 | 3.629444 | 628 | 90 | APO |
| 1140.01 | 8397675 | 66.44743 | 0.553259 | 1146 | 120 | Binary occultation |
| 1143.01 | 8312852 | 67.76284 | 7.440416 | 592 | 23 | APO |
| 1147.01 | 8299955 | 67.25617 | 2.682674 | 132 | 21 | APO |
| 1153.01 | 10351767 | 67.01944 | 0.635073 | 36772 | 273 | Binary |
| 1154.01 | 10295951 | 69.04584 | 6.810826 | 14832 | 713 | Binary occultation |
| 1155.01 | 10342041 | 66.38725 | 0.933744 | 2797 | 469 | APO |
| 1156.01 | 10514770 | 67.36536 | 1.872422 | 2964 | 112 | Binary occultation |
| 1157.01 | 10342065 | 66.38704 | 0.933747 | 1474 | 377 | Binary grazing binary |
| 1158.01 | 10352945 | 289.38614 | 6.471815 | 583 | 26 | APO |
| 1167.01 | 10485179 | 66.97753 | 0.445263 | 206 | 42 | APO |
| 1171.01 | 10485069 | 66.52966 | 0.445267 | 182 | 21 | Binary contact binary |
| 1172.01 | 10341913 | 67.31822 | 0.933753 | 122 | 55 | APO |
| 1173.01 | 10480921 | 66.616 | 2.037225 | 111 | 16 | APO |
| 1178.01 | 3869825 | 67.36948 | 4.800633 | 13791 | 132 | Binary phase linked variations |
| 1179.01 | 3655332 | 73.65674 | 15.066423 | 25123 | 163 | APO |
| 1180.01 | 4042026 | 96.01592 | 34.820008 | 18755 | 488 | Binary eccentric binary |
| 1181.01 | 3344419 | 66.11162 | 0.651782 | 4689 | 188 | APO |
| 1182.01 | 3865567 | 66.09875 | 11.116227 | 5904 | 127 | APO |
| 1183.01 | 3544689 | 67.65783 | 1.922868 | 2214 | 56 | APO |
| 1184.01 | 4037164 | 67.18271 | 0.635445 | 2951 | 122 | APO |
| 1185.01 | 3443790 | 66.86259 | 1.665782 | 1707 | 108 | Binary occultation |
| 1186.01 | 3966912 | 89.34976 | 55.659966 | 2702 | 54 | APO |
| 1188.01 | 3860441 | 66.26721 | 2.988155 | 845 | 37 | APO |
| 1189.01 | 3765771 | 66.22795 | 2.783865 | 898 | 43 | APO |
| 1190.01 | 3557341 | 111.58633 | 0.393729 | 727 | 34 | Binary contact binary |
| 1196.01 | 3348082 | 68.5425 | 3.981818 | 510 | 19 | APO |
| 1197.01 | 3853673 | 66.98158 | 0.643798 | 478 | 33 | APO |
| 1200.01 | 3557493 | 66.3066 | 0.393732 | 424 | 41 | Binary contact binary |
| 1211.01 | 3858704 | 68.64523 | 3.003592 | 139 | 38 | APO |
| 1213.01 | 3556220 | 66.08445 | 0.796712 | 132 | 35 | APO |
| 1217.01 | 3542588 | 66.75171 | 3.47124 | 105 | 15 | APO |
| 1223.01 | 6613006 | 66.93373 | 7.388831 | 11651 | 777 | APO |
| 1224.01 | 6606653 | 69.33242 | 2.698025 | 7480 | 50 | Binary |
| 1225.01 | 6620003 | 66.31535 | 1.714272 | 28197 | 668 | Binary |
| 1228.01 | 6387450 | 68.26916 | 3.661328 | 17248 | 376 | Binary occultation |
| 1229.01 | 6432059 | 66.29515 | 0.769738 | 10490 | 210 | Binary occultation |
| 1231.01 | 6462874 | 88.08445 | 22.342917 | 3039 | 45 | APO |
| 1232.01 | 6665223 | 165.64948 | 238.814686 | 18640 | 289 | Binary (large radius) |
| 1233.01 | 6545358 | 111.59514 | 1.171542 | 1366 | 48 | APO |
| 1234.01 | 6390824 | 66.49873 | 0.973544 | 1442 | 75 | APO |
| 1235.01 | 6546528 | 66.88287 | 3.053602 | 459 | 33 | APO |
| 1237.01 | 6531491 | 297.38226 | 14.325861 | 772 | 13 | APO |
| 1243.01 | 6677256 | 289.40143 | 3.126045 | 261 | 13 | APO |
| 1247.01 | 8801343 | 67.54977 | 2.739874 | 22189 | 455 | Binary (phased locked variations) |
| 1248.01 | 8488878 | 290.45296 | 5.801871 | 20730 | 495 | APO |
| 1250.01 | 8620565 | 67.02127 | 0.782044 | 14820 | 195 | APO |
| 1251.01 | 8616873 | 67.10443 | 0.576082 | 9222 | 383 | Binary |
| 1252.01 | 8737796 | 67.48073 | 0.885763 | 4247 | 281 | APO |
| 1253.01 | 8462258 | 290.24997 | 3.611524 | 2157 | 71 | APO |
| 1254.01 | 8454250 | 70.25924 | 5.082704 | 2168 | 58 | APO |
| 1256.01 | 8848271 | 70.44893 | 9.991579 | 3415 | 167 | APO |
| 1259.01 | 8823426 | 111.18663 | 1.506477 | 739 | 47 | APO |
| 1260.01 | 8766222 | 110.97695 | 5.296678 | 997 | 30 | APO |
| 1262.01 | 8703884 | 292.82951 | 14.170888 | 2374 | 45 | APO |
| 1263.01 | 8560840 | 66.88154 | 31.971634 | 1074 | 27 | Binary eccentric binary |
| 1265.01 | 8552583 | 66.84451 | 1.061949 | 407 | 39 | APO |
| 1267.01 | 8519253 | 69.53055 | 5.938123 | 834 | 35 | APO |
| 1269.01 | 8757910 | 66.17576 | 0.655003 | 382 | 87 | APO |
| 1272.01 | 8552498 | 66.84806 | 0.530968 | 201 | 60 | Binary 12 sigma odd-even |
| 1277.01 | 8552565 | 66.84725 | 1.061938 | 161 | 46 | APO |
| 1280.01 | 8509361 | 68.52649 | 6.099026 | 225 | 25 | APO |
| 1284.01 | 10960993 | 66.12539 | 1.558546 | 8533 | 426 | Binary |
| 1286.01 | 10879208 | 111.2802 | 0.668484 | 3332 | 126 | APO |
| 1289.01 | 10748393 | 287.53941 | 4.88778 | 3142 | 26 | APO |
| 1290.01 | 10874226 | 77.77651 | 11.973776 | 3485 | 78 | Binary |
| 1291.01 | 10661771 | 66.0339 | 1.231376 | 980 | 129 | APO |
| 1292.01 | 10924853 | 67.3941 | 2.102421 | 1783 | 78 | APO |
| 1293.01 | 10874926 | 66.72149 | 11.703074 | 3999 | 88 | Binary occultation |
| 1294.01 | 10549562 | 72.07875 | 9.089494 | 1023 | 64 | APO |
| 1295.01 | 10666230 | 66.30807 | 1.577794 | 684 | 51 | Binary occultation |
| 1296.01 | 10971674 | 68.32449 | 2.380863 | 34583 | 1371 | Binary occultation |
| 1297.01 | 10676923 | 66.9362 | 1.031112 | 384 | 39 | APO |
| 1313.01 | 10785538 | 66.74787 | 0.522466 | 184 | 40 | APO |
| 1318.01 | 4070376 | 66.96805 | 1.634614 | 19619 | 195 | Binary |
| 1319.01 | 4078157 | 300.69473 | 16.025273 | 20671 | 218 | APO |
| 1321.01 | 4480676 | 67.22337 | 0.711965 | 6016 | 245 | Binary |
| 1322.01 | 4079535 | 289.79691 | 17.726895 | 16147 | 276 | APO |
| 1324.01 | 4551328 | 66.1377 | 0.522059 | 1448 | 113 | Binary 24 sigma odd-even |
| 1326.01 | 4639868 | 68.97114 | 53.100926 | 13485 | 847 | Binary |
| 1327.01 | 4372768 | 292.04613 | 15.642622 | 1720 | 25 | APO |
| 1330.01 | 4150539 | 67.32308 | 8.65258 | 688 | 19 | APO |
| 1333.01 | 4285107 | 66.44838 | 2.24301 | 277 | 28 | APO |
| 1334.01 | 4150624 | 67.29188 | 8.653379 | 510 | 15 | APO |
| 1340.01 | 4386059 | 112.84431 | 2.900473 | 206 | 18 | APO |
| 1343.01 | 4570931 | 66.5491 | 1.54492 | 138 | 20 | APO |
| 1345.01 | 7284688 | 66.6039 | 0.32302 | 48165 | 262 | Contact Binary |
| 1346.01 | 7199774 | 70.07146 | 4.708125 | 52057 | 318 | Binary occultation |
| 1348.01 | 6866228 | 70.56335 | 7.702363 | 16949 | 93 | Binary |
| 1349.01 | 6847018 | 75.18696 | 16.662103 | 22623 | 107 | Binary |
| 1350.01 | 7220322 | 67.04433 | 0.752164 | 5591 | 130 | APO |
| 1352.01 | 6956233 | 291.16441 | 4.818807 | 3700 | 124 | APO |
| 1354.01 | 6891543 | 67.47279 | 1.752572 | 771 | 66 | APO |
| 1365.01 | 7174351 | 66.9749 | 1.487105 | 408 | 13 | APO |
| 1368.01 | 7357531 | 163.02483 | 251.059866 | 6255 | 89 | APO |
| 1371.01 | 6878167 | 66.6844 | 0.833971 | 216 | 28 | APO |
| 1373.01 | 6863839 | 66.7014 | 1.926129 | 185 | 19 | APO |
| 1374.01 | 7296086 | 67.56837 | 0.890732 | 218 | 28 | APO |
| 1380.01 | 7025526 | 66.84556 | 1.074081 | 65 | 28 | APO |
| 1381.01 | 9451127 | 67.58585 | 5.117403 | 64104 | 411 | Binary occultation |
| 1383.01 | 8953257 | 69.14077 | 3.221787 | 45922 | 317 | Binary occultation |
| 1384.01 | 8971432 | 66.04537 | 0.62438 | 38858 | 415 | Binary occultation |
| 1386.01 | 9143254 | 66.80499 | 1.137524 | 16933 | 537 | Binary |
| 1388.01 | 9346253 | 83.45355 | 34.064261 | 28776 | 684 | Binary |
| 1389.01 | 9002237 | 67.79932 | 4.350087 | 21642 | 162 | Binary |
| 1390.01 | 9288786 | 67.50781 | 1.744101 | 9692 | 124 | Binary |
| 1392.01 | 9040849 | 290.25992 | 4.118762 | 1838 | 57 | APO |
| 1394.01 | 8937021 | 66.88839 | 5.663653 | 2236 | 53 | APO |
| 1400.01 | 9157908 | 75.33007 | 9.414682 | 783 | 70 | Binary |
| 1414.01 | 8916492 | 69.24279 | 4.02365 | 34 | 20 | APO |
| 1415.01 | 11193447 | 66.80918 | 0.312943 | 34162 | 250 | Contact Binary |
| 1416.01 | 11517719 | 68.335 | 2.495801 | 26075 | 154 | Binary |
| 1443.01 | 11197126 | 67.77461 | 4.494499 | 232 | 28 | APO |
| 1446.01 | 12506351 | 66.78619 | 1.227759 | 44429 | 235 | Binary |
| 1447.01 | 7622486 | 94.30528 | 40.246662 | 148151 | 303 | V-shaped Binary |
| 1447.02 | 7622486 | 66.63923 | 2.279999 | 15260 | 205 | Binary Brown Dwarf |
| 1449.01 | 7802136 | 70.1369 | 10.980248 | 49724 | 929 | Binary |
| 1450.01 | 7532973 | 66.98404 | 2.144631 | 18255 | 225 | Binary |
| 1451.01 | 9632895 | 92.74682 | 27.322068 | 78850 | 1241 | Binary occultation |
| 1453.01 | 7842610 | 66.85321 | 0.971933 | 9596 | 344 | APO |
| 1454.01 | 7830637 | 70.3035 | 121.590891 | 11369 | 35 | Binary occultation |
| 1455.01 | 4760746 | 296.54861 | 15.068135 | 13227 | 143 | APO |
| 1460.01 | 7751571 | 296.97389 | 17.041841 | 5115 | 110 | APO |
| 1461.01 | 9579499 | 73.70776 | 7.946693 | 5706 | 68 | Binary |
| 1462.01 | 11913013 | 288.31141 | 3.747881 | 2747 | 25 | APO |
| 1464.01 | 7838655 | 289.32963 | 2.113152 | 1991 | 92 | APO |
| 1467.01 | 7770450 | 287.74883 | 1.157752 | 1040 | 47 | APO |
| 1469.01 | 7543649 | 288.51958 | 3.581956 | 1168 | 54 | APO |
| 1471.01 | 11858748 | 110.82595 | 1.780979 | 881 | 35 | APO |
| 1482.01 | 7812167 | 298.35756 | 17.792773 | 1545 | 30 | APO |
| 1485.01 | 9692345 | 66.66256 | 0.687895 | 517 | 38 | APO |
| 1487.01 | 12062667 | 66.72336 | 2.929223 | 427 | 50 | APO |
| 1490.01 | 9602514 | 66.51884 | 3.556617 | 324 | 40 | APO |
| 1492.01 | 12108312 | 66.80535 | 0.705449 | 490 | 57 | APO |
| 1497.01 | 11774387 | 111.15178 | 0.520205 | 430 | 26 | APO |
| 1500.01 | 9719634 | 67.27943 | 3.351587 | 486 | 34 | APO |
| 1504.01 | 9641018 | 67.1991 | 2.178178 | 346 | 41 | APO |
| 1509.01 | 9535080 | 86.04295 | 49.644312 | 650 | 64 | Binary phased locked variations |
| 1513.01 | 9784222 | 66.96052 | 1.197304 | 319 | 31 | APO |
| 1514.01 | 9520668 | 66.61153 | 1.399319 | 276 | 15 | APO |
| 1524.01 | 4826110 | 67.18301 | 1.333363 | 189 | 13 | APO |
| 1538.01 | 9963461 | 116.00779 | 10.581586 | 115934 | 982 | Eccentric Binary |
| 1539.01 | 8081482 | 66.8696 | 2.819448 | 72512 | 205 | Binary |
| 1542.01 | 8113154 | 66.81655 | 2.586873 | 26617 | 162 | Binary |
| 1548.01 | 9940565 | 68.06092 | 2.13933 | 5437 | 114 | Binary |
| 1550.01 | 8111381 | 67.72031 | 2.233799 | 2966 | 197 | APO Binary Secondary eclipse |
| 1551.01 | 5444549 | 348.94868 | 31.138459 | 9975 | 122 | APO |
| 1554.01 | 9899355 | 287.58427 | 1.332604 | 911 | 49 | APO Binary Secondary eclipse |
| 1555.01 | 12644774 | 312.12282 | 41.077414 | 4286 | 60 | APO |
| 1556.01 | 9902856 | 210.32841 | 135.913711 | 10827 | 95 | APO |
| 1559.01 | 9899280 | 66.37641 | 1.332583 | 621 | 66 | APO |
| 1562.01 | 5308663 | 288.58306 | 0.784473 | 905 | 33 | APO |
| 1565.01 | 5636648 | 66.97062 | 0.466743 | 837 | 72 | APO |
| 1566.01 | 5564247 | 288.36227 | 1.727256 | 649 | 14 | APO |
| 1568.01 | 5210475 | 287.85676 | 1.008933 | 432 | 19 | APO |
| 1571.01 | 5557821 | 289.29149 | 2.928799 | 473 | 19 | APO |
| 1575.01 | 5553652 | 74.04144 | 24.329742 | 1217 | 15 | APO Eccentric Binary |
| 1578.01 | 5629985 | 67.78972 | 2.272058 | 255 | 36 | APO |
| 1579.01 | 9898364 | 67.47018 | 7.132434 | 695 | 64 | APO |
| 1580.01 | 5193400 | 80.18576 | 21.382619 | 418 | 18 | APO |
| 1592.01 | 5217586 | 76.14359 | 26.06809 | 681 | 30 | APO |
| 1594.01 | 9895709 | 66.12217 | 1.818944 | 325 | 21 | APO |
| 1600.01 | 4860932 | 67.76764 | 3.091207 | 235 | 17 | APO |
| 1604.01 | 10033279 | 72.77578 | 72.491574 | 1267 | 49 | APO |
| 1607.01 | 5477805 | 67.53825 | 5.006818 | 232 | 20 | APO |
| 1610.01 | 5474733 | 66.60625 | 0.883781 | 122 | 17 | APO |

**Table 5.** Candidates in or near the Habitable Zone (sorted by $T_{eq}$)

| KOI | Kp (mag) | $R_p$ ($R_\oplus$) | Period (days) | $T_{eff}$ (°K) | $R_*$ ($R_\odot$) | $T_{eq}$ (°K) | *a* (AU) |
|---|---|---|---|---|---|---|---|
| 683.01 | 13.71 | 4.14 | 278.12 | 5624 | 0.78 | 239 | 0.84 |
| 1582.01 | 15.4 | 4.44 | 186.38 | 5384 | 0.64 | 240 | 0.63 |
| 1026.01 | 14.75 | 1.77 | 94.1 | 3802 | 0.68 | 242 | 0.33 |
| 1503.01 | 14.83 | 2.68 | 150.24 | 5356 | 0.56 | 242 | 0.54 |
| 1099.01 | 15.44 | 3.65 | 161.53 | 5665 | 0.55 | 244 | 0.57 |
| 854.01 | 15.85 | 1.91 | 56.05 | 3743 | 0.49 | 248 | 0.22 |
| 433.02 | 14.92 | 13.37 | 328.24 | 5237 | 1.08 | 249 | 0.94 |
| 1486.01 | 15.51 | 8.43 | 254.56 | 5688 | 0.83 | 256 | 0.8 |
| 701.03 | 13.73 | 1.73 | 122.39 | 4869 | 0.68 | 262 | 0.45 |
| 351.01 | 13.8 | 8.48 | 331.65 | 6103 | 0.94 | 266 | 0.97 |
| 902.01 | 15.75 | 5.66 | 83.9 | 4312 | 0.65 | 270 | 0.32 |
| 211.01 | 14.99 | 9.58 | 372.11 | 6072 | 1.09 | 273 | 1.05 |
| 1423.01 | 15.74 | 4.28 | 124.42 | 5288 | 0.66 | 274 | 0.47 |
| 1429.01 | 15.53 | 4.15 | 205.93 | 5595 | 0.86 | 276 | 0.69 |
| 1361.01 | 14.99 | 2.2 | 59.88 | 4050 | 0.59 | 279 | 0.24 |
| 87.01 | 11.66 | 2.42 | 289.86 | 5606 | 1.14 | 282 | 0.88 |
| 139.01 | 13.49 | 5.65 | 224.79 | 5921 | 0.9 | 288 | 0.74 |
| 268.01 | 10.56 | 1.75 | 110.37 | 4808 | 0.79 | 295 | 0.41 |
| 1472.01 | 15.06 | 3.57 | 85.35 | 5455 | 0.56 | 295 | 0.37 |
| 536.01 | 14.5 | 2.97 | 162.34 | 5614 | 0.84 | 296 | 0.59 |
| 806.01 | 15.4 | 8.97 | 143.18 | 5206 | 0.88 | 296 | 0.53 |
| 1375.01 | 13.71 | 17.88 | 321.22 | 6169 | 1.17 | 300 | 0.96 |
| 812.03 | 15.95 | 2.12 | 46.19 | 4097 | 0.57 | 301 | 0.21 |
| 865.01 | 15.09 | 5.94 | 119.02 | 5560 | 0.73 | 306 | 0.47 |
| 351.02 | 13.8 | 6 | 210.45 | 6103 | 0.94 | 309 | 0.71 |
| 51.01 | 13.76 | 4.78 | 10.43 | 3240 | 0.27 | 314 | 0.06 |
| 1596.02 | 15.16 | 3.44 | 105.36 | 4656 | 0.98 | 316 | 0.42 |
| 416.02 | 14.29 | 2.82 | 88.25 | 5083 | 0.75 | 317 | 0.38 |
| 622.01 | 14.93 | 9.28 | 155.05 | 5171 | 1.17 | 327 | 0.57 |
| 555.02 | 14.76 | 2.27 | 86.5 | 5218 | 0.78 | 331 | 0.38 |
| 1574.01 | 14.6 | 5.75 | 114.73 | 5537 | 0.85 | 331 | 0.47 |
| 326.01 | 12.96 | 0.85 | 8.97 | 3240 | 0.27 | 332 | 0.05 |
| 70.03 | 12.5 | 1.96 | 77.61 | 5342 | 0.7 | 333 | 0.35 |
| 1261.01 | 15.12 | 6.25 | 133.46 | 5760 | 0.9 | 335 | 0.52 |
| 1527.01 | 14.88 | 4.84 | 192.67 | 5470 | 1.31 | 337 | 0.67 |
| 1328.01 | 15.67 | 4.81 | 80.97 | 5425 | 0.72 | 338 | 0.36 |
| 564.02 | 14.85 | 4.97 | 127.89 | 5686 | 0.93 | 340 | 0.51 |
| 1478.01 | 12.45 | 3.73 | 76.13 | 5441 | 0.7 | 341 | 0.35 |
| 1355.01 | 15.9 | 2.81 | 51.93 | 5529 | 0.52 | 342 | 0.27 |
| 372.01 | 12.39 | 8.44 | 125.61 | 5638 | 0.95 | 344 | 0.5 |
| 711.03 | 13.97 | 2.62 | 124.52 | 5488 | 1 | 345 | 0.49 |
| 448.02 | 14.9 | 3.78 | 43.62 | 4264 | 0.71 | 346 | 0.21 |
| 415.01 | 14.11 | 7.7 | 166.79 | 5823 | 1.15 | 352 | 0.61 |
| 947.01 | 15.19 | 2.74 | 28.6 | 3829 | 0.64 | 353 | 0.15 |
| 174.01 | 13.78 | 2.52 | 56.35 | 4654 | 0.8 | 355 | 0.27 |
| 401.02 | 14 | 6.6 | 160.01 | 5264 | 1.4 | 357 | 0.59 |
| 1564.01 | 15.29 | 3.07 | 53.45 | 5709 | 0.56 | 360 | 0.28 |
| 157.05 | 13.71 | 3.23 | 118.38 | 5675 | 1 | 361 | 0.48 |
| 365.01 | 11.2 | 2.34 | 81.74 | 5389 | 0.86 | 363 | 0.37 |
| 374.01 | 12.21 | 3.33 | 172.67 | 5829 | 1.26 | 365 | 0.63 |
| 952.03 | 15.80 | 2.4 | 22.78 | 3911 | 0.56 | 365 | 0.12 |
| 817.01 | 15.41 | 2.1 | 23.97 | 3905 | 0.59 | 370 | 0.13 |
| 847.01 | 15.20 | 5.1 | 80.87 | 5469 | 0.88 | 372 | 0.37 |
| 1159.01 | 15.33 | 5.3 | 64.62 | 4886 | 0.91 | 372 | 0.30 |